\documentclass[brazil,12pt,a4paper,openany]{book}
\usepackage{anysize}
\usepackage[latin1]{inputenc}
\usepackage[english]{babel}
\usepackage[ruled,chapter]{algorithm}
\usepackage{makeidx}
\usepackage{longtable}
\usepackage{amsmath}
\usepackage{amssymb}
\usepackage{amsthm}
\usepackage{mathrsfs}
\usepackage{graphicx}
\usepackage{fancyhdr}
\usepackage{array}
\usepackage{geometry}
\usepackage{simplewick}
\usepackage{latexsym}
\usepackage[all]{xy}
\usepackage{eufrak}
\usepackage{euscript}
\usepackage{enumerate}
\usepackage{dsfont}
\usepackage{slashed}
\usepackage{cite}
\usepackage[plainpages=false,pagebackref=true]{hyperref}
\usepackage{scrextend}
\usepackage{pstricks}
\usepackage{pst-node}
\usepackage{pst-coil}
\usepackage{pst-grad}
\usepackage{multido}
\usepackage{chngcntr}

\makeindex
\marginsize{25mm}{25mm}{25mm}{25mm}

\counterwithout{footnote}{chapter}

\newcommand{\be}{\begin{equation}}
\newcommand{\ee}{\end{equation}}
\newcommand{\bea}{\begin{eqnarray}}
\newcommand{\eea}{\end{eqnarray}}

\hypersetup{pdfpagelabels,hyperindex,colorlinks=true,breaklinks=true,bookmarks=true,linkcolor=black,citecolor=black,
urlcolor=black}

\begin{document}

\setlength{\abovecaptionskip}{0.0cm}
\setlength{\belowcaptionskip}{0.0cm}
\setlength{\baselineskip}{24pt}

\pagestyle{fancy}
\lhead{}
\chead{}
\rhead{\thepage}
\lfoot{}
\cfoot{}
\rfoot{}

\fancypagestyle{plain}
{
	\fancyhf{}
	\lhead{}
	\chead{}
	\rhead{\thepage}
	\lfoot{}
	\cfoot{}
	\rfoot{}
}

\renewcommand{\headrulewidth}{0pt}

%%%%%%%%%%%%%%%%%%%%%%%%%%%%%%%%% Contra Capa %%%%%%%%%%%%%%%%%%%%%%%%%%%%%%%%%%

\frontmatter

\thispagestyle{empty}

\begin{figure}[!h]
\begin{minipage}[c]{0.2\linewidth}
\includegraphics[scale=0.5]{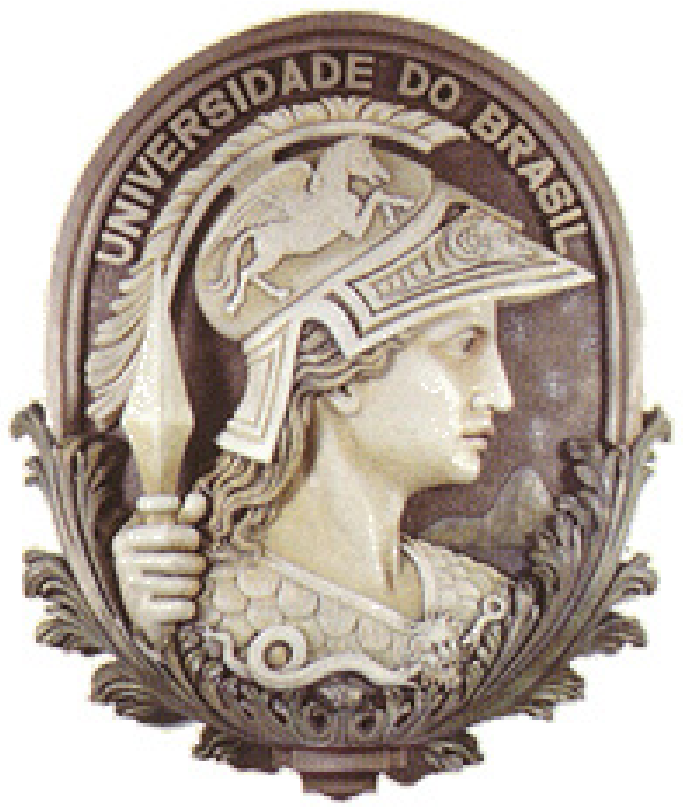}
\end{minipage}
\hspace{0.5cm}
\begin{minipage}[c]{0.8\linewidth}
\begin{center}
{ {\Large Universidade Federal do Rio de Janeiro}}\\
\vspace{0.3cm}
{\Large Centro de Ciências Matemáticas e da Natureza}\\
\vspace{0.3cm}
{\Large Instituto de Física}
\end{center}
\end{minipage}
\end{figure}
\vspace{40pt}
\begin{center}
{\Huge
 Novel approaches to tailor and tune

\vspace{20pt}
light-matter interactions at the nanoscale
}
\end{center}
%
%\vspace{20pt}
%
%\begin{figure}[h!]
%\centering
%\label{capa}
%	\includegraphics[scale=0.7]{capa.eps}
%\end{figure}
%
\vspace{100pt}
\begin{center}
{\Large
Wilton Júnior de Melo Kort-Kamp

\vspace{200pt}
2015
}
\end{center}

%%%%%%%%%%%%%%%%%%%%%%%%%%%%%
\newpage

\thispagestyle{empty}

\noindent

\phantom{Extra}

\clearpage

%%%%%%%%%%%%%%%%%%%%%%%%%%%%%%

\newpage

\thispagestyle{empty}

\begin{figure}[h]
	\includegraphics[scale=0.5]{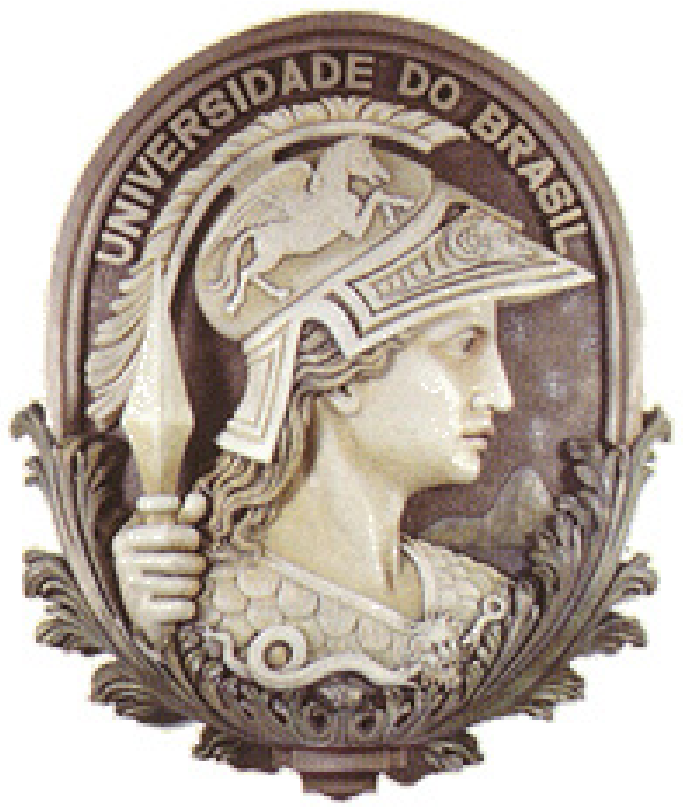}
\end{figure}

\vspace{15pt}

\begin{center}

\textbf{UNIVERSIDADE FEDERAL DO RIO DE JANEIRO}

\textbf{INSTITUTO DE FÍSICA}

\vspace{30pt}

{\Large \bf Novel approaches to tailor and tune light-matter interactions at the nanoscale}

\vspace{25pt}

{\large \bf Wilton Júnior de Melo Kort-Kamp}

\vspace{35pt}

\begin{flushright}
\parbox{10.3cm}{Ph.D. Thesis presented to the Graduate Program in Physics of the Institute of Physics of the Federal University of Rio de Janeiro - UFRJ, as part of the requirements to the obtention of the title of Doctor in Sciences (Physics).

\vspace{18pt}

{\large \bf Advisor: Carlos Farina de Souza}

\vspace{12pt}

{\large \bf Co-advisor: Felipe Arruda de Araújo \\ \phantom{Co-advisor::} Pinheiro}}
\end{flushright}

\vspace{90pt}

\textbf{Rio de Janeiro}

\textbf{February, 2015}

\end{center}

%%%%%%%%%%%%%%%%%%%%%%%%%%%%%%%%%%%%%%%%%%%%%%%%%%%%%%%%%%%%%%%%%%%%%%%%%

%%%%%%%%%%%%%%%%%%%% Folha de Assinaturas da Banca %%%%%%%%%%%%%%%%%%%%%%

%\thispagestyle{empty}

\noindent

\newpage

\thispagestyle{empty}

\noindent

\vspace{-20pt}
\begin{flushleft}
\begin{figure}[h!]
\centering
\includegraphics[scale=0.7]{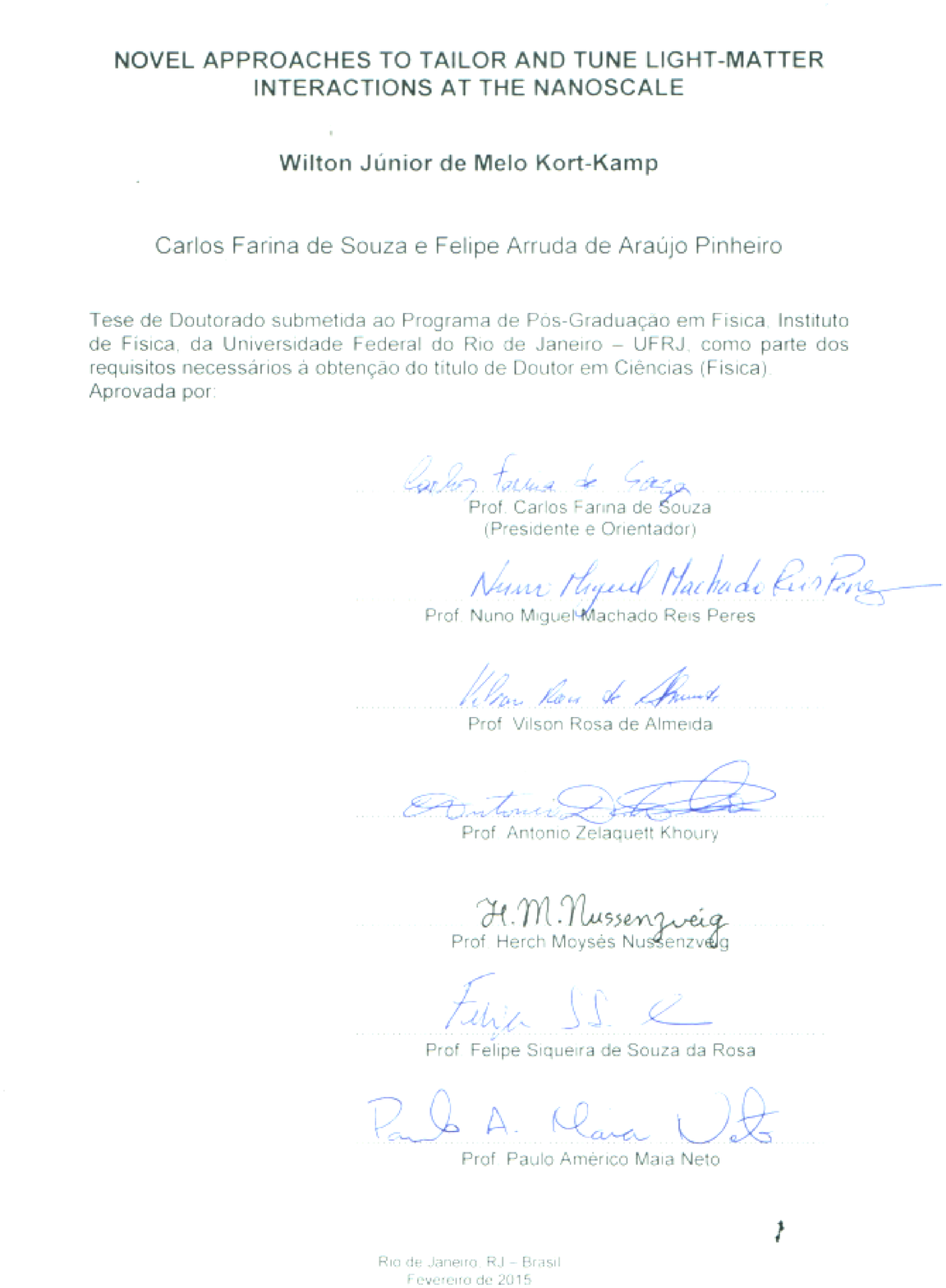}
\end{figure}
\end{flushleft}
\thispagestyle{empty}

\clearpage

%%%%%%%%%%%%%%%%%%%%%%%% Ficha Catalográfica %%%%%%%%%%%%%%%%%%%%%%%%%%%%

\newpage
\thispagestyle{empty}
\mbox{}\vspace{5cm}

\begin{center}
\begin{tabular}{|c|}
\hline\\
\begin{minipage}{15cm}
\rule{15cm}{0pt} \flushright

\begin{minipage}{12.5cm}
     Kort-Kamp, Wilton Júnior de Melo

\hspace{-2cm} K851 \hspace{2cm} Novel approaches to tailor and tune light-matter interactions at the nanoscale / Wilton Júnior de Melo Kort-Kamp. - Rio de Janeiro: UFRJ/IF, 2015.

\hspace{2cm} xlii, 231f.; il.; 31cm

\hspace{2cm} Orientador: Carlos Farina de Souza

\hspace{2cm} Coorientador: Felipe Arruda de Araújo Pinheiro

\hspace{2cm} Tese (doutorado) - UFRJ / Instituto de Física / Programa de Pós-graduação em Física, 2015.

\hspace{2cm} Referências Bibliográficas: f. 177-222.

\hspace{2cm} 1. Tunable invisibility cloaks. 2. Magneto-optical materials. 3. Spontaneous emission.  4. Dispersive interactions.  5. Near-field heat transfer. I.  Souza, Carlos Farina de. II. Pinheiro, Felipe Arruda de Araújo. III. Universidade Federal do Rio de Janeiro, Instituto de Física, Programa de Pós-graduação em Física. IV. Novel approaches to tailor and tune light-matter interactions at the nanoscale.

\bigskip
\end{minipage}
\end{minipage}
\\
\hline
\end{tabular}
\end{center}

%%%%%%%%%%%%%%%%%%%%%%%%%%%%%%% Abstract %%%%%%%%%%%%%%%%%%%%%%%%%%%%%%%%

\newpage

\noindent

\vspace*{20pt}
\begin{center}
{\LARGE\bf Abstract}\\
\vspace{15pt}
{\Large\bf  Novel approaches to tailor and tune  light-matter interactions at the nanoscale}\\
\vspace{6pt}
{\bf Wilton Júnior de Melo Kort-Kamp}\\
\vspace{12pt}
{\bf Advisor: Carlos Farina de Souza}\\
{\bf Co-advisor: Felipe Arruda de Araújo Pinheiro}\\
\vspace{20pt}
\parbox{14cm}{\emph{Abstract} of the Ph.D. Thesis presented to the Graduate Program in Physics of the Institute of Physics of the Federal University of Rio de Janeiro - UFRJ, as part of the requirements to the obtention of the title of Doctor in Sciences (Physics).}
\end{center}
\vspace*{35pt}

%The art of tailoring light-matter interactions at subwavelength scales has received renewed interest since the advent of metamaterials.
In this thesis we propose new, versatile schemes to control light-matter interactions at the nanoscale for both classical and quantum applications.

In the first part of the thesis,  we envisage a new class of plasmonic cloaks made of magneto-optical materials. We demonstrate that the application of a uniform magnetic field in these cloaks may not only switch on and off the cloaking mechanism but also mitigate the electromagnetic absorption. In addition,  we prove that the angular distribution of the scattered radiation can be effectively controlled by external magnetic fields, allowing for a swift change in the scattering pattern.

The second part of the thesis is devoted to the study of light-matter interactions mediated by fluctuations of the vacuum electromagnetic field. We present a novel application of plasmonic cloaking in atomic physics, demonstrating that the Purcell effect can be effectively suppressed even for small separations between an excited atom and a cloaking device. Our results suggest that the radiative properties of an excited atom or molecule could be exploited to probe the efficiency of plasmonic cloakings. Furthermore, the decay rate of a two-level quantum emitter near a graphene-coated wall under the influence of an external magnetic field is studied. We demonstrate that we can either increase or inhibit spontaneous emission in this system by applying an external magnetic field. We show that the magneto-optical properties of graphene strongly affect the atomic lifetime at low temperatures. We also demonstrate that the external magnetic field allows for an unprecedented control of the decay channels of the system.

Also in the second part of the thesis, we discuss the main features of dispersive interactions between atoms and arbitrary bodies. We present an appropriate method for calculating the non-retarded dispersive interaction energy between an atom and conducting objects of arbitrary shapes. In particular, we focus on the atom-sphere and atom-ellipsoid interactions. In addition, we study the dispersive interaction between an atom and suspended graphene in an external magnetic field. For large atom-graphene separations and low temperatures, we show that the interaction energy becomes a quantized function of the magnetic field. Besides, we show that at room temperature thermal effects must be taken into account even in the extreme near-field regime.

Finally, the third part of the thesis deals with the study of near-field heat transfer. We analyze the energy transfered from a semi-infinite medium to a composite sphere made of metallic spheroidal inclusions embedded in a dielectric host medium. Within the dipole approximation and using effective medium theories, we show that the heat transfer can be strongly enhanced at the insulator-metal phase transition. By demonstrating that our results apply for different effective medium models, and that they are robust against changing the inclusions' shape and materials, we expect that that enhancement of NFHT at the percolation threshold should occur for a wide range of inhomogeneous materials.

\vspace{15pt}

\textbf{Keywords:} 1. Tunable invisibility cloaks. 2. Magneto-optical materials. 3. Spontaneous emission. 4. Dispersive interactions.  5. Near-field heat transfer.

%%%%%%%%%%%%%%%%%%%%%%%%%%%%%%%% Resumo %%%%%%%%%%%%%%%%%%%%%%%%%%%%%%%%%

\newpage

\noindent

\vspace*{10pt}
\begin{center}
{\LARGE\bf Resumo}\\
\vspace{10pt}
{\Large\bf Novas abordagens para modelar e controlar a interação entre luz e matéria na nanoescala}\\
\vspace{6pt}
{\bf Wilton Júnior de Melo Kort-Kamp}\\
\vspace{8pt}
{\bf Orientador: Carlos Farina de Souza}\\
{\bf Coorientador: Felipe Arruda de Araújo Pinheiro}\\
\vspace{15pt}
\parbox{14cm}{Resumo da Tese de Doutorado apresentada ao Programa de Pós-Graduação em Física do Instituto de Física da Universidade Federal do Rio de Janeiro - UFRJ, como parte dos requisitos necessários à obtenção do título de Doutor em Ciências (Física).}
\end{center}
\vspace*{12pt}

Nesta tese propomos novos e versáteis dispositivos para controlar a interação entre luz e matéria na nanoescala tanto em aplicações clássicas quanto quânticas.

Na primeira parte da tese, introduzimos uma nova classe de capas de invisibilidade plasmônicas fabricadas com materiais magneto-óticos.  Mostramos que a aplicação de um campo magnético uniforme sobre o sistema permite não apenas ligar e desligar dispositivos de invisibilidade como também atenuar a absorção de energia eletromagnética.   Além disso, demonstramos que a distribuição angular da radiação eletromagnética espalhada pelo sistema pode ser efetivamente controlada por campos magnéticos externos, permitindo uma rápida mudança no padrão de espalhamento.

A segunda parte da tese é dedicada ao estudo das interações entre matéria e radiação mediadas pelas flutuações quânticas do campo eletromagnético de vácuo. Apresentamos uma nova aplicação de capas de invisibilidade plasmônicas ligada à física atômica: mostramos que o efeito Purcell pode ser suprimido mesmo para pequenas separações entre um átomo excitato e um dispositivo de camuflagem. Nossos resultados sugerem que as propriedades radiativas de átomos excitados poderiam ser exploradas para testar a eficiência de mantos plasmônicos.  Além disso, estudamos a taxa de decaimento de um emissor quântico próximo a uma parede revestida por grafeno sob a influência de um campo magnético. Mostramos que tanto redução quanto amplificação da taxa de decaimento podem ser obtidas.  Verificamos que as propriedades magneto-óticas do grafeno tem forte influência sobre o tempo de vida atômico no regime de baixas temperaturas. Demonstramos também que o campo magnético permite um controle sem precedentes sobre os canais de decaimento do sistema.

Ainda na segunda parte da tese, discutimos as principais características das interações dispersivas entre átomos e corpos arbitrários. Apresentamos um método apropriado para calcular a interação dispersiva não retardada entre um átomo e um objeto perfeitamente condutor. Em particular, investigamos as interações átomo-esfera e átomo-elipsóide. Além disso, estudamos a interação dispersiva entre um átomo e uma folha de grafeno na presença de um campo magnético. Mostramos que a energia de interação é uma função quantizada do campo magnético para grandes distâncias entre o átomo e o grafeno e baixas temperaturas. Demonstramos ainda que na temperatura ambiente os efeitos térmios devem ser levados em consideração mesmo no regime de campo próximo.

Finalmente, a terceira parte da tese trata a transferência de calor no campo próximo.  Analisamos a troca de energia térmica entre um meio semi-infinito e uma esfera feita de inclusões metálicas esferoidais embebidas em um meio hospedeiro dielétrico. Na aproximação de dipolo e utilizando teorias de meios efetivos mostramos que a transferência de calor pode ser fortemente amplificada na transição de fase  isolante-metal. Verificamos que nossos resultados aplicam-se para diferentes teorias de meios efetivos e são robustos quanto às mudanças na forma e nas propriedades materias das inclusões.

\vspace{5pt}

\textbf{Palavras-chave:} 1. Capas de invisibilidade sintonizáveis. 2. Materiais magneto-óticos. 3. Emissão espontânea. 4. Interações dispersivas. 5. Transferência de calor no campo próximo.

%%%%%%%%%%%%%%%%%%%%%%%%%%%%%%%%%%%%%%%%%%%%%%%%%%%%%%%%%%%%%%%%%%%%%%%%%%%%%%%%%%%%%%%%%%%%%%%%%%%%%%%%%%%%%%%%%%%%
\newpage

\noindent
\vspace{18cm}
\begin{flushright}
{\it
To my parents,\\ Nilton and Eliana.
}
\end{flushright}

\newpage

\phantom{bla}

\clearpage

%%%%%%%%%%%%%%%%%%%%%%%%%%%% Agradecimentos %%%%%%%%%%%%%%%%%%%%%%%%%%%%%

\newpage

\noindent

\vspace*{20pt}

\begin{center}

{\LARGE\bf Acknowledgments}

\end{center}

\vspace*{40pt}

My Ph.D. would not have been possible without the support of many people.

I am extremely grateful to ``the three musketeers" that have guide my Ph.D. research: my friends Prof. Carlos ``the malander" Farina, Prof. Felipe Pinheiro, and Prof. Felipe Rosa. Farina is one of the most generous person that I have the opportunity to meet.  He has been my advisor since I was a freshman at the university. I benefited a lot not only from his professional experience but also from his personal advices. His enthusiasm even with the small things surely has been the fuel that keeps several students still believing  that Physics may be a pleasant and exciting carrier. I thank Felipe Pinheiro for co-advising my Ph.D. work. He  has been a continuous source of ideas and many topics discussed throughout this thesis were directly inspired on his proposals, that we have extensively discussed in the last years. It was a pleasure each time I have gone to his office to either move physics forward or to have informal conversations (for instance, about soccer).  I have to mention that I really appreciated the level of freedom that Farina  and Pinheiro gave me during my Ph.D. work, allowing me to study some subjects that were not originally our main topic research. Finally, I must acknowledge my friend and ``under-the-table" co-advisor Felipe Rosa. He is the guy responsible for clarifying most of my questions and for correcting several of my mistakes. Certainly his dedication to understand as deep as possible all our results contributed to improve the level of our works. I am also indebted to him for revising  several projects, presentation letters and conference abstracts that I have written recently. Besides, I thank him for all advices in both personal and professional life.

I would like also to express my deepest gratitude to my wife Jessica. She has been my source of inspiration in the last 7 years. Without her I would not be able to overcome the life obstacles and keep on giving my best in everything. Paraphrasing Isaac Newton: if I have reached to this moment it is by standing on the shoulders of my giant wife. Her care, affection, love and support give me daily the guts to explore uncharted lands. To her, all my love and admiration.

I acknowledge my parents Nilton and Eliana for all their effort in providing me the better education I could have.  Surely all experiences I have shared with them were of fundamental relevance for instigating my curiosity and have encouraged me to look for some answers by means of Physics. In addition, I thank my brother Henrique for his friendship as well as my grandparents Nilton, Wilma, Danclares and Maria for \linebreak their inestimable advices.

I must thank those that have kindly hosted me in their own homes during my undergraduate and graduate studies: my wife's family Nininha, Almir, Cristiane, Minervina; my granduncle Alceu; my cousin Rosilma.

I acknowledge the patience of my psychologist Ivoneide for helping such a stubborn person like me. Her advices were extremely useful.

There are also several other people who played a prominent role in my academic life. I am grateful to the advices of Profs. Miriam Gandelman, Wania Wolff and Marcus Venicius. I thank Prof. Henrique Boschi for interesting discussions and for giving me some recommendation letters. I am glad to Prof. Paulo Américo Maia Neto for his critical remarks, incentive and for his help with my new Post Doctoral position.  Furthermore, I acknowledge Prof. Luiz Davidovich for presenting me several subtleties underlying the interaction between light and matter in his astounding courses of Quantum Mechanics and Quantum Optics I and II.  I thank also Prof. Nuno Peres for enlightening discussions on graphene's world that had assisted me in interpreting some of the results in this thesis. Finally, I am glad to Profs. Stenio Wulck, Marcelo Byrro, Angela Rocha, Marta Feijó, Felipe Acker, and Mario de Oliveira for obtaining to me some financial support in my early days in \linebreak Rio de Janeiro.

I acknowledge the members of the Quantum Vacuum Fluctuation group. Particularly, I thank my officemate Andreson Rego for putting up with me for 3 years and Reinaldo de Melo for valuable discussions on everything.  I am glad to Diney Ether for relaxing conversations, Guilherme Bastos for his contribution in Chapter \ref{cap6} and Marcos ``o rapaz" Bezerra for allowing me to co-advise his undergraduate research projects.

I could not forget of my friends and colleagues that have shared several moments with me. My gratitude to my first UFRJ's friend Ana Barbara Cavalcante, who is always ready to help no matter the problem. I thank Daniela Szilard for her friendship and for always providing me up to date solutions of exercise lists and tests. I must thank my friend and old office neighbor Marcio Taddei for enlightening discussions we have about almost everything in physics and also for futile conversations. I express my gratitude to Tarik Cysne and Diego Oliver for their help with the numerical calculations in Chapter \ref{cap7}. I am glad to  my colleagues Daniel Niemeyer, Daniel Kroff, Maurício Hippert, José Hugo, Elvis Soares, Tiago Mendes, Tiago Arruda, Camille Latune, Gabriel Bié, Wellison Peixoto, Sergio Abreu,  Leandro Nascimento, Danielle Tostes, Daniel Vieira, Oscar Augusto, Marcos Brum, Cleiton Silva, Vanderlei, Anderson Kendi, Carlos Zarro for nonsense and distractive conversations.

I acknowledge the Graduation Secretariat Carlos José and Pedro Ribeiro for working out all kind of bureaucracy involved in the process of obtaining the Ph.D. in \linebreak Sciences degree.

I thank the funding agencies that have partially supported this Ph.D. work, CNPq (Brazilian National Council for Scientific and Technological Development) and  FAPERJ (Foundation for Research Support of the State of Rio de Janeiro).

\newpage

\clearpage

\phantom{bla}

%%%%%%%%%%%%%%%%%%%%%%%%%%%%%%%%%%%%%%%%%%%%%%%%%%%%%%%%%%%%%%%%%%%%%%%%%

\newpage

\noindent
\vspace{18cm}
\begin{flushright}
{\it
Science, my lad, is made up of mistakes,\\
but they are mistakes which it is useful to make,\\
because they lead little by little to the truth.
}\\
{\sc Jules Verne}
\end{flushright}

\clearpage

%%%%%%%%%%%%%%%%%%%%%%%%%%%%%%%%%%%%%%%%%%%%%%%%%%%%%%%%%%%%%%%%%%%%%%%%%

\newpage

\phantom{Extra}

\clearpage
%%%%%%%%%%%%%%%%%%%%%%%%%%%%%%%%%%%%%%%%%%%%%%%%%%%%%%%%%%%%%%%%%%%%%%%%%

\newpage
\phantomsection
\addcontentsline{toc}{chapter}{Contents}
\tableofcontents

\newpage
\phantomsection
\addcontentsline{toc}{chapter}{List of Acronyms}
\chapter*{List of Acronyms}
\begin{longtable}[l]{c p{20pt}   p{260pt}}
\textbf{Acronym} &	& \textbf{Description} \\ \hline \\
BC  &    & Boundary condition\\
BEMT  &    & Bruggeman effective medium theory \\
CQED  &  & Cavity quantum electrodynamics\\
CP  &    & Casimir-Polder\\
DL & & Drude-Lorentz\\
EM  &    & Electromagnetic\\
LL  &    & Landau level\\
LSW  &    & Lossy surface wave\\
MO  &    & Magneto-optical\\
NFHT  &    & Near-field heat transfer\\
PE  &    & Purcell effect\\
QED  &   & Quantum electrodynamics\\
QS  &   & Quasi-static\\
SCT  &   & Scattering cancelation technique\\
SE  &    & Spontaneous emission\\
SPP  &    & Surface plasmon polariton\\
SPhP  &    & Surface phonon polariton\\
TE  &   & Transverse electric\\
TM  &   & Transverse magnetic\\
TIR  &   & Total internal reflection\\
TOM  &   & Transformation optics method\\
\end{longtable}

%%%%%%%%%%%%%%%%%%%%%%%%%%%%%%%%%%%%%%%%%%%%%%%%%%%%%%%%%%%%%%%%%%%%%%%%%%%%%%%%%%%%%%%%%%%%%%%%%%%%%%%%%%%%%%

\newpage
\phantomsection
\addcontentsline{toc}{chapter}{List of Principal Symbols}
\chapter*{List of Principal Symbols}
\begin{longtable}[l]{c p{20pt}   p{260pt}}
\textbf{Symbol} &	& \textbf{Description} \\ \hline \\
$a$  &    & Inner radius of a spherical (cylindrical) cloak\\
$\hat{a}, \, \hat{a}^{\dagger}$  &    & Electromagnetic field annihilation and creation operators. \\
${\bf A}_{\zeta}$  &    & Electromagnetic field modes\\
${A}_{21}$  &    & Einstein spontaneous emission coefficient\\
$b$  &    & Outer radius of a spherical (cylindrical) cloak\\
${\bf B}$  &    & Magnetic induction\\
$c_n^{\textrm{TE}}, \, c_n^{\textrm{TM}}$  &    & Mie's scattering coefficients\\
$C_{\textrm{sca}}\, , \ C_{\textrm{abs}}\, , \ C_{\textrm{ext}} $ & & Scattering, absorption and extinction cross sections \\
$dC_{\textrm{sca}}/d\Omega$ & & Differential scattering cross section\\
${\bf \hat{d}}$& & Electric dipole operator\\
${\bf D} $ &    & Electric displacement field\\
${\bf E} $ &	& Electric Field\\
$f$ & 	&  Frequency\\
$f_c$ & 	&  Percolation treshold\\
${\bf f}_{x}({\bf k},{\bf k}_0)\, , \ {\bf f}_{y}({\bf k},{\bf k}_0)$ & & Scattering amplitude for $x-$polarized and $y-$polarized incident plane waves\\
${\bf F}_0^{{\bf k}, p}, {\bf F}_{\textrm{sca}}^{{\bf k}, p}$ & & Free space and scattered electric vector potentials \\
$G$ & & Geometric cross-sectional area of a body\\
$G({\bf r}, {\bf r'})$ & & Electrostatic Green function\\
${\bf G}_0^{{\bf k}, p}, {\bf G}_{\textrm{sca}}^{{\bf k}, p}$ & & Free space and scattered magnetic vector potentials \\
$\mathbb{G} $ & & EM dyadic Green's function \\
$h^{(1)}_n(x)\, , h^{(2)}_n(x)$ & & Spherical Hankel functions of first and second kind\\
${\hat{\cal{H}}} $ &	& Hamiltonian operator\\
$H^{(1)}_n(x)\, , H^{(2)}_n(x)$ & & Cylindrical Hankel functions of first and second kind\\
${\bf H} $ &	& Magnetic field\\
$I$ & & Intensity\\
$\mathbb{I}$ & & Unit dyad\\
$j_n(x)$ & & Spherical Bessel function\\
$J_n(x)$ & & Cylindrical Bessel function\\
${\bf J}$ & & Current density \\
\mbox{{\mathversion{bold}${\cal J}$}} & & Coordinate transformation Jacobian \\
${\bf k}\ (k)$  &    & Wave vector (modulus of ${\bf k}$)\\
${\bf m}$  &    & Magnetic dipole moment\\
${\bf M}$  &    & Magnetization vector\\
$n_f(E)$ & & Fermi-Dirac distribution\\
$N_n(x)$ & & Cylindrical Neumann function\\
${\bf p}$  &    & Electric dipole moment\\
${\bf P}$  &    & Polarization vector\\
${\cal P}_{\textrm{abs}}^{E} (\omega)$, ${\cal P}_{\textrm{abs}}^{H} (\omega)$  &    & Electric and magnetic mean spectral absorbed power by a sphere\\
${\cal P}_{\textrm{abs}}^{\textrm{total}}$  &    & Total mean spectral absorbed power by a sphere\\
$Q_{\textrm{abs}}\, , \ Q_{\textrm{sca}}\, , \ Q_{\textrm{ext}}$  & & Absorption, scattering and extinction cross section efficiencies\\
$r^{\textrm{i, j}}$  &    & Reflection coefficients for a $i-$polarized incident wave and a $j-$polarized reflected wave\\
$\mathbb{R}$  &    & Reflection matrix of a flat surface\\
${\bf S}$  &    & Poynting vector\\
$T$  &    & Temperature\\
$u $ & & Electromagnetic energy density \\
$U^{\textrm{NR}} $ & & Dispersive interaction energy between an atom and an surface of arbitrary shape in the non-retarded regime\\
$U_{\textrm{T}} $ & & Dispersive interaction energy between an atom and a dispersive half-space at temperature $T$\\
$\langle W_{\textrm{abs}} \rangle\, ,\ \langle W_{\textrm{sca}} \rangle\, , \ \langle W_{\textrm{ext}} \rangle$    & & Mean rates at which energy is absorbed, scattered and extinct by a single particle\\
$y_n(x)$ & & Spherical Neumann function\\
$Z$      &    & Impedance\\ \hline
$\alpha$ & & Polarizability\\
$\gamma_s$ & & Out-of diagonal permittivity component in a magneto-optical media\\
$\Gamma({\bf r})$ & & Spontaneous emission rate\\
$\Delta({\bf r})$ & & Lamb shift\\
$\varepsilon (\mbox{{\mathversion{bold}${\varepsilon}$}})$ & & Electric permittivity (tensor)\\
$\epsilon $ & & Dielectric constant\\
$\hat{\mbox{{\mathversion{bold}${\varepsilon}$}}}_{\textrm{TE (TM)}}^{\pm}$ & & TE and TM polarization vectors\\
$\eta$      &    & Ratio between inner and outer radii of a cylindrical or spherical double-layered body\\
$\lambda$     & & Wavelength \\
$\mu (\mbox{{\mathversion{bold}${\mu}$}})$      &    & Magnetic permeability (tensor)\\
$\mu_c$      &    & Graphene's chemical potential\\
$\rho_e$ & & Charge density\\
$\rho$ & & Radial distance in cylindrical coordinates\\
$\sigma_{xx}\, , \ \sigma_{xy}$  &    & Graphene longitudinal and transversal conductivities\\
$\hat{\sigma}_{x}\, , \ \hat{\sigma}_{y}\, , \hat{\sigma}_{z}$  &    & Pauli's matrices\\
$1/\tau$  &    & Relaxation frequency in the Drude-Lorentz model  \\
$\Phi$, & & Rate at which energy crosses a closed surface\\
$\Phi({\bf r})$, & & Electrostatic potential\\
$\omega$  &    & Angular frequency\\
$\omega_c(B)$  &    & Cyclotron angular frequency\\
$\omega_0$  &    & Resonance frequency in the Drude-Lorentz model or transition frequency of a quantum emitter\\
$\Omega$  &    & Oscillating strength frequency in the Drude-Lorentz model \\ \hline\hline
$\hat{{\bf x}}\, ,\hat{{\bf y}}\, , \hat{{\bf z}} $  &    & Unit vectors of the cartesian basis \\
$\hat{\mbox{{\mathversion{bold}${\rho}$}}}\, ,\hat{\mbox{{\mathversion{bold}${\varphi}$}}}\, , \hat{{\bf z}} $  &    & Unit vectors of the cylindrical basis \\
$\hat{{\bf r}}\, ,\hat{\mbox{{\mathversion{bold}${\theta}$}}}\, , \hat{\mbox{{\mathversion{bold}${\varphi}$}}} $  &    & Unit vectors of the spherical basis \\
\end{longtable}

\newpage
\phantomsection
\addcontentsline{toc}{chapter}{List of Figures}
\listoffigures

\newpage
\phantomsection
\addcontentsline{toc}{chapter}{List of Tables}
\listoftables

\newpage
\phantom{bla}

\mainmatter
\phantomsection
\addcontentsline{toc}{chapter}{Introduction}

\begin{chapter}*{Introduction}
\label{cap1}

%\begin{flushright}
%{\it
%In physics, you don't have to go around making trouble for yourself.\\
%Nature does it for you.
%}

%{\sc F. Wilczek}
%\end{flushright}

\begin{flushright}
{\it
Science never solves a problem without creating ten more.
}

{\sc G. B. Shaw}
\end{flushright}

\hspace{5mm} In recent years, we have been privileged to witness spectacular advances in the art of harnessing the interactions between matter and the electromagnetic (EM) field.  In particular, the progress in the area of metamaterials and plasmonics~\cite{zheludev2012, EnghetaLivro, ShalaevLivro, SmithLivro, Maier_Plasmonics} has permitted not only the discovery of novel physical phenomena but also the development of new applications that allow for an unprecedented control of EM waves, far beyond to what can be achieved with natural media. Among a plethora of new results, we can mention the different types of electromagnetic cloaks \cite{AluReview2008, AluReview2014},  the ever-increasing degree of control of radiative properties of quantum emitters~\cite{Novotny, klimovreview, greffet2005}, the remarkable advances in tailoring dispersive interactions \cite{Lamoreaux05, CasimirLivro, Mostepanenko-Book-2009, Buhmann-Welsch-2007}, and the astonishing number of applications of near-field heat transfer \cite{Greffet2005, Volokitin2007, Basu2009, Dorofeyev2011,  Raschke2013, BiehsIR}.  However, despite the notable breakthroughs brought about by the fabrication of nanostructured materials, designing  tunable, versatile photonic devices at the nanoscale remains both a scientific and technological challenge. From the fundamental point of view, controlling the flow of light by tuning the material electromagnetic response using external agents would facilitate the investigation of optical transport phenomena in the meso and microscopic regimes. From the technological point of view, a dynamical tuning of light-matter interactions in integrated photonic systems would   have a drastic impact on the performance of meso-photonic devices.

\pagebreak

The purpose of the present thesis is to provide theoretical proposals of tunable, versatile  material platforms to control  electromagnetic radiation at subwavelength scales  for both classical and quantum applications\cite{kortkamp2013-2, kortkampPRL, kortkampJOSAA, KortKamp2013, WiltonAJP, Cysne-2014, WiltonPRB2014}. For the sake of clarity, we have divided the thesis in three parts. In the first part, we focus on applications that do not demand either the EM field or matter quantizations. In the second part, we focus on light-matter interactions mediated by fluctuations of the vacuum EM field. Finally, in the third part of the thesis, we are interested in studying radiative heat transfer in the near field, where the EM radiation can be treated within the scope of the stochastic electrodynamics.

Classical electrodynamics has received renewed interest from physicists and engineers since the advent of metamaterials ~\cite{zheludev2012, EnghetaLivro, ShalaevLivro, SmithLivro}, which consist essentially in nanostructured artificial materials with EM properties that can be controlled by manipulating the structural composition of their unit cell. Among several applications, the development of invisibility cloaking devices \cite{AluReview2008, AluReview2014} is possibly the most fascinating one. Nowadays, there are a number of modern approaches to cloaking \cite{AluReview2014}, proving that a properly designed metamaterial can strongly suppress EM scattering in a given frequency range.
%Among these approaches, the scattering cancellation technique (SCT) [REF] constitute the basis for fabrication of plasmonic cloaks and is the more suitable method for improving the performance of EM sensors. In this technique the invisibility cloak produces a scattered electromagnetic field that interferes destructively with the one generated by the cloaked object so that the whole system becomes almost undetectable.
In spite of the recent success in the experimental realizations of invisibility cloaking techniques \cite{AluReview2014}, the  vast majority of existing devices suffer from physical and/or practical limitations as, for instance, the narrow operation frequency bandwidth and the detrimental effect of losses. In addition, once the device is designed and fabricated, the cloaking mechanism works usually only around a restricted frequency band that cannot be freely modified after fabrication,  limiting the device applicability. The feasibility of controlling the operation of an invisibility cloak depends essentially on the fabrication of metamaterials whose electromagnetic properties can be modified by using external agents. Although there are some progresses in this direction ~\cite{peining2010_1,peining2010_2,peining2010_3,zharova2012,milton2009, milton2009-2,farhat2013,Schofield_2014_1,Rybin}, the tunable metamaterials proposed so far are either based on a passive mechanism of tuning or depend on a given range of intensities for the incident radiation. On the other hand, since the EM properties of some magneto-optical media, {\it e.g.} graphene-based materials, are strongly sensitive to external magnetic fields they may be excellent candidates for dynamically controlling the EM properties of invisibility cloaks.

In the first part of this thesis we investigate, for the first time, tunable plasmonic cloaks based on magneto-optical effects ~\cite{kortkamp2013-2,kortkampPRL,kortkampJOSAA}. We demonstrate that the application of an external magnetic field may not only switch on and off the cloaking mechanism but also mitigate the electromagnetic absorption, one of the major limitations of the existing plasmonic cloaks.
%Indeed, we show that the absorption cross section in the operation frequency band of the cloaking device may be substantially diminished by the application of a uniform magnetic field.
In addition,  we prove that the angular distribution of the scattered radiation can be effectively controlled by the magnetic field, allowing for a swift change in the scattering pattern. Our results suggest that magneto-optical materials enable a precise control of light scattering by a tunable mechanism and may be useful in disruptive photonic technologies.

In the second part of the thesis, we study some phenomena that can be interpreted as direct manifestations of quantum fluctuations of the EM field of vacuum. The quantum vacuum state, assumed to be the state with the lowest energy of the system, may be quite complex. In this state, the processes of creation an annihilation of particles may occur provided they do not violate the uncertainty principle \cite{Milonni-1994}. During their short lifetime, such particles can be affected by  external agents as, for instance, EM fields, gravitational fields, or cavities. In other words, quantum vacuum is far from being just an empty space devoided of matter. On the contrary, it can be seen as a ``sea of virtual particles" that continuously appear and disappear, behaving as if it were a material medium with macroscopic properties that may be affected by external agents. Among the several phenomena related to fluctuations of the quantum vacuum, spontaneous emission (SE) \cite{Novotny} and dispersive interactions \cite{Mostepanenko-Book-2009} are extremely relevant in the development of photonic and micro- and nano-electromechanical devices, respectively. Spontaneous emission corresponds to the case where an excited atom decays to the ground state emitting radiation without any apparent influence of external agents on the system. Dispersive interactions correspond to forces between neutral but polarizable bodies that do not have permanent electric or magnetic multipoles. %Dispersive forces and spontaneous emission are the focus of the second part of the thesis.

The possibility of tailoring and controlling light-matter interactions at a quantum level has been a sought-after goal in optics since the pioneer work of E. M. Purcell in 1946 \cite{Purcell-46}. To achieve this goal, several approaches have been proposed so far \cite{blanco2004, thomas2004, carminati2006, vladimirova2012, klimovreview, Sture-2007, Kien-2000, Klimov-2002, Milonni-2003, Kastel-2005, Klimov-2004, Biehs-2011, jacob2012,hyperbolicreview}. Nowadays, it is well known that whenever objects are brought to the vicinities of a quantum emitter, its lifetime is strongly affected by  the boundary conditions. However, we demonstrate that the SE rate of an emitter may not be necessarily  modified in the presence of objects coated by plasmonic cloaks \cite{KortKamp2013}. To the best of our knowledge, we show for the first time that, in the dipole approximation, the Purcell effect can be suppressed even for small separations between the excited atom and a cloak, with realistic material and geometrical parameters. This result suggests that the radiative properties of an excited atom could be exploited to quantically probe the performance of an invisibility cloaking. In addition to this problem,  we study the SE of an excited two-level emitter near a graphene sheet on an isotropic dielectric substrate under the influence of a uniform static magnetic field. This system is specially appealing since graphene possesses unique mechanical, electrical, and optical properties \cite{Graphene1, Graphene2, Graphene3}. Both inhibited and enhanced decay rates are predicted, depending on the emitter-wall distance and on the magnetic field strength. We show that the magnetic field allows for an extraordinary control of the atomic decay rate in the near-field regime. Besides, at low temperatures, the emitter's lifetime presents discontinuities as a function of the magnetic field strength  which is physically explained in terms of the discrete Landau energy levels in graphene. Furthermore, we demonstrate that the magnetic field allows us to tailor the decay channels of the system.

The second part of the thesis is  also devoted to dispersive interactions between atoms and nanostructures. Specifically, we present a proper method for calculating the non-retarded dispersive interaction between an atom and conducting objects of arbitrary shape. The atom-sphere and atom-ellipsoid interactions are investigated using this method \cite{WiltonAJP}. Moreover, we study the dispersive interaction between an atom and a suspended graphene sheet in an external magnetic field \cite{Cysne-2014}.  The possibility of varying the atom-graphene interaction without changing the physical system would be extremely appealing for both experiments and applications. We show that, just by changing the applied magnetic field, the atom-graphene interaction can be strongly reduced. Besides, we demonstrate that, at low temperatures, the Casimir-Polder energy exhibits sharp discontinuities at certain values of the magnetic field. As the distance between the atom and the graphene sheet increases, these discontinuities show up as a plateau-like pattern with quantized values for the dispersive energy.  We also show that, at room temperature, thermal effects must be taken into account even for considerably short distances. %Finally, at room temperature we demonstrate that thermal effects must be taken into account even in the extreme near-field regime.

In the third part of the thesis, we deal with radiative heat transfer in the near-field. For billions of years, thermal radiation has been of fundamental relevance for the development of life on Earth. Solar EM radiation is not only one of the main existing sources of energy for heating,  but it is also crucial for various biological processes, such as photosynthesis. In the modern history of thermal radiation, Planck's spectral distribution for EM radiation has accurately described far-field emission for more than a century. However, recent works on radiative transfer between bodies separated by sub-wavelength distances showed that Planck's law may fail in this case~\cite{Greffet2005, Basu2009, Dorofeyev2011,Raschke2013,BiehsIR,Volokitin2007}. On the other hand, since the seminal work by Polder and van Hove~\cite{PvH1971}, it is known that for two bodies separated by a distance much smaller than the typical thermal wavelengths, evanescent waves play a role in heat transfer, sometimes even surpassing the propagating contribution. Actually, it can be shown that this so called near-field heat transfer (NFHT) can vastly overcome the blackbody limit of radiative transfer by orders of magnitude, radically changing the landscape of possibilities in the arena of out-of-equilibrium phenomena~\cite{Greffet2005, Basu2009, Dorofeyev2011,Raschke2013,BiehsIR,Volokitin2007}. All the recent development in the area of NFHT has naturally led to investigations of possible applications~\cite{BiehsEtAl2008,Fan2010,Zwol2011,BiehsRosaAPL,BiehsPBA2013, Narayanaswamy2003,Laroche2006}, all of them taking advantage of the large increase of the heat flux brought forth by the near-field. As a result, enhancing the process of NFHT is crucial for the development of new and/or optimized applications. In this thesis, we introduce a new approach to enhance the heat transfer in the near-field, by exploiting the versatile material properties of composite media \cite{WiltonPRB2014}.  Within the formalism of the stochastic electrodynamics, we investigate the NFHT between a semi-infinite dielectric medium and metallic nanoparticles embedded in dielectric hosts. By using homogenization techniques, we demonstrate that the NFHT can be strongly enhanced in composite media if compared to the case where homogeneous media are considered. In particular, we show that NFHT is maximal precisely at the insulator-metal (percolation) transition. We also demonstrate that at the percolation threshold an increasingly number of modes effectively contribute to the heat flux, widening the frequency band where the  NHFT occurs.

The thesis is organized as follows. In Chapter {\bf \ref{cap2}} we describe the physics underlying the EM scattering by arbitrary objects. In Chapter {\bf \ref{cap3}} we discuss the physical mechanisms of invisibility cloaking devices designed based on the transformation optics method and the scattering cancelation technique. In Chapter {\bf \ref{cap4}} we investigate the electromagnetic radiation scattered by magneto-optical plasmonic cloaks. In chapter {\bf \ref{cap5}} we derive a convenient  formula for calculating the lifetime of two level quantum emitters in terms of EM field modes and dyadic Green's function. These expressions are applied in Chapter  {\bf \ref{cap6}} in order to investigate the spontaneous emission of a two-level atom near an invisibility cloak or a graphene-coated wall. Chapter {\bf \ref{cap7}} is devoted to an analysis of dispersive interactions between atoms and arbitrary bodies. Here, we also address our attention to the atom-graphene interaction. Finally, in Chapter {\bf \ref{cap8}}, we analyze the near-field heat transfer between a semi-infinite medium and a sphere made of randomly dispersed metallic inclusions embedded in a dielectric host medium.

\end{chapter}

\begin{chapter}{Light absorption and scattering by an arbitrary object}
\label{cap2}

\begin{flushright}
{\it
We have no knowledge of what energy is...\\
However, there are formulas for calculating some numerical quantity,\\
and when we add it all together it gives ``28''- always the same number.
}

{\sc R. P. Feynman}
%{\it
%Science shows us that the visible world is neither matter nor spirit,\\
% the visible world is the invisible organization of energy.
%}
%
%{\sc Heinz R. Pagels}
\end{flushright}

\hspace{5 mm}  Light is rarely seen directly from its source. Most of the light that comes to us is due to the scattering of radiation from different bodies. The characteristics of light scattered by a target, as well as the amount of absorbed radiation, depend on the size, shape, and materials of which the object is made. Although there are various distinct problems on electromagnetic scattering, they share some common features. In this chapter, we formulate the classical problem of scattering of EM radiation by an arbitrary object and discuss the general features involved in such process. Particularly, we use energy conservation to define the scattering, absorption and extinction cross sections. This establishes the main mathematical and physical tools underlying the specific problems regarding invisibility cloaks to be discussed in the next chapter.

\section{Introduction}
\label{Intro2}
\hspace{5 mm} Propagation of electromagnetic radiation through any medium depends on its optical properties. From the macroscopic point of view, we usually characterize the electromagnetic response of a given material by its refractive index $n$. If $n$ is isotropic and uniform, light will propagate throughout the material without being deflected. However, if inhomogeneities are present in the system, radiation will be scattered in all directions~\cite{BohrenHuffman, VanDeHulst,BornWolf, Ishimaru, Kerker, Jackson, Straton}. These inhomogeneities are due to spatial or temporal changes in the refractive index as, for instance, in a system of a particle embedded in a given medium or in a medium with density fluctuations.  In this way, all media scatter light since any material but vacuum presents imperfections when seen with a suitably fine probe. Whatever the heterogeneity, the physics underlying the scattering process is similar for all problems.

In order to describe the scattered, absorbed and extinct electromagnetic energy, let us consider a monochromatic EM wave that illuminates a body with an arbitrary shape embedded in a non-absorptive host medium, as shown in Fig.~\ref{ScatteringArbitraryParticle}. From the microscopic point of view, matter is made of discrete electric charges - electrons and protons. Hence, the applied field on the system will put the charges in the object into motion. The accelerated charges emit EM radiation in all directions and it is precisely this secondary radiation that we call scattered radiation by the body~\cite{BohrenHuffman, VanDeHulst, BornWolf}. Usually, scattering is followed by the transformation of part of the incident EM energy into other forms, such as thermal energy. This process is named absorption~\cite{BohrenHuffman, VanDeHulst, BornWolf} and it is a common process in every-day life. The leaves of a tree look green because they scatter green light more effectively than other colors. It means that blue, red and yellow light, for instance, are absorbed by the leaves and their energy may be converted into some other forms of energy, such as the energy used in photosythesis. Absorption, as well as scattering, removes energy from the incident radiation attenuating the impinging beam. \pagebreak This attenuation is called extinction~\cite{BohrenHuffman,VanDeHulst,BornWolf} and can be measured by a detector located after the obstacle (detector $D_1$ in Fig.~\ref{ScatteringArbitraryParticle}\footnote{\label{Note0} The figure is not in scale: the distance between the detectors and the scattering body is, actually, much larger than the size of the object.}). Consequently, \linebreak we can write
\begin{equation}
\label{DefinitionExtinction}
\textrm{extinction = scattering + absorption}\, .
\end{equation}
%Although scattering and absorption are distinct processes we shall be concerning here mainly with scattering.
%
     \vspace{10pt}
\begin{figure}[!ht]
\centering
	\includegraphics[scale=0.65]{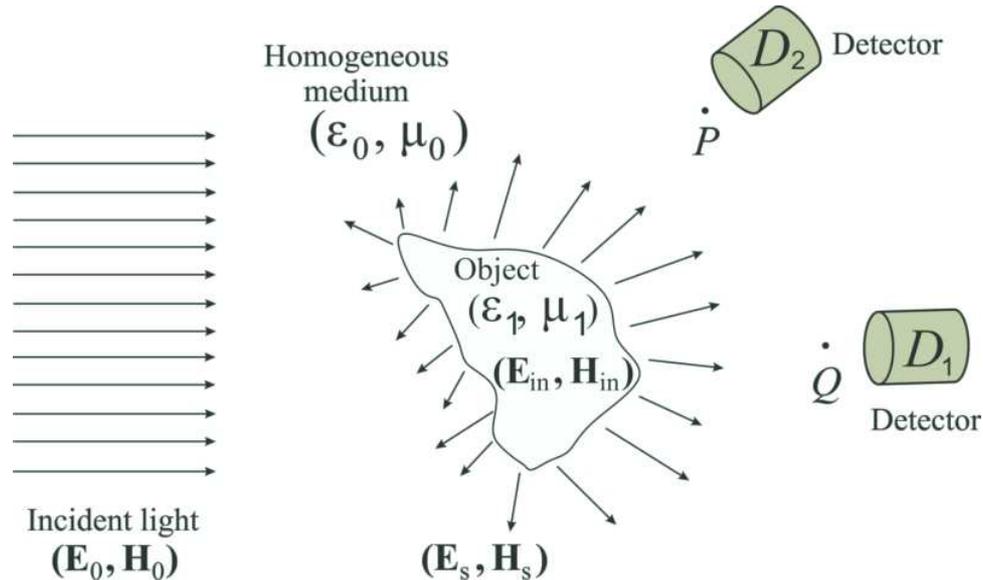}
    \vspace{10pt}
    \caption{Electromagnetic scattering by a single object. The impinging light $({\bf E}_0, {\bf H}_0)$ interacts with the obstacle and produces a field $({\bf E}_{\textrm{in}}, {\bf H}_{\textrm{in}})$ inside it and a scattered field $({\bf E}_{\textrm{s}}, {\bf H}_{\textrm{s}})$ in the host medium surrounding it$^{\textrm{\ref{Note0}}}$.}
    \label{ScatteringArbitraryParticle}
\end{figure}

Some qualitative information about the pattern of the scattered light can be acquired if we consider that the body is subdivided into very small regions. Each region is described by a dipole oscillating at the frequency of the incident wave, if we neglect the interaction between the dipoles~\cite{BohrenHuffman}. Therefore, the total scattered electromagnetic  field detected at a particular point $P$ very distant from the object is given by the superposition of the radiation emitted by all the dipoles on the body. The relative phases between the EM fields at point $P$ generated by distinct dipoles play a fundamental role in the problem. Apart from being functions of geometrical features such as the shape and size of the obstacle, the relative phases depend on the scattering direction. Hence, we expect that by varying the position of detection from $P$ to $Q$, the pattern of the detected field changes. Furthermore, since the strength and phase of the induced oscillating dipoles depend on the kind of material of which the object is composed, a full description of the scattering problem by a given particle must take into account the dispersive character of the material. The formal treatment of the problem is discussed in the next sections.

\section{Formulation of the problem}
\label{ScatteringProblem}

\hspace{5 mm} In this section, we study the interaction of monochromatic light of arbitrary frequency with a single object embedded in an homogeneous host medium, as shown in Fig.~\ref{ScatteringArbitraryParticle}. Here, homogeneous means that the length scale of the atomic inhomogeneities is much smaller than the wavelength of the propagating radiation. For the sake of simplicity, we do not consider scattering by fluctuations. Also, there are some considerations that simplify the analytical solution of the problem considerably and provide excellent results, when compared with numerical simulations~\cite{BohrenHuffman, VanDeHulst,Baber1990}. The general hypotheses are:
\begin{enumerate}
\item[(i)] Both electromagnetic radiation and matter will be treated classically. In other words, it is not necessary to take into account the interaction of photons with elementary quantum excitations of matter through quantum mechanics.
\item[(ii)] We assume that the optical properties of the obstacle and host medium can be well described in terms of linear, isotropic, and homogeneous frequency-dependent optical constants: permittivity, permeability, conductivity, refractive index, etc. Besides, the host medium (not necessarily  vacuum) will be considered nondispersive and nonabsorptive. 
\item[(iii)] The incident EM radiation is assumed to be monochromatic and generated by a source very distant from the target, in such a way that it can be treated as a plane wave. Since the materials are supposed to be linear, the solution of the problem of scattering of polychromatic radiation can obtained with the aid of Fourier series and transforms.
\item[(iv)] Only elastic scattering is considered: the frequency of the scattered light is the same as that of the incident wave. Hence, phenomena such as Raman scattering \cite{Raman} is not contemplated in the treatment.
\end{enumerate}

In addition, we are adopting the temporal dependence $e^{-i\omega t}$ for the electromagnetic field. It is worth mentioning that in Refs.~\cite{VanDeHulst, Kerker} the convention $e^{+i\omega t}$ is used and, as a result, the signal of the imaginary part of the optical constants characterizing the materials are opposite to that used in Refs.~\cite{BohrenHuffman, BornWolf, Jackson,  Straton, Baber1990}.

With the above assumptions, let us consider a linearly polarized EM plane wave with angular frequency $\omega$ and wave vector ${\bf k}_0$ propagating in a host medium with permittivity $\varepsilon_0$ and permeability $\mu_0$. The electric and magnetic fields of the wave are given by
\begin{eqnarray}
{\bf E}_0({\bf r}, t) =  {\bf {\cal{E}}}_0e^{i({\bf k}_0 \cdot {\bf r} - \omega t)}\, , \ \ \textrm{and} \ \
{\bf H}_0({\bf r}, t) =  {\bf {\cal{H}}}_0e^{i({\bf k}_0 \cdot {\bf r} - \omega t)}\, .
\label{IncidentElectromagneticField}
\end{eqnarray}
This wave impinges upon a single obstacle with dielectric constant $\varepsilon_1(\omega)$ and magnetic permeability $\mu_1(\omega) = \mu_0$, as depicted in Fig.~\ref{ScatteringArbitraryParticle}. The fields inside the object are denoted by $({\bf E}_{\textrm{in}}, {\bf H}_{\textrm{in}})$, whilst the fields outside consist of the superposition of the incident field $({\bf E}_0, {\bf H}_0)$ and the scattered field $({\bf E}_{\textrm{s}}, {\bf H}_{\textrm{s}})$. Assuming that there are no sources, the EM field dynamics is governed by the macroscopic Maxwell's equations~\cite{Jackson}
\begin{eqnarray}
\label{MaxwellEquations1}
\nabla \cdot {\bf D} &=& 0\, , \\
\label{MaxwellEquations2}
\nabla \cdot {\bf B} &=& 0\, , \\
\label{MaxwellEquations3}
\nabla \times {\bf E} &=& i\omega{\bf B}\, , \\
\label{MaxwellEquations4}
\nabla \times {\bf H} &=& -i\omega{\bf D}\, .
\end{eqnarray}
Taking the curl of Eqs. (\ref{MaxwellEquations3}) and (\ref{MaxwellEquations4}), and using the constitutive relations for the electric displacement field ${\bf D} = \varepsilon{\bf E}$ and  magnetic induction ${\bf B} = \mu_0{\bf H}$, we arrive at \pagebreak
\begin{eqnarray}
\label{WaveEquation1}
\!\!\!\!\!\!\nabla \times (\nabla \times {\bf E})\!\!\! &=&\!\!\! i\omega\nabla\times{\bf B} \Rightarrow \nabla\!\!\! \underbrace{(\nabla \cdot {\bf E})}_{=0,\ \textrm{Eq. (\ref{MaxwellEquations1})}}\!\!\!-\nabla^2{\bf E} = \mu_0\varepsilon\omega^2 {\bf E} \Rightarrow (\nabla^2 + k^2){\bf E} = {\bf 0}\, ,\\ \cr
\label{WaveEquation2}
\!\!\!\!\!\!\nabla \times (\nabla \times {\bf H})\!\!\! &=&\!\!\! -i\omega\nabla\times{\bf D} \Rightarrow \nabla\!\!\! \underbrace{(\nabla \cdot {\bf H})}_{=0,\ \textrm{Eq. (\ref{MaxwellEquations2})}}\!\!\!-\nabla^2{\bf H} = \mu_0\varepsilon\omega^2 {\bf H}  \Rightarrow (\nabla^2 + k^2){\bf H} = {\bf 0}\, ,
\end{eqnarray}
where $k^2=\mu_0\varepsilon\omega^2$. Therefore, the electric and magnetic fields satisfy the vector \linebreak Helmholtz equation~\cite{Jackson}.

In addition to the Maxwell's equations, there are boundary conditions that must be satisfied by the EM fields at the interface between the object and the host medium. With the usual approach of constructing closed curves and closed surfaces intersecting the boundary points and taking the appropriate limits, it is straightforward to show that the tangential components of {\bf E} and {\bf H} have to be continuous across the boundary. \linebreak Therefore, we have\footnote{\label{note1} Since we are assuming a $e^{-i\omega t}$ dependence, we will omit $t$ in the arguments of $({\bf E}, \, {\bf H})$ whenever convenient.}~\cite{Jackson} 
\begin{eqnarray}
\label{BoundaryCondition1}
\Big\{\left[{\bf E}_0({\bf r}) + {\bf E}_{\textrm{s}}({\bf r}) - {\bf E}_{\textrm{in}}({\bf r}) \right]\times \hat{{\bf n}}\Big\}_{{\bf r} \in S} = {\bf 0}\, , \\
\label{BoundaryCondition2}
\Big\{\left[{\bf H}_0({\bf r}) + {\bf H}_{\textrm{s}}({\bf r}) - {\bf H}_{\textrm{in}}({\bf r}) \right]\times \hat{{\bf n}}\Big\}_{{\bf r} \in S} = {\bf 0}\, ,
\end{eqnarray}
where $\hat{{\bf n}}$ is the unit vector normal to the surface $S$ of the object and is pointing outward.

As we shall see, these boundary conditions are crucially related to the energy conservation theorem. Consider two arbitrary closed surfaces: $S_1$ completely inside to the object and $S_2$ totally outside to the object, as shown in Fig.~\ref{EnergyAcrossSurface}. Both $S_1$ and $S_2$ can be made arbitrarily close to the surface $S$ of the body. The rate $\Phi_1$ at which energy crosses $S_1$ is given by the flow of the Poynting vector through this surface, namely,
\begin{eqnarray}
\Phi_1 = \oint_{S_1} {\bf S}_1 \cdot \hat{{\bf n}}\ dA = \oint_{S_1}({\bf E}_{\textrm{in}}\times {\bf H}_{\textrm{in}}) \cdot \hat{{\bf n}}\ dA
 = \oint_{S_1}(\hat{{\bf n}} \times {\bf E}_{\textrm{in}}) \cdot {\bf H}_{\textrm{in}}\ dA\, ,
\end{eqnarray}
where we used the well know identity $({\bf A} \times {\bf B}) \cdot {\bf C}=({\bf C} \times {\bf A}) \cdot {\bf B}=({\bf B} \times {\bf C}) \cdot {\bf A}$. When $S_1$ and $S_2$ approach $S$ we can use (\ref{BoundaryCondition1}) and (\ref{BoundaryCondition2}) to rewrite the previous equation as \pagebreak
\begin{eqnarray}
\label{EnergyConservationSurface}
\Phi_1 &=& \oint_{S}[\hat{{\bf n}} \times  ({\bf E}_{0}+{\bf E}_{\textrm{s}})] \cdot {\bf H}_{\textrm{in}}\ dA \cr
 &=& \oint_{S}[{\bf H}_{\textrm{in}} \times\hat{{\bf n}} ] \cdot ({\bf E}_{0}+{\bf E}_{\textrm{s}})\ dA \cr
 &=& \oint_{S}[ ({{\bf H}_{0}}+{\bf H}_{\textrm{s}}) \times\hat{{\bf n}} ] \cdot ({\bf E}_{0}+{\bf E}_{\textrm{s}})\ dA \cr
 &=& \oint_{S}[({\bf E}_{0}+{\bf E}_{\textrm{s}}) \times ({{\bf H}_{0}}+{\bf H}_{\textrm{s}})] \cdot \hat{{\bf n}}\ dA\cr
 &=& \Phi_2\, ,
\end{eqnarray}
where in the last step we used that in the absence of sources and sinks the rate at which EM energy crosses $S_2$ is given by
\begin{eqnarray}
\Phi_2 = \oint_{S_2} {\bf S}_2 \cdot \hat{{\bf n}}\ dA &=& \oint_{S_2}[({\bf E}_{0}+{\bf E}_{\textrm{s}}) \times ({{\bf H}_{0}}+{\bf H}_{\textrm{s}})] \cdot \hat{{\bf n}}\ dA\, .
\end{eqnarray}

Equation (\ref{EnergyConservationSurface}) puts in evidence  that the continuity of the tangential components of the electric and magnetic fields at the interface separating two media with different optical properties is a sufficient condition for energy conservation across such interface~\cite{BohrenHuffman}.
\vspace{10pt}
\begin{figure}[!ht]
\centering
	\includegraphics[scale=0.55]{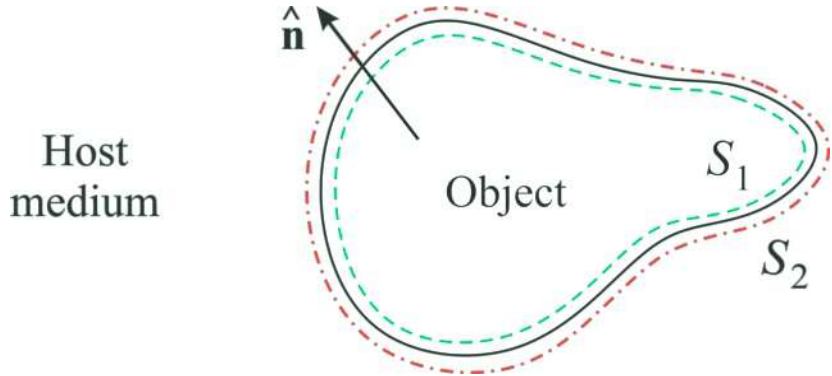}
    \vspace{10pt}
    \caption{$S_1$ (green dashed line) and $S_2$ (red  dot-dashed line) are closed surfaces arbitrarily near the interface between two materials with distinct optical properties. If there are no sources or sinks the energy flow through $S_1$ will be the same as that through $S_2$.}
    \label{EnergyAcrossSurface}
\end{figure}
\section{Absorption, extinction and scattering cross sections}
\label{CrossSection}

\hspace{5 mm} In Fig.~\ref{ScatteringArbitraryParticle}, the detector $D_1$ is after the object and pointing to the light source; it measures the rate at which electromagnetic energy $U$ arrives on it. If the scatter is absent, the detected power by $D_1$ will be $U_0 > U$. Once it is assumed that the host medium is nonabsorbent, the difference $U_0-U$ is due to absorption and scattering of the impinging wave by the object. Since this extinction (attenuation) of the incident beam might depend on the material composition of the body, its size and shape, as well on frequency and polarization of the EM radiation, the scattering problem by a single target can be quite complex. However, we can define some useful quantities, namely, the scattering, absorption, and extinction cross sections, that can help the analytical understanding of such problems.

Let us assume that a $x-$polarized incident plane wave that propagates in $z-$direction impinges on an arbitrary shape object, as represented in Fig. \ref{ScatteringArbitraryParticle2}. Since the homogeneous medium is lossless, the Poynting's theorem states that the absorbed energy rate $W_{\textrm{abs}}$ by the body ({\it e.g.} converted into mechanical or thermal energy) is~\cite{Jackson}
\begin{eqnarray}
W_{\textrm{abs}} =\int_{{\cal{R}}} {\bf J} \cdot {\bf E}\ dV =  -\left[\int_{{\cal{R}}}{\dfrac{\partial u}{\partial t}\ dV+ \oint_{\partial{\cal{R}}} {\bf S} \cdot \hat{{\bf r}}\ dA} \right]\, ,
\label{PoyntingTheorem}
\end{eqnarray}
where ${\cal{R}}$ is the spherical region of radius $R$ around the target, $\partial{\cal{R}}$ is the corresponding boundary and $\hat{{\bf r}}$ is the usual unit vector in the radial direction pointing outward the sphere. Also, $u$
%
%\begin{eqnarray}
%u = \dfrac{1}{2} ({\bf E}\cdot {\bf D} + {\bf B} \cdot {\bf H})\,
%\label{DensityElectromagneticEnergy}
%\end{eqnarray}
%
is the electromagnetic energy density~\cite{Jackson}, ${\bf J}$ is the current density and ${\bf S} $ is the Poynting vector~\cite{Jackson}. It is worth emphasizing that, in the steady regime, the time average of the first term on the right hand side of Eq. (\ref{PoyntingTheorem}) vanishes. Therefore, the mean rate at which EM energy is absorbed by the particle can be cast as~\cite{BohrenHuffman}
\begin{eqnarray}
\langle W_{\textrm{abs}} \rangle = -\oint_{\partial{\cal{R}}} \langle{\bf S}\rangle \cdot \hat{{\bf r}}\ dA\, ,
\label{AbsorbedEnergyParticle}
\end{eqnarray}
where $\langle...\rangle$ denotes temporal averaging. It should be noted that $\langle W_{\textrm{abs}}\rangle \geq 0$ since both the host and the scatter media are passive, {\it i.e.} there is no optical gain, and there are no sources within ${\cal{R}}$.
\vspace{10pt}
\begin{figure}[!ht]
\centering
	\includegraphics[scale=0.55]{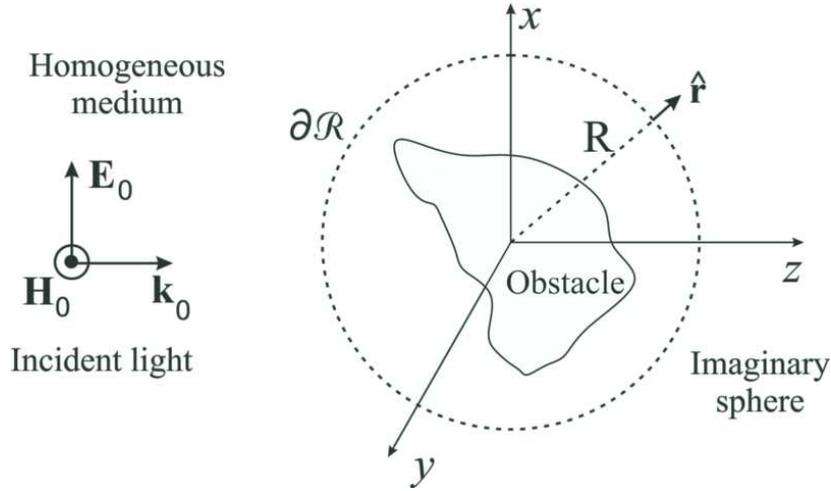}
    \vspace{10pt}
    \caption{Electromagnetic scattering of $x-$polarized plane wave by a single obstacle. The mean rate at which electromagnetic energy is absorbed by the object is given by the flow of the Poynting vector that crosses the boundary of the imaginary sphere.}
    \label{ScatteringArbitraryParticle2}
\end{figure}
\pagebreak

Since the incident and scattered EM fields are harmonic functions of time with the same frequency $\omega$ (elastic scattering), the time-averaged Poynting vector $\langle {\bf S} \rangle$ at the surface of the sphere is given by~\cite{BohrenHuffman}
\begin{eqnarray}
\langle {\bf S} \rangle = \langle ({\bf E}_{0}+{\bf E}_{\textrm{s}}) \times ({{\bf H}_{0}}+{\bf H}_{\textrm{s}})  \rangle =
\langle {\bf S}_{0} \rangle + \langle {\bf S}_{\textrm{s}}  \rangle + \langle{\bf S}_{\textrm{ext}} \rangle\, ,
\label{PoyntingVector1}
\end{eqnarray}
where
\begin{eqnarray}
\langle {\bf S}_{0} \rangle = &\dfrac{1}{2}&\!\!\! \textrm{Re}\left[{\bf E}_{0} \times {\bf H}_{0}^{*} \right]\, , \
\langle {\bf S}_{\textrm{\textrm{s}}} \rangle = \dfrac{1}{2} \textrm{Re}\left[{\bf E}_{\textrm{s}} \times {\bf H}_{\textrm{s}}^{*} \right]\, ,
\end{eqnarray}
and
\begin{eqnarray}
\langle {\bf S}_{\textrm{ext}} \rangle = \dfrac{1}{2} \textrm{Re}\left[{\bf E}_{0} \times {\bf H}_{\textrm{s}}^{*} + {\bf E}_{\textrm{s}} \times {\bf H}_{0}^{*} \right]\,
\label{PoyntingVector2}
\end{eqnarray}
are the Poynting vectors related to the incident wave, scattered field and interaction between impinging and scattered radiation. Substituting (\ref{PoyntingVector1}) in (\ref{AbsorbedEnergyParticle}) and using that ${\bf S}_0$ is independent of the position for a nonabsorbing homogeneous medium, we can \linebreak rewrite $\langle W_{\textrm{abs}} \rangle $ as~\cite{BohrenHuffman}
\begin{eqnarray}
\langle W_{\textrm{abs}} \rangle  = -\oint_{\partial{\cal{R}}} \langle{\bf S}_{\textrm{s}}\rangle \cdot \hat{{\bf r}}\ dA - \oint_{\partial{\cal{R}}} \langle{\bf S}_{\textrm{ext}}\rangle \cdot \hat{{\bf r}}\ dA
= - \langle W_{\textrm{sca}} \rangle + \langle W_{\textrm{ext}} \rangle\, ,
\label{AverageEnergy}
\end{eqnarray}
where  $\langle W_{\textrm{sca}} \rangle$ is the net rate at which scattered light crosses the boundary of the sphere and $\langle W_{\textrm{ext}} \rangle$ is, according to Eq. (\ref{DefinitionExtinction}), the extinct electromagnetic energy.

As the host medium is lossless, $\langle W_{\textrm{abs}} \rangle$, $\langle W_{\textrm{sca}} \rangle$ and $\langle W_{\textrm{ext}} \rangle$ are independent functions of the radius $R$ of the sphere so that we can choose it conveniently. On one hand, if $R\ll D^2k_0/2\pi$, where $D$ is a typical dimension of the object, the fields $({\bf E}_{\textrm{s}}, {\bf H}_{\textrm{s}})$ at the boundary of the imaginary sphere can be a complicated function due to interference effects that arise from contributions of different parts of the body to the scattered radiation. On the other hand, for $R \gg D^2k_0/2\pi$, we are in the radiation zone where the scattered field behaves as a spherical wave and can be written as~\cite{BohrenHuffman, Ishimaru}
\begin{eqnarray}
{\bf E}_{\textrm{s}}^{\textrm{rad}} = \dfrac{e^{ik_0r}}{-ik_0r} |{\bf E}_0| {\bf f}_x({\bf k},{\bf k}_0)\, , \ \textrm{and} \ \
{\bf H}_{\textrm{s}}^{\textrm{rad}} = \dfrac{k_0}{\omega\mu_0} \hat{{\bf k}} \times {\bf E}_{\textrm{s}}^{\textrm{rad}}\, ,
\label{RadiativeFields}
\end{eqnarray}
where ${\bf f}_x({\bf k},{\bf k}_0) |{\bf E}_0|$ describes the amplitude, phase and polarization of the scattered radiation in the far field in the $\hat{{\bf k}} = {\bf k}/k_0 = \hat{{\bf r}} $ direction for a $x-$polarized incident plane wave with wave vector ${\bf k}_0 = k_0\hat{{\bf z}}$. It should be noticed that the scattered light is in general elliptically polarized even though the impinging wave is linearly polarized. Also, ${\bf f}_x({\bf k},{\bf k}_0) \cdot \hat{{\bf k}} = 0 $ since the radiation field is transverse to the direction of energy propagation. Therefore, we might choose a sufficiently large $R$ so that Eq. (\ref{RadiativeFields}) can be used.

The scattering cross section $C_{\textrm{sca}}$ is defined as the ratio between the mean rate at which scattered light crosses the boundary of the sphere and the incident field intensity $I_0 = k_0|E_0|^2/2\omega\mu_0$~\cite{BohrenHuffman, VanDeHulst}. Thus, using (\ref{PoyntingVector1}), (\ref{PoyntingVector2}) and (\ref{RadiativeFields}), we get \cite{BohrenHuffman}
\begin{eqnarray}
C_{\textrm{sca}}:= \dfrac{\langle W_{\textrm{sca}} \rangle}{I_0}
= \int_0^{2\pi} d\varphi \int_0^{\pi} \dfrac{|{\bf f}_x({\bf k},{\bf k}_0)|^2}{k_0^2} \sin{\theta} \ d\theta\, .
\label{ScatteringCrossSectionDefinition}
\end{eqnarray}
The quantity
\begin{eqnarray}
\dfrac{dC_{\textrm{sca}}}{d\Omega}:=\dfrac{|{\bf f}_x({\bf k},{\bf k}_0)|^2}{k_0^2}\, ,
\label{DifferentialScatteringCrossSection}
\end{eqnarray}
is called differential scattering cross section~\cite{BohrenHuffman, VanDeHulst, Ishimaru} and gives the angular distribution of the scattered radiation.

Similarly to the scattering cross section, the extinction cross section is defined as~\cite{BohrenHuffman}
\begin{eqnarray}
C_{\textrm{ext}} &:=& \dfrac{\langle W_{\textrm{ext}} \rangle}{I_0}\cr\cr
&=& -\textrm{Re}\left\{ \dfrac{R}{ik_0}e^{-ik_0R}\int_0^{2\pi} d\varphi \int_0^{\pi} e^{ik_0R\cos{\theta}}
\{\hat{{\bf x}}\cdot{\bf f}_x^{*}({\bf k},{\bf k}_0)\} \sin{\theta} \ d\theta \right. \cr\cr
&-&\!\!\! \left. \dfrac{R}{ik_0}e^{ik_0R}\int_0^{2\pi} d\varphi \int_0^{\pi}  e^{-ik_0R\cos{\theta}}
\{\cos{\theta}[\hat{{\bf x}}\cdot{\bf f}_x({\bf k},{\bf k}_0)]\} \sin{\theta} \ d\theta \right. \cr\cr
&+&\!\!\!\left. \dfrac{R}{ik_0}e^{ik_0R}\int_0^{2\pi} d\varphi \int_0^{\pi}  e^{-ik_0R\cos{\theta}}
\{\sin{\theta}\cos{\varphi}[\hat{{\bf z}}\cdot{\bf f}_x({\bf k},{\bf k}_0)]\} \sin{\theta} \ d\theta \right\}\, .
\label{ExtinctionCrossSectionDefinition}
\end{eqnarray}
The integrals on $\theta$ are of the form
\begin{eqnarray}
\int_{-1}^{1}e^{\pm ik_0R\zeta}g(\zeta)d\zeta =
\dfrac{e^{\pm ik_0R}g(1)-e^{\mp ik_0R}g(-1)}{\pm ik_0R}+{\cal{O}}\left(\dfrac{1}{k^2R^2}\right)\, ,
\end{eqnarray}
where $\zeta = \cos{\theta}$ and we have integrated by parts and discarded terms of order $(k_0R)^{-2}$ provided that $dg/d\zeta$ is finite and $R$ is arbitrarily large. Thus, after some lengthy but straightforward algebraic manipulations equation (\ref{ExtinctionCrossSectionDefinition}) can be cast as \cite{BohrenHuffman}
\begin{eqnarray}
C_{\textrm{ext}} = \dfrac{4\pi}{k_0^2} \textrm{Re}\left[\hat{{\bf x}}\cdot{\bf f}_x({\bf k}_0,{\bf k}_0)\right]\, .
\label{ExtinctionCrossSectionFinal}
\end{eqnarray}
The above equation is one of the forms of the optical theorem~\cite{BohrenHuffman, VanDeHulst, BornWolf, Ishimaru,  Kerker, Jackson}, which physically expresses energy conservation in  the scattering process.  It shows that, in spite of the fact that extinction is a combined effect of absorption and scattering in all directions,  the extinction cross section depends only on the scattering amplitude  in the forward direction.

Equations (\ref{ScatteringCrossSectionDefinition}), (\ref{ExtinctionCrossSectionDefinition}) and (\ref{ExtinctionCrossSectionFinal}) can be combined to (\ref{AverageEnergy}) to calculate the absorption cross section~\cite{BohrenHuffman} of the object
\begin{eqnarray}
C_{\textrm{abs}} := \dfrac{\langle W_{\textrm{a}}\rangle}{I_0} = C_{\textrm{ext}} - C_{\textrm{sca}}\, .
\end{eqnarray}

The above equations for the cross sections were obtained under the assumption of a $x-$polarized impinging radiation. If the incident field is arbitrarily polarized, ${\bf E}_0 = E_{0x} \hat{{\bf x}} + E_{0y}\hat{{\bf y}}$, the previous equation will remain valid with the following changes: $\hat{{\bf x}} \rightarrow {\bf E}_0/|{\bf E}_0|$ and ${\bf f}_x({\bf k},{\bf k}_0) \rightarrow \left[E_{0x}{\bf f}_x({\bf k},{\bf k}_0) + E_{0y}{\bf f}_y({\bf k},{\bf k}_0)\right]/|{\bf E}_0|$, where ${\bf f}_y({\bf k},{\bf k}_0)$ is the scattering amplitude for $y-$polarized incident light.

It is convenient to work with the dimensionless absorption, scattering,  and extinction cross section efficiencies, defined as~\cite{BohrenHuffman}
\begin{eqnarray}
Q_{\textrm{abs}} := \dfrac{C_{\textrm{abs}}}{G}\, , \ \ Q_{\textrm{sca}} := \dfrac{C_{\textrm{sca}}}{G}\, , \ \textrm{and} \ \ Q_{\textrm{ext}} := \dfrac{C_{\textrm{ext}}}{G}\, ,
\label{EfficiencyCrossSections}
\end{eqnarray}
where $G$ is the geometric cross section of the particle.

\vspace{30pt}

\hspace{5 mm} In this chapter we have discussed the physics underlying the scattering of EM waves by arbitrarily shaped objects. Particularly, we have used Poynting's theorem to define the absorption, scattering and extinction cross sections, which  are expressed in terms of the scattering amplitude.
%, the only component necessary to be calculated so as to obtain the amount of absorbed and scattered electromagnetic energy.
%

\end{chapter}

\begin{chapter}{Physical mechanisms of invisibility cloaks}
\label{cap3}

\begin{flushright}
{\it
The true mystery of the world is the visible,\\
not the invisible.
}

{\sc O. WILDE}
\end{flushright}

\hspace{5 mm} The idea of rendering an object invisible in free space, which had been restricted to human imagination for many years, has become an important scientific and technological challenge since the advent of metamaterials. In this chapter we discuss the two main techniques to achieve invisibility, namely, the transformation optics method (TOM) and the scattering cancellation technique (SCT). In the former the invariance of Maxwell's equations under coordinate transformations is exploited to design materials with unique optical properties capable to bend light around an object, making it invisible. In the latter, covers made of materials with low positive or negative permittivities scatter electromagnetic waves which interfere destructively with the EM field scattered by the object to be cloaked. The mathematical formalism developed in the previous chapter will be used to treat the case of a plasmonic cloak with spherical symmetry.

\section{Introduction}
\label{Intro3}

\hspace{5 mm} Classical electrodynamics and its applications have experienced notable progress in the last decade after the introduction of metamaterials~\cite{zheludev2012, EnghetaLivro, ShalaevLivro, SmithLivro}. Indeed, metamaterials have permitted not only the discovery of novel physical phenomena but also the development of applications that led to an unprecedented control of electromagnetic waves, far beyond to what is achievable with natural media~\cite{zheludev2012, EnghetaLivro, ShalaevLivro, SmithLivro}. Among these applications, EM cloaking is arguably the most fascinating one, since the idea of rendering an object invisible has fueled human imagination for several centuries. In an ideal scenario, making an object invisible involves no electromagnetic energy absorption and complete suppression of the scattered light at all observation angles over a wide range of frequencies~\cite{AluReview2008, AluReview2014}. From a more realistic point of view, a perfect cloaking device might not be achieved, as intrinsic material losses are unavoidable~\cite{Ruan2007}. Indeed, passive cloaking suffers from some limitations~\cite{Miller2006, Greenleaf2007, Ruan2007, Yan2007, monticone2013_1} so that a perfect cloak may possibly never be constructed. However, the main goal of a cloaking mechanism is, under suitable conditions, to reroute the major fraction of the incident EM radiation around the object to be concealed, making it nearly undetectable.

Although studies of  realistic engineered passive invisible mantles are quite recent, the notion of non-radiating oscillating distributions has been studied since the beginning of the twentieth century~\cite{Schott1933}. Other related works in low-scattering antennas and invisible particles and sources in the quasi-static limit have already been extensively discussed in the literature~\cite{Kahn1965,Kerker1975,Kerker1976, Kerker82, Uzonoglu1978, Devaney1978,  Devaney1982, KiLdal1996, Hoenders1997, Greenleaf2003, Boardman2005,Goedecke1964}. Nowadays, there are a number of modern approaches to cloaking~\cite{alu2005_1,alu2005_2,alu2006,alu2007,alu2007_2,alu2008NJP,alu2008PRL,alu2010, alu2010_2, alu2008PRE,alu2009, alu2009-2,alu2009mantle,chen2010,chen2011,chen2012,Cummer2006,edwards2009,ergin2010,farhat2011,
farhat2013,filonov2012,kortkampPRL,kortkamp2013-2, kortkampJOSAA,leonhardt2006_1,leonhardt2006_2,leonhardt2006_3, yan2009, Yan2007,Ruan2007, leonhardt2008, milton2006,milton2007,milton2009,milton2009-2,monticone2013, Tretyakov,Schofield_2014_1, Rybin,
peining2010_1,peining2010_2, peining2010_3, pendry2006,rainwater2012,schurig2006,Schurig2007, Kundtz, silveirinha2008,soric2013,valentine2009,zharova2012, nicorovici1994, Zhang}, proving that a properly designed metamaterial can strongly suppress EM scattering around a given frequency. Among these approaches, we can highlight the transformation optics method~\cite{Cummer2006,leonhardt2006_1,leonhardt2006_2,leonhardt2006_3,pendry2006,Schurig2007, Kundtz, yan2009, Yan2007,Ruan2007, leonhardt2008, chen2010, schurig2006,ergin2010,valentine2009, Zhang} and the scattering cancellation technique~\cite{alu2005_1,alu2005_2,alu2006,alu2007,alu2007_2,alu2008NJP,alu2008PRL,alu2010, alu2010_2,alu2008PRE,silveirinha2008,alu2009,alu2009-2,chen2011,farhat2011,filonov2012,chen2012,
monticone2013,edwards2009,rainwater2012}.

The transformation optics method~\cite{Cummer2006,leonhardt2006_1,leonhardt2006_2,leonhardt2006_3,pendry2006,Schurig2007, Kundtz, yan2009, Yan2007,Ruan2007, leonhardt2008, chen2010, Zhang} is based on a coordinate transformation that stretches and squeezes the grid of space in such a way that light rays follow curvilinear trajectories. Due to the form-invariance of Maxwell's equations under any conformal coordinate transformation, it is possible to re-interpret the flow of the electromagnetic energy in the transformed space as a propagation in the original system with re-scaled anisotropic and inhomogeneous permittivity and permeability tensors. Therefore, provided we have the full control over $\mbox{{\mathversion{bold}${\varepsilon}$}}({\bf r})$ and $\mbox{{\mathversion{bold}${\mu}$}}(\bf r)$ profiles (as in a metamaterial) we can create a coordinate transformation that leads to a bending of the incoming EM radiation around a given region of space, rendering it invisible to an external observer. This effect is in some way similar to the mirage effect: in hot days, the density of the air near the surface of a road is smaller than in higher air layers so that there is a gradient in the refractive index of the atmosphere that distorts the optical rays.
%creating the illusion of a pool of water on the road.
An appropriate choice of material parameters in the cloaked region allows to guide the electromagnetic radiation through the system without distortion or light scattering. The TOM has firstly been realized experimentally for microwaves~\cite{schurig2006}  and has been later extended to infrared and visible frequencies~\cite{ergin2010,valentine2009}. This method to achieve invisibility will be examined in more details in the next section.

In the scattering cancellation technique~\cite{alu2005_1,alu2005_2,alu2006,alu2007,alu2007_2,alu2008NJP,alu2008PRL, alu2010, alu2010_2, alu2008PRE,silveirinha2008,alu2009,alu2009-2,chen2011,farhat2011,filonov2012,chen2012,
monticone2013}, a dielectric or conductor object can be effectively cloaked if it is coated by an isotropic and homogeneous material with a local electric permittivity smaller than that of the host medium (usually air/vacuum). In this kind of system, the multipoles induced on the coating shell are out of phase with respect to those induced on the object to be cloaked. For an appropriate set of material and geometric parameters, the net effect is that the dominant multipoles are such that the scattered EM radiation from the whole system becomes orders of magnitude smaller than that of the isolated object itself. In other words, in the SCT the invisibility cloak produces a scattered electromagnetic field that interferes destructively with the one generated by the object to be cloaked. As a matter of fact, since in the SCT the cloaked body interacts directly with the impinging wave,  this technique is more suitable than other cloaking methods, regarding  applications  for improving the performance of sensors~\cite{alu2009-2}. The SCT  was observed by the first time in two-dimensions for a cylindrical geometry in the microwave frequency range~\cite{edwards2009}. The first three-dimensional cloak based on the SCT was achieved in 2012~\cite{rainwater2012}. This technique will be treated in details for the case of spherical cloaks in Section ~\ref{ScatteringCancelation}.

It should be mentioned that other invisibility techniques based on anomalous localized resonances~\cite{nicorovici1994, milton2006, milton2007}, mantle cloaking~\cite{alu2009mantle,soric2013}, and transmission-line networks~\cite{Tretyakov} do exist. Although experimental demonstrations of some of these techniques have been performed in recent years, they are out of the scope of this thesis and will not be discussed.
%The interested reader can find more details about them in Refs. \cite{nicorovici1994, milton2006, milton2007,Rybin,Tretyakov,alu2009mantle,soric2013,AluReview2014} and references therein.

\section{The transformation optics method}
\label{TransformationOptics}

\hspace{5 mm} The transformation optics method to achieve invisibility is based on the form-invariance of the Maxwell's equations under coordinate transformations~\cite{Cummer2006,leonhardt2006_1,leonhardt2006_2,leonhardt2006_3,pendry2006,Schurig2007, Kundtz, yan2009, Yan2007,Ruan2007, leonhardt2008, chen2010, schurig2006,ergin2010,valentine2009, Zhang}.
%Even though the TOM in 4D Minkowski space is well established~\cite{leonhardt2008} most applications of this method are either in static or in slowly moving media (compared to the speed of light).
For the sake of simplicity we will not take into account temporal coordinate transformations but only spatial coordinate transformations. Let us start with the macroscopic Maxwell's equations in a flat 3D Euclidian space~\cite{Jackson}
\begin{eqnarray}
\label{MaxwellEqEinsteinNotation1}
\nabla \cdot {\bf D}\!\! &=&\!\! \rho_e\, , \ \ \ \nabla \times {\bf E} = -\dfrac{\partial {\bf B} }{\partial t} \, ,  \cr
 \nabla\cdot {\bf B}\!\! &=&\!\! 0\, , \ \ \ \nabla \times {\bf H} = {\bf J} + \dfrac{\partial {\bf D}}{\partial t} \, ,
\end{eqnarray}
where, for harmonic fields, we have
\begin{eqnarray}
\label{ConstitutiveRelations}
{\bf D} = \mbox{{\mathversion{bold}${\varepsilon}$}} \cdot {\bf E} \, , \ \ \ {\bf B} = \mbox{{\mathversion{bold}${\mu}$}} \cdot {\bf H}\, ,
\end{eqnarray}
with $\mbox{{\mathversion{bold}${\varepsilon}$}}$ and $\mbox{{\mathversion{bold}${\mu}$}}$ being the permittivity and permeability tensors, respectively. Consider now a coordinate transformation from the original cartesian coordinates $\{x_i\}, \, (i = 1\, , 2\, , 3)$  to a new set of coordinates $\{x'_i\}, \, (i = 1\, , 2\, , 3)$. Let $\mbox{{\mathversion{bold}${\cal J}$}}$ denote the $3 \times 3$ Jacobian \linebreak transformation matrix
\begin{equation}
{\cal{J}}_{ij} = \nabla_j x'_i\, ,
\end{equation}
where $\nabla_j = \partial/\partial x_j$. It is possible to show that, in the new coordinate system,  Maxwell's equations and the constitutive relations take exactly the same form as Eqs. (\ref{MaxwellEqEinsteinNotation1}) and (\ref{ConstitutiveRelations}), with $\nabla \rightarrow \nabla'$, provided that in the primed system the fields, sources and material parameters are given by~\cite{Kundtz, yan2009}
\begin{eqnarray}
\label{TransformedFields}
{\bf E}'\!\! &=&\!\! (\mbox{{\mathversion{bold}${\cal J}$}}^{T})^{-1}{\bf E}\, , \ \ \ {\bf H}' = (\mbox{{\mathversion{bold}${\cal J}$}}^{T})^{-1}{\bf H}\, , \\
\label{TransformedSources}
\rho_e'\!\! &=&\!\! \dfrac{\rho_e}{\textrm{det}  \mbox{{\mathversion{bold}${\cal J}$}} } \, , \ \ \ {\bf J}' =  \dfrac{\mbox{{\mathversion{bold}${\cal J}$}} {\bf J}}{ \textrm{det}  \mbox{{\mathversion{bold}${\cal J}$}}} \, , \\
\label{TransformedMaterials}
\mbox{{\mathversion{bold}${\varepsilon}$}}'\!\! &=&\!\! \dfrac{\mbox{{\mathversion{bold}${\cal J}$}} \mbox{{\mathversion{bold}${\varepsilon}$}} \mbox{{\mathversion{bold}${\cal J}$}}^T}{\textrm{det} \mbox{{\mathversion{bold}${\cal J}$}}} \, , \ \ \
\mbox{{\mathversion{bold}${\mu}$}}' = \dfrac{\mbox{{\mathversion{bold}${\cal J}$}} \mbox{{\mathversion{bold}${\mu}$}} \mbox{{\mathversion{bold}${\cal J}$}}^T}{\textrm{det} \mbox{{\mathversion{bold}${\cal J}$}}} \, .
\end{eqnarray}
In case we have a sequence of transformations $\mbox{{\mathversion{bold}${\cal J}$}}_1\, , \ \mbox{{\mathversion{bold}${\cal J}$}}_2\, , \ ... \mbox{{\mathversion{bold}${\cal J}$}}_n$, the previous equations are still valid, provided we take $\mbox{{\mathversion{bold}${\cal J}$}} = \mbox{{\mathversion{bold}${\cal J}$}}_n...\mbox{{\mathversion{bold}${\cal J}$}}_2\mbox{{\mathversion{bold}${\cal J}$}}_1$~\cite{yan2009, Kundtz}. It should be noticed that even if the permittivity and permeability are isotropic in the unprimed system, after the coordinate transformation the new material parameters will typically be anisotropic.

Equations (\ref{TransformedFields}), (\ref{TransformedSources}) and (\ref{TransformedMaterials}) constitute the basis of the transformation optics method. They show that, in general, light rays follow curvilinear trajectories in the primed coordinate system and primed EM fields appear distorted if they are reinterpreted in the original cartesian system, with the distortion depicted by the transformation Jacobian. Also, when the material parameters characterized by $\mbox{{\mathversion{bold}${\varepsilon}$}}'$ and $\mbox{{\mathversion{bold}${\mu}$}}'$ are interpreted in the original coordinate system, we obtain an effective medium, called transformation medium, that mimics the effect of curved spatial coordinates \cite{schurig2006}. Despite the existence in nature of materials that bend light, they can not do that in a complete controlled way. In this sense, the fabrication of man-made nanostructured materials  ({\it e.g.} metamaterials) would complete the final step to create complex media with proper material profiles capable to reroute EM radiation.
     \vspace{10pt}
\begin{figure}[!ht]
\centering
	\includegraphics[scale=0.65]{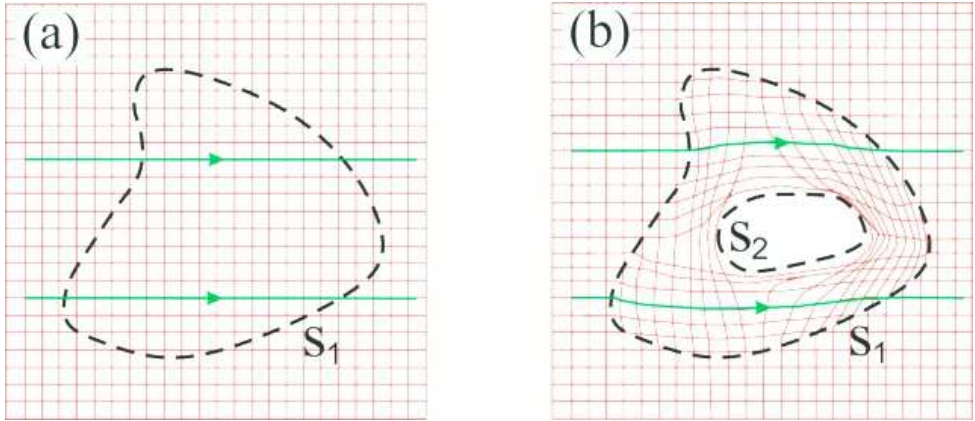}
    \vspace{10pt}
    \caption{Schematic representation of a coordinate transformation from the {\bf (a)} original cartesian system to  the {\bf (b)} new coordinate system that mimics a cloaking device of arbitrary shape. The coordinate lines are in red and  green arrows represent light rays in the original and new systems. The region bounded by $S_1$ and $S_2$ is the transformation medium whilst the voided region is within $S_2$. The materials outside $S_1$ are the same in both panels due to an identity transformation in that region.}
    \label{TransformationOptics1}
\end{figure}

The main application of TOM is in designing new devices that could present novel optical phenomena. Among these applications, EM cloaking is surely the most fascinating one. Within the TOM, the possibility to achieve invisibility cloaks occurs whenever the new set of coordinates present a voided region, as illustrated in Fig.~\ref{TransformationOptics1}. In Fig.~\ref{TransformationOptics1}(a) the gridlines of the original homogeneous, isotropic, flat space are shown in red. In this case, the light rays follow straight lines as  indicated by the green arrows. Assume now that a coordinate transformation is performed only in a finite region of space bounded by a surface $S_1$ of arbitrary shape. Consequently, the material parameters characterizing the region outside $S_1$ are the same in both panels~\ref{TransformationOptics1}(a) and \ref{TransformationOptics1}(b). Besides, suppose that the transformation preserves the shape of  $S_1$ ($\left.x'_i\right|_{S_1} = \left.x_i\right|_{S_1}$) and creates a region inside a surface $S_2$ in the coordinate system that is not crossed by any coordinate line as presented in Fig.~\ref{TransformationOptics1}(b).
%The void volume inner to $S_2$ corresponds to a point (line, in case of translational symmetry in the direction perpendicular to the page) in the original coordinates. On the other hand, despite having different material properties the regions enclosed by $S_1$ in \ref{TransformationOptics1}(a) and between $S_1$ and $S_2$ in Fig.~\ref{TransformationOptics}(b) are electromagnetically identical.
The fact that light rays follow paths parallel to the grid lines of space [green lines in Fig.~\ref{TransformationOptics1}(b)] implies that an object could be hidden in the voided region, since EM radiation will be unable to penetrate it. Moreover, note that the TOM gives not only the design of the transformation media but also determines the EM field profile inside (${\bf E}' = (\mbox{{\mathversion{bold}${\cal J}$}}^{T})^{-1}{\bf E}_0\, , \ {\bf H}' = (\mbox{{\mathversion{bold}${\cal J}$}}^{T})^{-1}{\bf H}_0$)  and outside (${\bf E}' = {\bf E}_0\, , \ {\bf H}' = {\bf H}_0$) the cloak, with $({\bf E}_0, {\bf H}_0)$ being the incident EM radiation. It is worth mentioning that, in spite of being solutions of Maxwell's equations in the primed system, it is not obvious that the primed fields completely describe the scattering of EM radiation by a TOM-based invisibility cloak. To show that Eq. (\ref{TransformedFields}) provides the full solution of the problem and no additional contribution to the fields exists, it is necessary to show that $(i)$ there is no reflection at the outer boundary $S_1$, and $(ii)$ there is no reflected light at $S_2$ and no light passes through the cloaked region. The proof that these conditions are indeed satisfied can be found in Ref.~\cite{yan2009}. 
% the interested reader the task of showing that these conditions are indeed satisfied~\cite{yan2009}.
%by showing that the tangential component of the transformed fields satisfy the appropriate boundary conditions at $S_1$ and $S_2$~\cite{yan2009}.

The TOM applied to invisibility cloaks with cylindrical geometry is largely the most studied case among all invisibility devices~\cite{pendry2006,Schurig2007, Kundtz, yan2009, leonhardt2008, chen2010, schurig2006,ergin2010,valentine2009}. A cylindrical cloak takes advantage of radial transformations where the fields within a cylinder of radius $b$ in the original coordinates are compressed into a shell with inner and outer radii $a$ and $b$, respectively. For instance, consider the following transformation:
\begin{eqnarray}
\rho\!\!\! &=&\!\!\! f(\rho')\, , \ \varphi = \varphi' \, , \ z = z'\, , \ \ \textrm{for}\ \ \rho \leq b\, , \\
\rho\!\!\! &=&\!\!\! \rho'\, , \ \varphi = \varphi' \, , \ z = z'\, , \ \ \textrm{for}\ \ \rho > b\, ,
\end{eqnarray}
with $f(a) = 0$ and $f(b) = b$. Note that the coordinates are kept unchanged for $\rho > b$. The transformed material parameters and EM fields can be obtained following appropriately the TOM prescription.
%following a track of transformations that transform from cartesian to cylindrical coordinates, where the above transformation is performed, and finally back to the original system so as to visualize the effects of the transformation.
According to Eq. (\ref{TransformedMaterials}),  the permittivity and permeability in the transformation media expressed in cylindrical coordinates are given by~\cite{schurig2006,Kundtz,yan2009}
\begin{eqnarray}
\label{TransformedPermittivityPermeability1}
\dfrac{\varepsilon'_{\rho \rho}(\rho')}{\varepsilon_0}&=& \dfrac{\mu'_{\rho \rho}(\rho')}{\mu_0}=  \dfrac{f(\rho')}{\rho' f'(\rho')}\, ,
\end{eqnarray}
\begin{eqnarray}
\label{TransformedPermittivityPermeability2}
\dfrac{\varepsilon'_{\varphi \varphi}(\rho')}{\varepsilon_0}  &=& \dfrac{\mu'_{\varphi\varphi}(\rho')}{\mu_0}=\dfrac{\rho'f'(\rho')}{f(\rho')}\, ,
\end{eqnarray}
and
\begin{eqnarray}
\label{TransformedPermittivityPermeability3}
\dfrac{\varepsilon'_{zz}(\rho')}{\varepsilon_0}  &=& \dfrac{\mu'_{zz}(\rho')}{\mu_0}  = \dfrac{f(\rho')f'(\rho')}{\rho'}\, .
\end{eqnarray}
Similarly, it follows from Eq. (\ref{TransformedFields}) that if $({\bf E}_0,{\bf H}_0)$ is the EM wave impinging on the cloak device the fields in the cloaked medium can be written as~\cite{schurig2006,Kundtz,yan2009}
\begin{eqnarray}
\!\!\!\!\!\!\!\!\!{{E'}}_{\rho} (\rho', \varphi', z')\!\!\! &=&\!\!\! f'(\rho')E_{0\rho}(f(\rho'),\varphi',z')\, ,
\ \ {{H'}}_{\rho} (\rho', \varphi', z') = f'(\rho')H_{0\rho}(f(\rho'),\varphi',z')\, ,\\
\!\!\!\!\!\!\!\!\!{{E'}}_{\varphi} (\rho', \varphi', z')\!\!\! &=&\!\!\! \dfrac{f(\rho')}{\rho'} E_{0\varphi}(f(\rho'),\varphi',z')\, ,
\ \ {{H'}}_{\varphi} (\rho', \varphi', z') = \dfrac{f(\rho')}{\rho'} H_{0\varphi}(f(\rho'),\varphi',z')\, ,\\
\!\!\!\!\!\!\!\!\!{{E'}}_{z} (\rho', \varphi', z')\!\!\! &=&\!\!\! E_{0z}(f(\rho'),\varphi',z')\, ,
\ \ \ \ \ \ \ \ \ {{H'}}_{z} (\rho', \varphi', z') = H_{0z}(f(\rho'),\varphi',z')\, .
\end{eqnarray}
%
%As mentioned before no reflection is excited at the inner boundary since the field components $E_{\theta}$, $H_{\theta}$, $D_r$ and $B_r$ are all zero and surface displacement currents make $E_z$ and $H_z$ vanish at $\rho' = a$ regardless the form of $f(\rho')$~\cite{Zhang}.

By choosing different functions $f$ we get various profiles for the EM parameters of the cloaking device. The simplest radial transformation relating the primed and unprimed coordinates is a linear function,
\begin{eqnarray}
\rho = f(\rho') = \dfrac{b}{(b-a)}(\rho'-a)\, , \ \ \ a\leq \rho' \leq b\, ,
\end{eqnarray}
for which the transformed permittivity tensor elements can be cast as~\cite{schurig2006,Kundtz,yan2009}
\begin{eqnarray}
\label{TransformedPermittivityPermeability4}
\dfrac{\varepsilon'_{\rho \rho}(\rho')}{\varepsilon_0}=   \dfrac{\rho'-a}{\rho'}\, , \ \ \
\dfrac{\varepsilon'_{\varphi \varphi}(\rho')}{\varepsilon_0}  &=& \dfrac{\rho'}{\rho'-a}\, , \ \ \
\dfrac{\varepsilon'_{zz}(\rho')}{\varepsilon_0}  =  \dfrac{b^2}{(b-a)^2} \dfrac{(\rho'-a)}{\rho'}\, ,
\end{eqnarray}
with similar expressions for $\mu'_{\rho \rho}(\rho')/\mu_0$, $\mu'_{\varphi\varphi}(\rho')/\mu_0$, and $\mu'_{zz}(\rho')/\mu_0$.

In Fig.~\ref{TransformationOptics2} we show the $E_z$ field distribution in the $xy$ plane for a monochromatic plane wave propagating in $x-$direction and polarized along the $z$-axis with wavelength $\lambda$. The left panel corresponds to free space whereas the middle one presents results for EM scattering by a gold cylinder of radius $a_c \simeq 0.2\lambda$. Note that the field patterns in Fig.~\ref{TransformationOptics2}(a) and Fig.~\ref{TransformationOptics2}(b) are quite different. This allow us to detect the presence of the gold cylinder. In Fig.~\ref{TransformationOptics2}(c) the conducting object is inside an electromagnetic cloak with $a \simeq 0.77 \lambda$, $b \simeq 1.67 \lambda $ and its material properties are given by (\ref{TransformedPermittivityPermeability4}). In this case the field distribution in the region outside the invisibility device is not disturbed neither by the inner cylinder nor by the cloak, so that the whole system is undetectable.
     \vspace{10pt}
\begin{figure}[!ht]
\centering
	\includegraphics[scale=0.78]{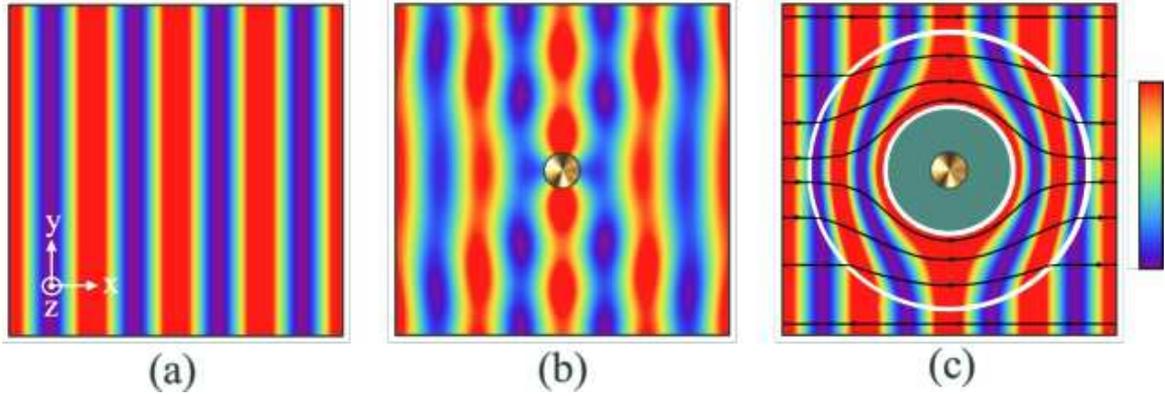}
    \vspace{10pt}
    \caption{Spatial distribution of the $E_z$ component of the electric field of a monochromatic plane wave {\bf (a)} propagating in free space, {\bf (b)} impinging on a gold cylinder of radius $a_p \simeq 0.2 \lambda$, and {\bf (c)} impinging on this same gold cylinder but now with it covered by an EM cloak of permittivity and permeability given by Eq. (\ref{TransformedPermittivityPermeability4}). The inner and outer radii of the cloak are $a \simeq 0.77 \lambda$ and $b \simeq 1.67 \lambda$, respectively. The arrows in (c) show the time average of the Poynting vector.}
    \label{TransformationOptics2}
\end{figure}

It should be noticed that the components $\varepsilon'_{\varphi\varphi}(\rho')$ and $\mu'_{\varphi\varphi}(\rho')$ of the permittivity and permeability tensors are singular at the inner surface since $f(a) = 0$. From the experimental point of view, such material parameters cannot be obtained. To circumvent this limitation, in Ref. \cite{pendry2006} simpler expressions for the permittivity and permeability tensors have been proposed. In the case of normal incidence of an electromagnetic wave with the electric field along the $z$ axis (TE wave) only $\varepsilon'_{zz}(\rho')$, $\mu'_{\rho\rho}(\rho')$ and $\mu'_{\varphi\varphi}(\rho')$ are relevant to describe the field dynamics. The authors of \cite{pendry2006} argue that rays trajectories of the EM field are completely determined by the dispersion relation and, therefore, the following equations
\begin{eqnarray}
\label{TransformedPermittivityPermeability5}
\dfrac{\varepsilon'_{zz}(\rho')}{\varepsilon_0}  =  \dfrac{b^2}{(b-a)^2}\, , \ \ \
\dfrac{\mu'_{\rho \rho}(\rho')}{\varepsilon_0}=   \dfrac{(\rho'-a)^2}{\rho'^2}\, , \ \ \textrm{and} \ \
\dfrac{\mu'_{\varphi \varphi}(\rho')}{\varepsilon_0}  &=& 1\,
\end{eqnarray}
would result in the same field dynamics than Eq. (\ref{TransformedPermittivityPermeability4}) with the cost of a small reflectance. Note that there are no more singularities in the permittivity (permeability) and  only $\mu'_{\rho\rho}(\rho')$ is inhomogeneous.
%, making easier the implementation of a device with such material properties.

The experimental verification of the invisibility cloaking device described above was performed in 2006 ~\cite{schurig2006} with a 2D apparatus for microwave frequencies ($\sim$ 8.5 GHz). The material properties described in Eq. (\ref{TransformedPermittivityPermeability5}) were engineered by means of an appropriate arrangement of split ring resonators in concentric rings, shown in Fig. \ref{TransformationOptics3}(a). In Figs. \ref{TransformationOptics3}(b) and \ref{TransformationOptics3}(c) the expected electric field distributions simulated for idealized [Eq. (\ref{TransformedPermittivityPermeability2})] and simplified [Eq. (\ref{TransformedPermittivityPermeability5})] material properties show that in the former a perfect invisibility could be obtained (nonrealistic situation) whereas in the latter a nonzero reflectance at the cloak surfaces still exists. The field profile in the region outside the cloak in Fig. \ref{TransformationOptics3}(b) corresponds to the EM radiation propagating in free space. Finally, Figs. \ref{TransformationOptics3}(d) and \ref{TransformationOptics3}(e) present the measured electric field patterns due to the scattering of TE radiation impinging perpendicularly on a bare and cloaked Copper disc, respectively. A comparison between these two plots reveals that, at least qualitatively, the microwave cloaking device partially restores the field distribution in free space. Similar results hold in the infrared and visible frequency ranges~\cite{ergin2010,valentine2009}.
     \vspace{10pt}
\begin{figure}[!ht]
\centering
	\includegraphics[scale=0.78]{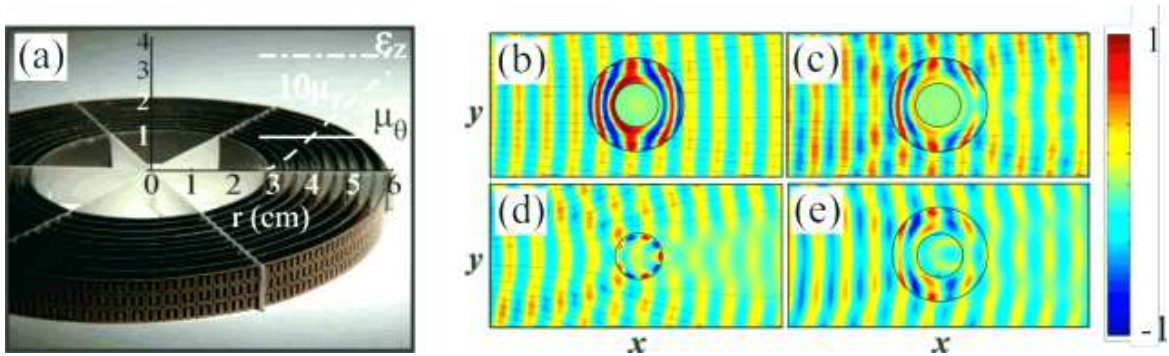}
    \vspace{10pt}
    \caption{{\bf (a)} Engineered 2D microwave cloaking device used in the experiment described in Ref. \cite{schurig2006}. The permittivity and permeabilities in Eq. (\ref{TransformedPermittivityPermeability5}) are plotted in the region of the cloak. Panels {\bf (b)} and {\bf (c)} present the simulated electric field patterns for ideal and simplified material properties, respectively. The measured electric field distributions due to scattering by a bare Copper disc {\bf (d)} and a cloaked Copper disc {\bf (e)} show qualitatively the performance of the cloaking device. Adapted from Ref.~\cite{schurig2006}.}
    \label{TransformationOptics3}
\end{figure}

It is worth mentioning that a quantitative analysis of the scattering by simplified electromagnetic cloaks obtained through TOM shows that they are inherently detectable~\cite{Yan2007,yan2009}. Indeed, the field dynamics is quite distinct for ideal and simplified cloaks,  since the methodology adopted for material simplification presupposes that $\mu'_{\varphi\varphi}$ is constant before determining the dispersion relation. As a consequence, the difference between these two cases is not only a small reflectance in the latter. Actually, it is possible to show that the zero-order cylindrical wave component of the impinging EM radiation always experiences high scattering
%(scattering coefficient $\sim 0.8$) 
due to the simplified cloaking device. In addition, the region inside to the cloak is not completely isolated from the exterior world since any object inside it will interact with the monopole component of the incident wave. Hence, scattering by the cloaked object will always occur, allowing its detection. A detailed discussion about these issues as well as alternative simplified material cloaking devices with a better performance than the one reported in Ref.~\cite{schurig2006} can be found in Refs.~\cite{Yan2007,yan2009}.

We would like to emphasize that the main advantage of invisibility cloaking devices designed through the transformation optics method lies on the fact that their operation do not depend on size, material composition, or even the geometry of the object to be cloaked. However, the efficiency of such devices is substantially affected by small perturbations~\cite{Ruan2007}, being strongly dependent on the inhomogeneous and anisotropic profiles of the used metamaterials. Moreover, the operation of TOM-based cloaks usually relies on the excitation of specific resonances of the metamaterials which in turn are functions of frequency, polarization and propagation direction of the incoming EM field~\cite{AluReview2008}. These facts make the system more sensitive to ohmic losses and changes in geometry. Finally, the fact that the impinging EM field must avoid the cloaked region precludes TOM to be applied in cloaking sensors, designed to detect the incident EM signal without disturbing their environment, and hence being detected. As we will see in the next section, the scattering cancellation technique may be used to circumvent these limitations~\cite{alu2009-2}.

\section{The scattering cancellation technique}
\label{ScatteringCancelation}

\hspace{5 mm} The scattering cancellation technique to achieve invisibility~\cite{alu2005_1,alu2005_2,alu2006,alu2007, alu2007_2,alu2008NJP,alu2008PRL, alu2010, alu2010_2, alu2008PRE,silveirinha2008,alu2009,alu2009-2,chen2011,farhat2011,filonov2012,chen2012,
monticone2013} is based on an approach entirely different from TOM. The aim of the SCT is not to make the EM fields completely vanish inside a given region of space, but rather minimize the scattered EM field generated by the dominant multipoles induced on the object to be hidden. The main idea is that an appropriate coating shell will respond to the impinging EM radiation in such way that their electric and magnetic multipoles will be out of phase with respect to those induced on the body to be cloaked. As a result, the overall scattering of the system will be reduced,  leading in this way to a low-observability of the setup by any detector. It is interesting to mention that the expression ``plasmonic cloaking" is often used as a synonym of SCT, since to cloak most objects, an invisibility device made of materials with a plasma-like dispersion relation, called plasmonic materials, is required~\cite{alu2005_1,alu2005_2,alu2006,alu2007,alu2008NJP,silveirinha2008,alu2009}. However, the reader should keep in mind that the SCT may work even without the utilization of low-permittivity coatings \cite{alu2005_2}.
     \vspace{10pt}
\begin{figure}[!ht]
\centering
	\includegraphics[scale=0.62]{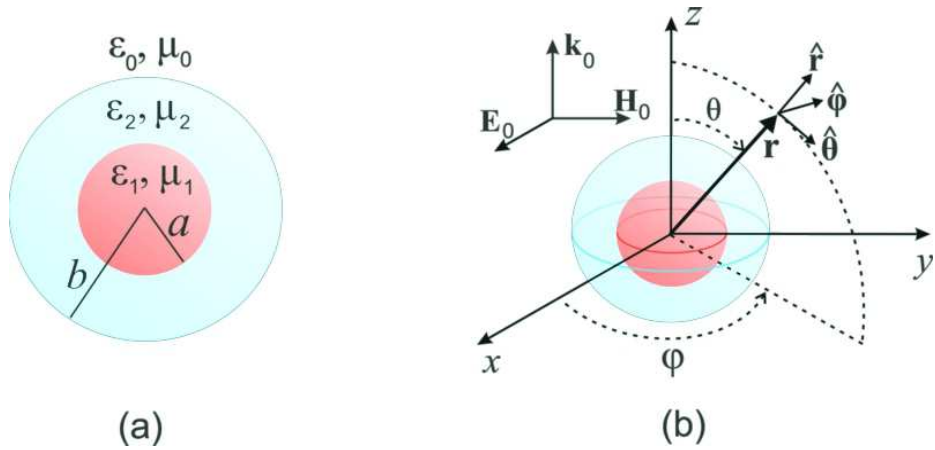}
    \vspace{10pt}
    \caption{ {\bf (a)} Cross section of the spherical object under study: both the inner sphere (radius $a$) and the covering layer (outer radius $b$) are assumed to be made of isotropic homogenous materials, with permittivities and permeabilities $\varepsilon_1(\omega)$,  $ \mu_1(\omega)$ and $\varepsilon_2(\omega)$, $\mu_2(\omega)$, respectively. {\bf (b)} Suitable spherical reference system used to calculate the EM field at position ${\bf r}$ due to scattering of a monochromatic plane wave propagating in vacuum along $z$-axis ($x$-polarized) by the particle described in (a).}
    \label{ScatteringCancelation1}
\end{figure}
%
%\vspace{-30pt}

Although the idea of using materials with low positive permittivities to obtain invisibility has already been suggested in Refs.~\cite{BohrenHuffman, Kerker82} in the Rayleigh (quasistatic) limit, the dynamical full-wave scattering was carried out only two decades later by A. Al\`u and N. Engheta for objects with spherical shape~\cite{alu2005_2}. In order to understand the physical mechanism behind the SCT and to establish some basic results to be used in Chapter {\bf \ref{cap6}} let us study plasmonic cloaking in the geometry depicted in Fig. ~\ref{ScatteringCancelation1}(a). A sphere of radius $a$ with permittivity $\varepsilon_1(\omega)$ and permeability $\mu_1(\omega)$ is covered by an isotropic and homogeneous spherical shell of outer radius $b$ with permittivity $\varepsilon_2(\omega)$ and permeability $\mu_2(\omega)$. This situation corresponds to the one discussed in Ref.~\cite{alu2005_2}. The system is illuminated by an EM monochromatic plane wave of angular frequency $\omega$ propagating in vacuum ($\varepsilon_0\, , \mu_0$) along the $z-$axis of a proper cartesian coordinate system with its electric field along the $x-$axis, as shown in Fig~\ref{ScatteringCancelation1}(b). Due to the spherical symmetry of the problem, it is suitable to expand the incident EM field into spherical harmonics as~\cite{BohrenHuffman}
%Eq: 4.47 e 4.38 do Bohren-Huffman
%
\begin{eqnarray}
\label{IncidentField1}
{\bf E}_0({\bf r}) &=& E_0 e^{ik_0z}\hat{{\bf x}} = \dfrac{E_0}{\varepsilon_0} \nabla \times {\bf F}_0^{k_z, x} ({\bf r})
- \dfrac{iE_0}{\omega \mu_0 \varepsilon_0} \nabla \times \nabla \times {\bf G}_0^{k_z, x}({\bf r})\, , \\ \vspace{10pt}
\label{IncidentField2}
{\bf H}_0({\bf r}) &=&\dfrac{E_0}{Z_0} e^{ik_0z} \hat{{\bf y}} = -\dfrac{E_0}{\mu_0} \nabla \times {\bf G}_0^{k_z, x} ({\bf r})
- \dfrac{iE_0}{\omega \mu_0 \varepsilon_0} \nabla \times \nabla \times {\bf F}_0^{k_z, x}({\bf r})  \, ,
\end{eqnarray}
where the electric ${\bf F}_0^{k_z, x} ({\bf r})$ and magnetic ${\bf G}_0^{k_z, x} ({\bf r})$ vector potentials\footnote{\label{note2} The upper indexes $k_z$ and $x$ in ${\bf F}^{k_z, x} ({\bf r})$ and ${\bf G}^{k_z, x} ({\bf r})$ indicate the directions of propagation and polarization of the impinging wave, respectively.}  can be written as a sum of TE$_r$ and TM$_r$ spherical waves, respectively ~\cite{BohrenHuffman}
\begin{eqnarray}
{\bf F}_0^{k_z, x} ({\bf r}) &=& \hat{{\bf r}} \dfrac{\sin \varphi}{\omega Z_0}\sum_{n=1}^{\infty} i^n \dfrac{2n+1}{n(n+1)}k_0rj_n(k_0r) P_n^1(\cos \theta)\, , \\ \vspace{10pt}
{\bf G}_0^{k_z, x} ({\bf r}) &=& \hat{{\bf r}} \dfrac{\cos \varphi}{\omega}\sum_{n=1}^{\infty} i^n \dfrac{2n+1}{n(n+1)}k_0rj_n(k_0r) P_n^1(\cos \theta)\, .
\end{eqnarray}
In the previous equations the coordinates $(r\, , \theta\, , \varphi)$ are defined in Fig.~\ref{ScatteringCancelation1}(b), $j_n(x)$ are the spherical Bessel functions~\cite{Abramowitz,Arfken,BohrenHuffman}, $P_n^1(x)$ are the associate Legendre Polynomials\footnote{\label{note3}In Ref.~\cite{Abramowitz} the associate Legendre Polynomials are defined as $P_n^m(x) = (-1)^m(1-x^2)^{m/2}d^mP_n(x)/dx^m$ whereas in Refs.~\cite{Arfken,BohrenHuffman} the factor $(-1)^m$ is missing. We use throughout the thesis the definition given by Bohren and Huffman~\cite{BohrenHuffman}.} of first degree and order $n$~\cite{Abramowitz,Arfken,BohrenHuffman}, $k_0 =\omega\sqrt{\mu_0\varepsilon_0} = \omega/c$ is the wave number of the incident plane wave, and $Z_0 = \sqrt{\mu_0/\varepsilon_0}$ is the vacuum impedance.

The scattered EM field can be similarly expanded in a superposition of vector \linebreak harmonics as~\cite{BohrenHuffman}
\begin{eqnarray}
\label{scatteredfield1}
{\bf E}_{\textrm{s}}({\bf r}) &=& -\dfrac{E_0}{\varepsilon_0} \nabla \times {\bf F}_{\textrm{s}}^{k_z, x} ({\bf r})
+ \dfrac{iE_0}{\omega \mu_0 \varepsilon_0} \nabla \times \nabla \times {\bf G}_{\textrm{s}}^{k_z, x}({\bf r})\, , \\ \vspace{10pt}
\label{scatteredfield2}
{\bf H}_{\textrm{s}}({\bf r}) &=&\dfrac{E_0}{\mu_0} \nabla \times {\bf G}_{\textrm{s}}^{k_z, x} ({\bf r})
+ \dfrac{iE_0}{\omega \mu_0 \varepsilon_0} \nabla \times \nabla \times {\bf F}_{\textrm{s}}^{k_z, x}({\bf r})  \, ,
\end{eqnarray}
with ${\bf F}_{\textrm{s}}^{k_z, x} ({\bf r})$ and ${\bf G}_{\textrm{s}}^{k_z, x} ({\bf r})$ given by
\begin{eqnarray}
{\bf F}_{\textrm{s}}^{k_z, x} ({\bf r}) &=& \hat{{\bf r}} \dfrac{\sin \varphi}{\omega Z_0}\sum_{n=1}^{\infty} i^n \dfrac{2n+1}{n(n+1)}c_n^{\textrm{TE}}k_0r h_n^{(1)}(k_0r) P_n^1(\cos \theta)\label{VectorPotentialFScattered}\, , \\ \vspace{10pt}
{\bf G}_{\textrm{s}}^{k_z, x} ({\bf r}) &=& \hat{{\bf r}} \dfrac{\cos \varphi}{\omega}\sum_{n=1}^{\infty} i^n \dfrac{2n+1}{n(n+1)}c_n^{\textrm{TM}}k_0r h_n^{(1)}(k_0r) P_n^1(\cos \theta)\label{VectorPotentialGScattered}\, ,
\end{eqnarray}
where $h_n^{(1)}(x) = j_n(x) + iy_n(x)$ are the spherical Hankel functions of the first kind~\cite{Abramowitz,Arfken,BohrenHuffman} and $c_n^{\textrm{TE}}$, and $c_n^{\textrm{TM}}$ are the Mie scattering coefficients\cite{alu2005_1,alu2005_2,BohrenHuffman,alu2006}.  It is interesting to mention that ${\bf F}_{\textrm{s}}^{k_z\, , x}({\bf r})$ and ${\bf G}_{\textrm{s}}^{k_z\, , x}({\bf r})$ are the potentials for the EM field due to a superposition of magnetic and electric multipoles, respectively. Due to the linearity of Maxwell's equations and orthogonality of the spherical harmonic waves, we can separately solve the problem for each multipole term in previous equations, with $c_n^{\textrm{TE}}$ and $c_n^{\textrm{TM}}$ fully determined by imposing the continuity of the tangential components of ${\bf E}$ and ${\bf H}$ at $r = a$ and $r = b$.  After some lengthy but straightforward algebra, it is possible to show that the analytical expressions for the Mie coefficients can be cast into the form \cite{alu2005_1,alu2005_2,alu2006}
\begin{eqnarray}
\label{miecoefficients}
c_n^{\textrm{TE}} = -\dfrac{U_n^{\textrm{TE}}}{U_n^{\textrm{TE}}+iV_n^{\textrm{TE}}}\, , \ \ \ \textrm{and} \ \ \ c_n^{\textrm{TM}} = - \dfrac{U_n^{\textrm{TM}}}{U_n^{\textrm{TM}}+iV_n^{\textrm{TM}}}\, .
\end{eqnarray}
Functions $U_n^{\textrm{TE, TM}}$ and $V_n^{\textrm{TE, TM}}$ can be written as
\begin{eqnarray}
\label{functionU}
U_n^{\textrm{TM}} =
\begin{vmatrix}
  j_n(k_1a) & j_n(k_2a)  & y_n(k_2a) & 0\\
 [k_1aj_n(k_1a)]'/\varepsilon_1 & [k_2aj_n(k_2a)]'/\varepsilon_2  & [k_2ay_n(k_2a)]'/\varepsilon_2  &  0\\
  0 & j_n(k_2b)  & y_n(k_2b) & j_n(k_0b) \\
  0&  [k_2bj_n(k_2b)]'/\varepsilon_2    & [k_2by_n(k_2b)]'/\varepsilon_2   & [k_0bj_n(k_0b)]'/\varepsilon_0
 \end{vmatrix}\, , \\ \vspace{100pt}
 \label{functionV}
 V_n^{\textrm{TM}} =
\begin{vmatrix}
  j_n(k_1a) & j_n(k_2a)  & y_n(k_2a) & 0\\
 [k_1aj_n(k_1a)]'/\varepsilon_1 & [k_2aj_n(k_2a)]'/\varepsilon_2  & [k_2ay_n(k_2a)]'/\varepsilon_2  &  0\\
  0 & j_n(k_2b)  & y_n(k_2b) & y_n(k_0b) \\
  0&  [k_2bj_n(k_2b)]'/\varepsilon_2    & [k_2by_n(k_2b)]'/\varepsilon_2   & [k_0by_n(k_0b)]'/\varepsilon_0
 \end{vmatrix}\, ,
\end{eqnarray}
where $k_1 = \omega\sqrt{\mu_1\varepsilon_1}$ and $k_2 = \omega\sqrt{\mu_2\varepsilon_2}$ are the wave numbers inside the sphere and shell, respectively, and $[...]'$ means derivative with respect to the argument. Owing to electromagnetic duality analogous expressions for $U_n^{\textrm{TE}}$ and $V_n^{\textrm{TE}}$ can be directly obtained from Eqs. (\ref{functionU}) and (\ref{functionV}) by exchanging $\varepsilon_{0,1,2} \leftrightarrow \mu_{0,1,2}$.

From the knowledge of $c_n^{\textrm{TE}}$ and $c_n^{\textrm{TM}}$ the scattering, absorption and extinction cross sections can be explicitly calculated. Indeed,  for $k_0r\gg1$  the asymptotic behavior of the spherical Hankel functions of the first kind $h_n^{(1)}(k_0r) \sim (-i)^n e^{ik_0r}/(ik_0r)$ ~\cite{Abramowitz, Arfken}, allows us to write Eqs. (\ref{scatteredfield1}) and (\ref{scatteredfield2}) in the form of  (\ref{RadiativeFields}) with \cite{BohrenHuffman}
\begin{eqnarray}
{\bf f}_x({\bf k},{\bf k}_0) = \sum_{n=1}^{\infty} \dfrac{2n+1}{n(n+1)}&&\!\!\!\!\!\!\!\!\!\!\!\! \{ \cos \varphi \left[c_n^{\textrm{TE}}\pi_n(\cos \theta)+c_n^{\textrm{TM}} \tau_n(\cos \theta)\right]\hat{\mbox{{\mathversion{bold}${\theta}$}}} \!\!\!\!\!\!\!\!\! \cr &-&\!\!\! \sin \varphi \left[c_n^{\textrm{TE}}\tau_n(\cos \theta)+c_n^{\textrm{TM}} \pi_n(\cos \theta)\right]\hat{{\mbox{{\mathversion{bold}${\varphi}$}}}} \}\, ,
\end{eqnarray}
where $\pi_n(\cos\theta) =  P_n^1(\cos\theta)/\sin\theta$ and $\tau_n(\cos\theta) = dP_n^1(\cos\theta)/d\theta$. If we use now Eqs. (\ref{ScatteringCrossSectionDefinition}) and (\ref{ExtinctionCrossSectionDefinition}), the orthonormality of the associate Legendre polynomials and the fact that $\pi_n(1) = \tau_n(1) = n(n+1)/2$ we can show that the scattering and extinction cross sections are given in terms of $c_n^{\textrm{TE}}$ and $c_n^{\textrm{TM}}$ respectively by~\cite{BohrenHuffman}
\begin{eqnarray}
C_{\textrm{sca}} &=& \dfrac{2\pi}{k_0^2}\sum_{n=1}^{\infty} (2n+1)\left(\left|c_n^{\textrm{TE}}\right|^2+\left|c_n^{\textrm{TM}}\right|^2\right)\, ,\\ \vspace{10pt}
C_{\textrm{ext}} &=&  \dfrac{2\pi}{k_0^2}\sum_{n=1}^{\infty} (2n+1)\textrm{Re}(c_n^{\textrm{TE}}+c_n^{\textrm{TM}})\, .
\end{eqnarray}

Generally speaking the dominant contributions to the scattering and extinction cross sections in the above expressions come from multipoles up to $n  \sim k_0a$~\cite{BohrenHuffman}. Consequently larger objects have wider cross sections since the number of terms that effectively contribute to $C_{\textrm{sca}}$ and $C_{\textrm{ext}}$ increases with the size of the target.  However, a suitable choice of the material properties of the cloaking shell may cancel the contributions of the dominant multipoles to the EM scattering, thus reducing the body detectability even if the overall diameter of the system is enlarged. To clearly show this statement let us focus on the dipole approximation, {\it i.e.}  $k_0b\, , k_1b\, , k_2b \ll1$, where the main contribution to the scattering cross section is due to $c_1^{\textrm{TM}}$. In this regime the spherical Bessel and Neumann functions can be expanded for small arguments and, by imposing $U_1^{\textrm{TM}}=0$, the condition that must be satisfied so that invisibility is achieved is given by~\cite{alu2005_1,alu2005_2,alu2007}
\begin{equation}
\label{invisibilitycondition_alu}
\dfrac{a}{b}={\eta_{\,}}_{\textrm{TM}} = \sqrt[3]{\dfrac{(\varepsilon_2-\varepsilon_0)(2\varepsilon_2+\varepsilon_1)}{(\varepsilon_2-\varepsilon_1)(2\varepsilon_2-\varepsilon_0)}}\, .
\end{equation}
Clearly,  ${\eta_{\,}}_{\textrm{TM}}$ must be a real function with values in the interval $0 < {\eta_{\,}}_{\textrm{TM}} < 1$.  Analogous equations for cancelling the contribution of the $n-$th electric or magnetic multipole to the EM scattered fields can be obtained by enforcing $U_n^{\textrm{TE, TM}} = 0$ \cite{alu2005_2}.

The conditions for achieving the cancellation of $c_1^{\textrm{TM}}$, obtained from Eq. (\ref{invisibilitycondition_alu}), are shown in Fig.~\ref{ScatteringCancelation2} as a function of the permittivities of the inner sphere and outer cloaking shell for lossless materials. Blue regions correspond to situations where $c_1^{\textrm{TM}}$ exactly vanishes, with bright colors corresponding to higher values of ratio $a/b$.  White regions correspond to the cases where invisibility is never achieved in the long wavelength limit.  Note that it is always possible to find a cloak capable to make a given dielectric sphere invisible under the considered approximations.
     \vspace{10pt}
\begin{figure}[!ht]
\centering
	\includegraphics[scale=0.5]{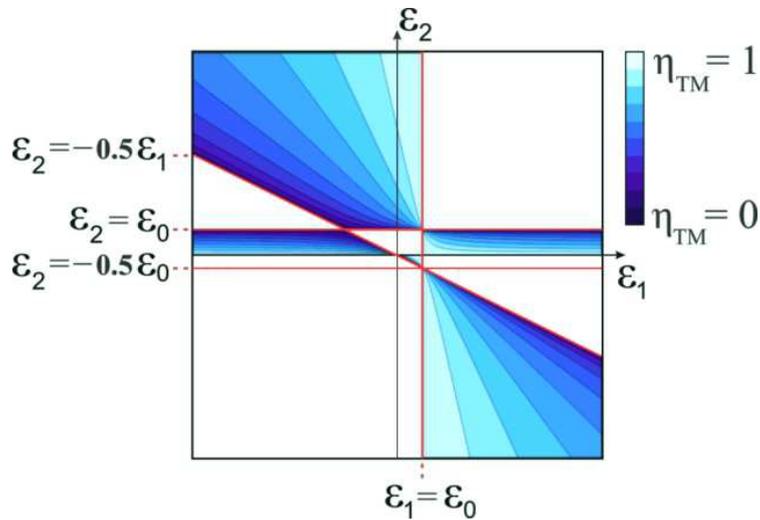}
    \vspace{10pt}
    \caption{Regions for which the invisibility condition in the dipole approximation is satisfied for a given value of ${\eta_{\,}}_{\textrm{TM}} = a/b$ between $0$ and $1$.  White regions correspond to unphysical results.}
    \label{ScatteringCancelation2}
\end{figure}

In order to understand the physical mechanism behind the SCT,  note that the allowed values of $\varepsilon_1$ and $\varepsilon_2$ for which the invisibility condition written in Eq. (\ref{invisibilitycondition_alu}) is satisfied, always predict opposite induced dipoles on the object and on the coating shell. For instance, suppose $\varepsilon_1 > \varepsilon_0 > \varepsilon_2$. In this case, the polarization vectors ${\bf P}_1 \propto (\varepsilon_1-\varepsilon_0) {\bf E}^{(1)}_{\textrm{local}}$ and ${\bf P}_2 \propto (\varepsilon_2-\varepsilon_0){\bf E}^{(2)}_{\textrm{local}}$ within the sphere and the cover, respectively, are out of phase as shown in Fig.~\ref{ScatteringCancelation3}. Therefore, in spite of being individually detectable, the spherical body and its cloak may have together a total electric dipole that vanishes provided an appropriate choice of their volumes is made. The correct relation between these volumes, that leads to a total destructive interference between the scattered fields of the sphere and the cloak, is precisely given by Eq. (\ref{invisibilitycondition_alu}).
     \vspace{10pt}
\begin{figure}[!ht]
\centering
	\includegraphics[scale=0.7]{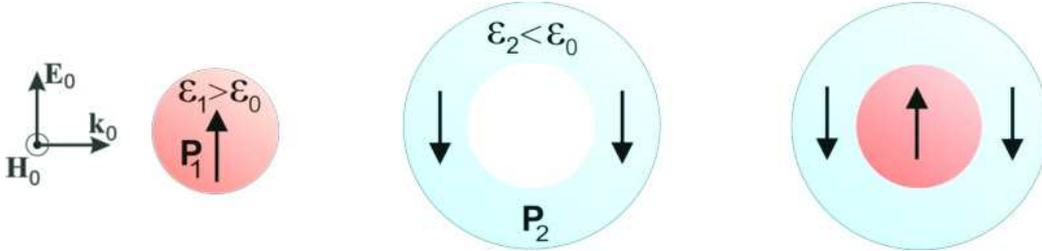}
    \vspace{10pt}
    \caption{Physical interpretation, in the dipole approximation, of the scattering cancellation technique mechanism to achieve invisibility: the incident radiation induces a local polarization on the shell which is out of phase with respect to that induced on the cloaked sphere. As a result, the total electric dipole of the system may vanish for a proper choice of the inner and outer radii.}
\label{ScatteringCancelation3}
\end{figure}

It is worth mentioning that the SCT is not well suited for making invisible objects with typical sizes of the order or larger than the radiation's vacuum wavelength, because the number of relevant multipoles to the EM scattering increases quickly with the size of the body.  For large targets it is not possible to achieve a strong reduction of $C_{\textrm{sca}}$ by using only one coating shell. Nevertheless, for dielectric spheres with radius $a\, \lesssim \lambda/5$, the scattering cross section can be considerably attenuated with the use of a single layer~\cite{alu2005_2,AluReview2008,chen2012,AluReview2014}. This can be seen in Fig.~\ref{ScatteringCancelation4} where we plot $C_{\textrm{sca}}$ for a coated sphere with $a = \lambda/5$, normalized by the scattering cross section of a bare sphere $C_{\textrm{sca}}^{(0)}$, as a function of $b/a$. Panel \ref{ScatteringCancelation4}(a) shows the results for $\varepsilon_1 = -6\varepsilon_0$, $\mu_1 = \mu_0$ and $\varepsilon_2 = 10\varepsilon_0$, $\mu_2 = \mu_0$. Note that $C_{\textrm{sca}}$ is reduced to $\sim30\%$ of $C_{\textrm{sca}}^{(0)}$ for $b \simeq 1.16 a$.  This value of $b/a$ does not correspond to the minimum of $c_1^{\textrm{TM}}$ since the dipole approximation is not valid in this case and higher order multipoles are also greatly affected by the shell radius, as can be inferred from Fig.~\ref{ScatteringCancelation4}(b). Notice for instance that two peaks in $C_{\textrm{sca}}$ occur at $b \simeq 1.06 a$ and $b \simeq 1.08 a$ which are due to the excitation of electric quadrupole and electric octopole resonances, respectively.  
The existence of such peaks near the invisibility region obviously limits the efficiency of the plasmonic cloaking, as they make the operation frequency range of the device narrower. Panels \ref{ScatteringCancelation4}(c) and \ref{ScatteringCancelation4}(d) present $C_{\textrm{sca}}/C_{\textrm{sca}}^{(0)}$ and Mie coefficients, respectively, as a function of $b/a$ for $\varepsilon_1 = 10\varepsilon_0$, $\mu_1 = \mu_0$, $\varepsilon_2 = 0.1\varepsilon_0$, and $\mu_2 = \mu_0$. In this situation the scattering cross section reaches $\sim 20 \%$ of  $C_{\textrm{sca}}^{(0)}$ for $b \simeq 1.12 a$.  The ratio $b/a \simeq 1.12$ coincides approximately with the position of the minimum of $c_1^{\textrm{TM}}$ inasmuch as the other Mie coefficients are barely altered by the cover. Actually, as $c_1^{\textrm{TE}}$ is almost constant in the region presented in the plot, the residual scattering cross section at the minimum in Fig. \ref{ScatteringCancelation4}(c) is essentially due to the magnetic dipole. However, $c_1^{\textrm{TE}}$ can also be cancelled at $b \simeq 1.12 a$ by a proper choice of the permeability of the shell. Indeed, panels \ref{ScatteringCancelation4}(e) and \ref{ScatteringCancelation4}(f) show similar plots to \ref{ScatteringCancelation4}(c) and \ref{ScatteringCancelation4}(d), respectively, but with $\mu_2 = 0.025\mu_0$. For these material parameters both contributions from electric and magnetic dipoles vanish at the same value of $b/a$ [see Fig. \ref{ScatteringCancelation4}(f)] and an attenuation of $C_{\textrm{sca}}$ larger than $95\%$ may be achieved.
     \vspace{10pt}
\begin{figure}[!ht]
\centering
	\includegraphics[scale=0.75]{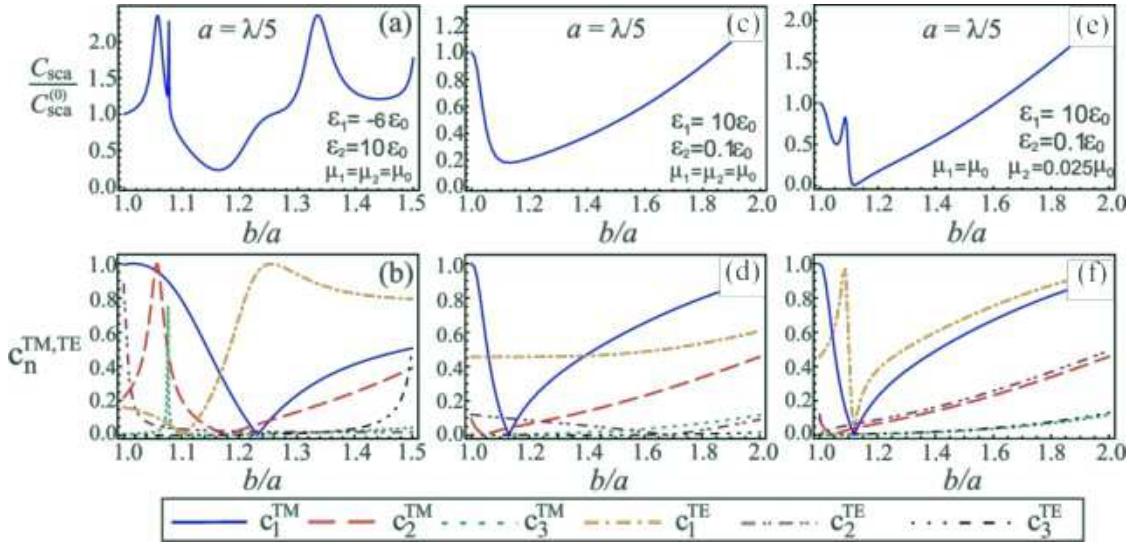}
    \vspace{10pt}
    \caption{{\bf (a)}  Scattering cross section $C_{\textrm{sca}}$ of a coated sphere (normalized by the scattering cross section $C_{\textrm{sca}}^{(0)}$ of a bare sphere) with $\varepsilon_1 = -6\varepsilon_0$, $\mu_1 = \mu_0$, $\varepsilon_2 = 10\varepsilon_0$, and $\mu_2 = \mu_0$ as a function of the ratio $b/a$ for $a = \lambda/5$. {\bf (b)} Scattering coefficients $c_n^{\textrm{TM}}$, $c_n^{\textrm{TE}}$ of the dominant electric and magnetic multipoles as functions of $b/a$ for the same material parameters used in the first panel. Panels {\bf (c)} and {\bf (d)} show the same plots as those appearing in panels (a) and (b), respectively, but for  $\varepsilon_1 = 10\varepsilon_0$, $\mu_1 = \mu_0$, $\varepsilon_2 = 0.1\varepsilon_0$, and $\mu_2 = \mu_0$. Panels {\bf (e)} and {\bf (f)} are identical to panels (c)  and (d), respectively, except from the fact that now $\mu_2 = 0.025 \mu_0$.}
    \label{ScatteringCancelation4}
\end{figure}
     \vspace{10pt}
\begin{figure}[!ht]
\centering
	\includegraphics[scale=0.7]{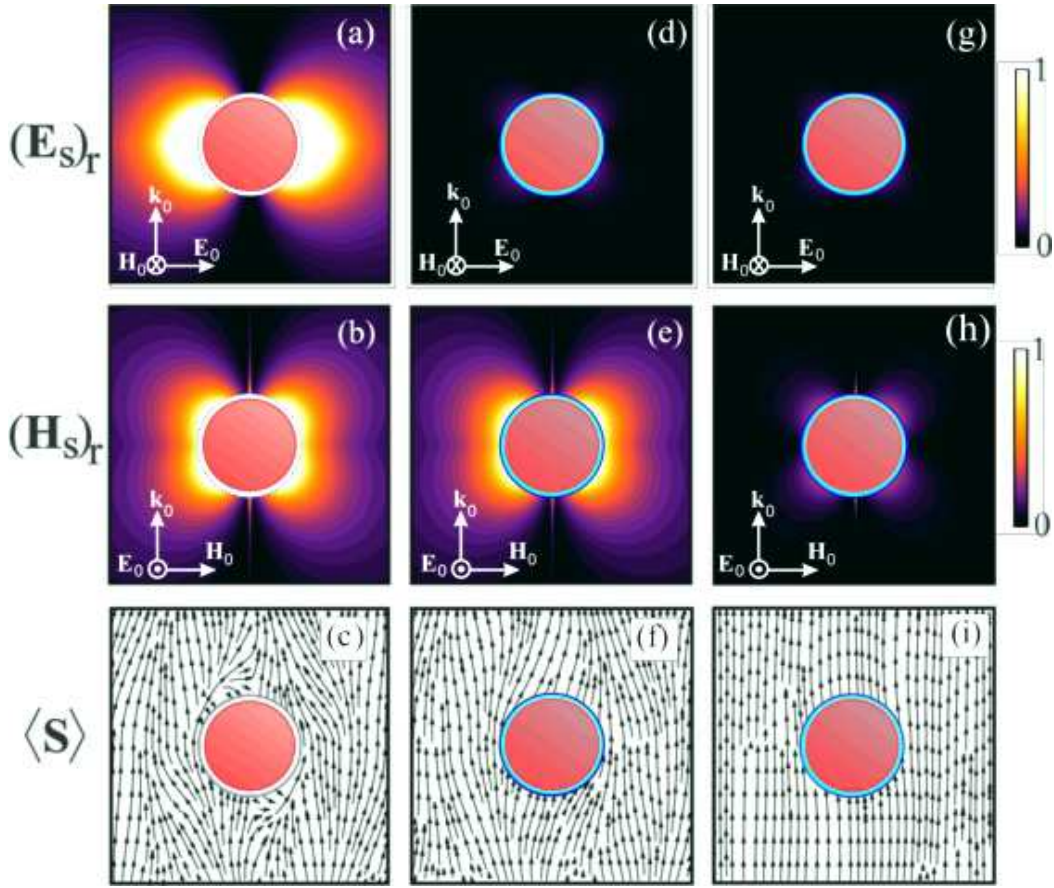}
    \vspace{10pt}
    \caption{{\bf Above:} Spatial distribution of the radial component $\left({\bf E}_{\textrm{s}}\right)_r$ of the scattered electric field in $xz$ plane due to scattering of a monochromatic plane wave propagating along $z$-axis ($x$-polarized) by {\bf (a)} a bare sphere of radius $a = \lambda/5$ and $\varepsilon_1 = 10\varepsilon_0$, $\mu_1 = \mu_0$; {\bf (d)} the same sphere but covered with a shell with $\varepsilon_2 = 0.1\varepsilon_0$, $\mu_2 = \mu_0$ and $b = 1.12 a$; and {\bf (g)} the same as (d) but with $\mu_2 = 0.025\mu_0$. {\bf Middle:} Spatial distribution of the radial component $\left({\bf H}_{\textrm{s}}\right)_r$ of the scattered magnetic field in $yz$ plane are shown in {\bf (b)}, {\bf (e)}, and {\bf (h)} for the same geometrical and material parameters as (a), (d), and (g), respectively. {\bf Bellow:} Panels {\bf (c)}, {\bf (f)}, and {\bf (i)} show the time-averaged Poynting vector in $xz$ plane for the same material parameters as in (a), (d), and (g), respectively.}
    \label{ScatteringCancelation5}
\end{figure}

In Fig.~\ref{ScatteringCancelation5} we can see the effects of the cancellation of the contributions of the electric and magnetic dipoles terms in the scattering pattern.  In the first column, panels (a) and (b) show  contour plots for the amplitude of the radial component of the electric and magnetic scattered fields in $xz$ and $yz$-planes, respectively, for the case of a bare sphere with $\varepsilon_1 = 10\varepsilon_0$, $\mu_1 = \mu_0$ [same values as those used in Figs.~\ref{ScatteringCancelation4}(c)-\ref{ScatteringCancelation4}(f)]. The spatial region that would be occupied by the plasmonic cloaking is indicated by the blue dashed line. As it is evident, the impinging EM wave induces strong electric and magnetic dipoles within the sphere. Consequently, the scattered fields have a dipole-like pattern.  Also, panel  \ref{ScatteringCancelation5}(c)  presents the time averaged Poynting vetor in the $xy$-plane, which is quite different from the one of a plane wave propagating in free space. In the second column, panels (d), (e) and (f) show the results for the same sphere, but covered with a shell with $b = 1.12a$,   $\varepsilon_2 = 0.1\varepsilon_0$, and $\mu_2 = \mu_0$ . This material and the geometric parameters precisely correspond to the situation where $c_1^{\textrm{TM}}$ vanishes, as previously seen in panels~\ref{ScatteringCancelation4}(c) and ~\ref{ScatteringCancelation4}(d). Figure \ref{ScatteringCancelation5}(d) reveals that once the electric dipole contribution is cancelled the remaining scattered electric field in ${\bf E}_0$-plane is mainly  due to the electric quadrupole term. Figure \ref{ScatteringCancelation5}(e) confirms the result exhibited in  ~\ref{ScatteringCancelation4}(d) that $c_1^{\textrm{TE}}$ is not affected by the low permittivity of the shell.  Figure \ref{ScatteringCancelation5}(f) shows that cancelling the electric dipole contribution to the scattering cross section may partially restore the free space energy flow pattern of a plane wave. The third column presents the results for the same material parameters as those used in Figs. ~\ref{ScatteringCancelation4}(e) and ~\ref{ScatteringCancelation4}(f). In other words, in this case both electric and magnetic dipole contributions to the scattered field vanish and the field profiles in ${\bf E}_0$-plane and ${\bf H}_0$-plane are dominated by the electric and magnetic quadrupoles, as shown in Figs. ~\ref{ScatteringCancelation5}(g) and ~\ref{ScatteringCancelation4}(h). Finally, Fig. ~\ref{ScatteringCancelation4}(i) highlights the fact that for this set of material and geometric parameters the flow of energy in free space remains practically unperturbed by the presence of the cloaked sphere.

It should be mentioned that the SCT has already been applied to more complex situations and  geometries. As selected examples (but by no means exhausting), we highlight studies including material losses and geometry imperfections~\cite{alu2007}, plasmonic cloaking of a collection of objects\cite{alu2007_2}, the use of several layers for achieving multifrequency invisibility as well as for cloaking larger objects \cite{alu2008NJP,alu2008PRL}, effects of dispersion \cite{alu2008PRE} and irregularly shaped anisotropic bodies \cite{alu2010}. In addition, for cylindric targets it was shown that plasmonic cloaking is robust against finite size effects even under oblique illumination\cite{alu2010_2}.
     \vspace{10pt}
\begin{figure}[!ht]
\centering
	\includegraphics[scale=0.7]{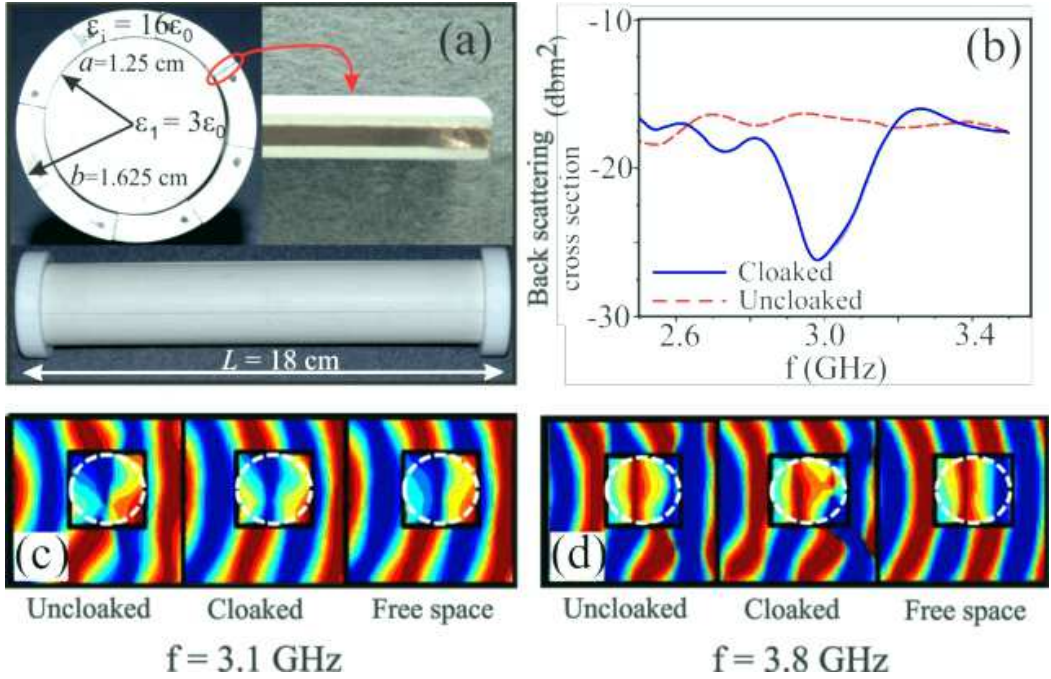}
    \vspace{10pt}
    \caption{{\bf (a)} (top left) Cross sectional view of a cylinder with $a = 1.25$ cm and $\varepsilon_1 = 3\varepsilon_0$, $\mu_1 = \mu_0$ covered by a shell with $b = 1.3a = 1.625$ cm made of $8$ copper strips (top right) embedded in a host medium with $\varepsilon_i = 16\varepsilon_0$, $\mu_i = \mu_0$; (bottom) picture of the full system showing its finite size. {\bf (b)} Measured back scattering cross section for normal incidence of a monochromatic plane wave as a function of frequency for the uncloaked (red dashed line) and cloaked (blue solid line) dielectric cylinder shown in (a). Measured near-field distribution of the electric field  is presented in {\bf (c)} and {\bf (d)} for $3.1$ GHz and $ 3.8$ GHz, respectively. Adapted from Ref.~\cite{rainwater2012}.}
    \label{ScatteringCancelation6}
\end{figure}

The experiments based on the SCT were performed in the cylindrical geometry in 2009 \cite{edwards2009} and in 2012 \cite{rainwater2012}, in the microwave range. Particularly, in the latter case it was reported the first 3D demonstration of an invisibility cloaking using metamaterials in free space. The authors showed that a dielectric cylinder with $a = 1.25$ cm, length \linebreak $L = 18$ cm, $\varepsilon_1 = 3\varepsilon_0$, and $\mu_1 = \mu_0$ can be camouflaged at microwave frequencies \linebreak ($\sim 3.0$ GHz) with an efficiency $\sim 70\%$ by using a coating shell made of several copper strips embedded in a host medium with $\varepsilon_i = 16\varepsilon_0$ and $\mu_i = \mu_0$, as shown in Fig. \ref{ScatteringCancelation6}(a). In this case it is possible to show that the effective permittivity of the layer is $\varepsilon_2 \simeq \varepsilon_i - N^2/(4k_0^2ab)$ where $N$ is the number of used strips (8 in the experiment). Note that $\varepsilon_2 < 0$ at the designed frequency, which allows a partial cancellation of the multipoles induced on the dielectric cylinder. Figure \ref{ScatteringCancelation6}(b) presents the measured back scattering cross section for normal incidence as a function of the frequency of the impinging EM wave. Note that a strong attenuation occurs around $3$ GHz.  Figures \ref{ScatteringCancelation6}(c) and \ref{ScatteringCancelation6}(d) show the measured near-field spatial distribution of the electric field in the uncloaked system, cloaked cylinder and free space at $3.1$ GHz  and $3.8$ GHz, respectively. Note that in the former the coating shell partially restores the field distribution in free space. In the latter, the invisibility device scatters more radiation than the bare cylinder since the frequency of the incident wave does not match the operation band of the  cloaking.

We would like to remark that the main advantage of the SCT relies on the fact that the performance of the designed devices does not depend on the excitation of material resonances in the system \cite{AluReview2008}. Therefore, such kind of cloaks usually work in a broader frequency band than invisibility cloaks based on TOM. Besides, in the SCT the necessary materials to construct the coating shell are isotropic and homogeneous and the effectiveness of the device is hardly affected by the polarization and propagation direction of the incident wave. In addition, as the impinging EM field penetrates the cloaked object without producing a substantial scattering, the SCT is the ideal method for applications in communication networks, particularly in cloaking sensors and antennas \cite{alu2009-2}. The idea that plasmonic cloaks are fabricated to hide only objects with a particular shape and do not work for other targets is not necessarily true. It is always possible to cloak an arbitrarily shaped body by covering it with a metallic surface that isolates its interior (where the cloaked object is located) and, then, use a plasmonic shell to cancel the scattered field by the conducting material. The disadvantages of the SCT are mainly related to its intrinsic limitation for camouflaging large objects, since the number of relevant multipoles increases with the size of the object to be cloaked.  Moreover, the difficulty in cancelling all the multipoles induced on the system always leads to a residual scattering, preventing perfect invisibility. However, this is not necessarily a problem, all depends on the efficiency of the detectors used to measure the scattered radiation.

\vspace{30pt}

In this chapter we have discussed the physical mechanisms underlying invisibility cloaks, with emphasis in devices based on the transformation optics method and scattering cancellation technique. While in the former the form-invariance of Maxwell's equations under coordinate transformations can be used to properly fabricate anisotropic and inhomogeneous metamaterials to bend light around a region of space, in the latter the EM scattering is almost suppressed by using plasmonic coatings capable to cancel the dominant multipoles induced in the cloaked object. Experiments on both techniques have already been performed in recent years and they were briefly commented in the text. Again, we would like to remark that other invisibility techniques do exist and the interested reader should refer to Refs. ~\cite{nicorovici1994, milton2006, milton2007, alu2009mantle,soric2013,Tretyakov} and references therein. Finally, although not considered in this chapter we should mention that invisibility devices have been developed very quickly for other types of waves. Among them, we have Refs.~\cite{acustico1,acustico2,acustico3,acustico4,acustico5,acustico6,acustico7,acustico8,acustico9,acustico10,acustico11,acustico12} about acoustic cloaking, Refs.~\cite{termico1,termico2,termico3,termico4,termico5,termico6,termico7,termico8,termico9,termico10} on thermal cloaking, and Refs.~\cite{materia1,materia2,materia3,materia4,materia5,materia6,materia7} on quantum matter-waves cloaking.
\end{chapter}

\begin{chapter}{Molding the flow of light with a tunable magneto-optical device}
\label{cap4}

\begin{flushright}
{\it
``This invisibility, however, \\means that the opportunities for creative research are infinite."
}

{\sc B. Smith}
\end{flushright}

\hspace{5 mm}  Despite the notable progress in all cloaking techniques, the development of a tunable cloaking device remains a challenge. In this chapter  we propose a mechanism to actively tune the operation of plasmonic cloaks with an external magnetic field. In our system electromagnetic waves are scattered by a cylinder coated with a magneto-optical (MO) shell. In the long wavelength regime we show that the application of an external magnetic field may not only modify the operation wavelength and switch on and off the cloaking mechanism but also mitigate losses, once the absorption cross section experiences a sharp decrease precisely in the cloaking operation frequency band.  We also show that the angular distribution of the scattered radiation can be effectively controlled by applying an external magnetic field, allowing for a swift change in the scattering pattern. By demonstrating that these results are feasible with realistic, existing magneto-optical materials, we suggest that magnetic fields could be used as an effective, versatile external agent to tune plasmonic cloaks and to dynamically control electromagnetic scattering in an unprecedented way.
\section{Introduction}
\label{Intro4}

\hspace{5 mm} The recent progress in micro- and nanofabrication have fostered the development of artificial metamaterials for designing EM cloaks. As discussed in the previous chapter, among all approaches to achieve invisibility, the transformation optics method and the scattering cancellation technique are the most promising ones. However, despite the success of the existing cloaking techniques, they generally suffer from practical physical limitations, namely, the detrimental effect of losses and the limited operation frequency bandwidth. In the particular case of the SCT the operation bandwidth relies on the plasmonic properties of the shell. As a result, once the cover is designed and constructed, the cloaking mechanism works only around a relatively narrow frequency band that cannot be freely modified after fabrication. Therefore, in case there is a need to modify the operation frequency band it is usually necessary to engineer a new device, with different geometric parameters and materials, limiting its applicability~\cite{AluReview2008, AluReview2014}.  Since practical applications often require more flexibility in the design and in the operational bandwidth of invisibility devices, proposals of tunable cloaks have been developed~\cite{peining2010_1,peining2010_2,peining2010_3,zharova2012,milton2009, milton2009-2,farhat2013,kortkamp2013-2,kortkampPRL,kortkampJOSAA,Schofield_2014_1,Rybin}. One route to create tunable cloaking devices involves the use of a graphene shell~\cite{farhat2013}.  Another possible implementation of tunable plasmonic cloaks is to introduce a nonlinear layer in multi-shell plasmonic cloaks, so that the scattering cross section can be controlled by changing the intensity of the incident EM field~\cite{zharova2012}. Core-shell plasmonic objects can also be designed to exhibit Fano resonances, allowing a fast switch from a completely cloaked situation to a strongly resonant scattering~\cite{monticone2013}. However, the disadvantage of this scheme is that its effectiveness depends on a given range of intensities for the incident excitation.

The feasibility of controlling the operation of an invisibility device depends on the fabrication of metamaterials whose electromagnetic properties can be altered by using external agents. Among several approaches, the use of magneto-optical materials arise as good candidates for tailoring light-matter interactions since their dispersion relations are affected whenever an external magnetic field is applied. Indeed, the material becomes optically anisotropic in the presence of the external magnetic field ${\bf B}$, so that when polarized light propagates through it, the polarization plane is rotated by an angle $\theta_{\textrm{rot}}$ proportional to the strength of ${\bf B}$  ~\cite{LivroMagnetoOptics1}. In addition, owing to the presence of the magnetic field, time-reversal symmetry is broken. As a consequence, if after accumulating a rotation angle $\theta_{\textrm{rot}}$ light is reflected from a mirror and start propagating in the opposite direction the angle $\theta_{\textrm{rot}}$ keeps on increasing\footnote{\label{Note4} This is one of the main differences between Faraday's effect and natural optical activity. In the latter, the value of $\theta_{\textrm{rot}}$ decreases after the reflection in the mirror.}. From a microscopic point of view, these properties are directly related to the Zeeman split of the atomic energy levels~\cite{LivroMagnetoOptics1}. Phenomenologically, magneto-optical effects are described in terms of an anisotropic permittivity tensor whose elements are strongly affected by the applied magnetic field. Therefore, the characteristics of a magneto-optical cloak, could be in principle significantly altered by adjusting the strength of the external magnetic field.

In the present chapter we investigate tunable plasmonic cloaks based on magneto-optical effects. We demonstrate that the application of an external magnetic field may not only switch on and off the cloaking mechanism but also reduce the electromagnetic absorption, one of the major limitations of the existing plasmonic cloaks. Indeed, we show that the absorption cross section in the operation frequency band of the cloaking device may be substantially diminished by the application of a uniform magnetic field. In addition,  we prove that the angular distribution of the scattered radiation can be effectively controlled by external magnetic fields, allowing for a swift change in the scattering pattern. We also discuss possible realistic implementations of magneto-optical shells, such as covers made of graphene. Our results suggest that magneto-optical materials allow a precise control of light scattering by a tunable  plasmonic cloak under the influence of an external magnetic field and may be useful in disruptive photonic technologies. Most of the results in the following sections were published in Refs.~\cite{kortkamp2013-2,kortkampPRL,kortkampJOSAA}

\section{Electromagnetic scattering by a cylinder coated with a magneto-optical shell}
\label{MOCloak}

\hspace{5 mm} Let us consider an infinitely long\footnote{\label{note5} This is a plausible idealization to study the amount of scattered radiation by a finite cylinder with its height several orders of magnitude larger than its diameter.} homogeneous cylinder with permittivity $\varepsilon_1$ and radius $a$ covered by a magneto-optical shell with outer radius $b>a$. The $z-$axis coincides with the symmetry axis of the cylinder. Both the inner cylinder and the shell have trivial permeabilities, namely, $\mu_1 = \mu_2 = \mu_0$. The whole system is subjected to a uniform static magnetic field ${\bf B}$ parallel to the $z$-axis, as depicted in Fig. \ref{MOC-Esquema}. %
 \vspace{10pt}
\begin{figure}[!ht]
  \centering
  \includegraphics[scale=0.4]{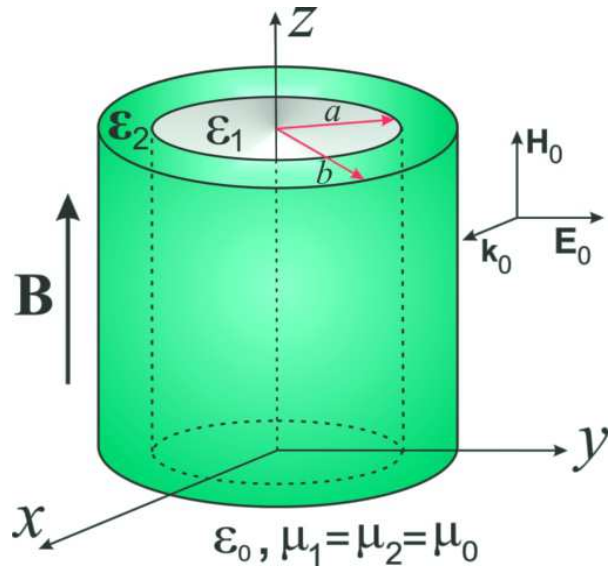}
   \vspace{10pt}
  \caption{The scattering system: an isotropic cylinder with permittivity  $\varepsilon_1$ and radius $a$ coated by a magneto-optical shell with permittivity tensor $\mbox{{\mathversion{bold}${\varepsilon}$}}_2$ and outer radius $b>a$ under the influence of a static magnetic field ${\bf B}$ and an incident TM$_{\rho}$ polarized monochromatic plane wave.}
  \label{MOC-Esquema}
\end{figure}

A monochromatic plane wave of angular frequency $\omega$ propagating in vacuum with its magnetic field parallel to the $z$-axis (TM$_{\rho}$ polarization) impinges on the cylinder perpendicularly to its symmetry axis,
\begin{eqnarray}
{\bf H}_{\rm 0} = H_{\rm 0}\, e^{- i \omega (t - x/c)} \hat{{\bf z}} \, , \ \ \ \textrm{and}\ \ \   {\bf E}_{\rm 0} = - (\mu_0 c) \, \hat{{\bf x}} \times {\bf H}_{\rm 0} = E_{\rm 0}\, e^{- i \omega (t - x/c)} \hat{{\bf y}}\, ,
\end{eqnarray}
where $E_0 = \mu_0 c H_0$.  In the absence of the external magnetic field the electromagnetic response of the shell to the impinging radiation may be well described by an isotropic permittivity tensor. However, for ${\bf B} \neq {\bf 0}$ the magneto-optical shell becomes anisotropic, and its permittivity tensor acquires non-diagonal elements.  For the geometry shown in Fig. \ref{MOC-Esquema}, the permittivity tensor $\mbox{{\mathversion{bold}${\varepsilon}$}}_2$  of the cloak can be cast into the form \cite{LivroMagnetoOptics1,Stroud1990}
\begin{eqnarray}
\label{TensoresEeM}
\mbox{{\mathversion{bold}${\varepsilon}$}}_2 =
\left(
  \begin{array}{ccc}
    \varepsilon_{xx} & \varepsilon_{xy} & 0 \\
    \varepsilon_{yx} & \varepsilon_{yy} & 0 \\
    0 & 0 & \varepsilon_{zz} \\
  \end{array}
\right)
= \left(
  \begin{array}{ccc}
    \varepsilon_{s} & i\gamma_s & 0 \\
    -i\gamma_s & \varepsilon_{s} & 0 \\
    0 & 0 & \varepsilon_{zz} \\
  \end{array}
\right)\, ,
\end{eqnarray}
where the specific expressions of $\varepsilon_s$, $\gamma_s$ and $\varepsilon_{zz}$ in terms of $\omega$ and $B = |{\bf B}|$ depend on the details of the materials under consideration. It is important to emphasize that the dependence of the dielectric tensor with $B$ plays a fundamental role in the scattering problem since the electromagnetic properties of the shell may be modified by controlling the strength and direction of the uniform magnetic field. In this way, the pattern of the scattered electromagnetic fields by the system can be drastically altered in comparison to the case where ${\bf B}$ is absent.

In order to determine the scattered electromagnetic fields we have to solve \linebreak Maxwell's equations,
\bea
\nabla \times {\bf E} = i\omega {\mu_0} {\bf H}\, , \ \ \ \textrm{and} \ \ \
\nabla \times {\bf H}= -i\omega  \mbox{{\mathversion{bold}${\varepsilon}$}} \cdot {\bf E}\, ,
\eea
in the regions $\rho\leq a$, $a< \rho \leq b$, and $b<\rho$ with the appropriate boundary conditions at $\rho=a$ and $\rho=b$ \cite{BohrenHuffman}. In spite of the anisotropic properties of the shell, for a TM$_{\rho}$ normally incident plane wave, Maxwell's equations can be decoupled in all three media \cite{Monzon}, and the only nonvanishing field components are $E_{\rho}(\rho,\varphi)$, $E_{\varphi}(\rho,\varphi)$ and $H_z(\rho,\varphi)$. More explicitly, the $z$-component of the magnetic field satisfies \cite{Monzon, kortkampJOSAA}
\begin{eqnarray}
 \!\!\!\!\!\!(\mbox{{\mathversion{bold}${\varepsilon}$}}_i)_{xx}  \dfrac{\partial^2H_z^{(i)}}{\partial x^2} +
(\mbox{{\mathversion{bold}${\varepsilon}$}}_i)_{yy}\dfrac{\partial^2H_z^{(i)}}{\partial y^2}  + [(\mbox{{\mathversion{bold}${\varepsilon}$}}_i)_{xy} + (\mbox{{\mathversion{bold}${\varepsilon}$}}_i)_{yx}] \dfrac{\partial^2H_z^{(i)}}{\partial x \partial y} + \omega^2 \alpha_i{\mu}_0 H_z^{(i)}=0\, ,
\label{EquacaoCampoMagnetico}
\end{eqnarray}
where $i = 0, \, 1, \, 2 $.  It can be shown that the remaining components of the electric field can be calculated in terms of $H_z^{(i)}(\rho,\varphi)$ as~\cite{Monzon, kortkampJOSAA}
\bea
\label{EquacaoCampoEletrico1}
E^{(i)}_{\rho}(\rho,\varphi)\!\!\! &=&\!\!\! \dfrac{i}{\omega\alpha_i}\left[(\mbox{{\mathversion{bold}${\varepsilon}$}}_i)_{xy}\dfrac{\partial H_z^{(i)}}{\partial \rho} +
\dfrac{(\mbox{{\mathversion{bold}${\varepsilon}$}}_i)_{xx}}{\rho} \dfrac{\partial H_z^{(i)}}{\partial \varphi} \right]\, , \\ \cr
E^{(i)}_{\varphi}(\rho,\varphi)\!\!\! &=&\!\!\! \dfrac{-i}{\omega\alpha_i}\left[(\mbox{{\mathversion{bold}${\varepsilon}$}}_i)_{xx} \dfrac{\partial H_z^{(i)}}{\partial \rho} -
\dfrac{(\mbox{{\mathversion{bold}${\varepsilon}$}}_i)_{xy}}{\rho} \dfrac{\partial H_z^{(i)}}{\partial \varphi} \right]\, ,
\label{EquacaoCampoEletrico2}
\eea
where $(\mbox{{\mathversion{bold}${\varepsilon}$}}_1)_{ij} = \varepsilon_1 \delta_{ij}$, $\mbox{{\mathversion{bold}${\varepsilon}$}}_2$ is given by Eq. (\ref{TensoresEeM}), and $(\mbox{{\mathversion{bold}${\varepsilon}$}}_0)_{ij} = \varepsilon_0 \delta_{ij}$. Also, we defined
\bea
\label{alpha}
\alpha_i := (\mbox{{\mathversion{bold}${\varepsilon}$}}_i)_{xx}(\mbox{{\mathversion{bold}${\varepsilon}$}}_i)_{yy}-
(\mbox{{\mathversion{bold}${\varepsilon}$}}_i)_{xy}(\mbox{{\mathversion{bold}${\varepsilon}$}}_i)_{yx}\, .
\eea
Note that Eqs. (\ref{EquacaoCampoMagnetico}), (\ref{EquacaoCampoEletrico1}), (\ref{EquacaoCampoEletrico2}), and (\ref{alpha}) simplify considerably in the isotropic regions $\rho\leq a$ and $b<\rho$.

Given the anti-symmetry of $\mbox{{\mathversion{bold}${\varepsilon}$}}_2$, the term containing the cross derivative in Eq. (\ref{EquacaoCampoMagnetico}) vanishes even within the shell, and therefore $H_z^{(i)}$ satisfies a Helmholtz equation,
\begin{equation}
(\nabla^2 + k_i^2)H_z^{(i)} = 0\,\, ,
\label{Helmholtz}
\end{equation}
with the modulus of the wave vector given by~\cite{kortkampJOSAA}
\begin{equation}
k_i = \omega \sqrt{\dfrac{{\mu}_0\alpha_i}{(\mbox{{\mathversion{bold}${\varepsilon}$}}_i)_{xx}}}\, .
\label{wavevector}
\end{equation}

The solution of the Helmholtz equation in cylindrical coordinates is well known and can be written as \cite{BohrenHuffman}
\begin{eqnarray}
H_z^{(1)}(\rho,\varphi)\!\!\! &=&\!\!\! H_0 \displaystyle{\sum_{m=-\infty}^{+\infty}} A_m J_m(k_1\rho)e^{im\varphi}\, ,\;\;\;\;  (\rho \leq a)\, ,
\label{CampoNoCilindroInterno}\\
H_z^{(2)}(\rho,\varphi)\!\!\! &=&\!\!\!H_0 \displaystyle{\sum_{m=-\infty}^{+\infty}} i^m\left[B_m J_m(k_2\rho) + C_m N_m(k_2\rho)\right]e^{im\varphi}\, ,\;\;\;\;  (a<\rho\leq b)\, ,
\label{CampoNaCasca}\\
H_z^{(0)}(\rho,\varphi)\!\!\! &=&\!\!\!H_0 \displaystyle{\sum_{m=-\infty}^{+\infty}} i^m\left[J_m(k_0\rho) + D_m H_m^{(1)}(k_0\rho)\right]e^{im\varphi}\, ,\;\;\;\; ( b < \rho)\, ,
\label{CampoEspalhado}
\end{eqnarray}
where we have used that the expansion of the incident plane wave $(e^{ik_0x})$ in cylindrical waves reads~\cite{BohrenHuffman,VanDeHulst,BornWolf}
\begin{eqnarray}
e^{ik_0x}= \displaystyle{\sum_{m=-\infty}^{+\infty}} i^mJ_m(k_0\rho)e^{im\varphi}\, .
\end{eqnarray}
In previous equations, $J_m(x)$, $N_m(x)$ and  $H_m^{(1)}(x)$ are the cylindrical Bessel, Neumann and Hankel (first kind) functions of $m-$th order, respectively \cite{Abramowitz, Arfken}. The coefficients $A_m$, $B_m$, $C_m$ and $D_m$ are entirely determined by imposing the usual electromagnetic boundary conditions on the transverse components of ${\bf E}$ and ${\bf H}$, namely,
\bea
&&\hspace{-10pt} H_z^{(1)}(a,\varphi) = H_z^{(2)}(a,\varphi) \;\; , \;\;\;
H_z^{(2)}(b,\varphi) = H_z^{(0)}(b,\varphi)  \\
&&\hspace{-10pt} E_{\varphi}^{(1)}(a,\varphi) = E_{\varphi}^{(2)}(a,\varphi)
\;\; , \;\;\;
E_{\varphi}^{(2)}(b,\varphi) = E_{\varphi}^{(0)}(b,\varphi) \, .
\label{CondicaoContorno1}
\eea

It follows from Eq. (\ref{CampoEspalhado}) that $D_m$ is the only coefficient necessary to calculate the EM field in region $b<\rho$. After a lengthy but straightforward calculation it is possible to show that $D_m$ can be put into the form~\cite{kortkampJOSAA}
\begin{equation}
D_m=\dfrac{{\cal{U}}_m}{{\cal{V}}_m}\, ,
\label{Dm}
\end{equation}
where
\bea
 {\cal{U}}_m =  \left|
  \begin{array}{cccc}
    J_m(k_1 a) 					&-J_m(k_2 a) 			& -N_m(k_2 a) 			& 0 \\
    \frac{k_1}{\varepsilon_1} J'_m(k_1 a) & -{\cal{J}}_m(k_2 a) 	& -{\cal{N}}_m(k_2 a) 	&0 \\
    0 							& J_m(k_2 b) 			&N_m(k_2 b) 			& J_m(k_0 b) \\
    0 							& {\cal{J}}_m(k_2 b) 		& {\cal{N}}_m(k_2 a) 		& \frac{k_0}{\varepsilon_0}J'_m(k_0 b) \\
  \end{array}
  \right|\, ,
  \label{Um}
\eea
and
\bea
 {\cal{V}}_m=  \left|
  \begin{array}{cccc}
    J_m(k_1 a) 						&-J_m(k_2 a) 			& -N_m(k_2 a) 			& 0 \\
    \frac{k_1}{\varepsilon_c} J'_m(k_1 a) 		& -{\cal{J}}_m(k_2 a) 	& -{\cal{N}}_m(k_2 a) 	&0 \\
    0 								& J_m(k_2 b) 			&N_m(k_2 b) 			& -H_m^{(1)}(k_0 b) \\
    0 								& {\cal{J}}_m(k_2 b) 		& {\cal{N}}_m(k_2 a) 		& -\frac{k_0}{\varepsilon_0}{H'}_m^{(1)}(k_0 b) \\
  \end{array}
\right|\, ,
\label{Vm}
\eea
with definitions
\bea
\hskip -0.5cm {\cal{J}}_m(x) &=& \dfrac{k_2}{\varepsilon_s^2-\gamma_s^2}\left[\varepsilon_s J'_m(x) + \dfrac{m\gamma_s}{x} J_m(x)\right]\, , \\ \cr
\hskip -0.5cm {\cal{N}}_m(x) &=& \dfrac{k_2}{\varepsilon_s^2-\gamma_s^2}\left[\varepsilon_s N'_m(x) + \dfrac{m\gamma_s}{x} N_m(x)\right]\, ,
\eea
where the primes in $J'_m(x)$, $N'_m(x)$, and ${H'}_m^{(1)}(x)$ denote differentiation with respect\linebreak  to the argument.

Obviously, the scattering, absorption and extinction cross sections of an infinitely long cylinder are infinite. Nevertheless, the amount of scattered, absorbed and extinct radiation per unit length are finite quantities. Therefore, similarly to the procedure adopted in Sec.~\ref{CrossSection}, we can calculate these cross sections per unit length of an infinite cylinder by evaluating the mean rates $\langle W_{\textrm{sca}} \rangle$, $\langle W_{\textrm{abs}} \rangle$, and $\langle W_{\textrm{ext}} \rangle$ at which energy crosses an imaginary cylindrical surface of length $L$ and radius $R$. By using the asymptotic expression of the Hankel functions of first kind \cite{Arfken, BohrenHuffman, Abramowitz}, namely,
\begin{equation}
 H_m^{(1)}(x) \sim (-i)^m \sqrt{\dfrac{2}{\pi x}}e^{i(x-\pi/4)}\, , \, x\gg1
\end{equation}
it is possible to show that the  differential scattering cross section efficiency $dQ_{\textrm{sca}}/d\varphi$, as well as the total scattering, extinction, and absorption cross sections efficiencies $Q_{\textrm{\textrm{sca}}}$,   $Q_{\textrm{ext}}$, and $Q_{\textrm{abs}}$ are given, respectively, by\footnote{\label{Note6} The difference in signal between our expression for $Q_{\textrm{\textrm{ext}}}$ and the one written in Refs.\cite{BohrenHuffman, VanDeHulst, BornWolf} is due to the fact that we have defined coefficients $A_m\, , B_m\, , C_m\, ,$ and $D_m$ through the expressions for the magnetic field whilst in Refs. \cite{BohrenHuffman, VanDeHulst, BornWolf} these coefficients are defined by means of the electric field.}~\cite{BohrenHuffman, VanDeHulst, BornWolf, kortkamp2013-2, kortkampPRL, kortkampJOSAA},
\begin{eqnarray}
\label{differential}
\dfrac{dQ_{\textrm{sca}}}{d\varphi}\!\!\! &=&\!\!\! \dfrac{dC_{\textrm{sca}}/d\varphi}{2bL} = \dfrac{1}{\pi k_0b}\left|\sum_{-\infty}^{\infty}D_m e^{im\varphi}\right|^2\, ,\\
\label{Qsc}
Q_{\textrm{\textrm{sca}}}\!\!\! &=&\!\!\! \dfrac{C_{\textrm{sca}}}{2bL} = \dfrac{2}{k_0b}\,\displaystyle{\sum_{-\infty}^{+\infty}|D_m|^2}\, , \\
\label{Qext}
Q_{\textrm{\textrm{ext}}}\!\!\! &=&\!\!\! \dfrac{C_{\textrm{ext}}}{2bL} =-\dfrac{2}{k_0b}\,\displaystyle{\sum_{-\infty}^{+\infty}{\textrm{Re}}(D_m)} \, , \\
\label{Qabs}
Q_{\textrm{\textrm{abs}}}\!\!\! &=&\!\!\! \dfrac{C_{\textrm{abs}}}{2bL}= Q_{\textrm{\textrm{ext}}} - Q_{\textrm{\textrm{sca}}}\, ,
\end{eqnarray}
where we used definitions written in Eq. (\ref{EfficiencyCrossSections}). It should be mentioned that despite the minus signal in Eq. (\ref{Qext})\footref{Note6} the extinction cross section is, of course, a positive quantity since $\textrm{Re}(D_m)<0$ (as can be checked numerically for magneto-optical dispersive media).
\vspace{20pt}
\section{Magneto-optical plasmonic cloaking}
\label{MOCloakResults}

\hspace{5mm} In order to analize the EM cloaking via SCT using a magneto-optical coating shell  let us consider the dipole approximation, {\it i.e.} $k_0b\ll 1\, , \ k_1b\ll 1\, , \ k_2b\ll1$. In this case the dominant scattering terms in Eqs. (\ref{differential}), (\ref{Qsc}), (\ref{Qext}), and (\ref{Qabs}) are $m = 0\, ,$ and  $m = \pm 1$. Besides,  by expanding the Bessel, Neumann and Hankel functions for small arguments it is possible to show that, up to terms $\sim (k_0b)^2$,  $D_0$ identically vanishes for nonmagnetic materials and $D_{\pm 1}$ can be cast as~\cite{kortkamp2013-2}
\be
D_{\pm 1} =\dfrac{i \pi (k_0b)^2}{4}\dfrac{(\gamma_s^2 \pm \gamma_s (\varepsilon_0 + \varepsilon_1)+\varepsilon_0\varepsilon_1-\varepsilon_{1}^2)(\eta^2-1 ) + \varepsilon_{s} (\varepsilon_1-\varepsilon_0)(\eta^2+1)}
{(\gamma_s^2 \pm \gamma_s (\varepsilon_1 - \varepsilon_0)-\varepsilon_0\varepsilon_1-\varepsilon_{s}^2)(\eta^2-1) + \varepsilon_{s} (\varepsilon_1 + \varepsilon_0)(\eta^2+1)}\, ,
\label{Dm_Aproximado}
\ee
where we defined $\eta = a/b$. Note that from Eqs. (\ref{Qsc}), (\ref{Qext}),  and (\ref{Dm_Aproximado}) the leading terms in the scattering and extinction cross section efficiencies are generally $\propto (k_0b)^3$ and $\propto k_0b$, respectively. However, for lossless materials Eqs. (\ref{Qext}) and (\ref{Dm_Aproximado}) yield $Q_{\textrm{ext}} = 0$. Surely, this result has no physical meaning for passive materials since it predicts a negative absorption cross section efficiency. This can be solved by expanding $D_0$ and $D_{\pm 1}$ up to the next nonvanishing order, which turns to be $\sim (k_0b)^4$. Hence, if losses are negligible, the leading terms in both $Q_{\textrm{sca}}$ and $Q_{\textrm{ext}}$ will be $\sim (k_0b)^3$.

For the purpose of discussing the SCT in our system, the approximate expressions for $D_{\pm 1}$ in Eq. (\ref{Dm_Aproximado}) are sufficient, because they will give the dominant contribution to the scattering cross section even if we neglect ohmic losses. In this case, the requirements for invisibility are determined by enforcing that each scattering coefficient vanishes separately.  Consequently, in the long wavelength approximation, Eq. (\ref{Dm_Aproximado}) shows that $D_{-1}$ and $D_{+1}$ are zero provided the ratio between the inner and outer radii of the cloak satisfies ~\cite{kortkamp2013-2, kortkampPRL, kortkampJOSAA}
\begin{equation}
\label{razaoraios}
{\eta_{\,}}_{\pm 1} = \sqrt{\dfrac{(\varepsilon_0\pm \gamma_s-\varepsilon_s)(\varepsilon_1 \pm \gamma_s+\varepsilon_s)}
{(\varepsilon_1 \pm \gamma_s-\varepsilon_s)(\varepsilon_0 \pm \gamma_s+\varepsilon_s)}},
\end{equation}
where ${\eta_{\,}}_{-1}$ and ${\eta_{\,}}_{+1}$ represent the values of $\eta = a/b$ that make $D_{-1}$ and $D_{+1}$ vanish, respectively. Note that when the external magnetic field is absent, {\it i.e.} $\gamma_s = 0$, Eq. (\ref{razaoraios}) yields ${\eta_{\,}}_{-1} =  {\eta_{\,}}_{+1}$ and the result for an isotropic cylindrical cloak is retrieved~\cite{alu2005_2}.

The conditions for achieving the cancellation of $D_{-1}$ and $D_{+1}$, obtained from Eq.~(\ref{razaoraios}), are shown in Fig.~\ref{MOC-ParametersSpace}, in terms of the permittivities of the inner core and of the outer cloaking shell for lossless materials and fixed values of $\gamma_s$. In Fig.~\ref{MOC-ParametersSpace}, blue regions correspond to situations where either $D_{-1}$ or $D_{+1}$ vanish for a given $\gamma_s$.  From Fig.~\ref{MOC-ParametersSpace} and Eq.~(\ref{razaoraios}) we conclude that it is not possible to obtain ${\eta_{\,}}_{-1} = {\eta_{\,}}_{+1}$ for $\gamma_s \neq 0$, {\em i.e.} the contributions of $D_{-1}$ and $D_{+1}$ to $Q_{\textrm{sca}}$ cannot be simultaneously canceled for the same $a/b$ in the presence of an external magnetic field. However, it is still possible to drastically reduce $Q_{\textrm{sca}}$ and, for a given $a/b$, to obtain values for the scattering cross section efficiency that are substantially smaller than those obtained in the absence of ${\bf B}$. Indeed, guided by Fig.~\ref{MOC-ParametersSpace} and by inspecting Eq.~(\ref{razaoraios}), we can choose a ratio $\eta_{-1} $ ($\eta_{+1}$) that makes the contribution of $D_{-1}$ ($D_{+1}$) to $Q_{\textrm{sca}}$ vanish without significant enhancement of $D_{+1}$ ($D_{-1}$). Another possibility is to choose a value of $a/b$ that minimizes the sum $|D_{-1}|^2 + |D_{+1}|^2$. As a result, a strong reduction of $Q_{\textrm{sca}}$ can be achieved since $D_{0}$ is already zero in the long wavelength approximation for nonmagnetic materials. In the following we will numerically analyze the effects of a nonzero magnetic field on the performance of a magneto-optical cloaking device.
 \vspace{10pt}
\begin{figure}[!ht]
  \centering
  \includegraphics[scale=0.75]{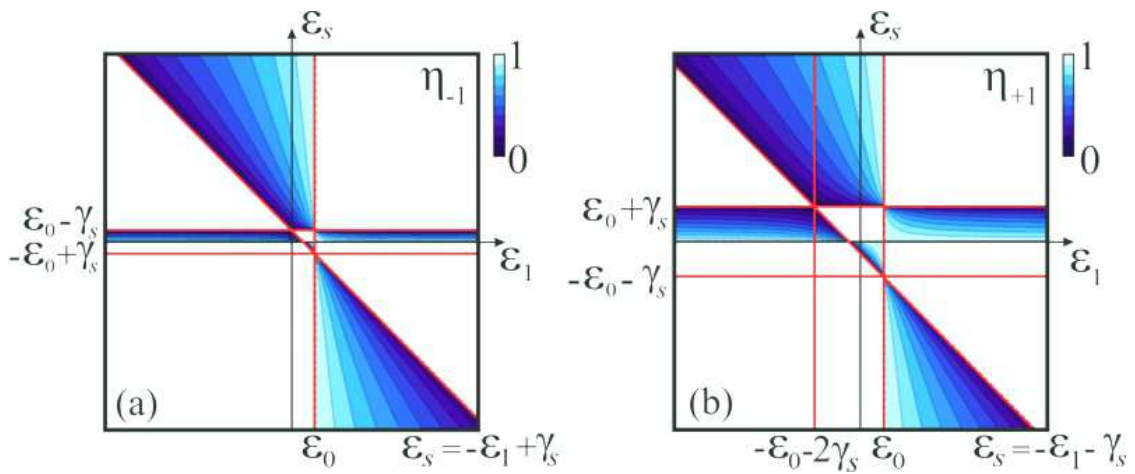}
   \vspace{10pt}
  \caption{Regions for which the invisibility condition (\ref{razaoraios}) is satisfied for the corresponding {\bf (a)} $\eta_{-1}$ and {\bf (b)} $\eta_{+1}$ between $0$ and $1$ in the presence of off-diagonal terms in the electric permittivity, proportional to the applied magnetic field, for lossless materials.}\label{MOC-ParametersSpace}
\end{figure}
\vspace{10pt}
\begin{figure}[!ht]
\centering
\includegraphics[scale=0.75]{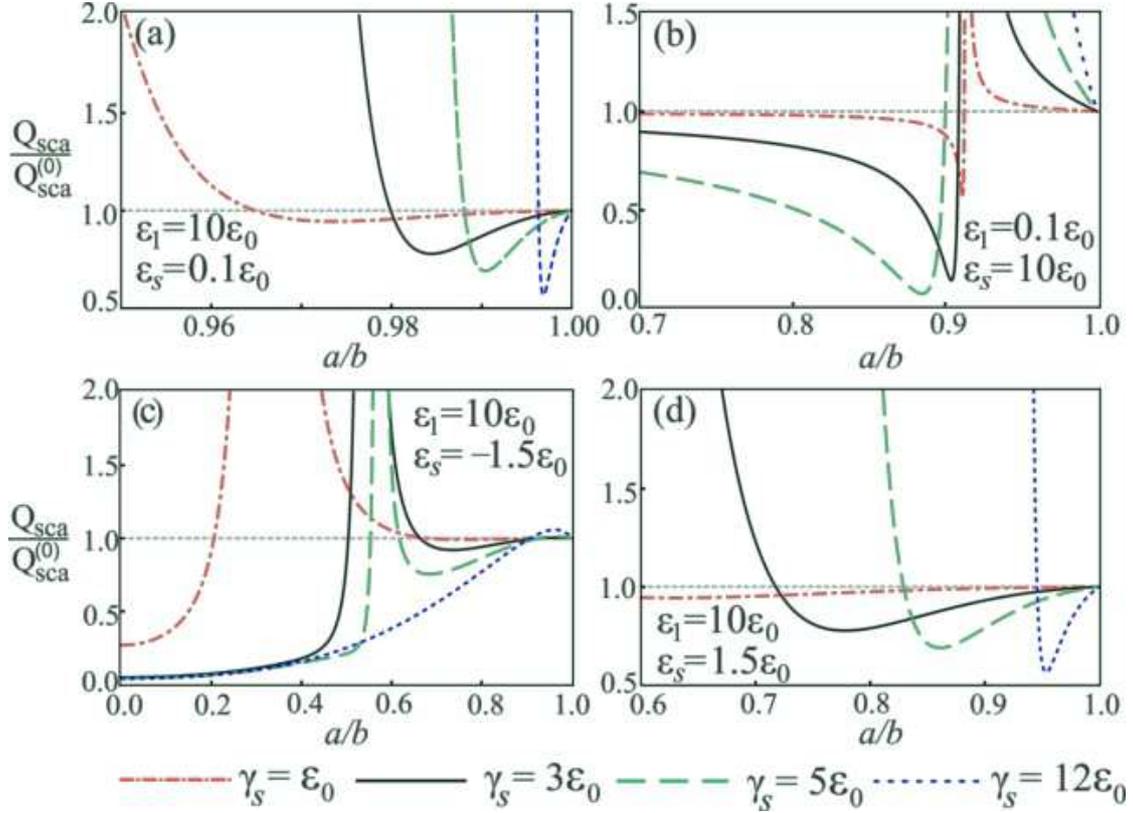}
 \vspace{10pt}
\caption{Scattering efficiency $Q_{\textrm{sca}}$ (normalized by its value in the absence of the external magnetic field, $Q_{\textrm{sca}}^{(0)}$) as a function of the ratio between the internal and external radii for {\bf (a)} $\varepsilon_1 = 10\varepsilon_0$ and $\varepsilon_s = 0.1 \varepsilon_0$, {\bf (b)} $\varepsilon_1 = 0.1\varepsilon_0$ and $\varepsilon_s = 10\varepsilon_0$, {\bf (c)} $\varepsilon_1 = 10\varepsilon_0$ and $\varepsilon_s = -1.5\varepsilon_0$, and {\bf (d)} $\varepsilon_1 = 10\varepsilon_0$ and $\varepsilon_s = 1.5\varepsilon_0$. Different curves correspond to different magnitudes of the off-diagonal term of the electric permittivity $\gamma_s$, proportional to $B$: $\gamma_s = \varepsilon_0$ (red dot-dashed curve), $\gamma_s = 3\varepsilon_0$ (black solid curve), $\gamma_s = 5\varepsilon_0$ (green dashed curve), $\gamma_s = 12\varepsilon_0$ (blue dotted curve).}
\label{MOC-SCS-1}
\end{figure}

In Fig.~\ref{MOC-SCS-1}, the scattering cross section efficiency in the presence of a uniform magnetic field ($\gamma_s \neq 0$) is calculated as a function of $a/b$ for different $\varepsilon_1$ and $\varepsilon_s$. We normalized $Q_{\textrm{sca}}$ by the scattering efficiency in the absence of the external magnetic field, $Q_{\textrm{sca}}^{(0)}$, and set $b= \lambda/100$. Figure~\ref{MOC-SCS-1}(a) corresponds to the situation where, given the parameters $\varepsilon_1 = 10\varepsilon_0$ and $\varepsilon_s = 0.1 \varepsilon_0$, it is possible to achieve invisibility for $a/b \simeq 0.92$ in the absence of the magnetic field \cite{alu2005_2}. To see this it suffices to put $\gamma_s =  0$ and substitute the set of material parameters used in Fig.~\ref{MOC-SCS-1}(a) in Eq.~(\ref{razaoraios}). The application of ${\bf B}$ shifts the values of $\eta = a/b$ that strongly reduce of $Q_{\textrm{sca}}$ to higher values. This in turn makes the ratio $Q_{\textrm{sca}}/Q^{(0)}_{\textrm{sca}}$ so large in the vicinities of $\eta = 0.92$ that it is impossible to plot it and still have enough resolution to see the reduction for $\eta > 0.96$ as we change the magnetic field. For this reason there is a cut-off in our plot at $\eta=0.95$.
Interestingly, by increasing $\gamma_s$ we can not only shift the operation range of the device for $\eta$ close to $1$, but also significantly decrease $Q_{\textrm{sca}}$. Indeed, for $0.98 \lesssim \eta < 1$ the application of ${\bf B}$ leads to a reduction of the order of $50\%$ in the scattering cross section efficiency if compared to the case where $B = 0$. This result indicates that, for this set of parameters, an optimal performance of the cloak could be achieved for very thin magneto-optical films.

In Fig.~\ref{MOC-SCS-1}(b), for the chosen set of parameters ($\varepsilon_1 = 0.1\varepsilon_0$ and $\varepsilon_s = 10\varepsilon_0$) EM transparency also occurs for $a/b \simeq 0.92$ in the absence of ${\bf B}$. In the presence of an external magnetic field, perfect invisibility (vanishing scattering cross section efficiency) is no longer possible for $\eta \simeq 0.92$, since a peak in $Q_{\textrm{sca}}$ emerges precisely at this value of $\eta$. Physically, this peak in $Q_{\textrm{sca}}$ can be explained by the fact that the EM response of the shell is no longer isotropic in the presence of ${\bf B}$. Hence, the induced electric polarization within the shell is not totally out of phase with respect to the one induced inside the inner cylinder. The net effect is that these two induced electric polarizations are not capable to cancel each other anymore in the presence of ${\bf B}$, leading to a nonvanishing scattered field. On the one hand, the fact that the application of a static magnetic field induces a peak in $Q_{\textrm{sca}}$ in a system originally conceived to achieve perfect invisibility (in the absence of ${\bf B}$) suggests that a magnetic field could be used as an external agent to reversibly switch on and off the operation of the cloak. On the other hand, for $\eta \lesssim 0.9$ (and the same set of parameters) the application of ${\bf B}$ always leads to a reduction of $Q_{\textrm{sca}}$ with respect to $Q_{\textrm{sca}}^{(0)}$, increasing the efficiency of the cloak. For example, for $\gamma_s = 5\varepsilon_0$ and $\eta = 0.88$, it is possible to achieve a reduction for $Q_{\textrm{sca}}$ of the order of $93\%$ with respect to $Q_{\textrm{sca}}^{(0)}$; for $\gamma_s = 3\varepsilon_0$ and $\eta \simeq 0.9$ this reduction is of the order of $80\%$. These results demonstrate that the reduction of the scattering cross section efficiency is robust against the variation of $\gamma_s(B)$. For $\eta \lesssim 0.9$ the increase of $\gamma_s$ not only improves the efficiency of the cloak (by decreasing $Q_{\textrm{sca}}/Q_{\textrm{sca}}^{(0)}$), but also shifts the operation of the device to lower values of $a/b$. This result contrasts to the behavior of $Q_{\textrm{sca}}$ shown in Fig.~\ref{MOC-SCS-1}(a) for different $\varepsilon_1$ and $\varepsilon_s$, where an increase in $\gamma_s$ shifts the operation range of the cloak to higher values of $a/b$. This means that, for fixed $\eta$, the scattering pattern of the system, and hence the cloaking operation functionality, can be drastically modified by varying either $B$ or the frequency, demonstrating the versatility of the magneto-optical cloak.

In Fig.~\ref{MOC-SCS-1}(c), the scattering cross section efficiency is calculated as a function of $a/b$ for $\varepsilon_1 = 10\varepsilon_0$ and $\varepsilon_s = -1.5\varepsilon_0$, values of permittivity that preclude the possibility of invisibility for any $a/b$ and ${\bf B} = {\bf 0}$. Figure \ref{MOC-SCS-1}(c) reveals that for $\gamma_s \neq  0$ a very strong reduction in $Q_{\textrm{sca}}$ can occur for a wide range of values of $\eta$, showing that we can drastically reduce the scattered field by applying a magnetic field to a system that is not originally designed to operate as an EM cloak. Indeed, for $\gamma_s = 3\varepsilon_0, 5\varepsilon_0, \ \ \textrm{and}\ \ 12\varepsilon_0$, the reduction in $Q_{\textrm{sca}}$ ranges from $80\%$ to $95\%$ for $0 < \eta \lesssim 0.5$. This result complements the one depicted in Fig.~\ref{MOC-SCS-1}(b), where the presence of ${\bf B}$ switched off the functionality of the system as a cloak. The crossover between these distinct scattering patterns could be achieved by varying the incident wave frequency.

Finally, Figure~\ref{MOC-SCS-1}(d) also corresponds to the case where invisibility cannot occur for $\varepsilon_1 = 10\varepsilon_0$,  $\varepsilon_s = 1.5\varepsilon_0$ and for zero magnetic field. For $\gamma_s\neq 0$ there is again a reduction of $Q_{\textrm{sca}}$ for $ \eta > 0.8$, which can be of the order of $50\%$ for $\gamma_s = 12\varepsilon_0$, further confirming that the application of ${\bf B}$ to a cylinder coated with a magneto-optical shell can switch on the functionality of the device as an invisibility cloak.

\section{Effects of dispersion in magneto-optical cloaks}
\label{MOC_Dispersion}

\hspace{5mm} In order to discuss a possible experimental realization of the magneto-optical invisibility device proposed here we must consider a more realistic situation where effects of dispersion and losses are taken into account in the performance of the cloak. With this purpose in mind, let us assume that the coating shell is made of a magneto-optical dielectric metamaterial described by the Drude-Lorentz (DL) model\footnote{\label{note7} A derivation of these expressions can be found in {\bf Appendix \ref{apendicea}}.}~\cite{Stroud1990, King2009}
\begin{eqnarray}
\label{DrudeLorentz1}
\varepsilon_s(\omega,B)\!\!\! &=&\!\!\! \varepsilon_0\left[1 - \dfrac{\Omega^2(\omega^2+i\omega/\tau - \omega_0^2)}{(\omega^2+i\omega/\tau - \omega_0^2)^2 - \omega^2\omega_c(B)^2} \right]\, , \\
\label{DrudeLorentz2}
\gamma_s(\omega,B)\!\!\! &=&\!\!\! \varepsilon_0 \dfrac{\Omega^2\omega\omega_c(B)}{(\omega^2+i\omega/\tau - \omega_0^2)^2 - \omega^2\omega_c(B)^2}\, ,
\end{eqnarray}
where $\omega_0$, $1/\tau$, and $\Omega$ are the resonance, damping, and oscillating strength frequencies, respectively. Also, $\omega_c(B) = eB/m$ where $e$ and $m$ are the electron's charge and mass. Note that if the external magnetic field vanishes, $\gamma_s(\omega,0) = 0$ and $\varepsilon_s(\omega,0)$ will yield the usual expression for the diagonal permittivity in the DL model~\cite{Jackson}. Moreover, it should be mentioned that Drude-Lorentz permittivities are extensively used to describe a wide range of media, including magneto-optical materials. A particularly recent example is monolayer graphene epitaxially grown on SiC~\cite{Crassee2011, Crassee2012}, where the Faraday rotation is very well described by an anisotropic DL model. Also, it is worth mentioning that the magneto-optical response of typical materials, which will ultimately govern the tuning speed of the magneto-optical cloak, is usually very fast. Indeed, magneto-optical effects manifest themselves in a time scale related to the spin precession; for typical paramagnetic materials this time is of the order of nanoseconds~\cite{kirilyuk2010}. Hence, the tuning mechanism induced by magneto-optical activity is expected to be almost instantaneous after the application of ${\bf B}$.

Hereafter we set $\Omega/2\pi = 3.0$ THz, $\omega_0/2\pi = 1.5$ THz, and $1/2\pi\tau = 0.03$ THz \footnote{These are proposed parameters for a magneto-optical metamaterial capable to work as a plasmonic  cloak. However, they do not correspond necessarily to an existing magneto-optical material.}. For concreteness, we choose the inner and outer radii to be $a = 0.6b$ and $b = \lambda/100$ such that the device operates around the frequency $f = 2.93$ THz. Also, we assume that the core is made of a dielectric with $\varepsilon_1 = 10\varepsilon_0$, and negligible losses.
 \vspace{10pt}
\begin{figure}[!ht]
\centering
\includegraphics[scale=0.75]{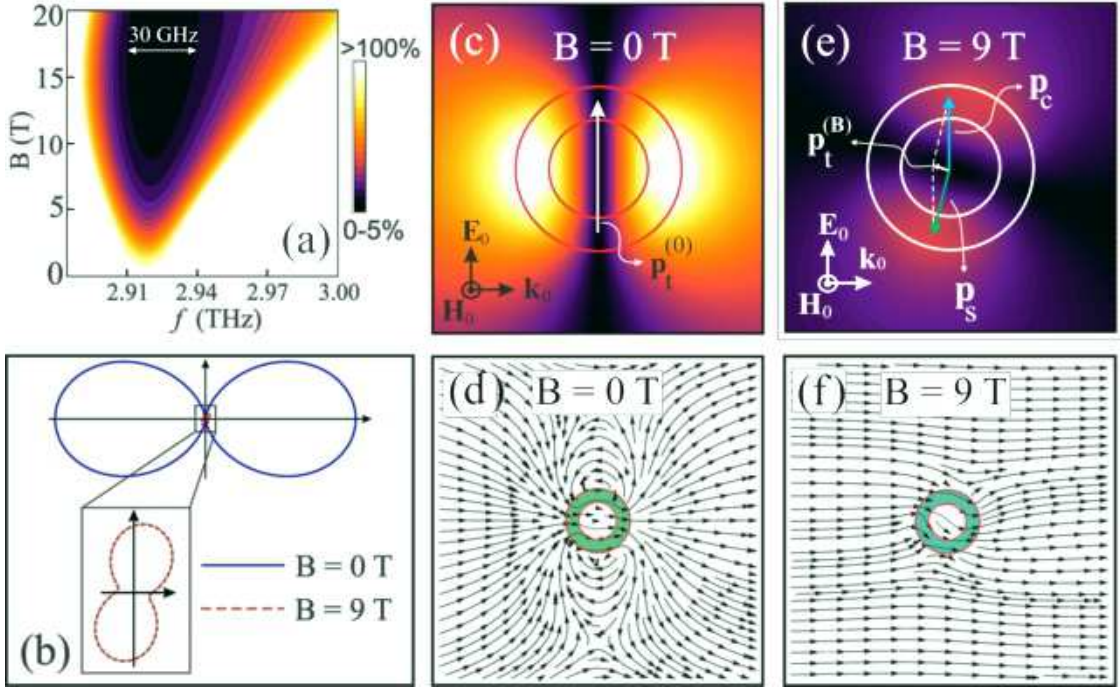}
 \vspace{10pt}
\caption{{\bf (a)} Scattering efficiency $Q_{\textrm{sca}}$ (normalized by $Q_{\textrm{sca}}^{(0)}$) as a function of the frequency $f$ and applied magnetic field strength $B$. The colored area indicates the regions in the parameters space where EM scattering is reduced due the presence of the external magnetic field. {\bf (b)} Differential scattering cross section showing the scattered pattern in the far-field for $B = 0$ T (blue solid line) and $B = 9$ T (red dashed line). {\bf (c)} Near-field spatial distribution of the $z-$component of the scattered magnetic field and {\bf (d)} time average of the Poynting vector in $xy-$ plane for $B = 0$ T and incident frequency $f = 2.93$ THz. Panels {\bf (e)} and {\bf (f)} show the same as (c) and (d), respectively, but for an external magnetic field of $9$  T. In all panels $\varepsilon_1 = 10\varepsilon_0$,  $a/b= 0.6$,  $b = \lambda/100$. The magneto-optical shell is described by the DL model given by Eqs. (\ref{DrudeLorentz1}) and (\ref{DrudeLorentz2}).}
\label{MOC-ScatterdField1}
\end{figure}

In Fig. \ref{MOC-ScatterdField1}(a)  we show a contour plot of the scattering efficiency $Q_{\textrm{sca}}$ normalized by $Q_{\textrm{sca}}^{(0)}$ as a function of both $B$ and the incident wave frequency $f$. From this plot we can see that a reduction of $Q_{\textrm{sca}}$ as large as $95$\% (if compared to the case without ${\bf B}$) can be achieved for $f\simeq 2.93$ THz and magnetic fields of the order of $10$ T. Besides, Fig. \ref{MOC-ScatterdField1}(a) demonstrates that for  $B \simeq 15$ T a attenuation of $Q_{\textrm{sca}}$ of the order of $95$\% occurs for a relatively broad band of frequencies, of the order of $30$ GHz. It should be noticed that the effect of increasing $B$ around the design operation frequency is to broaden the frequency bandwidth for which the reduction of $Q_{\textrm{sca}}$ induced by the magnetic field is larger than $95$\%. Also, note that even for smaller values of $B$ ($\sim 5$ T) the efficiency of the device can be as high as $80$\%.
%We emphasize that these findings are robust against material losses, illustrated by the fact that we are allowing for quite typical dissipation parameters and the cloak works at the levels discussed above.
Furthermore, we checked that even for much larger dissipative systems, characterized by $b=0.1\lambda$ and $b=0.2\lambda$, the cross section reduction (calculated including high order multipoles) can be as impressive as $92\%$ and $83\% $ (for $B = 10$ T), respectively. Finally, our results suggest that the efficiency of the proposed system is comparable to state-of-the-art existing cloaking apparatuses~\cite{edwards2009,rainwater2012} with the advantage of being highly tunable in the presence of magnetic fields.

In order to investigate the effects of ${\bf B}$ on the far field scattering pattern, we plot in  Fig. \ref{MOC-ScatterdField1}(b) the differential scattering efficiency for $B = 0$ T (blue solid line) and \linebreak $B = 9$ T (red dashed line) for an incident monochromatic plane wave with frequency $f = 2.93$ THz. On the one hand, in the absence of the external magnetic field the scattered radiation has a typical dipolelike profile since we are working in the long wavelength approximation. On the other hand,  the scattered radiation is strongly reduced in all directions whenever $ B  = 9$ T is applied on the system, even in the presence \linebreak of material losses.

Figure \ref{MOC-ScatterdField1}(c) exhibits the near-field spatial distribution of the $z-$component of the scattered magnetic field in the $xy-$plane for $f = 2.93$ THz. For the set of material and geometric parameters chosen invisibility cannot occur for $B= 0$ T. In this case, the spatial distribution of $H_z$ is dipolelike, as expected since for $b\ll \lambda$ the electric dipole term is dominating. Also, for zero magnetic field both the cylinder and the shell permittivities are isotropic and the total electric dipole ${\bf p}_t^{(0)}$  induced on the whole system is parallel to the impinging electric field, as shown in Fig.~\ref{MOC-ScatterdField1}(c). Furthermore, Figure \ref{MOC-ScatterdField1}(d) confirms that for a vanishing external magnetic field the cylinder can be easily observed since the time average of the Poynting vector in this case is quite distinct from the one of a plane wave propagating in free space (this would correspond to horizontal arrows pointing to the right in the whole space). On the other side, Figure \ref{MOC-ScatterdField1}(e) shows that the presence of a magnetic field with magnitude $B = 9$ T strongly reduces the scattered field intensity for all observation angles in the near-field, indicating that {\bf B} could play the role of an external agent to switch on and off the cloaking device operation. Besides, Fig. \ref{MOC-ScatterdField1}(e) unveils the physical mechanism behind the scattering suppression: since the optical properties of the inner cylinder do not depend on {\bf B}, the dipole ${\bf p}_c$ induced on the cylinder is always parallel to the impinging electric field; the shell, however, is optically anisotropic in the presence of an external magnetic field in such a way that the dipole ${\bf p}_s$ induced within the cover, is not parallel to the electric field anymore. The resulting electric dipole ${\bf p}_t^{(B)} = {\bf p}_c + {\bf p}_s$  of the whole system is therefore much smaller than ${\bf p}_t^{(0)}$ which explains why the scattered field is largely reduced for ${\bf B} \neq {\bf 0}$. Note also that the scattering pattern in Fig.~\ref{MOC-ScatterdField1}(e) is tilted in relation to the case shown in Fig~\ref{MOC-ScatterdField1}(c). The modification in the direction of maximum scattering is a direct consequence from the fact that ${\bf p}_t^{(B)}$ is not collinear with ${\bf E}_{0}$ when ${\bf B}$ is applied, as it will be elaborated in Section {\bf \ref{DirectionalScattering}}. Finally, Fig. ~\ref{MOC-ScatterdField1}(f) shows the same as panel ~\ref{MOC-ScatterdField1}(d) but for $B = 9$ T. As it can be seen, the application of an external magnetic field makes the system almost undetectable even in the near-field since the energy flow of the incident wave remains practically undisturbed in the presence of the cylinder.

It is worthwhile mentioning that there are dielectric materials that exhibit strong magneto-optical activity that could be used in the design of a magneto-optical cloak. Single- and multilayer graphene are promising candidates, as they exhibit giant Faraday rotations for moderate magnetic fields~\cite{Crassee2011}. Garnets are paramagnetic materials with huge magneto-optical response, with Verdet constants as high as $10^{4}$ deg./[T.m.] for visible and infrared frequencies~\cite{barnes1992}. There are composite materials made of granular magneto-optical inclusions that show large values of $\gamma_s$ for selected frequencies and fields in the 10-100 T range~\cite{Stroud1990, Reynet2002}. These materials offer an additional possibility of tuning EM scattering by varying the concentration of inclusions, and have been successfully employed in plasmonic cloaks~\cite{farhat2011}.

\section{Graphene-based magneto-optical coating shell}
\label{GrapheneOnSiCResults}

\hspace{5mm}  In addition to electromagnetic cloaking, in this section  we discuss the effects of ${\bf B}$ on the absorption cross section and  show that the scattered field profile can be actively controlled by the application of an external magnetic field on a polystyrene cylinder coated by a magneto-optical shell made of graphene grown on silicon carbide (SiC). Besides, we show that ohmic losses are one of the main obstacles to improve the performance of invisibility devices.

It has been recently shown that graphene epitaxially grown on SiC naturally exhibits nanoscale inhomogeneities that lead to a pronounced terahertz plasmonic resonance~\cite{Crassee2011, Crassee2012}. The excellent plasmonic properties and strong magneto-optical activity make graphene grown on SiC an excellent material platform to investigate the effects of an external magnetic field on plasmonic cloakings~\cite{kortkampPRL, kortkampJOSAA}. It is possible to show that the effective electric permittivity tensor of such magneto-optical material can be well described by the following expressions for the diagonal and off-diagonal matrix elements  \cite{Alaee2012,kortkampJOSAA, Lim2013}
\begin{eqnarray}
\varepsilon_s(\omega, B) = \varepsilon_0+\dfrac{i\sigma_{xx}^{2D}(\omega,B)}{\omega (b-a)}\, , \ \ \ \textrm{and} \ \ \
\gamma_s(\omega, B) = \dfrac{\sigma_{xy}^{2D}(\omega,B)}{\omega(b-a)},
\end{eqnarray}
where \cite{Crassee2012}
\begin{eqnarray}
\sigma_{xx}^{2D}(\omega,B)\!\!\! &=&\!\!\! 3.5\sigma_0+ \dfrac{\sigma_{+}^{2D}(\omega,B)+\sigma_{-}^{2D}(\omega,B)}{2}\, ,\\
\sigma_{xy}^{2D}(\omega,B)\!\!\! &=&\!\!\! \dfrac{\sigma_{+}^{2D}(\omega,B)-\sigma_{-}^{2D}(\omega,B)}{2i}.
\end{eqnarray}
with $\sigma_0=e^2/4\hbar$ being the universal conductivity and
\begin{eqnarray}
\sigma_{\pm}^{2D}(\omega,B) = \dfrac{2di/\pi}{\omega \mp \omega_c^{\textrm{exp}}(B)-\omega_0^2/\omega+i/\tau_g}.
\end{eqnarray}

In the following calculations we employ parameters directly fitted from experimental results \cite{Crassee2012}, namely, $\hbar d/\sigma_0 = 0.52 $ eV, $\omega_0 =  9.9\times10^{12}$ rad/s, $1/\tau_g = 18.3\times10^{12}$ rad/s, and $\omega_c^{\textrm{exp}} = 3.2\times10^{12}B\textrm{(T)}$ rad/s. It is worth mentioning that in the experiment described in Ref.~\cite{Crassee2012} the magnetic field is applied perpendicularly to the sample of graphene epitaxially grown on SiC. Here, since we are considering effective material parameters, the values extracted from the experimental data provide reasonable estimates to be used in the following numerical calculations.

For the inner polystyrene cylinder we use a Drude-Lorentz-like model \cite{Hough1980}
\begin{equation}
\dfrac{\varepsilon_{1}(\omega)}{\varepsilon_0} = 1 + \dfrac{\omega_{p1}^2}{\omega_{r1}^2 - \omega^2 - i\omega/\tau_{1}} +\dfrac{\omega_{p2}^2}{\omega_{r2}^2 - \omega^2 - i\omega/\tau_{2}}\, ,
\label{Poly}
\end{equation}
with material parameters, including losses, given by  $\omega_{p1} = 1.11\times10^{14}\  \textrm{rad/s}\, ,$ \linebreak $\omega_{r1} = 5.54\times10^{14}\  \textrm{rad/s}\, , \ \ \  \omega_{p2} = 1.96\times10^{16}\  \textrm{rad/s}\, , \ \ \omega_{r2} = 1.35\times10^{16}\  \textrm{rad/s}\, , \ \ \ \textrm{and}$ \linebreak $1/\tau_1 = 1/\tau_2 = 0.1 \times 10^{12}\  \textrm{rad/s} $ \footnote{\label{note8} In Ref. \cite{Hough1980}, $1/\tau_1 = 1/\tau_2 = 0$. Here, however, we decided to include small ohmic losses so as to make the model more realistic.}.

We also emphasize that in the following calculations we set $b=0.1\lambda$  and $a=0.6b$. We have verified that this condition guarantees the validity of the dipole approximation and that the dominant scattering coefficients are indeed $D_0$ and $D_{\pm 1}$.

\subsection{Plasmonic cloaking and enhanced scattering}
\label{PlasmonicCloaking_Graphene}

\hspace{5mm} Let us briefly discuss magneto-optical effects on plasmonic cloaking in this system. We show in Fig.~\ref{MOC-SCS-Graphene}(a) the scattering efficiency $Q_{\textrm{sca}}$  in the presence of an external magnetic field ${\bf B}$ (normalized to its value in the absence of ${\bf B}$, $Q^{(0)}_{\textrm{sca}}$) as a function of the frequency of the impinging wave for a polystyrene cylinder coated with graphene  grown on silicon carbide. Figure~\ref{MOC-SCS-Graphene}(a) shows that the application of ${\bf B}$ drastically reduces $Q_{\textrm{sca}}$, increasing the plasmonic cloaking performance in comparison to the case without the magnetic field treated so far; this reduction can achieve 80\% for $B = 20$ T, relative to the case where  $B = 0$ T. The analysis of Fig.~\ref{MOC-SCS-Graphene}(a) also reveals that by increasing $B$ it is possible to broaden the frequency band where cloaking occurs, which is particularly appealing for applications. Furthermore Fig.~\ref{MOC-SCS-Graphene}(a) shows that the application of ${\bf B}$ modifies considerably the scattering pattern. Indeed, for a fixed frequency we can readily move, by applying ${\bf B}$, from a situation where cloaking occurs to one in which the system scatters considerably more radiation than in the case ${\bf B}= {\bf 0}$. Figure~\ref{MOC-SCS-Graphene}(b) shows a contour plot of $Q_{\textrm{sca}}/Q^{(0)}_{\textrm{sca}}$ as a function of both $B$ and frequency in the terahertz range for the same system. Figure~\ref{MOC-SCS-Graphene}(b) confirms that the application of ${\bf B}$ can significantly reduce the scattering cross section for a broad frequency band in the terahertz even for modest magnetic fields. Altogether, Figs.~\ref{MOC-SCS-Graphene}(a) and \ref{MOC-SCS-Graphene}(b) demonstrate that the application of an external magnetic field can not only reduce the scattering cross section of a realistic system composed by existing magneto-optical materials (graphene) in a broad frequency range in the terahertz, but also tune its scattering properties.
 \vspace{10pt}
\begin{figure}[!ht]
  \centering
  \includegraphics[scale=0.7]{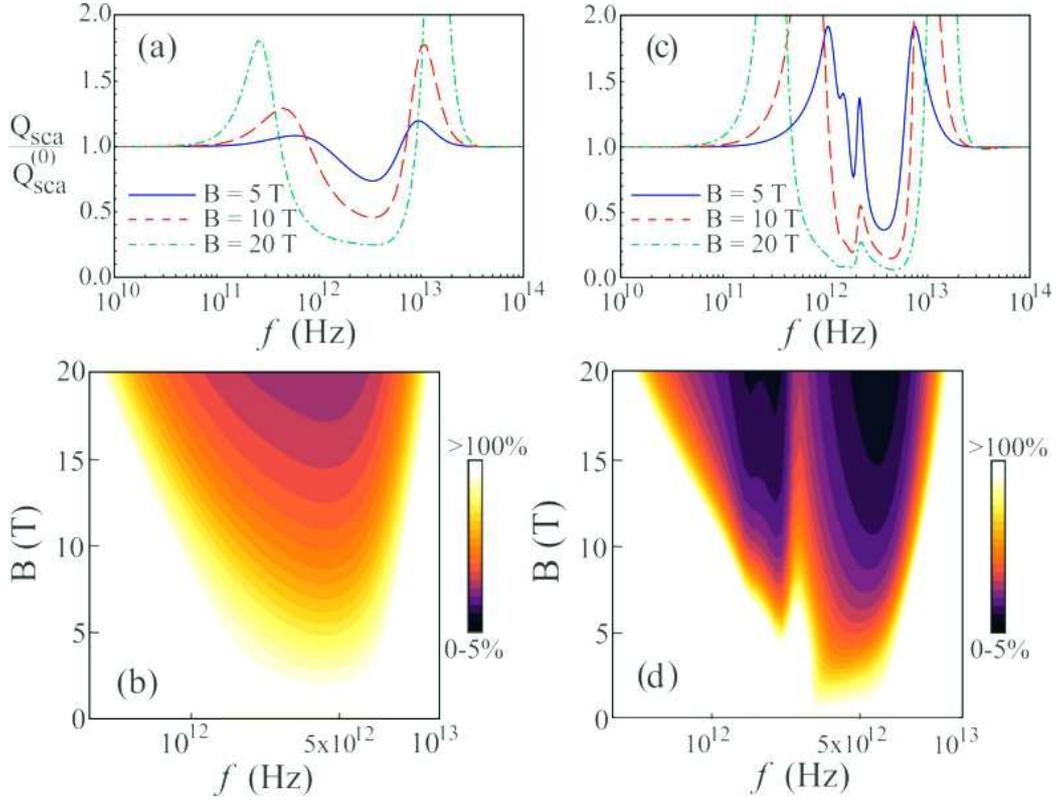}
   \vspace{10pt}
  \caption{Scattering efficiency $Q_{\textrm{sca}}$ (normalized by its value in the absence of  {\bf B}, $Q_{\textrm{sca}}^{(0)}$) as a function of the frequency of the impinging wave for $B = 5$ T (blue solid line), $10$ T (red dashed line) and $20$ T (green dot-dashed line) for {\bf (a)} the same losses as in graphene grown on SiC ($1/\tau_g = 18.3\times 10^{12}$ rad/s), and {\bf (c)} with losses ten times smaller ($1/\tau_g = 1.83\times 10^{12}$ rad/s). Contour plots of $Q_{\textrm{sca}}/Q_{\textrm{sca}}^{(0)}$ as a function of both frequency $f$ and magnetic field $B$  are shown in {\bf (b)} and {\bf (d)} for the same set of parameters chosen in (a) and (c), respectively.}
  \label{MOC-SCS-Graphene}
\end{figure}

With the purpose of analyzing how to improve the tuning mechanism and the plasmonic cloaking performance, in Figs.~\ref{MOC-SCS-Graphene}(c) and \ref{MOC-SCS-Graphene}(d) we show $Q_{\textrm{sca}}/Q^{(0)}_{\textrm{sca}}$ for the same system but with $1/\tau_g = 1.83\times 10^{12}$ rad/s, {\it i.e.} with values of ohmic losses ten times smaller than in the experiment of Ref.~\cite{Crassee2012}. From Figs.~\ref{MOC-SCS-Graphene}(c) and \ref{MOC-SCS-Graphene}(d) we can see that electromagnetic scattering can be almost totally suppressed in the presence of ${\bf B}$; this reduction attains up to 90\% for $B=15$ T in the frequency range from 1.2 THz to \linebreak 5.6 THz (except for a frequency band of 0.6 THz centered at 2.5 THz). Even for moderate magnetic fields the scattering reduction is quite large, approximately 70\% (typical efficiency in plasmonic cloaking experiments \cite{edwards2009, rainwater2012}) for $B=9$ T in the frequency range from 1.4 THz to 5.6 THz (except for a frequency bandwidth of 0.4 THz centered at \linebreak 2.4 THz). The effect of increasing the magnetic field is twofold: to further reduce $Q_{\textrm{sca}}$ and to broaden the frequency band where this reduction occurs. In both cases reducing material losses contributes to enhance these effects.  Here again it is possible to drastically modify the scattering properties by varying the magnitude of the magnetic field, changing from a highly scattering situation to cloaking,  while keeping the frequency constant.
 \vspace{10pt}
\begin{figure}[!ht]
  \centering
  \includegraphics[scale=0.75]{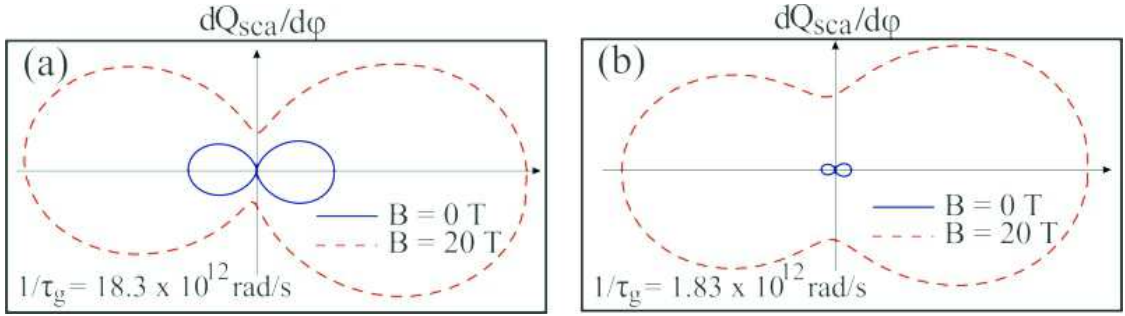}
   \vspace{10pt}
  \caption{Differential scattering cross section efficiency for $B = 0$ T (blue solid line) and $B = 20$ T (red dashed line) for an incident wave of frequency $14$ THz and two different values of loss parameters: {\bf (a)} $1/\tau_g = 18.3\times 10^{12}$ rad/s, and {\bf (b)} $1/\tau_g = 1.83\times 10^{12}$ rad/s.}
  \label{MOC-DSCS-Graphene}
\end{figure}

So far we have discussed possibilities of scattering attenuation, but an external magnetic field also allows for scattering enhancement. Indeed, from Fig. \ref{MOC-SCS-Graphene}  we can easily see that for frequencies slightly below or above the invisibility cloaking band there exists a strong increase in the scattering cross section efficiency. So as to observe the effects of this enhancement on the spatial distribution of the scattered radiation, in Fig. \ref{MOC-DSCS-Graphene} we show the polar representation of the differential scattering cross section efficiency given by Eq. (\ref{differential}) at frequency $14$ THz for two different values of the external magnetic field, namely, $B=0$ T and $B=20$ T. In panel \ref{MOC-DSCS-Graphene}(a) the dissipation parameter in the shell corresponds precisely to the one obtained from experimental data,\linebreak  $1/\tau_g = 18.3 \times 10^{12}$ rad/s, whereas in panel \ref{MOC-DSCS-Graphene}(b) it is ten times smaller. It is clear that Figs.~\ref{MOC-DSCS-Graphene}(a) and \ref{MOC-DSCS-Graphene}(b) highlight a situation where EM scattering is strongly enhanced omnidirectionally when the magnetic field is present, regardless of the values of $B$ and $\tau_g$. For $B = 20$ T and the set of parameters used in Fig.~\ref{MOC-DSCS-Graphene}(a) the total scattering cross section efficiency can be four times larger than in the case where $B$ is absent, while for the same value of $B$ and smaller losses [Fig.~\ref{MOC-DSCS-Graphene}(b)] the enhancement in the field intensities is almost twenty times greater. As discussed above, the effects of the external magnetic field on the scattering pattern are more pronounced for smaller losses. These results reinforce that magneto-optical materials can be explored as a new platform for tuning the operation of plasmonic cloaking devices.

\subsection{The role of absorption}
\label{AbsorptionCrossSection}

\hspace{5mm} It is interesting to mention that in some cases the reduction of the scattering is obtained at the expense of an enhancement of the absorption cross section in the operation frequency band of the device \cite{mullig2013Exp, miller2013, mullig2013}. It means that even though the detectability of the cloaked body is reduced in oblique directions it can be easily observed in the forward direction due to the shadows produced by the system.

In the following we discuss the effects of the external uniform magnetic field on the absorption of a magneto-optical cloaking made of  graphene grown on SiC. In Fig.~\ref{MOC-Absorption1} we present the absorption cross section $Q_{\textrm{abs}}$, given by Eq. (\ref{Qabs}), normalized by its value in the absence of ${\bf B}$, $Q^{(0)}_{\textrm{abs}}$, as a function of frequency for three values of the magnitude of ${\bf B}$. In Fig.~\ref{MOC-Absorption1}(a) the material parameters of the graphene layer are exactly the same as in the experiments of Ref.~\cite{Crassee2012} whereas in Fig.~\ref{MOC-Absorption1}(b) the loss parameter in such layer is ten times smaller. Comparing Fig.~\ref{MOC-Absorption1} and Fig.~\ref{MOC-SCS-Graphene} one can see that frequency regions where absorption and scattering minima occur approximately coincide when ${\bf B}$ is applied. This result is in contrast to what typically occurs in many cloaking devices: being based on resonant effects, cloaking mechanisms are usually associated with enhanced absorption, which degrades their performance. Figure~\ref{MOC-Absorption1} indicates that absorption would not be a nuisance in this implementation of the magneto-optical tunable cloak. On the contrary, absorption is expected to be less significative precisely in the regions where\linebreak cloaking occurs.
 \vspace{0pt}
\begin{figure}[!ht]
  \centering
  \includegraphics[scale=0.7]{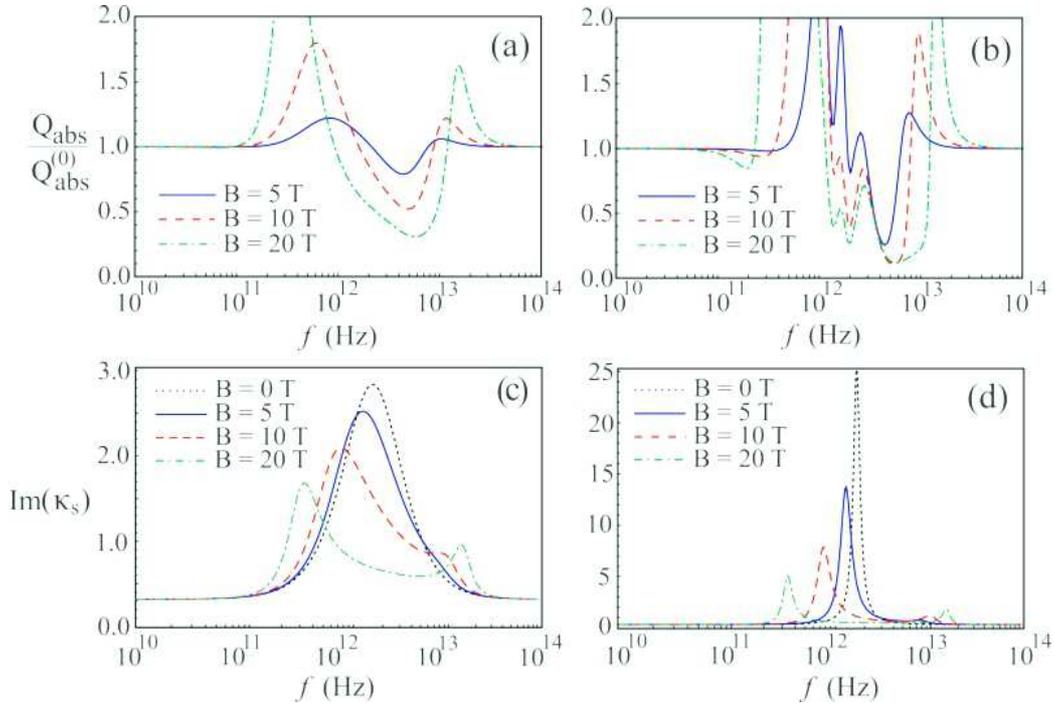}
   \vspace{5pt}
  \caption{Absorption efficiency $Q_{\textrm{abs}}$ (normalized by its value in the absence of  {\bf B}, $Q_{\textrm{abs}}^{(0)}$) as a function of the frequency of the impinging wave for $B = 5$ T (blue solid line), $10$ T (red dashed line) and $20$ T (green dot-dashed line) for {\bf (a)} the same losses as in graphene epitaxially grown on SiC ($1/\tau_g = 18.3\times 10^{12}$ rad/s), and {\bf (b)} ten times smaller losses ($1/\tau_g = 1.83\times 10^{12}$ rad/s). Panels {\bf (c)} and {\bf (d)} show the behavior of the imaginary part of $\kappa_{s} \equiv (\varepsilon_s^2-\gamma_s^2)/\varepsilon_s$ as a function of frequency for $B = 0$ T (black dotted line), $B = 5$ T (blue solid line), $10$ T (red dashed line) and $20$ T (green dot-dashed line) for  $1/\tau_g = 18.3\times 10^{12}$ rad/s and $1/\tau_g = 1.83\times 10^{12}$ rad/s, respectively.}
  \label{MOC-Absorption1}
\end{figure}

In order to bring to light the physical mechanism that governs the reduction of the absorption cross section,  it is necessary to investigate the behavior of both $\varepsilon_s$ and $\gamma_s$ when the external magnetic field $B$ is applied. It follows from  Eqs. (\ref{Helmholtz}) and (\ref{wavevector}) that $k_s^2 = (\varepsilon_s^2-\gamma_s^2)\omega^2/\varepsilon_s c^2$, so that $\kappa_{s} \equiv (\varepsilon_s^2-\gamma_s^2)/\varepsilon_s$ plays the role of the electric permittivity of the magneto-optical shell. This result, combined with the fact that $Q_{\textrm{abs}}$ is proportional to the imaginary part of the permittivity of the material \cite{BohrenHuffman}, has motivated us to calculate $\textrm{Im} (\kappa_{s})$ as a function of frequency for different values of $B$ and $1/\tau_g$ in Figs.~\ref{MOC-Absorption1}(c) and~\ref{MOC-Absorption1}(d). For $B=0$ T, $\textrm{Im} (\kappa_{s})$ has a peak exactly at the resonance frequency, $\omega_0/2\pi \simeq 1.6 $ THz. When $B \neq 0$, $\textrm{Im} (\kappa_{s})$ decreases in the range $\sim1$ THz to $\sim10$ THz for all values of $B$, which explains the reduction in $Q_{\textrm{abs}}$ in this same frequency bandwidth, as shown in Figs.~\ref{MOC-Absorption1}(a) and \ref{MOC-Absorption1}(b). Besides, as $B$ increases, the original peak in $\textrm{Im} (\kappa_{s})$ splits into two peaks  above and below $\omega_0/2\pi$, separated from each other by $\sim \omega_c(B)$~\cite{Crassee2012}. These two peaks are responsible for the enhancement of $Q_{\textrm{abs}}$ for frequencies below $\sim1$ THz and above $\sim10$ THz when compared to the case with ${\bf B}={\bf 0}$.

\subsection{Directional scattering}
\label{DirectionalScattering}

\hspace{5mm} As shown in Fig.~\ref{MOC-ScatterdField1}(e) the direction of the scattered radiation can be modified by the application of a magnetic field {\bf B}. The rotation of the scattered pattern can be quantified by the tilt angle $\varphi_r$ [see the inset of Fig.~\ref{MOC-RotationAngle1}(a)] between the direction of maximum scattering and the $x-$axis (which is the direction of maximum scattering in the long wavelength approximation for $B = 0$ T). In the case without losses and for nonmagnetic materials the coefficient $D_0$ vanishes and it follows from Eq. (\ref{differential}) that the maxima of $dQ_{\textrm{sca}}/d\varphi$ are given by
\begin{equation}
\varphi_r = \dfrac{1}{2} \tan^{-1}\left[\dfrac{\textrm{Im}(D_{-1}D_1^{*})}{\textrm{Re}(D_{-1}D_1^{*})}\right]\, .
\label{RotationAngle}
\end{equation}
Notice that $\varphi_r=0^o\, , \ 180^o$ for $B = 0$ T, as expected. Even with realistic losses taken into account, we have numerically verified that for the set of materials we have chosen the above equation gives an excellent estimation for the rotation angle. It is worth mentioning that by symmetry $\varphi_r$ is an odd function of $B$. In the dipole approximation and for nonmagnetic materials this can be analytically verified using Eq. (\ref{Dm_Aproximado}). 

To investigate how an external magnetic field can modify and tailor the direction of the scattered radiation in magneto-optical systems we show in Fig.~\ref{MOC-RotationAngle1}(a) the tilt angle $\varphi_{r}$ as a function of frequency $f$ for distinct values of ${\bf B}$. The parameters of the graphene layer are chosen from experimental data ~\cite{Crassee2012}.  It is evident from Figure~\ref{MOC-RotationAngle1}(a) that the application of an external magnetic field induces a rotation of the scattering pattern. This rotation increases as ${\bf B}$ is increased and $\varphi_{r}$ exhibits a pronounced maximum in the terahertz frequency range. Increasing ${\bf B}$ has the effect of broadening the frequency band for which the rotation occurs and of reducing the value of the frequency for which $\varphi_{r}$ has a local maximum; the greatest rotation can be as high as $\varphi_{r} \simeq 30^{o}$ for $B=20$ T and $f \simeq 0.6$ THz. It is worth mentioning that $\varphi_{r}$ changes its sign for higher frequencies ($f \gtrsim10$ THz) and this happens for all values of ${\bf B}$, as it can be seen from Fig.~\ref{MOC-RotationAngle1}(a). In Fig.~\ref{MOC-RotationAngle1}(b), $\varphi_{r}$ is plotted as a function of the strength of ${\bf B}$ for three distinct frequencies in the terahertz range. Figure~\ref{MOC-RotationAngle1}(b) further demonstrates that the radiation pattern can be significantly rotated by applying an external ${\bf B}$ for a broad frequency band in the terahertz range. In particular, the tilt angle $\varphi_{r}$ increases from $0^{o}$ to $30^{o}$ as the magnetic field increases up to approximately $15$ T for $f = 1$ THz.
 \vspace{10pt}
\begin{figure}[!ht]
  \centering
  \includegraphics[scale=0.75]{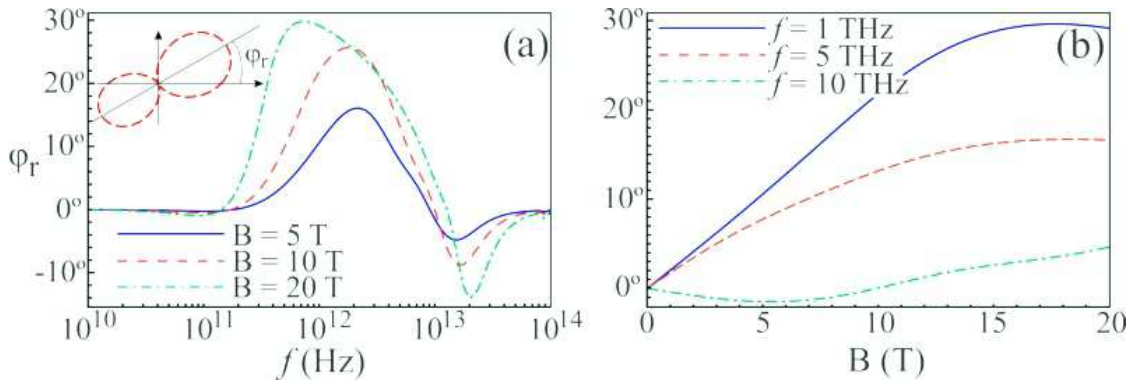}
   \vspace{10pt}
  \caption{{\bf (a)} Rotation angle $\varphi_r$ as a function of frequency for $B = 5$ T (blue solid line), $B = 10$ T (red dashed line), and $B = 20$ T (green dot-dashed line). The inset shows the definition of $\varphi_r$.  {\bf (b)} Angle $\varphi_r$ as a function of the external magnetic field strength $B$ for three different frequencies of the impinging wave, $f = 1$ THz (blue solid line), $f = 5$ THz (red dashed line) and $f = 10$ THz (green dot-dashed line).}
  \label{MOC-RotationAngle1}
\end{figure}

In Fig.~\ref{MOC-RotationAngle2} a contour plot of $\varphi_{r}$ as a function of both magnitude of ${\bf B}$ and frequency is shown. Note  the large frequency band (from $1$ THz to approximately $10$ THz) for which a substantial rotation in the radiation pattern takes place ($\varphi_{r} \simeq 30^{o}$), when we apply a magnetic field of the order of $9$ T. This frequency band is broadened as we increase the magnetic field. Even for modest magnetic fields (of the order of $5$ T) a rotation of the order of $10^{o}$ can be achieved in the frequency range of $1$ THz to $4$ THz.
 \vspace{10pt}
\begin{figure}[!ht]
  \centering
  \includegraphics[scale=0.42]{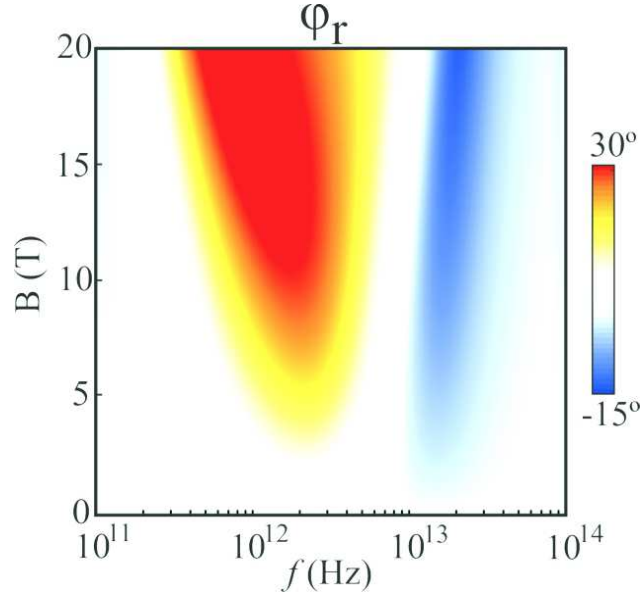}
   \vspace{10pt}
  \caption{Rotation angle given by Eq. (\ref{RotationAngle}) as a function of both frequency $f$ and external magnetic field strength $B$.}
  \label{MOC-RotationAngle2}
\end{figure}

In Fig.~\ref{MOC-RotationAngle3}(a) a polar representation of the scattering cross section $dQ_{\textrm{sca}}/d\varphi$ is calculated for three different values of ${\bf B}$ and with the parameters of the magneto-optical cover taken from the experimental data of Ref~\cite{Crassee2012}. The frequency of the incident electromagnetic plane wave is $5$ THz. Here one can appreciate the effects of reversing the direction of the magnetic field while keeping its magnitude constant: the angular distribution of the scattered radiation drastically changes its pattern, once $\varphi_{r}$ experiences a variation of 60$^o$ during this process. This effect of reorienting the scattering pattern by applying ${\bf B}$ will be  even more pronounced if one reduces losses in the system, as demonstrated in Fig.~\ref{MOC-RotationAngle2}(b), where the dissipation parameter in the magneto-optical shell is ten times smaller than that considered in Fig.~\ref{MOC-RotationAngle3}(a) ($1/\tau_g = 1.83 \times 10^{12}$ rad/s). In this case $\varphi_{r}$ varies 90$^o$ as $B$ changes from $20$ T to $-20$ T. Together with Figs.~\ref{MOC-RotationAngle1} and \ref{MOC-RotationAngle2}, these results demonstrate that we can achieve a high degree of tunability and control of the scattered radiation pattern by applying an external magnetic field in existing magneto-optical materials.
 \vspace{10pt}
\begin{figure}[!ht]
  \centering
  \includegraphics[scale=0.75]{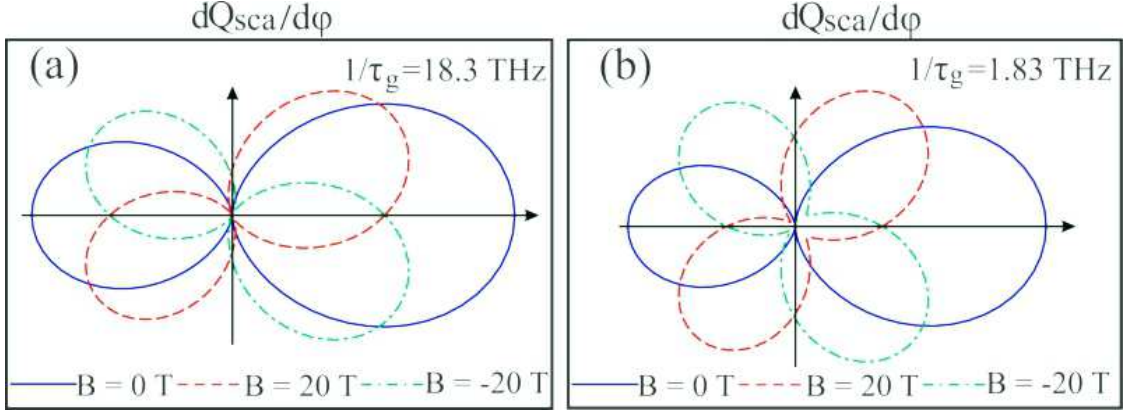}
   \vspace{10pt}
  \caption{Differential scattering cross section for an incident wave of frequency $5$ THz and {\bf (a)} $1/\tau_g = 18.3\times 10^{12}$ rad/s, and {\bf (b)} $1/\tau_g = 1.83\times 10^{12}$ rad/s. As the direction of the magnetic field is reversed whilst keeping constant its strength, the scattered pattern rotates up to (a) $60^o$, and (b) to $90^o$.}
  \label{MOC-RotationAngle3}
\end{figure}
\vspace{30pt}
In conclusion, we theoretically investigated EM scattering by a dielectric cylinder coated with a magneto-optical shell and showed that it is possible to actively tune the operation of plasmonic cloaks with an external magnetic field ${\bf B}$. In the long wavelength limit we showed that the application of ${\bf B}$ may drastically reduce the scattering cross section for all observation angles. The presence of the magnetic field can also largely modify the operation range of the proposed magneto-optical cloak in a dynamical way. Indeed, for a fixed $a/b$, we demonstrated that invisibility can be achieved by applying a magnetic field for situations where the condition for transparency cannot be satisfied without ${\bf B}$. Conversely, a magnetic field can suppress invisibility in a system originally designed to act as an invisibility cloak. Together, these results suggest that one can dynamically switch on and off the magneto-optical cloak by applying ${\bf B}$. In terms of frequency, the application of a uniform magnetic field in a system with fixed $a/b$ can largely shift the cloak operation range, both to higher and lower frequencies. Moreover, this tuning mechanism allows not only for a switchable cloak but also to a broadening of the operation frequency band of cloaking devices. In the particular case of a magneto-optical coating shell made of graphene grown on silicon carbide, we demonstrated that cloaking typically shows up in frequency ranges where the absorption cross section is minimal. This result suggests that magneto-optical cloaks could circumvent one of the major problems of many cloaking devices, whose performance is typically decreased due to unavoidable material losses. In addition to cloaking we also showed that it is possible to achieve a large enhancement of scattering by applying an external magnetic field. The interchange between these two distinct and opposite scenarios (enhanced scattering and cloaking) can indeed be tuned by varying the magnetic field, allowing for a high degree of external control of the scattering properties. This control is also possible for the angular distribution of scattered radiation, as the differential scattering radiation is shown to be highly tunable under the influence of an external magnetic field. As a result, the angular distribution of the scattered radiation can be significantly rotated by applying a magnetic field. Altogether, our findings may pave the way for the utilization of magneto-optical tunable cloaking devices with versatile functionalities, such as directional scattering, in disruptive photonic technologies.

\end{chapter}

\begin{chapter}{Quantum theory of spontaneous emission}
\label{cap5}

\begin{flushright}
{\it
No point is more important than this, \\ that empty space is not empty.\\
It is the seat of the most violent physics.
}

{\sc John A. Wheeler}
\end{flushright}

\hspace{5 mm} We cannot imagine the world without light. Light it is not only a fundamental source of energy for living creatures on Earth but also is crucial for scientific advances in physics, telecommunication, medicine, data storage, and so forth. Ultimately, most of the light around us is due to the spontaneous emission (SE) process. In this chapter we discuss the main concepts behind the SE of quantum emitters near arbitrary bodies.  In the context of the Cavity Quantum Electrodynamics (CQED), we establish  appropriate expressions to calculate the SE rate in terms of either EM field modes or Green's dyadic. As selected examples, we study the SE rate of a two-level quantum emitter near perfect and realistic metallic mirrors.

\pagebreak

\section{Introduction}
\label{Intro5}

\hspace{5mm} Quantum Electrodynamics (QED) corresponds to the relativistic quantum field theory of electrodynamics and, in essence, describes light-matter interactions through quantum mechanics and special relativity formalisms. Within this theory several phenomena, such as spontaneous emission, dispersive interactions, Lamb shift, and electron's anomalous moment can be interpreted as direct or indirect manifestations of quantum fluctuations of the QED vacuum (for a detailed discussion on these topics see Ref.~\cite{Milonni-1994} and references therein). The theory that describes the influence of the neighborhood of an atomic system in its properties is known as Cavity Quantum Electrodynamics. The purpose of this kind of study is not only to obtain more precise results concerning the radiative properties of a system, but also to control them. %as, for example, the atomic transition frequencies or the natural line widths of a quantum emitter. %Among the several phenomena related to fluctuations of the quantum (QED) vacuum,
Particularly, we shall be concerned in this thesis with spontaneous emission~\cite{Milonni-1994,Milonni-WhySpontaneousEmission, Eberly} of quantum emitters under the influence of neighboring bodies and dispersive interactions~\cite{CohenEtAl-1973, Milton-Book-2004,Mostepanenko-Book-2009,Buhmann-Welsch-2007} between atomic species and arbitrary objects. Spontaneous emission corresponds to the case where an excited atom decays to the ground state emitting radiation without any apparent influence of external agents on the system. This process is directly related to the natural line widths of the atomic levels and it may be enhanced, weakened or even suppressed depending on appropriately arranged objects in the vicinity of the quantum emitter. Dispersive interactions correspond to forces between neutral but polarizable bodies that do not have permanent electric or magnetic multipoles. They are connected to the energy level shifts of the interacting atoms (molecules). Dispersive interactions will be discussed in more detail in Chapter {\bf \ref{cap7}}.

The spontaneous emission phenomenon is a typical manifestation of vacuum fluctuations. Indeed, even in the case of an emitter far away from any existing body, it will not be absolutely isolated since generally it always interacts with the vacuum EM radiation. This interaction is due to field fluctuations that take place even if the EM field is in its lowest energy state. Consequently, although the initial excited state of an atom is a stationary state of the atomic Hamiltonian it is not a steady state of the full \linebreak atom-field Hamiltonian.

The concepts of both spontaneous and stimulated emission were introduced by Einstein in 1917 \cite{Einstein-1917}. In his seminal paper Einstein intended, among other things, to reobtain Planck's spectral distribution by analyzing the absorption and emission of EM radiation of two-level molecules in thermal equilibrium with radiation at temperature $T$. According to Einstein the molecules at ground state could be excited by the absorption of a photon. The molecules at the excited state might decay to the ground state through either spontaneous or stimulated emission. Using simple statistical and thermodynamical arguments Einstein was able to calculate the ratio between the rates $A_{21}$ and $B_{12}$ at which photons are emitted via SE or absorbed by the molecules. It should be mentioned, however, that Einstein did not compute explicitly the spontaneous emission rate of atoms and molecules in free space, but only the ratio $A_{21}/B_{12}$. The first formal calculation of Einstein's $A_{21}$ coefficient was accomplished in 1927 by Dirac \cite{Dirac27} in a work that is considered for several authors as the initial landmark of QED.

As a consequence of the coupling between the atom and the quantized EM field, the SE process may be affected by the presence of objects in the vicinity of the emitter. Indeed, since the EM field modes are modified by boundary conditions, the presence of bodies in the neighborhood of an atomic system will change its radiative properties. The fact that the environment can strongly modify the spontaneous emission rate of a system is called Purcell effect since it was theoretically predicted by Purcell in 1946 \cite{Purcell-46}. Approximately two decades later, several experimental works  \cite{DrexhageEtAl-68, Drexhage-70, Drexhage-70-2, Drexhage-74} verified the influence of dispersive mirrors in the lifetime of excited Eu$^{3+}$, presenting excellent agreement with theoretical predictions \cite{Morawitz-69}, at least for large atom-wall separations. Also, Kuhn \cite{Kuhn-1970} and Chance {\it et al} \cite{chance-1978} have discussed the necessity of including the contribution of surface plasmon-polaritons to the SE rate in the near-field. Furthermore, different experiments \cite{Rikken-1995, Snoeks-1995, Lukosz-1979, Lukuoz-1987, Yablonovitch-1988} have also measured the decay rate of quantum emitters within a medium, demonstrating the importance of taking into account not only the effects of boundary conditions on SE but also the effects due to the refractive index of the surrounding environment. Generalizations for an atom between two parallel metallic plates were done by many authors \cite{Barton-70, Barton-87, Stehle-70, Milonni-Knight-73, Philpott-73, Alves-Farina-Tort-2000, MendesEtAl-08}, and the possibility of modifying the emission properties of atoms and molecules in a controlled way has stimulated the investigation of SE in different systems and geometries \cite{KortKamp2013, blanco2004, thomas2004, carminati2006, vladimirova2012, klimovreview, Sture-2007, Kien-2000, Klimov-2002, Milonni-2003, Kastel-2005, Klimov-2004, Biehs-2011, hyperbolicreview, jacob2012}.  Even in simple situations, as those with an atom between two parallel mirrors, quite interesting phenomena may occur. For instance, in the case of a transition dipole moment parallel to the mirrors, the SE rate may be suppressed if the distance between the plates is shortened below a critical value~\cite{Barton-70, Barton-87}. Suppression of SE rate was observed by the first time in 1985 by Hulet {\it et al} \cite{Hulet-Hilfer-Kleppner-85} by using a beam of excited Cesium atoms passing between  two parallel mirrors. Later on suppresion was also observed by other groups~\cite{JheEtAl-87,DeMartiniEtAl-87,HeinzenEtAl-87}. There are excellent reviews on how the environment can modify the SE rate of an emitter (such as atoms, molecules, or quantum dots), see, for instance, Refs. \cite{klimovreview, kuhnreview, Milonni-1994, Haroche-Kleppner-89, Hinds-90, Berman-94, Weber-1984, Barnes-1998, Haroche-1992}.

In the last decade the research on the radiative properties of atoms and molecules has received renewed interest due to the progress in near-field optics. Indeed, advances in nano-optics have not only allowed the increase of the spectroscopical resolution of molecules in complex environments~\cite{betzig1993} but also have led to the use of nanometric objects (e.g. nanoparticles and nano-tips) to modify the lifetime and to enhance the fluorescence of single molecules~\cite{bian1995,sanchez1999, blanco2004, thomas2004, carminati2006, vladimirova2012, klimovreview, Klimov-2002, Milonni-2003, Kastel-2005, Klimov-2004}. Another important example is the development of nanoantennas that can enhance the local optical field and modify the fluorescence of single-emitters~\cite{greffet2005,muhlschlegel2005}. The advent of plasmonic devices has also opened new possibilities for tailoring the emission properties of atoms and molecules. When a quantum emitter is located near a plasmonic structure it may experience an enhancement of the local field due to the excitation of a plasmonic resonance, which affects the lifetime of an excited state. This effect has been exploited in the development of important applications in nanoplasmonics, such as surface-enhanced fluorescence and surface-enhanced Raman scattering~\cite{jackson2004,wei2008,li2010} and the modification of two-level atom resonance fluorescence~\cite{vladimirova2012}.
%In the next Subsections we discuss the physics underlining the SE process in both near- and far-field.

The main goal of this chapter is to establish a convenient formalism to compute spontaneous emission rates of atomic systems. In Sections {\bf \ref{SETwoLevelAtoms}} and {\bf \ref{SEGreenFunction}} , we study the influence of boundary conditions (BC) in the radiative properties of quantum emitters, known as the Purcell effect. In particular, we obtain a suitable expression for the spontaneous emission rate in terms of the EM field modes in Sec. {\bf \ref{SETwoLevelAtoms}}. In Sec. {\bf \ref{SEGreenFunction}} we show that the SE rate can be conveniently written in terms of the Green's dyadic function.

\section{Nonrelativistic theory of two-level quantum emitters in vacuum}
\label{SETwoLevelAtoms}

\hspace{5mm} In order to establish basic concepts and notation, as well as a convenient expression for the SE rate of a two-level atom in the presence of an arbitrary arrangement of bodies in its surroundings, we start discussing the atom-field dynamics. For simplicity, we consider an atom that can be well described by two of its eigenstates. In the absence of interaction we assume that the lowest state $\vert g \rangle$, with energy $E_{\textrm{g}} = -\hbar\omega_0/2$, has a very long lifetime and a well-defined parity, while the highest energy state $\vert e \rangle$, with energy $E_{\textrm{e}} = \hbar\omega_0/2$, has a nonvanishing electric dipole coupling to $\vert g \rangle$. Besides, the influence of the aforementioned bodies on the atomic radiative properties are taken into account in this model by the boundary conditions (BC). Bearing this in mind, let us assume that the system can be described by the well-known hamiltonian \cite{Eberly, Milonni-2005}
\begin{equation}
\hat{{\cal H}} = \hat{{\cal H}}_{\textrm{at}} + \hat{{\cal H}}_{\textrm{f}} + \hat{{\cal H}}_{\textrm{int}} ,
\label{HamiltonianoSistema}
\end{equation}
where $\hat{{\cal H}}_{\textrm{at}}$, $\hat{{\cal H}}_{\textrm{f}}$, and $\hat{{\cal H}}_{\textrm{int}}$  are the atomic, the electromagnetic field, and the interaction hamiltonians, respectively. More specifically,
\begin{equation}
\hat{{\cal H}}_{\textrm{at}} = \dfrac{1}{2} \hbar \omega_0 \big(| e \rangle \langle e| - | g\rangle \langle g|\big)
= \dfrac{1}{2} \hbar \omega_0 \hat{\sigma}_{z}\,
\label{HamiltonianoAtomo}
\end{equation}
is the pure hamiltonian describing the atomic internal dynamics, $\omega_0$ is the associated transition frequency and $\hat{\sigma}_x, \hat{\sigma}_y,$ and $\hat{\sigma}_z$ are the Pauli operators \cite{Eberly}. The second term in (\ref{HamiltonianoSistema}) is given by \cite{Milonni-2005}
\begin{equation}
\hat{{\cal H}}_{\textrm{f}} = \dfrac{1}{4} \int_V\left(\varepsilon_0\hat{{\bf E}}^2 + \mu_0\hat{{\bf H}}^2\right)d{\bf r},
\label{HamiltonianoCampo}
\end{equation}
where $\hat{{\bf E}}({\bf r})$ and $\hat{{\bf H}}({\bf r})$ are the quantum electric and magnetic field operators. Furthermore, the atom is assumed to be much smaller than all other relevant length scales of the problem, so that the electric dipole approximation is valid. As a consequence the third term in (\ref{HamiltonianoSistema}) can be written as \cite{Milonni-1994, Eberly, Milonni-2005}
\begin{equation}
\hat{{\cal H}}_{\textrm{int}} = - \hat{{\bf d}}\cdot \hat{{\bf E}}\, ,
\label{HamiltonianoInteracao}
\end{equation}
where $\hat{{\bf d}}$ is the atomic electric dipole operator and, in the previous equation, $\hat{{\bf E}}$ must be evaluated at the position of the atom ({\it i.e.} its center of mass, since in the dipole approximation we neglect the variation of the field along the atom).  We assume that the atom has no permanent electric dipole moment, thus $\langle g\vert\hat{{\bf d}}\vert g\rangle = \langle e\vert\hat{{\bf d}}\vert e\rangle = {\bf 0}$. Consequently, in the two-level approximation $\hat{{\bf d}}$ takes the simple form $\hat{{\bf d}} = {\bf d}_{\textrm{eg}}\hat{\sigma}_x$, where ${\bf d}_{\textrm{eg}} = \langle e\vert\hat{{\bf d}}\vert g\rangle$ is the transition electric dipole moment that can be made real by an appropriate choice of the relative phase between the ground and the excited states. 

The electromagnetic and interaction Hamiltonians may be recast into more convenient forms by expanding the EM field in their normal modes ${\bf A}_{\zeta}({\bf r})$, namely \cite{Milonni-1994,Milonni-2005},
\begin{eqnarray}
\hat{{\bf E}}({\bf r})\!\!\! &=&	\!\!\! i\sum_{\zeta}\sqrt{\dfrac{\hbar\omega_{\zeta}}{2\varepsilon_0}}
\ [\hat{a}_{\zeta} {\bf A}_{\zeta}({\bf r})-\hat{a}_{\zeta}^{\dagger} {\bf A}_{\zeta}^{*}({\bf r})],
\label{campoquantizadogeral0}\\\cr
\hat{{\bf H}}({\bf r})\!\!\! &=&	 \!\!\! \sum_{\zeta} \sqrt{\dfrac{\hbar c^2}{2\mu_0\omega_{\zeta}}}
\ \nabla \!\times\! [\hat{a}_{\zeta}{\bf A}_{\zeta}({\bf r}) - \hat{a}_{\zeta}^{\dagger}{\bf A}_{\zeta}^*({\bf r})],
\label{campoquantizadogeral}
\end{eqnarray}
where the label $\zeta$ represents an arbitrary complete set of quantum numbers, $\omega_{\zeta}$ are the EM mode frequencies, and $\hat{a}_{\zeta}, \hat{a}_{\zeta}^{\dagger}$ are the annihilation and creation operators that contain all quantum properties of the field and satisfy the following bosonic commutation relations:
\begin{eqnarray}
[\hat{a}_{\zeta},\hat{a}_{\zeta'}^{\dagger}] = \delta_{\zeta\zeta'}\, , \ \ [\hat{a}_{\zeta},\hat{a}_{\zeta'}] = 0\, ,\ \ \textrm{and}\ \ [\hat{a}_{\zeta}^{\dagger},\hat{a}_{\zeta'}^{\dagger}] =0\, .
\end{eqnarray}
In addition, the EM modes ${\bf A}_{\zeta}({\bf r})$ are classical functions determined by the Coulomb gauge $\nabla \cdot {\bf A}_{\zeta}  ({\bf r})= 0$ and homogeneous Helmholtz equation
\begin{equation}
\label{HelmholtzModes}
\nabla \times \nabla \times {\bf A}_{\zeta}({\bf r}) - \dfrac{\omega_{\zeta}^2}{c^2} {\bf A}_{\zeta}({\bf r})=0\, ,
\end{equation}
with the appropriate boundary conditions. Substituting Eqs. (\ref{campoquantizadogeral0}) and (\ref{campoquantizadogeral}) into (\ref{HamiltonianoCampo}) and choosing an orthonormal set of functions ${\bf A}_{\zeta}({\bf r})$, {\it i.e.} $\int d{\bf r} {\bf A}_{\zeta}({\bf r}) \cdot {\bf A}_{\zeta'}^{*}({\bf r}) = \delta_{{\bf \zeta\zeta}'}$, it can be shown that
\begin{equation}
\hat{{\cal H}}_{\textrm{f}} = \sum_{\zeta} \hbar\omega_{\zeta}\hat{a}^{\dagger}_{\zeta}\hat{a}_{\zeta}\, ,
\label{HamiltonianoCampo2}
\end{equation}
where we neglected the zero-point contribution since it does not contribute to the atom-field dynamics. Analogously, the interaction hamiltonian can be cast as\cite{Eberly,Milonni-1994,Milonni-2005}
\begin{eqnarray}
\hat{{\cal H}}_{\textrm{int}} = -i\hbar\sum_{\zeta}\hat{\sigma}_{x}\left[g_{\zeta}\hat{a}_{\zeta}
 -g^*_{\zeta}\hat{a}_{\zeta}^{\dagger} \right] ,
\label{HamiltonianoInteracao2}
\end{eqnarray}
where we defined
\begin{equation}
g_{\zeta} := \sqrt{\dfrac{\omega_{\zeta}}{2\varepsilon_0\hbar}}\; {\bf d}_{\textrm{eg}} \cdot
{\bf A}_{\zeta}\, .
\label{gzeta}
\end{equation}

In order to obtain to the SE rate, we have to determine the time evolution of the expectation value of $\hat{{\cal H}}_{\textrm{at}}$. Formally, this can be accomplished in the Heisenberg picture by calculating the expectation value of $\hat{\sigma}_z(t)$ in the initial state of the system, where the atom is assumed to be in the highest energy level $|e\rangle$ and the field in the vacuum state $|\textrm{vacuum}\rangle$. Applying the Heisenberg equation, it is straightforward to show that the atomic and field operators satisfy the following equations\cite{Eberly, Milonni-1994, Milonni-2005}
\begin{eqnarray}
\dot{{\hat{\sigma}}}^{\dagger}(t)\!\!\! &=&\!\!\! i\omega_0\hat{\sigma}^{\dagger}(t)\! -\!\!\! \sum_{\zeta}
%&&
\!\!\left[g_{\zeta}\hat{\sigma}_{z}(t)\hat{a}_{\zeta}(t) - g_{\zeta}^{*}\hat{a}_{\zeta}^{\dagger}(t)\hat{\sigma}_{z}(t)\!\right]\!\! , \label{EqSigmaDagger} \\
\dot{{\hat{\sigma}}}_z(t)\!\!\! &=&\!\!\! - 2 i  \sum_{\zeta} \left[ g_{\zeta}\hat{\sigma}_y(t)\hat{a}_{\zeta}(t)
- g_{\zeta}^{*}\hat{a}_{\zeta}^{\dagger}(t)\hat{\sigma}_y(t)\right]\! ,\label{EqSigmaZ} \\
\dot{{\hat{a}}}_{\zeta}(t)\!\!\! &=&\!\!\! -i\omega_{\zeta} \hat{a}_{\zeta}(t) + g_{\zeta}^{*}\hat{\sigma}_{x}(t) ,
\label{EqOp_a}
\end{eqnarray}
where $\hat{\sigma} = (\hat{\sigma}_x + i\hat{\sigma}_y)/2$, and  $\hat{\sigma}^{\dagger} = (\hat{\sigma}_x - i\hat{\sigma}_y)/2$ play the role of the Pauli lowering and raising operators. They are related to $\hat{\sigma}_z$ through the following commutation relations
\begin{eqnarray}
[\hat{\sigma},\hat{\sigma}_{z}] = 2 \hat{\sigma} , \;
[\hat{\sigma}^{\dagger},\hat{\sigma}_{z}] = -2 \hat{\sigma}^{\dagger}\; , \ \ \textrm{and} \ \
\left[ \hat{\sigma},\hat{\sigma}^{\dagger} \right]= -{\hat{\sigma}}_{z} .
\label{CommutationRelations}
\end{eqnarray}
Besides, since equal-time atomic and field operators commute, we have written the Heisenberg equations in the normal ordering in which annihilation (creation) operators appear at the right (left) in operator products.

The coupled Eqs. (\ref{EqSigmaDagger}), (\ref{EqSigmaZ}), (\ref{EqOp_a}) are quite difficult to solve, even for the simple system considered here. Besides, a direct integration in Eq. (\ref{EqOp_a}) shows that the relationship between $\hat{a}_\zeta(t)$  and $\hat{\sigma}_x(t)$ is non-local in time, with the field (atomic) operators depending on the atomic (field) operators in earlier times. Fortunately, considering that the atom is weakly coupled to the field, it is possible to simplify considerably these equations by performing a Markovian approximation; after this we arrive at \cite{Milonni-1994, Eberly}
\begin{eqnarray}
\hat{a}_{\zeta}(t) \simeq \hat{a}_{\zeta}^{v}(t) - i g_{\zeta}^{*}
\left[\hat{\sigma}^{\dagger}(t) \xi^{*}(\omega_{\zeta}+\omega_{0}) \right.
+ \left. \hat{\sigma}(t)\xi^{*}(\omega_{\zeta}-\omega_{0}) \right],
\label{Solucaoadagger}
\end{eqnarray}
where $\hat{a}_{\zeta}^{v}(t) =  \hat{a}_{\zeta}(0) e^{-i\omega_{\zeta} t}$ is the free field homogeneous solution of (\ref{EqOp_a}) and the term in brackets is the part due to the source, which in the present case is a two-level quantum emitter. Also, $\xi(x)$ is defined as
\begin{eqnarray}
\xi(x):= {\cal {P}}\left(\dfrac{1}{x}\right)-i\pi\delta(x),
\label{funcaoxi}
\end{eqnarray}
with ${\cal{P}}$ denoting the principal Cauchy value and $\delta(x)$ being the Dirac delta function. Using the last two equations we can rewrite (\ref{EqSigmaZ}) in a simpler way,
\begin{eqnarray}
\dot{\hat{\sigma}}_z(t) = -2i\sum_{\zeta}  \left[ g_{\zeta}\hat{\sigma}_y(t)
\hat{a}_{\zeta}^{v}(t) - g_{\zeta}^{*}\hat{a}_{\zeta}^{v \dagger}(t)\hat{\sigma}_y(t) \right]
-2\pi \left[ \mathds{1} + \hat{\sigma}_{z}(t)\right] \sum_{\zeta} |g_{\zeta}|^2
\delta(\omega_{\zeta}-\omega_{0}).
\label{EqSigmaZ2}
\end{eqnarray}

Taking now the expectation value of Eq. (\ref{EqSigmaZ2}) in the initial state $\vert i \rangle = \vert e \rangle \otimes \vert\textrm{vacuum} \rangle$ and using that $\hat{a}_{\zeta}^{v}(t)\vert\textrm{vacuum} \rangle = 0$, we obtain
\begin{eqnarray}
\langle\hat{\sigma}_{z}(t)\rangle_e = -1 + \left[ \langle\hat{\sigma}_{z}(0)\rangle_e + 1 \right]
e^{- \Gamma t}  = -1 + 2 e^{- \Gamma t},
\label{EqSigmaZ3}
\end{eqnarray}
where $\langle(...)\rangle_\textrm{e} = \langle e\vert(...)\vert e\rangle$, we used that $\langle\hat{\sigma}_{z}(0)\rangle_{\textrm{e}} =1$, and the SE rate $\Gamma({\bf r})$ is \linebreak identified as \cite{Milonni-1994,Eberly}
\begin{eqnarray}
\Gamma({\bf r}) = 2\pi  \sum_{\zeta} |g_{\zeta}|^2 \delta(\omega_{\zeta}-\omega_{0})= \dfrac{\pi \omega_0}{\varepsilon_0 \hbar} \sum_{\zeta} |{\bf d}_{\textrm{eg}} \cdot {\bf A}_{\zeta}({\bf r})|^2\delta(\omega_{\zeta}-\omega_{0})\, .
\label{TaxaEmissaoEspontanea}
\end{eqnarray}

Equation (\ref{TaxaEmissaoEspontanea}) is exactly the same as that obtained via the Weisskopf-Wigner theory \cite{Milonni-1994} for the SE rate of a two-level quantum emitter in the presence of bodies of arbitrary shape. The influence of such bodies is encoded in the functions ${\bf A}_{\zeta}({\bf r})$ that, as mentioned before, are classical solutions of the Helmholtz equation. This is extremely convenient, as it makes the problem amenable to various well-developed analytical and \linebreak numerical techniques.

Finally, it is interesting to mention that by analyzing the time evolution of $\langle \sigma(t) \rangle$ it is possible to obtain the shift in the oscillation frequency of the expectation value of the electric dipole operator. Indeed, making use of a Markovian and a rotating-wave approximations it is straightforward to show that
\begin{eqnarray}
\langle \dot{\hat{\sigma}}(t) \rangle = -i(\omega_0+\Delta) \langle {\hat{\sigma}}(t) \rangle - \dfrac{\Gamma}{2} \langle {\hat{\sigma}}(t) \rangle\, ,
\end{eqnarray}
where \cite{Milonni-1994,Eberly}
\begin{equation}
\label{LambShift}
\Delta({\bf r}) = -\dfrac{\omega_0}{\varepsilon_0\hbar} {\cal {P}} \left\{\sum_{\zeta} |{\bf d}_{\textrm{eg}} \cdot
{\bf A}_{\zeta}^{\textrm{(sca)}}({\bf r})|^2\ \dfrac{\omega_{\zeta}}{\omega_\zeta^2-\omega_0^2}\right\}\,
\end{equation}
corresponds to the contribution of the neighboring bodies to the Lamb shift. It is important to emphasize that in the last equation ${\bf A}_{\zeta}^{\textrm{(sca)}}$ are the scattered electromagnetic field modes and the divergent contribution to the frequency shift owing to the interaction between the emitter and the free space EM radiation is supposed to be already included in the definition of $\omega_0$.

\subsection{Free space spontaneous emission}

\hspace{5mm} As an easy but important check, and also for future convenience, we shall use Eq.~(\ref{TaxaEmissaoEspontanea}) to obtain the SE rate of a two-level system in free space. In this case, field modes are simply given by \cite{Milonni-1994, Milonni-2005}
\begin{equation}
{\bf A}_{\zeta}({\bf r}) \, \Longrightarrow \, {\bf A}_{{\bf k}p}^{(0)}({\bf r}) = \dfrac{e^{i{\bf k}\cdot {\bf r}}}{\sqrt{V}} \mbox{{\mathversion{bold}${\epsilon}$}}_{{\bf k} p},
\label{ModesFreeSpace}
\end{equation}
where $p = 1,\ 2$,  $\boldsymbol{\epsilon}_{{\bf k} 1}$ and $\boldsymbol{\epsilon}_{{\bf k} 2}$ are two orthogonal polarization vectors, and $V$ is the quantization volume. Inserting Eq. (\ref{ModesFreeSpace}) into Eq. (\ref{TaxaEmissaoEspontanea}) and using that
\begin{equation}
\displaystyle{\sum_{\zeta}} \rightarrow \displaystyle{\sum_{{\bf k}, \ p}} \rightarrow \dfrac{V}{8\pi^3}\displaystyle{\sum_{p=1, \ 2}}\ \displaystyle{\int} d{\bf k}\, ,
\end{equation}
we obtain
\begin{eqnarray}
\Gamma_{0} &=& \dfrac{\pi \omega_0}{\varepsilon_0\hbar} \dfrac{V}{8\pi^3} \dfrac{|{\bf d}_{\textrm{eg}}|^2}{V}  \sum_{p=1,2}
\int d\Omega \cos^2 \theta \sin\theta \int_0^{\infty}  dk\ k^2  \,\delta(\omega_{\bf k}-\omega_{0})\cr
&=& \dfrac{|{\bf d}_{21}|^2 \omega_{0}^3}{3\pi\varepsilon_0\hbar c^3},
\label{GammaFreeSpace}
\end{eqnarray}
which is the well-known result firstly obtained by P.A.M. Dirac in 1927 \cite{Dirac27} for the Einstein's $A_{21}$ coefficient.

\subsection{Quantum emitter near a perfect mirror}

\hspace{5mm} As a second example, let us consider the influence of a flat perfect mirror at $z = 0$ on the SE rate of a two-level atom located a distance $z$ above the wall. In this case, the appropriate field mode functions in the half-space $z\geq 0$ that satisfy the Helmholtz equation and the boundary condition ${\bf\hat{z}} \times {\bf E}|_{z = 0} = 0$ are given by \cite{Milonni-1994, Hinds-90}
\begin{eqnarray}
\label{FieldModes1Wall1}
{\bf A}_{{\bf k}\textrm{TE}}({\bf r})\!\!\! &=&\!\!\! \sqrt{\dfrac{2}{V}} ({\bf \hat{k}}_{||} \times {\bf \hat{z}}) \sin(k_z z) e^{i{\bf k}_{||}\cdot {\bf r}}\, , \\
\label{FieldModes1Wall2}
{\bf A}_{{\bf k}\textrm{TM}}({\bf r})\!\!\! &=&\!\!\! \dfrac{1}{k}\sqrt{\dfrac{2}{V}}
\left[k_{||} {\bf \hat{z}} \cos(k_z z) - i k_z {\bf \hat{k}}_{||} \sin(k_z z) \right]
e^{i{\bf k}_{||}\cdot {\bf r}} \, ,
\end{eqnarray}
where we have decomposed the wave vector ${\bf k}$ in its components parallel ${\bf k}_{||}$ and perpendicular $k_z\hat{{\bf z}}$ to the conducting surface.

In this case, if we write  ${\bf d}_{\textrm{eg}} = d_{\perp} {\bf \hat{z}} + {\bf d}_{||}$ it is possible to investigate the lifetime of the quantum emitter when it is properly prepared to decay with its electric dipole vector parallel or perpendicular to the metallic wall. Indeed, using (\ref{FieldModes1Wall1}) and (\ref{FieldModes1Wall2}) we can show that the SE rate of a two-level atom near a conducting plane can be written as \cite{Milonni-1994,Hinds-90}
\begin{eqnarray}
\label{SEPerfectWallPerp}
\dfrac{\Gamma_{\perp}(z)}{\Gamma_0}\!\!\! &=&\!\!\! \dfrac{d_{\perp}^2}{|{\bf d}_{\textrm{eg}}|^2} \left\{1 - \dfrac{3\cos(2k_0z)}{(2k_0z)^2} +  \dfrac{3\sin(2k_0z)}{(2k_0z)^3}\right\}\, , \\
\label{SEPerfectWallPar}
\dfrac{\Gamma_{||}(z)}{\Gamma_0}\!\!\! &=&\!\!\! \dfrac{d_{||}^2}{|{\bf d}_{\textrm{eg}}|^2} \left\{1 - \dfrac{3\sin(2k_0z)}{2(2k_0z)} - \dfrac{3\cos(2k_0z)}{2(2k_0z)^2} + \dfrac{3\sin(2k_0z)}{2(2k_0z)^3}\right\}\, ,
\end{eqnarray}
where $k_0 = \omega_0/c$, $\Gamma_{\perp}(z)$ and $\Gamma_{||}(z)$ are the contributions of ${\bf d}_{\perp}$ and ${\bf d}_{||}$ to the decay rate of the atomic system. If the atomic initial state is prepared so that all electric dipole transition orientations are equally likely to occur, the SE rate is given by summing the above equations and taking an orientational average~\cite{Milonni-1994,Hinds-90}, namely, $ \langle d_{\perp}^2 \rangle = |{\bf d}_{\textrm{eg}}|^2/3 $ and $\langle d_{||}^2 \rangle = 2|{\bf d}_{\textrm{eg}}|^2/3$.

In Fig. \ref{SEPerfectWall} we plot $\Gamma_{\perp}/\Gamma_0$, $\Gamma_{||}/\Gamma_0$, and $\Gamma_{\textrm{iso}}/\Gamma_0 = (\Gamma_{\perp}/\Gamma_0)/3 + 2(\Gamma_{||}/\Gamma_0)/3$  as functions of $k_0 z$. It should be noticed that for $k_0z \rightarrow 0$ the SE rate is canceled in the parallel configuration and is twice the value as the free-space $\Gamma_0$ in the perpendicular configuration. This result can be easily interpreted in terms of the image method if we consider the two-level quantum emitter as a classical radiating dipole \cite{Haroche-1992}. Indeed, in the perpendicular configuration the real and image dipoles [see inset \ref{SEPerfectWall}(a)]  add up for  $k_0z \rightarrow 0$,  giving a overall electric dipole that is twice the real dipole\footnote{\label{note9} It should be mentioned that if the amplitude of an oscillating dipole is doubled the power radiated in a $4\pi$ solid angle will be multiplied by four. Here the SE rate increases only by a factor two because the real and image dipoles radiate in a $2\pi$ solid angle.}. On the other hand, in the parallel configuration the real and image dipoles oscillate completely out of phase [see inset \ref{SEPerfectWall}(b)] so that in the limit $k_0z \rightarrow 0$ they cancel each other resulting in a total suppression of the dipole radiation. Furthermore, for $k_0z \rightarrow \infty$ the SE rate approaches $\Gamma_0$, as expected. In the intermediary regime of distances, the decay rate presents oscillations as a function of $k_0z$ with a spacial period of $\pi/k_0$. This behavior is a direct consequence of interference effects between the two distinct ways that a photon can be emitted into the same mode: in the first one, the atom emits the photon into the mode ({\bf k}, p) without reflection at the surface whereas in the second one, the photon ends in mode ({\bf k}, p) after a reflection at the mirror. Finally, it is  interesting to note that $\Gamma_{||}$ presents stronger oscillations than $\Gamma_{\perp}$. This is a consequence of the fact that the transverse electric modes have the electric field parallel to the $xy-$plane, thus not coupling with a transition electric dipole perpendicular to the mirror. Therefore, while ${\bf d}_{||}$ may see the influence of the mirror in both TE- and TM-modes $d_{\perp}$ detects the presence of the conductor half-space just through TM-modes.
 \vspace{10pt}
\begin{figure}[!ht]
  \centering
  \includegraphics[scale=0.6]{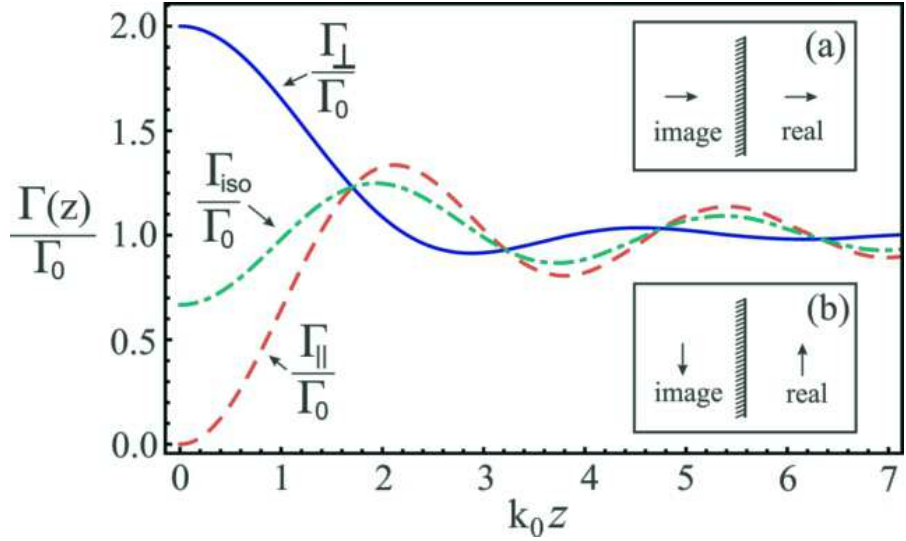}
   \vspace{10pt}
  \caption{Spontaneous emission rates $\Gamma_{\perp}/\Gamma_0$ (solid blue curve), $\Gamma_{||}/\Gamma_0$ (dashed red curve), and $\Gamma_{\textrm{iso}}/\Gamma_0$ (dot-dashed green curve) as a function of $k_0z$. Insets {\bf (a)} and {\bf (b)} show the classical analogue of a quantum emitter with its transition electric dipole parallel and perpendicular to the flat mirror, respectively.}
  \label{SEPerfectWall}
\end{figure}

\section{Spontaneous emission and the Green's dyadic}
\label{SEGreenFunction}

\hspace{5mm} In some cases, depending on the geometry and/or material properties of the bodies in the vicinities of the quantum emitter, the calculation of all EM field modes ${\bf A}_{\zeta}({\bf r})$ may be a daunting task. However, there are several analytical and numerical methods for calculating dyadic Green's functions. Our goal in this section is to derive a convenient expression for the SE rate in terms of the Green's function.

For further reference, it is convenient to rewrite the expression for the SE rate given by Eq. (\ref{TaxaEmissaoEspontanea}) as
\begin{equation}
\label{TaxaEmissaoEspontanea2}
\Gamma({\bf r}) = \dfrac{\pi \omega_0}{\varepsilon_0 \hbar} {\bf d}_{\textrm{eg}}\cdot \left[\sum_{\zeta} {\bf A}_{\zeta}^{*}({\bf r}) {\bf A}_{\zeta}({\bf r}) \delta(\omega_{\zeta}-\omega_{0})\right] \cdot {\bf d}_{\textrm{eg}}\, ,
\end{equation}
where  ${\bf A}_{\zeta}({\bf r}) {\bf A}_{\zeta}^{*}({\bf r})$ denotes the outer product ({\it i.e.} a $3 \times 3$ matrix). We will show  that, apart from a multiplicative factor,  the term in brackets corresponds to the imaginary part of the EM dyadic Green's function $\mathbb{G}({\bf r}, {\bf r}'; \omega)$, which satisfies the inhomogeneous Helmholtz equation with a punctiform source at position ${\bf r}'$, namely,
\begin{equation}
\label{HelmholtzGreenFunction}
\nabla \times \nabla \times \mathbb{G}({\bf r}, {\bf r}'; \omega) - \dfrac{\omega^2}{c^2} \mathbb{G}({\bf r}, {\bf r}'; \omega) = \mathbb{I} \delta({\bf r} - {\bf r}')\, ,
\end{equation}
with $\mathbb{I}$ being the unit dyad.

Once the EM field modes defined by Eq. (\ref{HelmholtzModes}) and boundary conditions form a set of orthonormal vectorial functions we can expand $\mathbb{G}({\bf r}, {\bf r}'; \omega)$ in terms of ${\bf A}_{\zeta}({\bf r})$ as follows
\begin{equation}
\label{GreenExpansion1}
\mathbb{G}({\bf r}, {\bf r}'; \omega) = \sum_{\zeta} {\bf C}_{\zeta}({\bf r}',\omega){\bf A}_{\zeta}({\bf r})
\end{equation}
where the coefficients $ {\bf C}_{\zeta}({\bf r}',\omega)$ are yet to be determined. By substituting (\ref{GreenExpansion1}) in (\ref{HelmholtzGreenFunction}) we obtain
\begin{equation}
\sum_{\zeta} {\bf C}_{\zeta}({\bf r}',\omega) \left[\nabla \times \nabla \times {\bf A}_{\zeta}({\bf r}) - \dfrac{\omega^2}{c^2} {\bf A}_{\zeta}({\bf r}) \right] = \mathbb{I} \delta({\bf r} - {\bf r}')\, .
\end{equation}
Using Eq. (\ref{HelmholtzModes}) the last equation can be cast as
\begin{equation}
\sum_{\zeta} {\bf C}_{\zeta}({\bf r}',\omega) (\omega_{\zeta}^2 - \omega^2){\bf A}_{\zeta}({\bf r}) = c^2 \mathbb{I} \delta({\bf r} - {\bf r}')\, .
\end{equation}
Performing the inner product by ${\bf A}_{\zeta'}^{*}({\bf r})$ on the right of both sides of the above equation, integrating over the whole space, and using the orthonormality of the EM field modes, it is straightforward to show that
\begin{equation}
{\bf C}_{\zeta'}({\bf r}', \omega) = c^2 \dfrac{{\bf A}_{\zeta'}^{*}({\bf r'})}{\omega_{\zeta'}^2 - \omega^2}\, .
\end{equation}
Therefore, the dyadic Green's function can be written in terms of the vectorial functions ${\bf A}_{\zeta}({\bf r})$ as \cite{Novotny}
\begin{equation}
\label{GreenExpansion2}
\mathbb{G}({\bf r}, {\bf r}'; \omega) = c^2\sum_{\zeta}\dfrac{{\bf A}_{\zeta}^{*}({\bf r'}){\bf A}_{\zeta}({\bf r})}{\omega_{\zeta}^2 - \omega^2}\, .
\end{equation}
Taking the imaginary part of Eq. (\ref{GreenExpansion2}) and using the well known result \cite{Novotny}
\begin{equation}
\lim_{\eta \to 0} \textrm{Im}\left[ \dfrac{1}{\omega_{\zeta}^2 - (\omega_0 + i\eta)^2} \right] =
\dfrac{\pi}{2\omega_{\zeta}} [\delta(\omega_{\zeta} - \omega_0) - \delta(\omega_{\zeta} + \omega_0)]\, ,
\end{equation}
which can be demonstrated by complex contour integration, we can show that
\begin{equation}
\label{ImaginaryPartGreenFunction}
\textrm{Im}\ \mathbb{G}({\bf r}, {\bf r}; \omega_0) = \dfrac{\pi c^2}{2\omega_0} \sum_{\zeta} {\bf A}_{\zeta}^{*}({\bf r}){\bf A}_{\zeta}({\bf r}) \delta(\omega_{\zeta} - \omega_0)\, ,
\end{equation}
where we have neglected the $\delta(\omega_{\zeta} + \omega_0)$ term since we are assuming that $\omega_0$ is a positive frequency. Hence, using Eq. (\ref{ImaginaryPartGreenFunction}) in (\ref{TaxaEmissaoEspontanea2}) we finally arrive at the desired result \cite{Novotny}
\begin{equation}
\label{TaxaEmissaoEspontanea3}
\Gamma({\bf r}) = \dfrac{2\omega_0^2|{\bf d}_{\textrm{eg}}|^2}{\varepsilon_0 \hbar c^2} \textrm{Im} \left[ \mbox{{\mathversion{bold}${\hat{\eta}}$}} \cdot \mathbb{G}({\bf r}, {\bf r}; \omega_0) \cdot \mbox{{\mathversion{bold}${\hat{\eta}}$}} \right]\, ,
\end{equation}
where $\mbox{{\mathversion{bold}${\hat{\eta}}$}}$ is the unitary vector pointing in the same direction as ${\bf d}_{\textrm{eg}}$. The last equation is the main result of this section. It allows us to calculate the decay rate of a two-level quantum emitter in the presence of an arbitrary arrangement of bodies in its neighborhood provided the dyadic Green's function is known.

For completeness we now derive an appropriate expression for the Lamb shift in terms of the Green's dyadic \cite{Dung-2001, Yao-2009, Sokhoyan-2013, Angelatos-2014}
\begin{eqnarray}
\Delta({\bf r})\!\!\! &=&\!\!\! -\dfrac{\omega_0}{\varepsilon_0\hbar}
{\bf d}_{\textrm{eg}}\cdot {\cal {P}} \left\{\sum_{\zeta} {{\bf A}_{\zeta}^{\textrm{(sca)}}}^{*}({\bf r}) {\bf A}_{\zeta}^{\textrm{(sca)}}({\bf r}) \ \dfrac{\omega_{\zeta}}{\omega_\zeta^2-\omega_0^2}\right\} \cdot {\bf d}_{\textrm{eg}}\cr
\!\!\!&=&\!\!\! -\dfrac{\omega_0}{\varepsilon_0\hbar}
{\bf d}_{\textrm{eg}}\cdot {\cal {P}} \left\{\sum_{\zeta} {{\bf A}_{\zeta}^{\textrm{(sca)}}}^{*}({\bf r}) {\bf A}_{\zeta}^{\textrm{(sca)}}({\bf r}) \ \int_0^{\infty}d\omega_{\zeta'}\dfrac{\omega_{\zeta'}}{\omega_{\zeta'}^2-\omega_0^2}\delta(\omega_{\zeta'}-\omega_{\zeta}) \right\}
\cdot {\bf d}_{\textrm{eg}}\cr
\!\!\!&=&\!\!\! -\dfrac{\omega_0}{\varepsilon_0\hbar}
{\bf d}_{\textrm{eg}}\cdot {\cal {P}} \left\{ \int_0^{\infty}d\omega_{\zeta'}  \dfrac{\omega_{\zeta'}}{\omega_{\zeta'}^2-\omega_0^2} \sum_{\zeta} {{\bf A}_{\zeta}^{\textrm{(sca)}}}^{*}({\bf r}) {\bf A}_{\zeta}^{\textrm{(sca)}}({\bf r}) \ \delta(\omega_{\zeta'}-\omega_{\zeta})\right\}
\cdot {\bf d}_{\textrm{eg}} \cr
\!\!\!&=&\!\!\! -\dfrac{\omega_0^2}{\varepsilon_0\hbar c^2}
{\bf d}_{\textrm{eg}}\cdot \dfrac{2}{\pi}  {\cal {P}}\! \int_0^{\infty}d\omega_{\zeta'}  \dfrac{\omega_{\zeta'} \textrm{Im}\ \mathbb{G}^{\textrm{(sca)}}({\bf r}, {\bf r}; \omega_{\zeta'})}{\omega_{\zeta'}^2-\omega_0^2} \cdot {\bf d}_{\textrm{eg}}\cr
\!\!\!&=&\!\!\! -\dfrac{\omega_0^2 |{\bf d}_{\textrm{eg}}|^2}{\varepsilon_0\hbar c^2}
\textrm{Re}\ [\mbox{{\mathversion{bold}${\hat{\eta}}$}} \cdot \mathbb{G}^{\textrm{(sca)}}({\bf r}, {\bf r}; \omega_0) \cdot \mbox{{\mathversion{bold}${\hat{\eta}}$}}]\, ,
\end{eqnarray}
where in the third step we used Eq. (\ref{ImaginaryPartGreenFunction}) and in the last one we applied the Kramers-Kr\"onig relations \cite{Jackson}.

\subsection{Spontaneous emission near a dispersive half-space}

\hspace{5mm} In this section we will discuss the spontaneous emission rate of a two-level quantum emitter located a distance $z$ above a semi-infinite dispersive homogeneous medium with flat surface at $z = 0$. For convenience, let us explicitly split the Green's dyadic as
\begin{eqnarray}
\label{GreenFunctionSplit}
\mathbb{G}({\bf r}, {\bf r'}; \omega_0)  = \mathbb{G}^{\textrm{(0)}}({\bf r}, {\bf r'}; \omega_0) +\mathbb{G}^{\textrm{(sca)}}({\bf r}, {\bf r'}; \omega_0)\, ,
\end{eqnarray}
where $\mathbb{G}^{\textrm{(0)}}({\bf r}, {\bf r'}; \omega_0)$  and $\mathbb{G}^{\textrm{(sca)}}({\bf r}, {\bf r'}; \omega_0)$ are the free space and scattered dyadic Green's functions. For $z\geq z'$ the free space contribution to the Green's function reads \cite{Novotny}
\begin{equation}
\label{FreeSpaceGreenFunction}
\mathbb{G}^{\textrm{(0)}}({\bf r}, {\bf r'}; \omega) = \dfrac{i}{2} \int \dfrac{d{\bf k}_{||}}{(2\pi)^2} \dfrac{\mathbb{M} e^{i[{\bf k}_{||}\cdot ({\bf r} - {\bf r}') + {k_z}_0(z-z')]}}{{k_{z}}_0}\, ,
\end{equation}
where
\begin{equation}
{k_z}_0 = \left\{ \!\!\!
\begin{tabular}{c}
$\sqrt{k_0^2-k_{x}^2 - k_{y}^2} = \sqrt{k_0^2-k_{||}^2} := \xi\, , \ \ \ k_{||} \leq k_0$ \\
									   \\
$i \sqrt{k_{x}^2 + k_{y}^2-k_0^2} = i\sqrt{k_{||}^2 - k_0^2}  := i \zeta\, , \ \ \ k_{||}> k_0$
\end{tabular}
\right.
\label{kz0}
\end{equation}
and tensor $\mathbb{M}$ is given by
\begin{equation}
\mathbb{M} = \mbox{{\mathversion{bold}${\epsilon}$}}_{\textrm{TE}}^{+} \otimes  \mbox{{\mathversion{bold}${\epsilon}$}}_{\textrm{TE}}^{+} +  \mbox{{\mathversion{bold}${\epsilon}$}}_{\textrm{TM}}^{+} \otimes  \mbox{{\mathversion{bold}${\epsilon}$}}_{\textrm{TM}}^{+}\, ,
\end{equation}
with
\begin{eqnarray}
\mbox{{\mathversion{bold}${\epsilon}$}}_{\textrm{TE}}^{-} = \mbox{{\mathversion{bold}${\epsilon}$}}_{\textrm{TE}}^{+} = \dfrac{-k_y{\bf \hat{x}} + k_x {\bf \hat{y}}}{k_{||}}\, , \ \  \textrm{and} \ \  \mbox{{\mathversion{bold}${\epsilon}$}}_{\textrm{TM}}^{\pm} = \dfrac{\pm{k_z}_0(k_x{\bf \hat{x}} + k_y{\bf \hat{y}}) - k_{||}^2{\bf \hat{z}}}{k_{||}k_0}\,
\end{eqnarray}
being the polarization vectors for TE- and TM-polarized waves. Notice that these vectors are orthogonal, but they are normalized only for propagating modes ($k_{||} < k_0$). Besides, at the coincidence (${\bf r}' = {\bf r}$) the only nonzero components of  $\mathbb{G}^{\textrm{(0)}}({\bf r}, {\bf r}; \omega)$ are $\mathbb{G}^{\textrm{(0)}}_{xx}({\bf r}, {\bf r}; \omega) = \mathbb{G}^{\textrm{(0)}}_{yy}({\bf r}, {\bf r}; \omega)$ and $\mathbb{G}^{\textrm{(0)}}_{zz}({\bf r}, {\bf r}; \omega)$.

Similarly, the reflected Green's dyadic can be conveniently expressed as \cite{Novotny, Felipe-2011}
\begin{equation}
\label{ReflectedGreenFunction}
\mathbb{G}^{\textrm{(sca)}}({\bf r}, {\bf r'}; \omega) = \dfrac{i}{2} \int \dfrac{d{\bf k}_{||}}{(2\pi)^2} \dfrac{\mathbb{R} e^{i[{\bf k}_{||}\cdot ({\bf r} - {\bf r}') + {k_z}_0(z+z')]}}{{k_{z}}_0}\, ,
\end{equation}
where matrix $\mathbb{R}$ is given in terms of the reflection coefficients as follows
\begin{equation}
\label{ReflectionMatrix}
\mathbb{R}\ \ =\!\!\! \sum_{i\, , \ j = \{\textrm{TE, TM}\}}\!\!\!\!\!\!\!\! r^{\textrm{i, j}}   \mbox{{\mathversion{bold}${\epsilon}$}}_{\textrm{i}}^{+} \otimes \mbox{{\mathversion{bold}${\epsilon}$}}_{\textrm{j}}^{-}\, .
\end{equation}

We should remark that, although at the coincidence $\mathbb{G}^{\textrm{(sca)}}_{xy}({\bf r}, {\bf r}; \omega)= - \mathbb{G}^{\textrm{(sca)}}_{yx}({\bf r}, {\bf r}; \omega) \neq 0$ these terms do not contribute to the decay rate since they give opposite sign contributions to Eq. (\ref{TaxaEmissaoEspontanea3}). Furthermore, as $\mathbb{G}^{\textrm{(sca)}}_{xz}({\bf r}, {\bf r}; \omega) = \mathbb{G}^{\textrm{(sca)}}_{zx}({\bf r}, {\bf r}; \omega)=\mathbb{G}^{\textrm{(sca)}}_{yz}({\bf r}, {\bf r}; \omega) = \mathbb{G}^{\textrm{(sca)}}_{zy}({\bf r}, {\bf r}; \omega) = 0$, the only relevant components of the Green's dyadic to the SE rate are the diagonal ones. In addition, note that we are allowing the semi-infinite medium to present anisotropic properties so that a TE (TM) electromagnetic wave impinging on the flat surface may have a TM (TE) reflected component. Nevertheless, it is possible to demonstrate that all terms involving cross polarization reflection coefficients
%($r^{\textrm{TE, TM}}$ and $r^{\textrm{TM, TE}}$) are either zero or cancel each other in the expression for the SE rate.
do not contribute to the SE rate. This, however, does not mean that anisotropy plays no role in the quantum emitter lifetime, as the diagonal reflection coefficients may have explicit dependence on the parameters characterizing the optical anisotropy (for instance, see Section \ref{SEGrapheneWall}). Clearly, in the case where the half-space $z<0$ is made of an isotropic material $r^{\textrm{TE, TE}}$ and $r^{\textrm{TM, TM}}$ are given by the usual Fresnel equations \cite{Jackson}.

Using Eqs. (\ref{GreenFunctionSplit}), (\ref{FreeSpaceGreenFunction}), and (\ref{ReflectedGreenFunction}) in (\ref{TaxaEmissaoEspontanea3}) we can show that the SE rate of a quantum emitter with its transition electric dipole parallel and perpendicular to the wall plane are given, respectively, by
\begin{eqnarray}
\label{SEPerp}
\dfrac{\Gamma_{\perp}(z)}{\Gamma_0} = \dfrac{d_{\perp}^2}{|{\bf d}_{\textrm{eg}}|^2}
\bigg\{ 1 \!\!\! &+&\!\!\! \dfrac{3}{2} \int_0^{k_0}\dfrac{k_{||}^3}{k_{0}^3\xi} \textrm{Re}[r^{\textrm{TM, TM}} e^{2i\xi z}]  dk_{||} \cr\cr
 \!\!\!&+&\!\!\!  \dfrac{3}{2} \int_{k_0}^{\infty} \dfrac{k_{||}^3}{k_0^3\zeta}  \textrm{Im}[r^{\textrm{TM, TM}}] e^{-2\zeta z} dk_{||}  \bigg\}\, ,
\end{eqnarray}
\begin{eqnarray}
\label{SEPar}
\dfrac{\Gamma_{||}(z)}{\Gamma_0} = \dfrac{d_{||}^2}{|{\bf d}_{\textrm{eg}}|^2}
\bigg\{ 1 \!\!\! &+&\!\!\! \dfrac{3}{4} \int_0^{k_0}\dfrac{k_{||}}{k_{0}^3\xi} \textrm{Re}[(k_0^2 r^{\textrm{TE, TE}}  - \xi^2r^{\textrm{TM, TM}}) e^{2i\xi z}]  dk_{||} \cr\cr
 \!\!\!&+&\!\!\!  \dfrac{3}{4} \int_{k_0}^{\infty}\dfrac{k_{||}}{k_{0}^3\zeta} \textrm{Im}[k_0^2 r^{\textrm{TE, TE}}  + \zeta^2r^{\textrm{TM, TM}}] e^{-2\zeta z} dk_{||} \bigg\}\, .
\end{eqnarray}

Some comments are in order here: $(i)$ in the simple limit of absence of the wall ($r^{\textrm{TE, TE}} = r^{\textrm{TM, TM}} = 0$) we recover the SE rate in vacuum given by Eq. (\ref{GammaFreeSpace}); (ii) in the case of a perfect conducting wall ($r^{\textrm{TE, TE}} = -1$ and $r^{\textrm{TM, TM}}  = 1$) the last equations lead to the same results as in (\ref{SEPerfectWallPerp}) and (\ref{SEPerfectWallPar}); (iii) in Eqs. (\ref{SEPerp}) and (\ref{SEPar}) we have explicitly separated the contributions from propagating  ($0 < k_{||} < k_0$) and evanescent modes ($k_0 < k_{||} < \infty$). While the former dominates the SE rate for distances $z \gtrsim 2\pi/ k_0 = \lambda_0 $ the latter becomes more relevant as the emitter-wall distance decreases ($z\ll \lambda_0$) \cite{Weber-1984,Barnes-1998, Novotny}.

\subsubsection{Spontaneous emission rate near a dispersive metallic wall}

\hspace{5mm} Let us apply Eqs.  (\ref{SEPerp}) and (\ref{SEPar}) to discuss the effects of the finite conductivity of an isotropic metallic wall in the lifetime of a two-level emitter. In this case the reflection coefficients of the semi-infinite medium are given by the well known \linebreak Fresnel expressions\cite{Jackson},
\begin{eqnarray}
\label{FresnelCoefficients}
r^{\textrm{TE, TE}}_{\textrm{iso}}(\omega, k_{||}) \!\!\! &=&\!\!\!  \dfrac{{k_z}_0(\omega, k_{||}) - {k_z}_1(\omega, k_{||})}{{k_z}_0(\omega, k_{||}) + {k_z}_1(\omega, k_{||})}\, , \\
r^{\textrm{TM, TM}}_{\textrm{iso}}(\omega, k_{||})\!\!\! &=&\!\!\! \dfrac{\varepsilon_1(\omega){k_z}_0(\omega, k_{||}) - \varepsilon_0{k_z}_1(\omega, k_{||})}{\varepsilon_1(\omega){k_z}_0(\omega, k_{||}) + \varepsilon_0{k_z}_1(\omega, k_{||})}\, ,
\end{eqnarray}
where $\varepsilon_1(\omega)$ is the wall permittivity and ${k_z}_1 = \sqrt{\mu_0\varepsilon_1(\omega)\omega^2 - k_{||}^2}$.

In Fig. \ref{SEGoldWall} we show the radiative decay rate as a function of $k_0z$ of a quantum emitter with transition frequency $\omega_0 \simeq 4.2 \times 10^{12}$ rad/s above a metallic wall. The half-space is supposed to be made of gold and described by a Drude-model \cite{Ordal}
\begin{equation}
\dfrac{{\varepsilon_{\! }}_{\textrm{Au}}(\omega)}{\varepsilon_0} = 1 - \dfrac{{\omega_{\! }}_{\textrm{Au}}^{\ 2}}{\omega^2 + i\omega/{\tau_{\! }}_{\textrm{Au}}}\, ,
\end{equation}
where ${\omega_{\! }}_{\textrm{Au}} = 1.37 \times 10^{16}$ rad/s and $1/{\tau_{\! }}_{\textrm{Au}} = 4.05\times 10^{13}$ rad/s. There are three main features to be highlighted in Fig. \ref{SEGoldWall}: $(i)$ for large separations between the wall and the emitter the SE rate oscillates as a function of distance and the results are quite similar to those depicted in Fig.~\ref{SEPerfectWall} for a perfect mirror. $(ii)$ As the atom-wall distance decreases bellow $1/k_0$, $\Gamma_{\perp}/\Gamma_0$ and $\Gamma_{||}/\Gamma_0$ approach asymptotically to two and zero, respectively, until $z \sim 1/(10k_0)$. (iii) For separations $z \ll \lambda_0$ the SE rate is highly enhanced due to the coupling between the emitter and evanescent modes. In other words, in this regime of distances the amount of available channels into which the photon can be emitted is greatly increased due to the near-field. Finally, in the near-field regime the decay rate diverges as $z^{-3}$, a behavior that has its origins in the coulombian interaction between the atomic transition electric dipole and the bulk.
 \vspace{10pt}
\begin{figure}[!ht]
  \centering
  \includegraphics[scale=0.6]{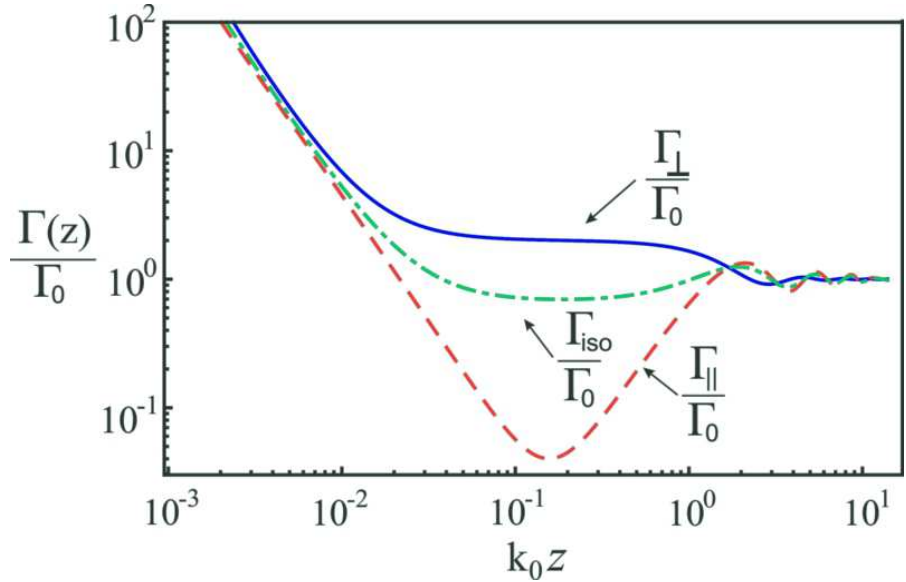}
   \vspace{10pt}
  \caption{$\Gamma_{\perp}/\Gamma_0$ (blue solid line), $\Gamma_{||}/\Gamma_0$ (red dashed line), and $\Gamma_{\textrm{iso}}/\Gamma_0$ (green dot-dashed line) as a function of $k_0z$ for a two-level quantum emitter with transition frequency $\omega_0 \simeq 4.2 \times 10^{12}$ rad/s near a gold half-space.}
  \label{SEGoldWall}
\end{figure}

\vspace{30pt}
In this chapter we have studied an important effect owing to the quantum fluctuations of the vacuum EM field, namely, the spontaneous emission rate of a two-level quantum emitter. We obtained analytical expressions for the emitter's lifetime in terms of both EM field modes and dyadic Green's function. The formalism developed here will be used in the next chapter.

\end{chapter}

\begin{chapter}{Novel strategies to tailor, tune, and suppress the Purcell effect}
\label{cap6}

\begin{flushright}
{\it
A splendid light has dawned on me\\
about the absorption and emission of radiation.
}

{\sc A. Einstein}

\end{flushright}

\hspace{5mm} Controlling radiative properties of quantum emitters is a topic that attracts strong interest nowadays. In this chapter we first show that the Purcell effect (PE) - the influence of a neighboring object in the SE rate of a quantum emitter - can be suppressed if the object is properly covered with a plasmonic cloaking device. Considering the current status of plasmonic invisibility cloaks, we show that the conditions for quenching the PE may be fulfilled for existing quantum emitters. We suggest that atoms with a sufficiently strong transition could be used as quantum, local probes to test the efficiency of plasmonic cloaks. As a second application, we investigate the SE rate of a quantum emitter near a graphene sheet on a dielectric substrate under the influence of an external magnetic field. We demonstrate that the application of the magnetic field may induce a variation in the quantum emitter lifetime as large as 99$\%$ if compared to the case where the magnetic field is absent.
%This effect is due to the discrete Landau levels in graphene.
Our findings suggest that an external magnetic field could act as an efficient external agent to dynamically tune and tailor light-matter interactions in graphene at a quantum level.

\section{Introduction}
\label{Intro6}

\hspace{5mm} The possibility of tailoring and controlling light-matter interactions at a quantum level has been a sought-after goal in optics since the pioneer work of Purcell in 1946, where it was first demonstrated that the environment can strongly modify the spontaneous emission rate of a quantum emitter ({\em e.g.} atom, molecule, quantum dot)\cite{Purcell-46}. To achieve this objective, several approaches have been proposed so far \cite{blanco2004, thomas2004, carminati2006, vladimirova2012, klimovreview, Sture-2007, Kien-2000, Klimov-2002, Milonni-2003, Kastel-2005, Klimov-2004, Biehs-2011, MendesEtAl-08, jacob2012,hyperbolicreview}. Advances in nanofabrication techniques have allowed not only the increase of the spectroscopical resolution of molecules in complex environments~\cite{betzig1993}, but also have led to the use of nanometric objects, such as antennas and tips, to modify the lifetime and to enhance the fluorescence of single molecules~\cite{bian1995,sanchez1999, blanco2004, thomas2004, carminati2006, vladimirova2012, klimovreview, Klimov-2002, Milonni-2003, Kastel-2005, Klimov-2004}.  The presence of metamaterials, artificial structures with engineered electromagnetic response, may also substantially affect quantum emitters' radiative processes. For instance, the impact of negative refraction~\cite{Klimov-2002} and of the hyperbolic dispersion of a certain class of metamaterials~\cite{jacob2012,hyperbolicreview} on the SE of atoms and molecules have been investigated.

Nowadays it is well-known that whenever objects are brought to the neighborhood of a quantum emitter its lifetime is affected due to the boundary conditions imposed by these bodies on the quantum vacuum EM modes. However, here we focus on a novel application of plasmonic cloaking in atomic physics, demonstrating that the spontaneous emission rate of an emitter may not be always modified in the presence of objects coated by plasmonic cloaks. To the best of our knowledge we show for the first time that, in the dipole approximation, the Purcell effect can be strongly suppressed even for small separations between the excited atom and a realistic (including losses) cloak. This result not only proves that SE decay may not be sensitive to the boundary conditions on the EM field but also suggests that the radiative properties of an excited atom or molecule could be exploited to probe the efficiency of an invisibility cloaking.

It is also important to point out here that most proposed schemes to control the radiative properties of excited atoms use materials whose EM properties are hardly modified by the application of external agents. As a consequence, tuning the lifetime of quantum emitters may be a difficult task that unavoidably limit their application in photonic devices. In this chapter, we show that such limitations can be circumvented by using graphene-based materials under the influence of external magnetic fields.  Particularly, we study the SE of an excited two-level emitter above a graphene sheet on an isotropic dielectric substrate under the influence of an uniform static magnetic field. Both inhibited and enhanced decay rates are predicted depending on the emitter-wall distance and on the magnetic field strength. We demonstrate that the application of an external magnetic field can induce a suppression in the Purcell effect as impressive as 99 $\%$ if compared to the case where the magnetic field is absent, for distances on the order of few microns. Besides, we show that at low temperatures the atomic lifetime presents discontinuities as a function of the magnetic field strength  which is physically explained in terms of the discrete Landau levels in graphene under the influence of a magnetic field. Altogether, our results suggest that an external magnetic field could be an efficient external agent to tune and control light-matter interactions in graphene at a quantum level.

This chapter is organized as follows. In section {\bf~\ref{SECloak}} we investigate the Purcell effect in the presence of plasmonic cloaks. Section {\bf~\ref{SEGrapheneWall}} is devoted to the study of the SE rate of a quantum emitter near a graphene-coated wall.

\section{Suppressing the Purcell effect with invisibility cloaks}
\label{SECloak}

\hspace{5mm} In this section we investigate the emission properties of an atom in the presence of a cloaking device and show that if the atomic transition frequency is within the operation band of the invisibility cloak the quantum emitter lifetime will not be affected by the presence of cloaked objects in its neighborhood. Most of the results presented here were published in Ref.~\cite{KortKamp2013}. For the sake of simplicity we assume that all materials involved in the problem are isotropic so that only cloaks based on the SCT are considered here. Particularly, we treat the case of a quantum emitter placed in vacuum at a distance $r$ from a cloak composed by a dielectric sphere of radius $a$ and permittivity $\varepsilon_1(\omega)$ covered with a spherical shell of outer radius $b > a$ and permittivity $\varepsilon_2(\omega)$, as shown in Fig. \ref{SistemaEixos}. Both the core and the shell are nonmagnetic,  $\mu_1(\omega) = \mu_2(\omega) =  \mu_{0}$.
\vspace{10pt}
\begin{figure}[!ht]
\centering
\includegraphics[scale=0.4]{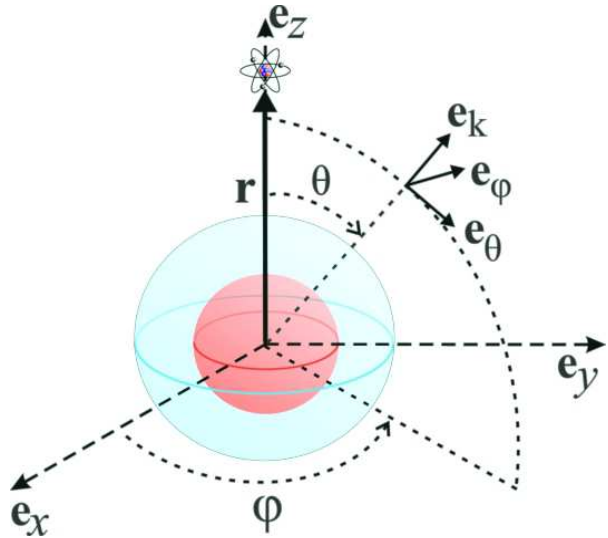}
\vspace{10pt}
\caption{Quantum emitter placed at a distance $r$ from the center of the spherical plasmonic cloak  depicted in Fig. \ref{ScatteringCancelation1}.}
\label{SistemaEixos}
\end{figure}

In order to calculate the spontaneous emission rate we just need the EM modes so as to use Eq. (\ref{TaxaEmissaoEspontanea}).  The field modes ${\bf A}_{{\bf k} p}({\bf r})$ in the region outside the cloak can be determined by using the standard Mie approach for the scattering by the coated sphere of a superposition of two linearly polarized plane waves with orthogonal polarizations. For the case of a plane wave propagating along the $z-$axis and $x-$polarized the resulting EM field at a position ${\bf r}$ is given by Eqs. (\ref{IncidentField1}), (\ref{IncidentField2}), (\ref{scatteredfield1}), and (\ref{scatteredfield2}). In this case the associated modes are
\begin{eqnarray}
{\bf A}_{{k_z} x}({\bf r}) = {\bf A}_{{k_z} x}^{\textrm(0)}({\bf r}) + {\bf A}_{{k_z} x}^{\textrm{(s)}}({\bf r}),
\label{ModosCampo}
\end{eqnarray}
where ${\bf A}_{{k_z} x}^{\textrm(0)}({\bf r})$ refer to the free space EM modes for the considered polarization while the scattered modes can be written as [see Eq. (\ref{scatteredfield1})]
\begin{eqnarray}
{\bf A}_{{k_z} x}^{\textrm{(s)}}(r, \theta, \varphi)  =  \dfrac{1}{\sqrt{V}}\left[-\dfrac{1}{\varepsilon_0} \nabla \times {\bf F}_{\textrm{s}}^{k_z, x} (r, \theta, \varphi)
+ \dfrac{i}{\omega \mu_0 \varepsilon_0} \nabla \times \nabla \times {\bf G}_{\textrm{s}}^{k_z, x}(r, \theta, \varphi)\right]\, ,
\label{CampoEspalhado11}
\end{eqnarray}
where $V$ is the quantization volume introduced here just for consistency with Eq. (\ref{campoquantizadogeral0}) and functions ${\bf F}_{\textrm{s}}^{k_z, x} (r, \theta, \varphi)$ and ${\bf G}_{\textrm{s}}^{k_z, x} (r, \theta, \varphi)$ are given by Eqs. (\ref{VectorPotentialFScattered}) and (\ref{VectorPotentialGScattered}), respectively. In the case of a impinging plane wave propagating in the $z-$direction and polarized along the $y-$axis the scattered EM field modes can be obtained from the previous one simply by making $\varphi \rightarrow \varphi + \pi/2$ \cite{BohrenHuffman}, to wit
\begin{eqnarray}
{\bf A}_{{k_z} y}^{\textrm{(s)}}(r, \theta, \varphi)  =  {\bf A}_{{k_z} x}^{\textrm{(s)}}(r, \theta, \varphi + \pi/2)\, .
\label{CampoEspalhado12}
\end{eqnarray}

Note that although Eqs. (\ref{CampoEspalhado11}) and (\ref{CampoEspalhado12}) give the EM field modes for two orthogonal polarizations they can not be directly used to calculate the SE rate. Indeed, while Eqs. (\ref{CampoEspalhado11}) and (\ref{CampoEspalhado12})  give the scattered field in a generic position of space for impinging waves propagating along a fixed direction (see Fig. \ref{ScatteringCancelation1}), we need the modes at a fixed position but for an incident field propagating in an arbitrary direction (see Fig. \ref{SistemaEixos}) to evaluate the sum in Eq. (\ref{TaxaEmissaoEspontanea}). However, a simple inspection in the reference frames of Figs. \ref{ScatteringCancelation1} and \ref{SistemaEixos} shows that the latter can be obtained by rotating the former by $\theta$ around the $y-$axis followed by a rotation of $\varphi$ around the $z-$direction. Formally, this can be carried out in Eqs. (\ref{VectorPotentialFScattered}) and (\ref{VectorPotentialGScattered}) by rotating the spherical harmonics with the aid \linebreak of the expression \cite{Morrison}
\bea
\label{RotationSphericalHarmonics}
 Y_n^{\pm 1}(\theta',\varphi') = \sum_m e^{-im\alpha_k} d^l_{m, \pm 1} (\beta_k) Y_n^m(\theta,\varphi)\, ,
\eea
where $Y_n^{m}(\theta,\varphi) = (-1)^{m} \sqrt{(2l+1)(n-m)!/4\pi(n+m)!}P_n^m(\cos \theta) e^{i m\varphi}$ are the spherical harmonics \cite{Abramowitz}, $d^l_{m, \pm 1} (\beta_k)$ are the so-called Wigner d-matrices and $\alpha_k = \varphi$, $\beta_k = \theta$ (and also $\gamma_k = 0$) are the respective Euler angles. With these remarks in mind we can show after a lengthy and tedious calculation that the proper EM field modes to be used in Eq. (\ref{TaxaEmissaoEspontanea}) may be cast into the form \cite{KortKamp2013}
\begin{eqnarray}
{\bf A}_{{\bf k}1}^{\textrm{(s)}}({\bf r})\!\!\! &=&\!\!\! -\dfrac{3  c_1^{\textrm{TM}}}{2\sqrt{V}}
\left[\dfrac{h_1^{(1)}(kr)}{kr}
+ {h'}_1^{(1)}(kr) \right] {\bf e}_{\varphi},
\label{DipoleFields1}\\ \cr
{\bf A}_{{\bf k}2}^{\textrm{(s)}}({\bf r})\!\!\! &=&\!\!\!  \dfrac{3  c_1^{\textrm{TM}}}{2{\sqrt{V}}} \left[
 \dfrac{\sin 2\theta}{2}
\left({h'}_1^{(1)}(kr) -\dfrac{h_1^{(1)}(kr)}{kr}\right) {\bf e}_{k} \right. \cr
\!\!\!&+&\!\!\! \left. \left([1+\sin^2\theta]\dfrac{h_1^{(1)}(kr)}{kr}-{h'}_1^{(1)}(kr)\cos^2\theta  \right)
{\bf e}_{\theta} \right],
\label{DipoleFields2}
\end{eqnarray}
where the prime in ${h'}_1^{(1)}(x)$ denotes differentiation with respect to the argument, the unitary vectors ${\bf e}_{k},\ {\bf e}_{\varphi},\ {\bf e}_{\theta}$ in ${\bf k}$-space and the angle $\theta$ are defined in Fig. \ref{SistemaEixos}. We have also assumed the long wavelength approximation. It is instructive to comment here that we have checked that the above formal procedure is equivalent to evaluate the curls in Eqs. (\ref{CampoEspalhado11}) and (\ref{CampoEspalhado12}), then to rewrite the old unit vectors in terms of the new ones and to make the changes $\theta \rightarrow -\theta$ and $\varphi \rightarrow 0$. This  simpler way to obtain Eqs. (\ref{DipoleFields1}) and (\ref{DipoleFields2})  is a consequence of the fact that the rotation angles in Eq. (\ref{RotationSphericalHarmonics}) are precisely the relative angles $\theta$ and $\varphi$ between ${\bf r}$, ${\bf k}$ and the polarization vectors.

At this point it is worth emphasizing that we have two different approximations here: on the one hand, we have the (electric) dipole approximation for the atom, meaning that the transition wavelength $\lambda_0 = 2 \pi c /\omega_0$ is much larger than the Bohr radius $a_0$; on the other hand, we have the dipole approximation for the cloaked sphere, which means that $\lambda_0 \gg b$. The former allows us to simplify the interaction part of the Hamiltonian as in Eq. (\ref{HamiltonianoInteracao}), while the latter allows us to neglect all the scattering coefficients except $c_1^{\textrm{TM}}$. The dipole approximation for the coated sphere also implies that $|c_1^{\textrm{TM}}| \ll 1$ and therefore  $|{\bf A}_{{\bf k} p}^{\textrm{(s)}}({\bf r})| \ll |{\bf A}_{{\bf k} p}^{(0)}({\bf r})|$, so that by substituting Eq. (\ref{ModosCampo}) into Eq. (\ref{TaxaEmissaoEspontanea}) we may keep only linear terms in ${\bf A}_{{\bf k} p}^{\textrm{(s)}}({\bf r})$, obtaining \cite{KortKamp2013}
\begin{eqnarray}
\label{SECloakDipoleApp}
\Gamma({\bf r}) \simeq \dfrac{\pi \omega_0}{\varepsilon_0 \hbar} \sum_{{\bf k}p}\! \Bigg\{ |{\bf d}_{\textrm{eg}} \cdot
{\bf A}^{(0)}_{{\bf k}p}({\bf r})|^2
+  2 \textrm{Re}\! \left[\!\left\{{\bf d}_{\textrm{eg}} \cdot
{{\bf A}_{{\bf k}p}^{(0)}}^{*}\!({\bf r})\right\}\!\! \left\{{\bf d}_{\textrm{eg}}
\cdot{\bf A}_{{\bf k}p}^{\textrm{(s)}}({\bf r}) \right\} \right]\!\!\!\Bigg\} \delta(\omega_k-\omega_0).
\end{eqnarray}

The first term in the previous equation gives the free space contribution (\ref{GammaFreeSpace}), while the second one is precisely the correction due to the presence of the coated sphere. A direct substitution of Eqs. (\ref{DipoleFields1}) and (\ref{DipoleFields2}) into Eq. (\ref{SECloakDipoleApp}) yields \cite{KortKamp2013} the following result for the relative SE rate  \cite{KortKamp2013},
\begin{eqnarray}
\label{emissionfinal}
\dfrac{\Delta\Gamma({\bf r})}{\Gamma_{0}}\!\!\! &\simeq&\!\!\! \dfrac{3}{2}|c_1^{\textrm{TM}}|
\left\{\dfrac{y_1(k_{0}r)}{k_{0}r}\left[-\dfrac{\sin(k_{0}r)}{(k_{0}r)^3}+\dfrac{2\cos(k_{0}r)}{(k_{0}r)^2}-
 \dfrac{\sin(k_{0}r)}{k_{0}r} \right]\right. \cr
 \!\!\! &+& \!\!\! \left. 2{y'}_1(k_{0}r)\left[-\dfrac{\sin(k_{0}r)}{(k_{0}r)^3}+
\dfrac{\cos(k_{0}r)}{(k_{0}r)^2} + \dfrac{\sin(k_{0}r)}{k_{0}r} \right]\right\}\, ,
\end{eqnarray}
where $\Delta\Gamma({\bf r}) = \Gamma({\bf r}) -  \Gamma_{0}$, $k_0 = 2\pi/\lambda_0$ and $y_1(x)$ is the spherical Neumann function~\cite{Abramowitz}. In order to obtain the previous result we considered an isotropic atom, $|{{\bf d}_{\textrm{eg}}}_x|^2 = |{{\bf d}_{\textrm{eg}}}_y|^2 = |{{\bf d}_{\textrm{eg}}}_z|^2 = |{\bf d}_{\textrm{eg}}|^2/3$, and neglected losses by assuming real permittivities at the transition frequency (this means that $c_1^{\textrm{TM}}$ is a pure imaginary number in the dipole approximation).

Equation (\ref{emissionfinal}) is the central result of this section and it demonstrates that the SE rate of a two-level quantum emitter placed in the vicinities of a spherical plasmonic cloak is proportional to the first TM scattering coefficient of the coated sphere, $c_1^{\textrm{TM}}$.  As we have discussed in Chapter {\bf \ref{cap3}} this coefficient, which corresponds to the electric dipole radiation, can vanish by a judicious choice of material parameters. As a consequence, the scattering cross section corresponding to the coated sphere will be greatly reduced since $c_1^{\textrm{TM}}$ largely dominates the scattering pattern in the long wavelength approximation, making the sphere practically invisible to the atom (see Section \ref{ScatteringCancelation}). The complete set of material parameters for which invisibility can occur in the dipole approximation for the coated sphere has been  discussed in Fig. \ref{ScatteringCancelation2}. In the following,  we explore the \linebreak case where $0<\varepsilon_{2}(\omega_0)< 1 $.
\vspace{10pt}
\begin{figure}[!ht]
\centering
\includegraphics[scale=0.65]{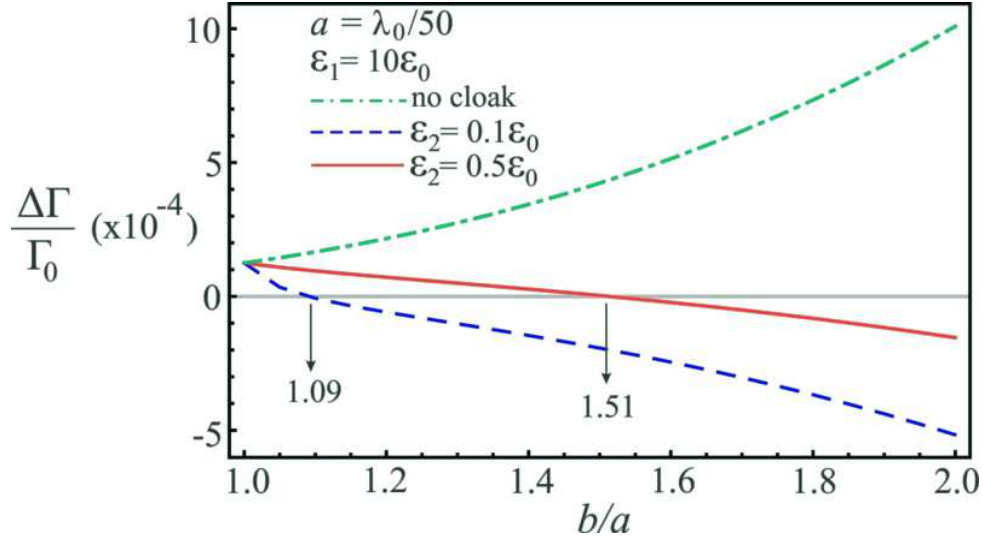}
\vspace{10pt}
\caption{Relative spontaneous emission rate $\Delta\Gamma({\bf r})/\Gamma_{0}$ as a function of $b/a$ for a fixed distance between the sphere and the atom. The sphere parameters have the following fixed values: $a = \lambda_{0}/50$ and $\varepsilon_1(\omega_0) = 10\varepsilon_0$. The blue dashed line and the red solid line show the results for $\varepsilon_2(\omega_0) = 0.1\varepsilon_0$ and $\varepsilon_2(\omega_0) = 0.5\varepsilon_0$,  respectively. Vertical arrows indicate the values of $b/a$ for which the SE rate reduces to its value in vacuum, {\it i.e.}, complete suppression of the Purcell effect. For comparison, the green dot-dashed line shows the result for the case where the covering layer is made of the same material as the inner sphere (no cloak).}
\label{EmissaoEspontaneaVersusRaioCloak}
\end{figure}

In Figure \ref{EmissaoEspontaneaVersusRaioCloak} we show $\Delta\Gamma({\bf r})/\Gamma_{0}$, the difference between the SE rate of a two-level atom placed near a coated sphere and its vacuum value normalized by $\Gamma_{0}$, as a function of $b/a$ for a given distance between the atom and the sphere. We fixed the physical parameters characterizing the core by setting the electric permittivity of the inner sphere as $\varepsilon_1(\omega_0) = 10\varepsilon_0$ and, in order to be consistent with the dipole approximation, we chose the radius $a = \lambda_0/50$. For comparison, in Fig. \ref{EmissaoEspontaneaVersusRaioCloak} it is also plotted the result for $\Delta\Gamma({\bf r})/\Gamma_{0}$ in the case where there is no cloak at all (green dot-dashed line), {\it i.e.}, the relative SE rate in the presence of a single sphere with $\varepsilon_1(\omega_0)=10\varepsilon_0$ as a function of its radius. In this case, as the radius of the sphere grows the SE rate increases monotonically from its value in vacuum $\Gamma_{0}$. In contrast, when the shell is taken into account, $\Delta\Gamma({\bf r})$ approaches zero, and hence $\Gamma({\bf r})\rightarrow \Gamma_{0}$,  as the outer radius $b$ of the cloak is increased. More interestingly, the SE rate  $\Gamma({\bf r})$ is identical to its value in vacuum for $b/a \simeq 1.09$ when $\varepsilon_2(\omega_0) = 0.1\varepsilon_0$ and for $b/a \simeq 1.51$ when $\varepsilon_2(\omega_0) = 0.5\varepsilon_0$,  precisely the cases where condition (\ref{invisibilitycondition_alu}) is satisfied for the respective choices of $\varepsilon_2(\omega_0)$.  It is worth mentioning that as the contribution due to the electric dipole of the cloaked sphere vanishes, the next order multipoles, namely the magnetic dipole and  the electric quadrupole, become relevant to the scattering. However, this does not preclude a strong reduction of $\Delta\Gamma({\bf r})$ since the contributions of $c_1^{\textrm{TE}}$ and $c_2^{\textrm{TM}}$ of a cloaked sphere to the atomic radiative properties are still much smaller than the one of a bare sphere [proportional to $(c_1^{\textrm{TM}})_{\!\!\textrm{bare} \atop \textrm{sphere}}$].  Indeed, we have verified that, for the material parameters chosen, a reduction of $\Delta\Gamma({\bf r})$ up to 98\% compared to the case of an uncloaked sphere can be obtained when (\ref{invisibilitycondition_alu}) is fulfilled and taking into account terms until octopole (inclusive) contributions.  This result means that the Purcell effect may be almost  suppressed if the cloaking condition is satisfied. Finally,  note that for values of $b/a$ larger than the condition of invisibility, the contribution of the shell to the scattered field is dominant and the system becomes visible, so that the relative SE rate increases again as it can be seen from Fig. \ref{EmissaoEspontaneaVersusRaioCloak}.

In order to investigate the dependence of the SE rate on the distance between the atom and the spherical plasmonic cloak, in Fig. \ref{EmissaoEspontanea3D} we exhibit a three-dimensional plot of $\Delta\Gamma({\bf r})/\Gamma_{0}$ as a function of both $b/a$ and $k_{0} r$. The parameters characterizing the inner sphere are the same as in Fig.~\ref{EmissaoEspontaneaVersusRaioCloak} and the electric permittivity of the shell is $\varepsilon_2(\omega_0)=0.5\varepsilon_0$. It is important to emphasize that in general the SE rate exhibits an oscillatory behavior with the distance $r$ except for $b/a = \sqrt[3]{38/11} \simeq 1.51$ (highlighted by the red line in Fig. \ref{EmissaoEspontanea3D}), where the invisibility condition (\ref{invisibilitycondition_alu})  is fulfilled and the SE rate is always equal to $\Gamma_{0}$ for any value of $k_0r$. As far as we know, this is the first situation where the SE rate of a quantum emitter is unaffected by the presence of a surrounding body for a broad range of separation distances  \cite{KortKamp2013}.
\vspace{10pt}
\begin{figure}[!ht]
\centering
\includegraphics[scale=0.6]{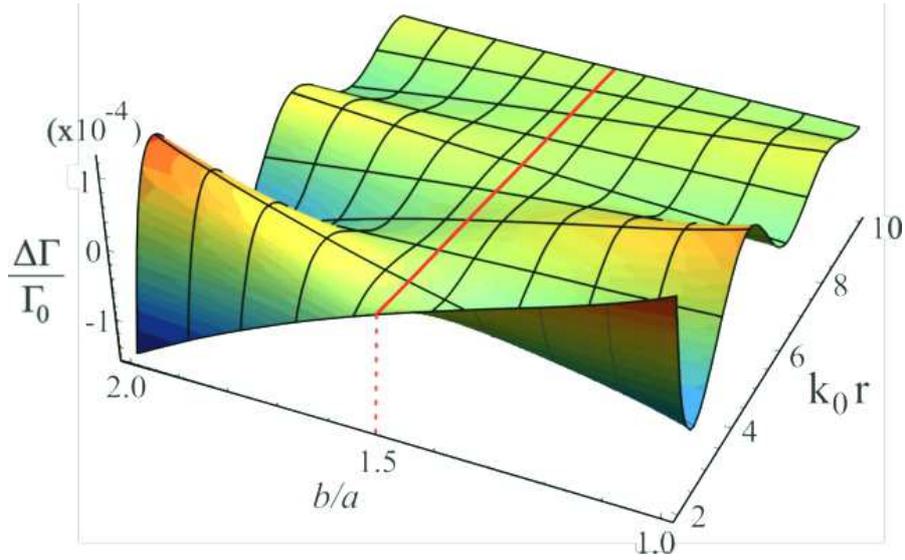}
\vspace{10pt}
\caption{Relative spontaneous emission rate as function of $b/a$ and $k_{0} r$. Parameters of the inner sphere are $a=\lambda_0/50$ and $\varepsilon_1(\omega_0)=10\varepsilon_0$, while the covering layer has $\varepsilon_2(\omega_0)=0.5\varepsilon_0$. The red solid line highlights the value $b/a=\sqrt[3]{38/11}\simeq 1.51$, for which the invisibility condition  (\ref{invisibilitycondition_alu}) is fulfilled. We note also that $b/a = 1$ corresponds to the SE rate of an atom in the presence of a dielectric sphere without cloak.}
\label{EmissaoEspontanea3D}
\end{figure}

To further explore the dependence of the SE rate upon the outer shell radius, in Fig.  ~\ref{EmissaoEspontaneaVersusDistancia} we plot $\Delta\Gamma({\bf r})/\Gamma_{0}$ for different values of the ratio $b/a$; all other parameters being the same as those used in Fig.~\ref{EmissaoEspontanea3D}. Again, it is clearly seen that the presence of the covering layer significantly reduces the amplitude of oscillation with respect to SE rate for a single sphere. The black dashed line corresponds to $b/a = 1.40$ and demonstrates that a reduction of about 80\% for $\Delta\Gamma({\bf r})/\Gamma_{0}$ (compared to the case without the shell) can be obtained even with plasmonic cloaks for which the invisibility condition (\ref{invisibilitycondition_alu}) is not strictly satisfied. The green shaded area highlights the interval of possible values of $b/a$ for which the relative SE rate is reduced  by at least 90\% due to the inclusion of the shell. More specifically, the results show that for $1.46\leq b/a \leq 1.56$, the influence of the cloaked sphere on $\Gamma({\bf r})/\Gamma_{0}$ is smaller than 10\% of the uncloaked case regardless of the distance between the atom and the center of the cloak. This shows that a tolerance of 10\% in the efficiency of the plasmonic cloak is possible in a relatively wide range of ratios $b/a$, highlighting the serendipitous robustness of our results.
\vspace{10pt}
\begin{figure}[!ht]
\centering
\includegraphics[scale=0.60]{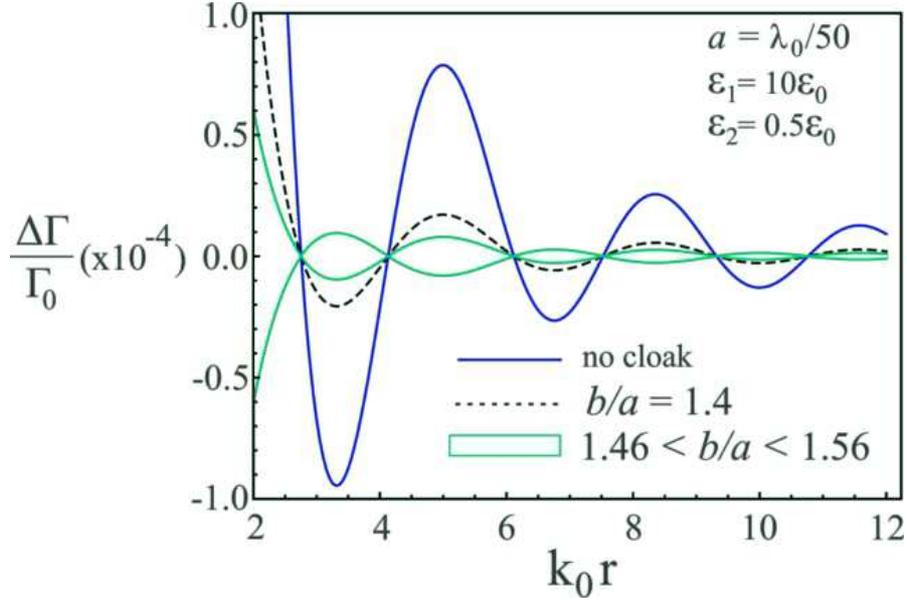}
\vspace{10pt}
\caption{Relative SE rate  as a function of $k_{0} r$ for different values of  $b/a$. The blue solid line corresponds to the absence of a cloaking shell, namely, $b/a=1$. The black dashed line corresponds to $b/a = 1.40$ and shows that the covering layer significantly reduces the amplitude of oscillation. The red dotted-dashed line shows the case where the condition (\ref{invisibilitycondition_alu})  is fulfilled, $b/a = \sqrt[3]{38/11} \simeq 1.51$. The shaded area corresponds to the interval $1.46<b/a<1.56$ and highlights the range of values of $b/a$ for which $\Delta\Gamma({\bf r})/\Gamma_{0}$ is reduced to 10\% or less than its value in the presence of a single sphere.}
\label{EmissaoEspontaneaVersusDistancia}
\end{figure}

Let us now consider some realistic parameters in order to assess the experimental viability to test the results discussed here. A well-suited quantum emitter is a Rubidium atom prepared in a Rydberg state with principal number $n=51$ and magnetic number $m=50$, since such an atom can be well described by a two-level system and therefore our previous discussion can be applied \cite{Brune1994}. Moreover, the corresponding transition frequency is $f_0 = \omega_0/2\pi \simeq 51.099$ GHz, is near the range of recent experiments with plasmonic cloaking devices~\cite{edwards2009, rainwater2012}. Suppose the Rubidium atom is placed near a spherical cloak composed of a non-magnetic sphere of radius $a$ and permittivity $\varepsilon_1(\omega_0) = 10\varepsilon_0$  covered by a spherical shell with outer radius $b$ and permittivity $\varepsilon_2(\omega_0) = 0.5\varepsilon_0$, being both permittivities evaluated at the transition frequency $\omega_0$. As before, let us take $a = \lambda_0/50 \simeq 117.4\ \mu$m. Therefore, the ideal value for $b$ for a perfect cloaking is $b=\sqrt[3]{38/11}\times117.4 \simeq 177.3\ \mu$m. If, instead of perfect cloaking, we require only that the SE rate is suppressed by at least $90\%$,  the possible values of $b$ lie in the interval $171.3\ \mu \textrm{m}\lesssim b\lesssim 183.3\ \mu \textrm{m}$. In other words, one could vary $b$ by $\sim \pm 6\ \mu$m around the ideal value and would still have a very efficient reduction of the Purcell effect ($\ge 90\%$). Had we been interested in a $95\%$ of suppression in $\Delta\Gamma({\bf r})/ \Gamma_0$, the allowed range for $b$ would have been narrower, namely, $ 174.1\ \mu \textrm{m}\lesssim b\lesssim 180.5\ \mu \textrm{m}$ (in this case, $b$ could be changed by $\sim \pm 3.2\ \mu$m around the ideal value). These numbers show that, at least in principle, our result could be tested with current apparatuses and techniques.
\vspace{10pt}
\begin{figure}[!ht]
\centering
\includegraphics[scale=0.77]{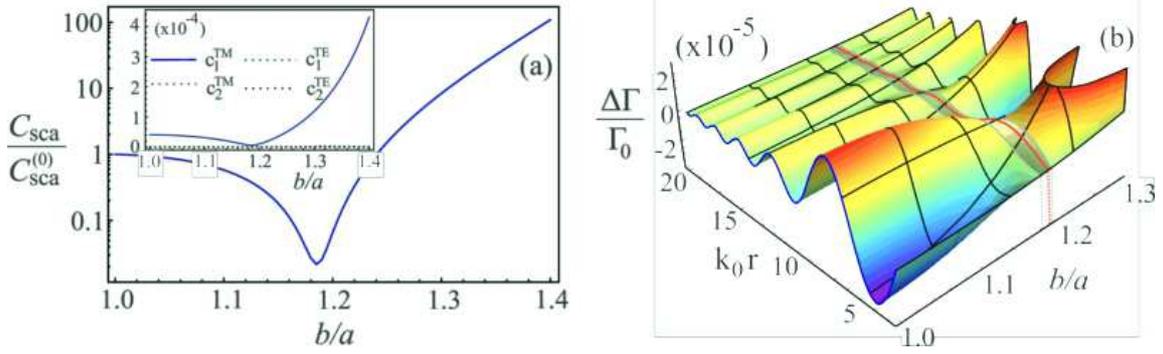}
\vspace{10pt}
\caption{{\bf(a)} Normalized scattering cross section of a cloaked sphere as a function of $b/a$ for $a = 203.7$ nm. The inner sphere and the outer shell are made of Polystyrene and Silicon Carbide, respectively. The frequency of the impinging wave is $\omega_0 = 168 \times 10^{12}$ rad/s ($\lambda_0 \simeq 55a = 11.2\ \mu\textrm{m}$). The inset shows the Mie coefficients $c_1^{\textrm{TM}}$, $c_2^{\textrm{TM}}$, $c_1^{\textrm{TE}}$, and $c_2^{\textrm{TE}}$ as a function of $b/a$. {\bf (b)} Plot of the relative SE rate $\Delta\Gamma({\bf r})/\Gamma_0$ as a function of both $b/a$ and $k_0r$ for a quantum emitter with transition frequency  $\omega_0 = 168 \times 10^{12}$ rad/s. The blue and red curves show the expected behavior of  $\Delta\Gamma({\bf r})/\Gamma_0$ as a function of $k_0r$ for a two-level atom near a bare sphere and a cloaked one exactly at the electric dipole invisibility condition, respectively. The gray shaded area presents the allowed values of $b/a$ for which a suppression of the Purcell effect by at least 90\% of its bare value is achieved.}
\label{EmissaoEspontaneaSiCPoly}
\end{figure}

Finally, we investigate the robustness of our results against (small) material losses. For concreteness we consider a Polystyrene sphere of radius $a = 203.7$ nm described by the model in Eq. (\ref{Poly}) and a coating shell made of Silicon Carbide with its electric permittivity given by \cite{Palik}
\begin{equation}
\dfrac{\varepsilon_{\textrm{SiC}}(\omega)}{\varepsilon_0} = \epsilon_{\infty}\left(1 + \dfrac{\omega_L^2 - \omega_T^2}{\omega_T^2 - \omega^2 - i \omega/{\tau_{\,}}_{\textrm{SiC}}}\right),
\label{sic}
\end{equation}
where $\varepsilon_{\infty} = 6.7$,  $\omega_L = 182.7\times10^{12}$ rad/s, $\omega_T = 149.5\times10^{12}$ rad/s, and $1/{\tau_{\,}}_{\textrm{SiC}} = 0.9\times10^{12}$ rad/s. Besides, we design our cloak to obtain invisibility around $\omega_0 = 168 \times 10^{12}$ rad/s ($\lambda_0 \simeq 55a = 11.2\ \mu\textrm{m}$). At this frequency $\varepsilon_1(\omega_0)/\varepsilon_0 \simeq 3.138$ and $\varepsilon_2(\omega_0)/\varepsilon_0\simeq -5.668 + 0.314i$ (note that $|\textrm{Im}[\varepsilon_2(\omega_0)]/\textrm{Re}[\varepsilon_2(\omega_0)]| \ll 1$) so that cancelling the electric dipole contribution to the scattered field might be possible. Indeed, in Fig. \ref{EmissaoEspontaneaSiCPoly}(a)  we plot the scattering cross section $C_{\textrm{sca}}$  of the cloaked sphere (normalized by the uncloaked result $C_{\textrm{sca}}^{(0)}$) as a function of $b/a$ for the aforementioned material parameters. As it can be seen $C_{\textrm{sca}}$ drops sharply near $b \simeq 1.185a \simeq 241.4$ nm, exactly the same result obtained via Eq. (\ref{invisibilitycondition_alu}) by neglecting the imaginary part of $\varepsilon_2(\omega_0)$. Moreover, the inset in \ref{EmissaoEspontaneaSiCPoly}(a) presents the first TM and TE Mie coefficients and highlights that  at the considered frequency the scattering is largely dominated by the electric dipole contribution. Hence, the analytical results for $\Delta\Gamma({\bf r})/\Gamma_0$ obtained in the long wavelength approximation remain valid provided we consider the following mapping $|c_1^{\textrm{TM}}| \rightarrow -c_1^{\textrm{TM}}$, $y_1(x) \rightarrow h_1^{(1)}(x)$  and then take the real part of Eq. (\ref{emissionfinal}) so as to take into account materials losses. In Fig.  \ref{EmissaoEspontaneaSiCPoly}(b) we show a 3D plot of  $\Delta\Gamma({\bf r})/\Gamma_0$ as a function of both $b/a$ and $k_0r$. The blue solid curve shows the behavior of the spontaneous emission rate  with $k_0r$ for a two-level quantum emitter with transition frequency $\omega_0 = 168 \times 10^{12}$ rad/s in the presence of a bare Polystyrene sphere. The red curve corresponds to $b/a = 1.185$, the situation where the electric dipole invisibility condition is fulfilled for real permittivities. Note that if small material losses are taken into account $\Delta\Gamma({\bf r})/\Gamma_0$ cannot be made zero for all $r$ at a fixed ratio $b/a$. However, a strong suppression of the Purcell effect is still possible. Actually, the gray shaded area (that includes $b/a = 1.185$) highlights the values of $b/a$ for which the relative SE rate can be reduced by at least 90\% of its bare value regardless of $k_0 r$. Although the previous plot does not show the (extreme) near-field regime to avoid confusion, we have verified that similar results hold even for distances much smaller than $\lambda_0$.

Altogether, our findings unambiguously demonstrate that quantum emitters could be explored as local quantum probes to test the effectiveness of plasmonic cloaking devices. Besides, our results show that it is possible to bring objects to the vicinity of excited atoms without affecting their lifetime provided the transition frequency of the quantum emitter is within the operation frequency band of the invisibility cloak.

\section{Controlling the Purcell effect with graphene}
\label{SEGrapheneWall}

\hspace{5mm} Control of spontaneously emitted radiation is one of the most important problems in quantum optics. It is a fundamental process for several applications such as light-emitting diodes \cite{Yablonovitch-87, Comlombelli-2003, Smith-1999, Painter-1999, Nod-2001} and single-photon sources for quantum information \cite{Kuhn-2002, Michler-2000}. However, designing a proper environment whose EM properties could be manipulated by external agents so as to tailor radiative properties of quantum emitters is still an obstacle.  In this section we show that graphene-based materials might be used in order to\linebreak circumvent  this limitation.

\subsection{Graphene in a magnetic field}

\hspace{5mm} Graphene is a 2-D monoatomic material made of Carbon atoms distributed in a honeycomb lattice that presents extraordinary mechanic, thermal, electronic, and optical properties \cite{Graphene1, Graphene2, Graphene3, Gusynin1, Gusynin2, grigorenko2012,bao2012,bludov2013,abajoreview}. It is the building block of graphite and it was first successfully isolated in 2004 \cite{Novoselov-2004}. From the fundamental point of view it is a special material for testing some Quantum Field Theory models since near the Dirac point the electrons in graphene behave as massless relativistic particles described by the Dirac equation.
%Hence, graphene could be the link between high energy physics and condensed matter physics.
From a practical point of view graphene is a promising material for  fabrication of smaller, faster and tunable electronic and photonic nanodevices owing to the high level of control over its EM properties. Actually, due to its band structure graphene presents a metallic behavior and can be usually well described in terms of its two-dimensional longitudinal and transverse conductivities. Among other properties, these conductivities are strongly sensitive to variations in the density of charge carriers, environment temperature, and applied static magnetic fields \cite{Graphene1, Graphene2, Graphene3, Gusynin1, Gusynin2, grigorenko2012,bao2012,bludov2013,abajoreview}.  As a consequence, altering graphene characteristics by means of a suitable choice of external agents acting on the system would allow us to modify the boundary conditions which could in principle allow for tailoring quantum vacuum fluctuations.  Indeed, graphene hosts extremely confined plasmons to volumes smaller than the diffraction limit, which facilitates strong light-matter interactions~\cite{grigorenko2012,bao2012,bludov2013,abajoreview}, specially in the THz frequency range~\cite{bludov2013}. At the quantum level, the strong spatial confinement of surface plasmons in graphene has been shown to induce high decay rates of quantum emitters in its vicinities~\cite{koopens2013, Vasilios-2014}. The EM field pattern excited by quantum emitters near a graphene sheet further demonstrates the huge field enhancement due to the excitation of surface plasmons ~\cite{hanson2012}. Also, graphene sheet has shown to be able to mediate sub- and superradiance between two quantum emitters~\cite{huidobro2012}.

Here we are mainly interested in studying the SE rate of a two-level quantum emitter near graphene-based materials under the influence of external magnetic fields. It is well known that when graphene is placed in a magnetic field ${\bf B}$ it presents optical anisotropy, with a nonvanishing transverse (Hall) conductivity that gives rise to a strong magneto-optical response. In addition,  the charge carriers can assume only discrete values of energy given by the so called Landau levels (LL) \cite{Graphene3, Gusynin1, Gusynin2}. Consequently, in the low temperature regime, and for strong enough magnetic fields, the Hall conductivity of graphene becomes a quantized function of $B$.
%and may be expressed as an integer multiple of the universal conductivity $\sigma_0 = e^2/4\hbar$ \cite{Graphene3}. This effect corresponds to the integer quantum Hall effect in graphene.

Closed-forms of the longitudinal $\sigma_{xx}(\omega, B)$ and transverse $\sigma_{xy}(\omega, B)$ conductivities have been obtained in \cite{Gusynin2} through a quantum mechanical study using the Kubo's formula. Within the Dirac-cone approximation and for the case of a magnetic field applied perpendicularly to the graphene sheet the conductivities read \cite{Gusynin2}
\vskip -0.4cm
\begin{eqnarray}
\sigma_{xx}(\omega,B) =\dfrac{e^3v_F^2 B\hbar(\omega+i\tau^{-1})}{i\pi} \sum\limits_{n=0}^\infty &&\!\!\!\!\!\!\!\!\!\!\!\! \Bigg[\dfrac{n_F(M_n)-n_F(M_{n+1})+n_F(-M_{n+1})-n_F(-M_n)}    {(M_{n+1}-M_n)[(M_{n+1}-M_n)^2-\hbar^2(\omega+i\tau^{-1})^2]} \cr
&+& (M_n\to-M_n)\Bigg]\, , \label{Conductivity1}\\
\hspace{-85pt}\sigma_{xy}(\omega, B)= \dfrac{e^3v_F^2B}{-\pi}\sum\limits_{n=0}^\infty [n_F(M_n)\!\!\! &-&\!\!\!n_F(M_{n+1})-n_F(-M_{n+1})+n_F(-M_n)] \cr
&& \hspace{-85pt}\times \Bigg[\dfrac{1}{(M_{n+1}-M_n)^2-\hbar^2(\omega+i\tau^{-1})^2} +(M_n\to-M_n) \Bigg]\, ,
\label{Conductivity2}
\end{eqnarray}
where $1/\tau$ is a phenomenological scattering rate, $n_F(E) = 1/[1+e^{(E-\mu_c)/k_BT}]$ is the Fermi-Dirac distribution, $\mu_c$ is the chemical potential, $M_n=\sqrt {n} M_1$ with $n = 0, \, 1, \, 2, \, 3, ... $ are the Landau energy levels, $M_1^2 = 2 \hbar e B v_F^2$ is the Landau energy scale, and $v_F \simeq 10^6$ m/s is the Fermi velocity. We can see from the above equations that whereas $\sigma_{xx}$ is an even function of the chemical potential,  $\sigma_{xy}$ is an odd function of $\mu_c$.
\begin{figure}[!ht]
\vspace{10pt}
  \centering
  \includegraphics[scale = 0.45]{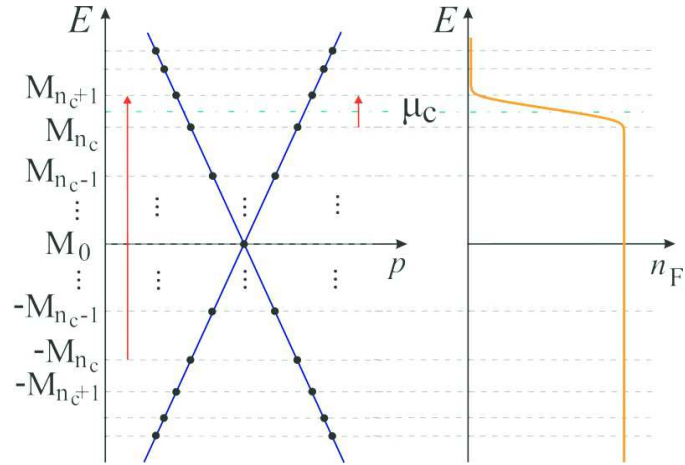}
  \vspace{10pt}
  \caption{{\bf Left:} energy-momentum dispersion diagram of graphene in a magnetic field. The blue lines show the usual linear dispersion relation of graphene whilst the Landau levels brought forth by the introduction of $B$ are represented by the black dots. The long (short) vertical arrow shows the lowest energy interband (intraband) transition crossing the chemical potential (dot-dashed green line). {\bf Right:} Fermi-Dirac distribution at temperature $T$.}
    \label{DispersionDiagram}
\end{figure}
\pagebreak

In Fig. \ref{DispersionDiagram} we present a scheme that helps us understand the behavior of the graphene's conductivity in the presence of an external magnetic field. The left panel depicts the energy-momentum dispersion diagram of graphene in the presence of a static magnetic field. The blue solid line shows the usual linear dispersion relation of graphene. However, due to the applied magnetic field the charge carriers in graphene can occupy only discrete values of energy, given by the Landau levels (LL) $M_n$ represented by black dots. The allowed transitions between two LL give rise to all terms of the summations in Eqs. (\ref{Conductivity1}) and (\ref{Conductivity2}).  There are two kinds of transitions: interband transitions, that connect levels at distinct bands and intraband transitions that involve levels of the same band. The possibility of occurrence of a specific transition is related to the difference between the probabilities of having the initial and final levels full and empty, respectively. Ultimately, these probabilities are given by the Fermi-Dirac distribution, illustrated by the solid orange line on the right of Fig. \ref{DispersionDiagram}. Clearly, the transitions that most contribute to $\sigma_{xx}$ and $\sigma_{xy}$ are the ones crossing the chemical potential $\mu_c$ (dot-dashed green line) and those that require smaller amounts of energy.  In Fig. \ref{DispersionDiagram} the long arrow between $-M_{n_c}$ and $M_{n_c+1}$ and short arrow between $M_{n_c}$ and $M_{n_c+1}$ show the most probable inter- and intraband transition, respectively.
\vspace{10pt}
\begin{figure}[!ht]
\centering
\includegraphics[scale=0.6]{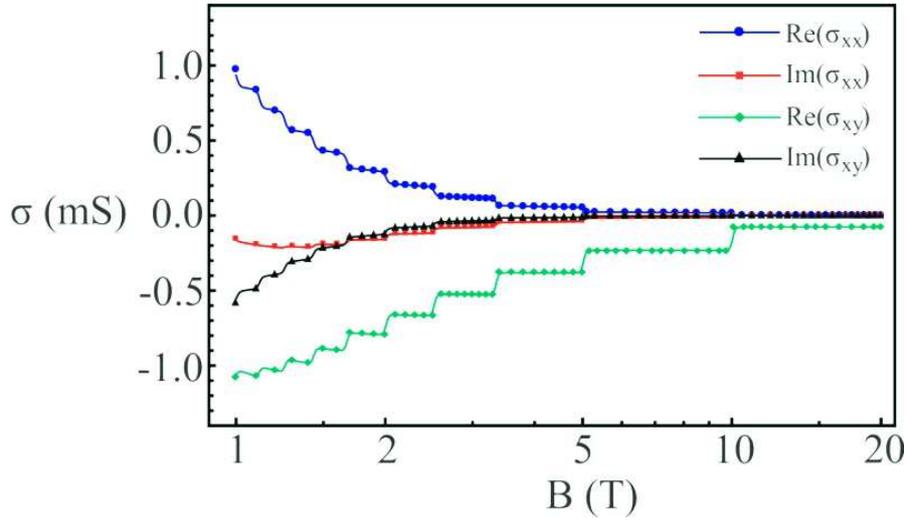}
\vspace{10pt}
\caption{Real and imaginary parts of graphene's longitudinal and transversal conductivities as a function of $B$ for $\omega = 4.2 \times 10^{12}$ rad/s. The solid lines correspond to the results obtained through Eqs. (\ref{Conductivity1}) and (\ref{Conductivity2}) whereas the circle, square, diamond, and triangle dotted ones represent the conductivities calculated only taking into account the intraband contribution in Eqs. (\ref{LongitudinalIntra}) and (\ref{TransversalIntra}). All calculations were performed for $T = 4$ K and $\mu_c = 115$ meV.}
\label{ConductivityFig1}
\end{figure}

For very-low temperatures, $k_BT \ll \mu_c$, the Fermi-Dirac distribution is a quasi-step-function of energy, so that all levels above (below) $\mu_c$ are empty (filled).  In this case, whenever a given LL, whose position in energy depends on $B$, crosses upwards (downwards) the chemical potential of the graphene sheet, it gets immediately depopulated (populated). Therefore the crossing of the $n$-th level sharply quenches the $M_{n} \leftrightarrow M_{n+1}$ transition at the same time that it induces the $M_{n-1} \leftrightarrow M_{n}$ one \cite{Gusynin1, Gusynin2}, in a process that changes the conductivity discontinuously. In Fig. \ref{ConductivityFig1} we show these effects by plotting the real and imaginary parts of $\sigma_{xx}$ and $\sigma_{xy}$ as a function of $B$ for $T = 4$ K and $\mu_c = 115$ meV. The discontinuities in the conductivities at specific values of the magnetic field correspond exactly to the situation where the energy of a given LL reaches $\mu_c$. The fact that, for strong magnetic fields,  the intensities of both $\sigma_{xx}$ and $\sigma_{xy}$ drop down at a discontinuity as we increase $B$ is a consequence of the behavior of the relativistic Landau levels with $\sqrt{n}$.  Indeed, after the crossing of a given Landau-level (say level $n$) through the chemical potential the energy gaps between the dominant electronic transitions are wider than in the case where $M_n < \mu_c$. Hence, the contribution of the dominant transitions to  $\sigma_{xx}$ and $\sigma_{xy}$ becomes smaller since they need a larger amount of energy to occur. As a consequence, graphene conductivity  is reduced (at least for $B\gtrsim 1$ T in our case).

In the regime of low temperatures Eqs. (\ref{Conductivity1}) and (\ref{Conductivity2}) can be put in a simpler form by separating the intraband and interband contributions to the conductivity. For $k_BT\ll \mu_c$ the intraband contribution to $\sigma_{xx}$ and $\sigma_{xy}$ is largely dominated by the $n_c \rightarrow n_c+1$ transition, where $n_c = \textrm{int}(\mu_c^2/M_1^2)$ denotes the number of occupied Landau levels. On the other side, the interband terms are dominated by transitions between levels with $n \geq n_c$. Therefore, by approximating the Fermi-Dirac distribution by a step function, the longitudinal and Hall conductivities can be conveniently written as $\sigma_{xx}(\omega, B) = \sigma_{xx}^{\textrm{intra}}(\omega, B) + \sigma_{xx}^{\textrm{inter}}(\omega, B)$ and $\sigma_{xy}(\omega, B) = \sigma_{xy}^{\textrm{intra}}(\omega, B) + \sigma_{xy}^{\textrm{inter}}(\omega, B)$, where \cite{Nuno-2012-MagnetoPlasmons}
\begin{eqnarray}
\sigma_{xx}^{\textrm{intra}}(\omega, B)\!\!\! &\simeq&\!\!\! \dfrac{e^3v_f^2\hbar B (\omega + i\tau^{-1})(1+\delta_{0,n_c})}{i\pi\Delta_{\textrm{intra}}[\Delta_{\textrm{intra}}^2 - \hbar^2(\omega + i\tau^{-1})^2]}\, ,
\label{LongitudinalIntra}\\ \cr
\sigma_{xx}^{\textrm{inter}}(\omega, B)\!\!\! &\simeq&\!\!\! \sum_{n\geq n_c}^{\infty}\dfrac{e^3v_f^2\hbar B (\omega + i\tau^{-1})(2-\delta_{n,n_c})(1-\delta_{0,n})}{i\pi\Delta_{\textrm{inter}}^{(n)}[{\Delta_{\textrm{inter}}^{(n)}}^2 - \hbar^2(\omega + i\tau^{-1})^2]}\, ,\label{LongitudinalInter} \\ \cr
\sigma_{xy}^{\textrm{intra}}(\omega, B)\!\!\! &\simeq&\!\!\! -\dfrac{e^3v_f^2 B (1+\delta_{0,n_{\textrm{c}}})}{\pi[\Delta_{\textrm{intra}}^2 - \hbar^2(\omega + i\tau^{-1})^2]}\, ,
\label{TransversalIntra} \\ \cr
\sigma_{xy}^{\textrm{inter}}(\omega, B)\!\!\! &\simeq&\!\!\! -\dfrac{e^3v_f^2 B (1-\delta_{0,n_c})}{\pi[{\Delta_{\textrm{inter}}^{(n_c)}}^2 - \hbar^2(\omega + i\tau^{-1})^2]}\, ,
\label{TransversalInter}
\end{eqnarray}
where $\Delta_{\textrm{intra}} = M_{n_c+1} - M_{n_c} $ and $\Delta_{\textrm{inter}}^{(n)} = M_{n+1} + M_{n}$. In order to obtain the above expressions we have considered transitions involving the zero energy LL as intraband-like.

\subsection{Spontaneous emission near a graphene-coated plane}

\hspace{5mm} Once we have discussed the basic features of graphene's conductivity under the influence of an external magnetic field, let us now investigate the radiative properties of a quantum emitter in the situation depicted in Fig.~\ref{SEGraphene1}. The half-space $z<0$ is made of an isotropic and homogeneous dielectric material of permittivity $\varepsilon_s(\omega)$. The substrate is coated by a flat graphene sheet ($z = 0$) held at temperature $T$ and with chemical potential $\mu_c$. The whole system is under the influence of an external uniform magnetic field  ${\bf B} = B\hat{{\bf z}}$ applied perpendicularly to the graphene plane. The upper medium $z>0$ is vacuum and an excited two-level quantum emitter is at a distance $z$ above the interface.
\vspace{10pt}
\begin{figure}[!ht]
\centering
\includegraphics[scale=0.45]{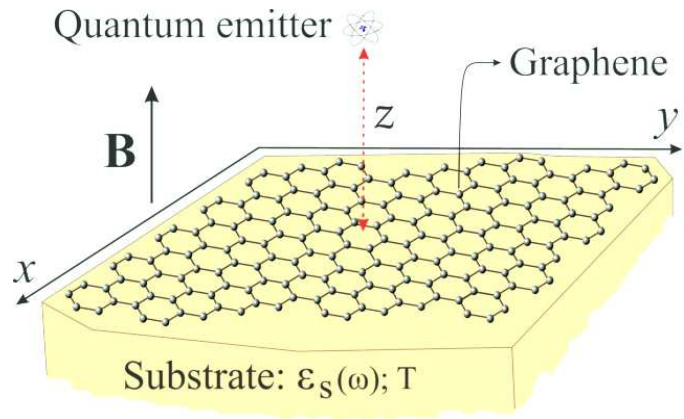}
\vspace{10pt}
\caption{Excited two-level quantum emitter at a distance $z$ above a graphene sheet on an homogeneous and isotropic substrate of permittivity $\varepsilon_s(\omega)$ held at temperature T. The whole system is under the influence of a uniform static magnetic field ${\bf B}$ applied perpendicularly to the graphene sheet.}
\label{SEGraphene1}
\end{figure}

In order to calculate the SE rate of the quantum emitter we can use  Eqs. (\ref{SEPerp}) and (\ref{SEPar}), provided we have the reflection coefficients $r^{\textrm{TE, TE}}$ and $r^{\textrm{TM, TM}}$. In our problem, these Fresnel coefficients can be obtained by modeling graphene as a surface density current ${\bf K} = \mbox{{\mathversion{bold}${\sigma}$}} \cdot {\bf E}|_{z=0}$ and applying the appropriate boundary conditions at $z = 0$,
\begin{eqnarray}
\label{CC1Graphene}
{\bf \hat{z}} \times \left[{\bf E}_T - {\bf E}_R - {\bf E}_0 \right]\!\!\! &=&\!\!\! {\bf 0}\, , \\
\label{CC2Graphene}
{\bf \hat{z}} \times \left[{\bf H}_T - {\bf H}_R - {\bf H}_0 \right]\!\!\! &=&\!\!\! {\bf K} = \mbox{{\mathversion{bold}${\sigma}$}}\cdot{{\bf E}_T}\, .
\end{eqnarray}
where ${\bf E}_0$ is an arbitrarily polarized wave impinging on the interface $z = 0$ wheras ${\bf E}_R$ and ${\bf E}_T$ are the corresponding reflected and transmitted EM fields, respectively. After straightforward algebraic manipulations we can demonstrate that the diagonal reflection coefficients for graphene on a non magnetic substrate can be written as\footnote{\label{note11} A short derivation of these equations is provided in Appendix {\bf \ref{apendiceb}}.}\cite{Macdonald_PRL}
\begin{eqnarray}
\label{ReflectionCoefficients_SS_Exact}
\!\!\!\!\!\!\!\!\!\!r^{\textrm{TE, TE}}(k_{||},\omega,B) \!\!\! &=& \!\!\!-\dfrac{\Delta_{+}^{E}(k_{||},\omega,B)\Delta_{-}^{H}(k_{||},\omega,B)+Z_0^2k_{z_0}k_{z_1} \sigma_{xy}(\omega,B)^2}{\Delta_{+}^{E}(k_{||},\omega,B)\Delta_{+}^{H}(k_{||},\omega,B)+Z_0^2k_{z_0}k_{z_1} \sigma_{xy}(\omega,B)^2}\, ,\\ \cr
\label{ReflectionCoefficients_PP_Exact}
\!\!\!\!\!\!\!\!\!\!r^{\textrm{TM, TM}}(k_{||},\omega,B) \!\!\!&=& \!\!\! \dfrac{\Delta_{-}^{E}(k_{||},\omega,B)\Delta_{+}^{H}(k_{||},\omega,B)+Z_0^2k_{z_0}k_{z_1} \sigma_{xy}(\omega,B)^2}{\Delta_{+}^{E}(k_{||},\omega,B)\Delta_{+}^{H}(k_{||},\omega,B)+Z_0^2k_{z_0}k_{z_1} \sigma_{xy}(\omega,B)^2}\, ,
\end{eqnarray}
with
\begin{eqnarray}
\Delta_{\pm}^{i}(k_{||},\omega,B) = (k_{z_1}\delta_{iE}+k_{z_0}\delta_{iH})(\mathfrak{S}^i  \sigma_{xx}(\omega,B)\pm 1) + [\varepsilon_s(\omega)/\varepsilon_0]k_{z_0} \delta_{iE} + k_{z_1} \delta_{iH} \, ,
\end{eqnarray}
where $i = (E, \, H)$,  $\mathfrak{S}^H = \omega \mu_0 / k_{z0}$, $\mathfrak{S}^E = k_{z0}/(\omega \varepsilon_0)$, $Z_0=\sqrt{\mu_0/\varepsilon_0}$ is the vacuum impedance, $k_{z0}$ is given by Eq. (\ref{kz0}), and ${k_z}_1 = \sqrt{\mu_0\varepsilon_{s}(\omega)\omega^2 - k_{||}^2}$. It should be noticed that if the graphene layer is absent,  $\sigma_{xx} = \sigma_{xy} = 0$,  we will reobtain the usual expressions for the Fresnel reflection coefficients.

With these reflection coefficients in hand we are able to evaluate Eqs. (\ref{SEPerp}) and (\ref{SEPar}). In the following calculations we assume that the semi-infinite medium is made of Silicon Carbide (SiC) with its dispersive permittivity given by Eq. (\ref{sic}). Besides, we consider that the quantum emitter's transition frequency is $\omega_0 = 4.2 \times 10^{12}$ rad/s.  It is worthwhile to mention that depending on the characteristics of the emitter, the Lamb shift and the Zeeman effect may play an important role in the problem. Here, however, we will neglect the dependence of $\omega_0$ on the distance $z$ as well as on the strength of the external magnetic field ${\bf B}$. Also, it is possible to check that for the graphene material parameters used throughout the text the value of the transition frequency is much smaller than $\mu_c$.
%and $M_1/\hbar$ provided $B \geq 1 T$.
As a consequence, the intraband transitions in graphene will be widely dominant in the conductivities expressions, as it can be seen from Fig. \ref{ConductivityFig1}.
\vspace{10pt}
\begin{figure}[!ht]
\centering
\includegraphics[scale=0.77]{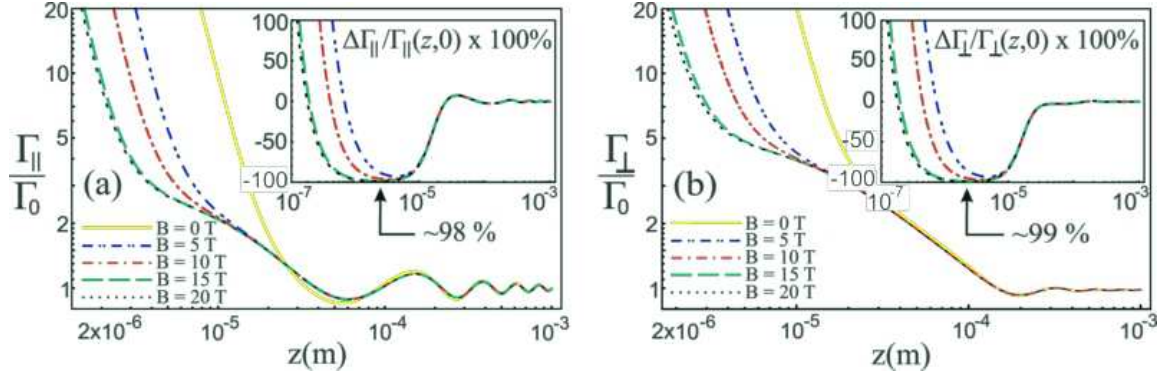}
\vspace{10pt}
\caption{Spontaneous emission rate (normalized by the free space rate of decay $\Gamma_0$) as a function of distance $z$ between the emitter and a graphene coated half-space made of Silicon Carbide for $B = 0$ T (yellow solid line), $B = 5$ T (blue dash-double dotted line), $B = 10$ T (red dash-dotted line), $B = 15$ T (green dashed line), and $B = 20$ T (black dotted line). Panels {\bf (a)} and {\bf (b)} correspond to transition dipole moments parallel and perpendicular to the wall, respectively. The insets in (a) and (b) present the relative SE rates $\Delta\Gamma_{||,\perp} = [\Gamma_{||,\perp}(z,B)-\Gamma_{||,\perp}(z,0)]/\Gamma_{||,\perp}(z,0)$ as functions of distance for the aforementioned magnetic fields. The transition frequency of the quantum emitter is set at $\omega_0 \simeq 4.2 \times 10^{12}$ rad/s, the temperature of the system is held at $T =4$ K and graphene's chemical potential is $\mu_c = 115$ meV.}
\label{SEGraphene2}
\end{figure}

In Fig. \ref{SEGraphene2} we plot the normalized spontaneous emission rates $\Gamma_{||}/\Gamma_0$ and $\Gamma_{\perp}/\Gamma_0$ as functions of the mutual distance $z$ between the emitter and the half-space. Different curves correspond to distinct values of $B$. Graphene's chemical potential has been set at $\mu_c = 115$ meV and the system is supposed to be at temperature $T = 4$ K. Notice that for $z \gtrsim 100\ \mu$m both $\Gamma_{||}$ and $\Gamma_{\perp}$ exhibit an oscillatory behavior as a function of $z$. In this regime of distances the SE process is said to be radiative and the coupling between the emitter and the graphene-coated wall is mediated by propagating modes ($k_{||} \leq \omega_0/c$) of the vacuum EM field. Indeed, we have verified that the last integrals (evanescent contributions) in Eqs. (\ref{SEPerp}) and (\ref{SEPar}) are negligible in this regime. Besides, note that the amplitude of the oscillations are stronger in the  parallel configuration [Fig. \ref{SEGraphene2}(a)] than in the perpendicular one [Fig. \ref{SEGraphene2}(b)]. This is due to the fact that whereas ${\bf d}_{\perp}$ couples only to TM-modes ${\bf d}_{||}$ may interact with both TE- and TM-modes. Furthermore, it is worth noticing that in the far-field regime the emitter's lifetime is barely affected by the applied magnetic field. This is a consequence of the fact that for large emitter-graphene separations the phase $e^{2ik_{z0}z}$ gives a highly oscillatory integrand in the first integrals of Eqs. (\ref{SEPerp}) and (\ref{SEPar}) except for $k_{z0} \sim 0$. Consequently, the propagating modes that most contribute to $\Gamma_{||}$ and $\Gamma_{\perp}$ are those with $k_{||}  = \sqrt{\omega_0^2/c^2 - k_{z0}^2} \sim \omega_0/c$. Hence, by expanding Eqs. (\ref{ReflectionCoefficients_SS_Exact}) and (\ref{ReflectionCoefficients_PP_Exact}) for $k_{||}$ around $\omega_0/c$ it is easy to show that $r^{\textrm{TE, TE}} \sim r^{\textrm{TM, TM}} \sim -1 + {\cal{O}}(ck_{z0}/\omega_0)$. This result is valid even for an uncoated half-space and shows that the reflectivity of the dielectric semi-infinite medium is almost saturated ($|r^{\textrm{TE, TE}}|^2 \approx 1$, $|r^{\textrm{TM, TM}}|^2\approx 1$) in this case. Therefore, it is not surprising that modifying graphene's optical properties would not affect significantly the reflectivity of the substrate in the considered regime.

In Fig. \ref{SEGraphene2}  we can also note that for  $z \lesssim 10\ \mu$m the SE rate is largely enhanced as the emitter-wall separation decreases.  In this regime of distances the emission is dominated by the evanescent ($k_{||} > \omega_0/c$) part of the EM field. Note that in this case changing the strength of the applied magnetic field strongly affects the lifetime of the two-level quantum emitter. Particularly, in the $1-10\ \mu$m range the Purcell effect is significantly suppressed when compared to the case where $B = 0$ T even for moderate magnetic fields. Indeed, for $z = 3\ \mu$m the influence of the graphene-coated wall on both  $\Gamma_{||}$ and $\Gamma_{\perp}$ can be reduced up to $\sim 90\%$ for $B = 5$ T, $\sim 96 \%$ for $B = 10$ T, and $\sim 98 \%$ for $B = 15$ T. These results are highlighted in the insets of Figs. \ref{SEGraphene2}(a) and \ref{SEGraphene2}(b) where we plot the relative SE rates $\Delta\Gamma_{||,\perp} = [\Gamma_{||,\perp}(z,B)-\Gamma_{||,\perp}(z,0)]/\Gamma_{||,\perp}(z,0)$ as functions of $z$ for different values of $B$. In the far-field regime $\Delta\Gamma_{||,\perp} \approx 0$ as discussed previously. For $1\ \mu$m $\lesssim z \lesssim 10\ \mu$m the decay rate is  reduced for any $B \gtrsim 5$ T. For even smaller atom-wall separations the insets show that the external magnetic field may not suppress the PE. On the contrary, for $B = 5$ T and $z = 200$ nm  the SE rate is enhanced by a factor $\sim 6$ ($\sim 500\%$).

Once the main effects of the magnetic field on the atomic lifetime occur in the near-field regime, let us concentrate in this case. In this regime, three different processes (in addition to the propagating modes) may contribute to the SE rate, namely: $(i)$ the photon can be emitted into an evanescent total internal reflection (TIR) mode, $(ii)$  a surface plasmon polariton (SPP) can be excited, and $(iii)$ the emitter energy can be lost without involving the emission of a photon\cite{Weber-1984,Barnes-1998}. Total internal reflection modes may show up if $\textrm{Re}(\varepsilon_s/\varepsilon_0) > 1$ and correspond to $k_0<k_{||}<k_0\sqrt{\varepsilon_s(\omega_0)/\varepsilon_0}$. If losses in the substrate are negligible we can show that $k_{z0}$ is purely imaginary whereas $k_{z1}$ is real in this range of values of $k_{||}$. It means that TIR modes propagate within the substrate but evanesce in vacuum. Their contribution to $\Gamma_{||}$ and $\Gamma_{\perp}$ can be cast into the form
\begin{eqnarray}
\label{SEPerp_TIR}
\dfrac{\Gamma_{\perp}^{\textrm{TIR}}(z)}{\Gamma_0}\!\!\! &=&\!\!\! \dfrac{3d_{\perp}^2}{2|{\bf d}_{\textrm{eg}}|^2} \displaystyle{\int_{k_0}^{k_0\sqrt{\frac{\varepsilon_s(\omega_0)}{\varepsilon_0}}}} \dfrac{k_{||}^3}{k_0^3\zeta}  \textrm{Im}[r^{\textrm{TM, TM}}] e^{-2\zeta z} dk_{||} \, , \\ \cr
\label{SEPar_TIR}
\dfrac{\Gamma_{||}^{\textrm{TIR}}(z)}{\Gamma_0}\!\!\! &=&\!\!\! \dfrac{3d_{||}^2}{4|{\bf d}_{\textrm{eg}}|^2} \displaystyle{\int_{k_0}^{k_0\sqrt{\frac{\varepsilon_s(\omega_0)}{\varepsilon_0}}}}\dfrac{k_{||}}{k_{0}^3\zeta} \textrm{Im}[k_0^2 r^{\textrm{TE, TE}}  + \zeta^2r^{\textrm{TM, TM}}] e^{-2\zeta z} dk_{||}\, ,
\end{eqnarray}
where we have used that $\textrm{Re}[\varepsilon_s(\omega_0)/\varepsilon_0] >0$ and that losses are negligible in our case, since $\textrm{Im}[\varepsilon_s(\omega_0)]/\textrm{Re}[\varepsilon_s(\omega_0)] \sim 10^{-5}$.

Surface plasmon polaritons correspond to the coupled oscillations of the EM field and surface charges. They are EM waves that propagate along the substrate-vacuum interface with amplitudes decaying exponentially along $z$-direction (both $k_{z0}$ and $k_{z1}$ are purely imaginary for lossless materials). The dispersion relation of SPP can be obtained by calculating the poles of the reflection coefficients. For graphene in a magnetic field two different kinds of surface waves may appear, namely, magnetoplasmon polaritons and weakly damped waves called quasi-transverse-electric modes  \cite{Nuno-2012-MagnetoPlasmons}.  We have checked that in our problem only the latter exist for the parameters chosen and magnetic fields considered ($B \gtrsim 1$ T). Nevertheless, it is possible to show that these modes do not give significant contribution to the emitter's lifetime in the extreme near-field regime.
% throughout the thesis.

Finally, the emitter can de-excite in a process where its energy is transferred directly to the half-space giving origin to a lossy surface wave (LSW) with high transverse wave vector ($k_{||} \gg \omega_0/c$). Among other processes, LSW could be related to the creation of an electron-hole pair within the medium \cite{Weber-1984,Barnes-1998}. Such waves are quickly damped and their energy is useless for any practical application, being converted into heat. They are called ``lossy" due to the fact that at very short distances the SE rate is proportional to the imaginary part of the bulk permittivity (in the case without graphene). Hence, the absorption of the material governs the atomic decay. This is a non-radiative process and it is one of the main responsible for the so called quenching effect of fluorescence \cite{chance-1978, Weber-1984,Barnes-1998}. LSW has been discussed in literature for molecules with short transition wavelengths near metallic substrates. In this situation they are relevant only for distances $z \lesssim 5$ nm. However, in our case, we will see in the following discussion that they may play a fundamental role in the emitter's fluorescence even at larger separations ($\sim 1\ \mu$m) since the transition wavelength we are considering is $\lambda_0 \simeq 450\ \mu$m. Since the contributions of LSW to $\Gamma_{\perp}$ and $\Gamma_{||}$ come from high values of $k_{||}$ we can perform a quasi-static (QS) approximation ($c\rightarrow \infty$) in Eqs. (\ref{SEPerp}) and (\ref{SEPar}). The decay rate due to LSW is approximately given by
\begin{eqnarray}
\label{SEPerp_NF}
\dfrac{\Gamma_{\perp}^{\textrm{LSW}}}{\Gamma_0}\!\!\! &\simeq&\!\!\! \dfrac{3d_{\perp}^2}{2|{\bf d}_{\textrm{eg}}|^2} \int_{k_0\sqrt{\frac{\varepsilon_s(\omega_0)}{\varepsilon_0}}}^{\infty} \left\{\dfrac{k_{||}^2 e^{-2k_{||} z}  }{k_0^3} \textrm{Im}[r^{\textrm{TM, TM}}_{\textrm{QS}}]\right\} dk_{||} \, , \\ \cr
 \label{SEPar_NF}
 \dfrac{\Gamma_{||}^{\textrm{LSW}}}{\Gamma_0}\!\!\! &\simeq&\!\!\! \dfrac{3d_{||}^2}{4|{\bf d}_{\textrm{eg}}|^2}\int_{k_0\sqrt{\frac{\varepsilon_s(\omega_0)}{\varepsilon_0}}}^{\infty} \left\{\dfrac{e^{-2k_{||} z} }{k_{0}}  \textrm{Im}[r^{\textrm{TE, TE}}_{\textrm{QS}}] +  \dfrac{k_{||}^2e^{-2k_{||} z} }{k_{0}^3}  \textrm{Im}[r^{\textrm{TM, TM}}_{\textrm{QS}}] \right\}dk_{||}\, ,
\end{eqnarray}
where
\begin{eqnarray}
\label{ReflectionCoefficients_SS_Approx}
\!\!\!\!\!\!\!\!\!\!\!\!\!\!\!r^{\textrm{TE, TE}}_{\textrm{QS}} \!\!\! &\simeq&\!\!\! \dfrac{i\left[Z_0^2\varepsilon_s+  \mu_0\right] \omega\sigma_{xx}/k_{||} - Z_0^2[\sigma_{xx}^2 + \sigma_{xy}^2]}{2[\varepsilon_s/\varepsilon_0 + ik_{||}\sigma_{xx}/\varepsilon_0\omega +1]+Z_0^2\sigma_{xy}^2}\, , \\ \cr
\label{ReflectionCoefficients_PP_Approx}
\!\!\!\!\!\!\!\!\!\!\!r^{\textrm{TM, TM}}_{\textrm{QS}} \!\!\!&\simeq& \!\!\!\dfrac{2[\varepsilon_s/\varepsilon_0 + ik_{||}\sigma_{xx}/\varepsilon_0\omega - 1] +Z_0^2\sigma_{xy}^2}{2[\varepsilon_s/\varepsilon_0 + ik_{||}\sigma_{xx}/\varepsilon_0\omega +1]+Z_0^2\sigma_{xy}^2} \, ,
\end{eqnarray}
are the QS reflection coefficients, valid for $B \neq 0$ T. For $B = 0$ T some approximations employed fail and Eq. (\ref{ReflectionCoefficients_SS_Approx}) do not hold. 
\vspace{10pt}
\begin{figure}[!ht]
\centering
\includegraphics[scale=0.77]{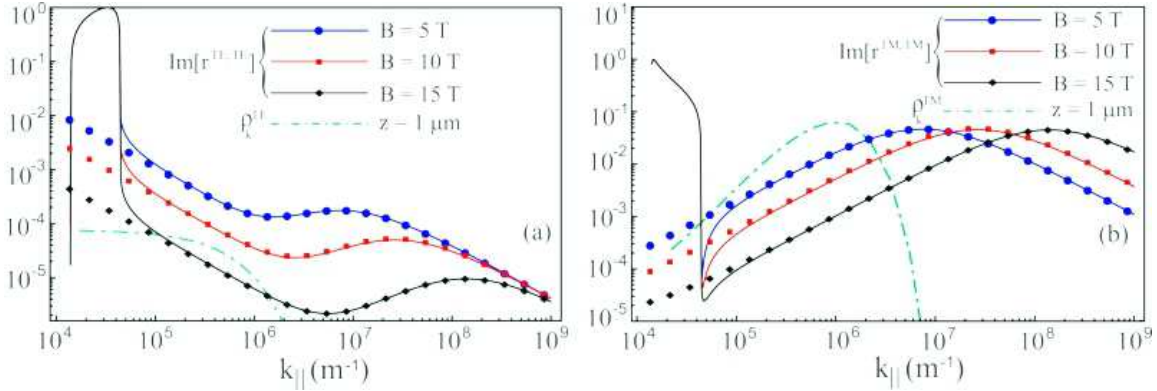}
\vspace{10pt}
\caption{Imaginary part of {\bf (a)} $r^{\textrm{TE, TE}}$ and {\bf (b)} $r^{\textrm{TM, TM}}$ as functions of $k_{||}$ for $k_{||} \geq k_0$. The blue, red and black lines are for $B = 5$ T, $B = 10$ T, and $B = 15$ T, respectively. The solid curves are obtained from Eqs. (\ref{ReflectionCoefficients_SS_Exact}) and (\ref{ReflectionCoefficients_PP_Exact}) whilst the circle, square, and diamond dotted ones correspond to the near-field approximated results achieved through Eqs. (\ref{ReflectionCoefficients_SS_Approx}) and (\ref{ReflectionCoefficients_PP_Approx}). The green dot-dashed lines show (a) $\rho_k^{\textrm{TE}} = e^{-2k_{||}z}dk_{||}/k_0$, and (b) $\rho_k^{\textrm{TM}} =k_{||}^2e^{-2k_{||}z}dk_{||}/k_0^3$  for $z = 1\ \mu$m. The chemical potential and temperature are $\mu_c = 115$ meV and $T = 4$ K.}
\label{SEGraphene3}
\end{figure}

It should be mentioned that, for the material parameters chosen in this thesis, the Hall conductivity is negligible in Eqs. (\ref{ReflectionCoefficients_SS_Approx}) and (\ref{ReflectionCoefficients_PP_Approx}), except in the numerator of $r^{\textrm{TE, TE}}_{\textrm{QS}}$. In Fig. \ref{SEGraphene3} we verify the accuracy of the QS approximation for the reflection coefficients for high values of $k_{||}$. Panel  \ref{SEGraphene3}(a) presents the imaginary part of $r^{\textrm{TE, TE}}$ as a function of $k_{||}$ calculated through Eq. (\ref{ReflectionCoefficients_SS_Exact}) [solid curves] and Eq. (\ref{ReflectionCoefficients_SS_Approx}) [dotted curves] for different values of $B$.  The accuracy of the QS approximation is clear for \linebreak $k_{||} \gtrsim 10^5 > k_0\sqrt{\varepsilon_s(\omega_0)}/\varepsilon_0$. Similar results are shown in panel \ref{SEGraphene3}(b) for $r^{\textrm{TM, TM}}$. In Fig. \ref{SEGraphene3}(a) and \ref{SEGraphene3}(b) we also plot the functions $\rho_k^{\textrm{TE}} := e^{-2k_{||}z}dk_{||}/k_0$ and $\rho_k^{\textrm{TM}} :=k_{||}^2e^{-2k_{||}z}dk_{||}/k_0^3$  for $z = 1\ \mu$m, respectively. These are the weight functions that multiply $\textrm{Im}[r^{\textrm{TE, TE}}_{\textrm{QS}}]$ and $\textrm{Im}[r^{\textrm{TM, TM}}_{\textrm{QS}}]$ in the integrals of Eqs. (\ref{SEPerp_NF}) and (\ref{SEPar_NF}). Note that $\rho_k^{\textrm{TE}}$ is much smaller than $10^{-4}$ for high values of $k_{||}$.  Therefore, the contribution of TE-polarized waves (even for shorter distances) to $\Gamma_{||}^{\textrm{LSW}}$ can be neglected.  On the other hand, $\rho_k^{\textrm{TM}}$ is a non-monotonic function of $k_{||}$ and has a maximum $\sim 0.1$ (for $z = 1\ \mu$m) at ${k_{||}^{\textrm{max}}}_{\!\!\!\!\!\!\!\rho}\ = 1/z$. Hence, TM-modes dominate the nonradiative energy transfer from the emitter to LSW.  Besides, note that apart from a factor $1/2$ the contributions of TM-modes for $\Gamma_{\perp}^{\textrm{LSW}}$ and $\Gamma_{||}^{\textrm{LSW}}$ are the same.
\vspace{10pt}
\begin{figure}[!ht]
\centering
\includegraphics[scale=0.77]{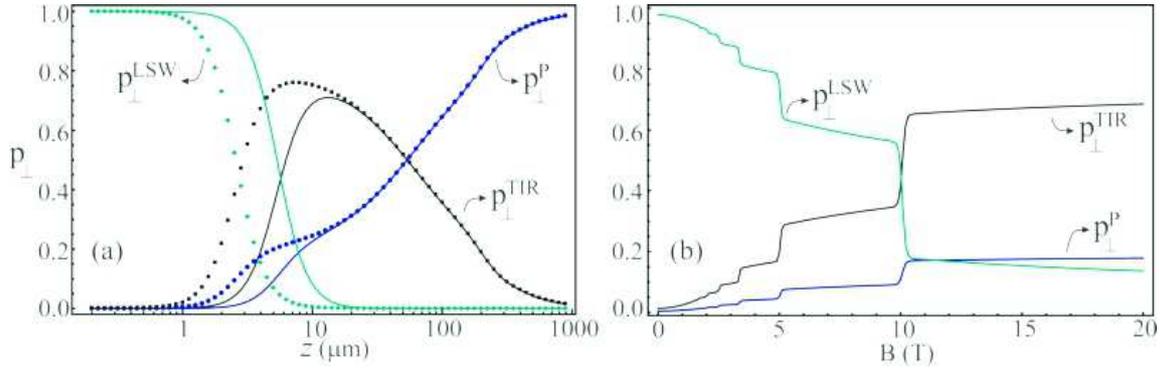}
\vspace{0pt}
\caption{Probabilities $p^{\textrm{P}}_{\perp}$ (blue curves), $p^{\textrm{TIR}}_{\perp}$ (black curves), and $p^{\textrm{LSW}}_{\perp}$ (green curves) of exciting a propagating mode, TIR mode or a LSW as functions of {\bf (a)} $z$ and {\bf (b)} $B$. In panel (a) the values of the magnetic field are $B = 5$ T (solid curves) and $B = 15$ T (dotted curves).  In panel (b) the emitter-graphene distance is $z = 4\ \mu$m. The material parameters are the same as in Fig. \ref{SEGraphene2}.}
\label{SEGraphene9}
\end{figure}
\pagebreak

Once the emitter de-excites, its energy can be transfered to the coated substrate through a LSW or a photon may be found in a propagating or TIR mode.  In Fig. \ref{SEGraphene9}(a) the probabilities ($p^{\textrm{LSW}}_{\perp}$, $p^{\textrm{P}}_{\perp}$, $p^{\textrm{TIR}}_{\perp}$) associated to each of these processes are shown as functions of $z$ for a transition electric dipole moment perpendicular to the wall.  All material parameters are the same as in Fig. \ref{SEGraphene2}. For $B = 5$ T (solid curves) we notice that propagating modes are the main decay channel for $ z\gtrsim 50\ \mu$m and are negligible for $z \lesssim 1\ \mu$m. Total internal reflection modes are also negligible for distances below $1\ \mu$m and present an antisymmetric peak around $10\ \mu$m spread in the interval $5\ \mu\textrm{m} \lesssim z \lesssim 100\ \mu\textrm{m}$. Moreover, LSW dominate the SE rate for distances smaller than $1\ \mu$m. If the magnetic field is increased up to $B = 15$ T we will observe that none of these probabilities are significantly modified for $z \gtrsim 20\ \mu$m and $z\lesssim 1\ \mu$m. However, for $1\ \mu\textrm{m} \lesssim z \lesssim 20\ \mu\textrm{m}$ the emission channels are strongly affected by changes in the applied magnetic field. Interestingly, the energy lost related to LSW drops sharply in this range of distances whereas $p^{\textrm{P}}_{\perp}$ and $p^{\textrm{TIR}}_{\perp}$ increase. For instance, around $z = 4\ \mu$m  $p^{\textrm{LSW}}_{\perp}$ decreases from $\sim 75\%$  ($B = 5$ T) to $\sim 15\%$ ($B = 15$ T).  On the other hand, the probabilities of observing the emission of a photon into a propagating or TIR waves change from  $\sim 5\%$ and $\sim20\%$ to $\sim18\%$ and $\sim67\%$, respectively. Furthermore, note that the peak in $p^{\textrm{TIR}}_{\perp}$ is now around $6\ \mu$m.  In order to further analyze the effects of the magnetic field on the decay routes we show $p^{\textrm{P}}_{\perp}$, $p^{\textrm{TIR}}_{\perp}$, and $p^{\textrm{LSW}}_{\perp}$ as functions of $B$ for $z = 4\ \mu$m  in Fig. \ref{SEGraphene9}(b). This plot complements Fig. \ref{SEGraphene2} and shows that the magneto-optical effects of graphene not only allow us to manipulate the total SE rate in the near-field regime but also enable us to have a substantial control on the partial probabilities of emission. Note that $p^{\textrm{P}}_{\perp}$, $p^{\textrm{TIR}}_{\perp}$, and $p^{\textrm{LSW}}_{\perp}$ present discontinuities at specific values of $B$. Such discontinuities are directly linked to the discrete LL brought about by the application of a magnetic field. They occur exactly at the same values of $B$ for which graphene's longitudinal and transverse conductivities show abrupt changes (see Fig. \ref{ConductivityFig1}).

%Let us continue investigating the effects of the external magnetic field on the emitters lifetime.
Figure \ref{SEGraphene5} presents $\Gamma_{\perp}(z, B)$ normalized by $\Gamma_{\perp}(z, 0)$ as a function of $B$ for two values of the chemical potential, namely, $\mu_c = 115$ meV (blue solid/circle dotted curves) and $\mu_c = 150$ meV (red dashed/square dotted curves). The temperature of the system is held at $T = 4$ K. Panels (a) and (b) are for $z = 100$ nm and $z = 1\ \mu$m, respectively. The solid curves present the results calculated using the exact Eqs. (\ref{SEPerp}) and (\ref{SEPar}) whilst the dotted ones were obtained using the approximated Eqs. (\ref{SEPerp_NF}) and (\ref{SEPar_NF}). In both panels the QS approximation describes accurately the atomic lifetime. In addition, the SE rate presents sharp discontinuities whenever a given LL energy reaches $\mu_c$. It should be noticed that there exists a critical magnetic field given by $B_c = \mu_c^2/(2\hbar e v_F^2)$ above which the discontinuities are not present any longer. This is related to the fact that for $B \geq B_c$ all LL in the conduction band are above $\mu_c$, except from $M_0$. As a consequence, graphene's conductivities are dominated by the transition $M_0 \rightarrow M_1$. Furthermore, note that the final plateau in $\Gamma_{\perp}(z, B)/\Gamma_{\perp}(z, 0)$ has always the same behavior regardless of the value of the chemical potential. Indeed, for magnetic fields above the critical value $\Delta_{\textrm{intra}} = \sqrt{2\hbar e B v_F^2}$ independently of $\mu_c$. Therefore, both $\sigma_{xx}$ and $\sigma_{xy}$ are approximately independent of the chemical potential provided $k_BT \ll \mu_c$ and $B \geq B_c$.
\vspace{10pt}
\begin{figure}[!ht]
\centering
\includegraphics[scale=0.82]{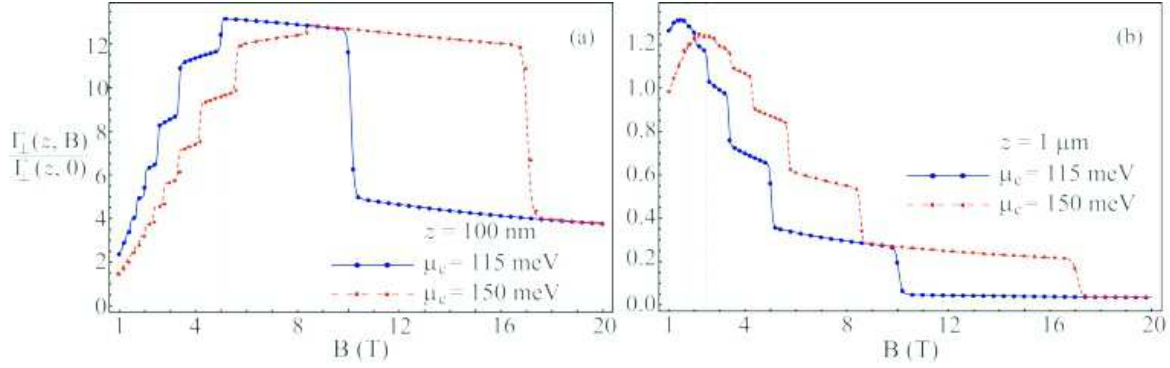}
\vspace{0pt}
\caption{Spontaneous emission rate $\Gamma_{\perp}(z, B)$ normalized by $\Gamma_{\perp}(z, 0)$ as a function of $B$ for {\bf (a)} $z = 100$ nm, and {\bf (b)} $z = 1\ \mu$m. The blue solid line and the red dashed one correspond to $\mu_c = 115$ meV and $\mu_c = 150$ meV, respectively. The circle and square dotted curves are obtained using Eq. (\ref{SEPerp_NF}), the near-field approximated result. The gray vertical lines show the values of $B_m$ calculated through Eq. (\ref{MaximoCampoMagnetico}).}
\label{SEGraphene5}
\end{figure}
\pagebreak

Another interesting point is that as we crank up the magnetic field the SE rate starts increasing, passes by a maximum and then decreases with $B$. As we can see from Figs. \ref{SEGraphene5}(a) and \ref{SEGraphene5}(b) the value of $B$ at which $\Gamma_{\perp}(z, B)/\Gamma_{\perp}(z, 0)$ achieves its maximum depends on both $\mu_c$ and $z$. This behavior can be understood in the QS regime if we look again at Fig. \ref{SEGraphene3}(b) and remember that the value of the integral in Eq. (\ref{SEPerp_NF}) depends on the product $\rho_k^{\textrm{TM}} \times \textrm{Im}[r^{\textrm{TM, TM}}]$ integrated over $k_{||}$. On one side, the function $\rho_k^{\textrm{TM}}$ has a maximum at ${k_{||}^{\textrm{max}}}_{\!\!\!\!\!\!\!\rho}\ = 1/z$. On the other side, it is possible to show that $\textrm{Im}[r^{\textrm{TM, TM}}]$ presents a peak near
%
%\begin{eqnarray}
%\label{ReflectionCoefficients_Im}
%\!\!\!\!\!\!\!\!\!\!\!\!\!\!\!\textrm{Im}[r^{\textrm{TM, TM}}_{\textrm{QS}}] \!\!\!&\simeq& \!\!\!\dfrac{2k_{||}\textrm{Re}[\sigma_{xx}]/\varepsilon_0\omega}
%{\left[\dfrac{\varepsilon_s}{\varepsilon_0}+1\right]^2 -\dfrac{2k_{||}\textrm{Im}[\sigma_{xx}]}{\varepsilon_0\omega}\left[\dfrac{\varepsilon_s}{\varepsilon_0}+1\right]+\dfrac{k_{||}^2 |\sigma_{xx}|^2}{\varepsilon_0^2\omega^2}} \, ,
%\end{eqnarray}
%
\begin{equation}
{k_{||}^{\textrm{max}}}_{\!\!\!\!\!\!\!r}\ \simeq \dfrac{\varepsilon_0 \omega_0
[\varepsilon_{s}(\omega_0)/\varepsilon_0+1]}{|\sigma_{xx}(\omega_0, B)|}\, ,
\label{MaximoIntegrando}
\end{equation}
where we took into account that $\textrm{Im}[\varepsilon_s(\omega_0)]\simeq 0$ and that the Hall conductivity can be neglected in Eq. (\ref{ReflectionCoefficients_PP_Approx}). Since $|\sigma_{xx}(\omega_0, B)|$ decreases with $B$ (for the considered values of magnetic field) we note that ${k_{||}^{\textrm{max}}}_{\!\!\!\!\!\!\!r}\ \ $ moves to higher values of $k_{||}$ as we enhance $B$. Therefore, for a fixed atom-graphene separation the overlap between $\rho_{k}^{\textrm{TM}}$ and  $\textrm{Im}[r^{\textrm{TM, TM}}]$ grows with the magnetic field intensity until ${k_{||}^{\textrm{max}}}_{\!\!\!\!\!\!\!r}\ \ \sim {k_{||}^{\textrm{max}}}_{\!\!\!\!\!\!\!\rho}\ $. After that, the superposition between $\textrm{Im}[r^{\textrm{TM, TM}}]$ and $\rho_{k}^{\textrm{TM}}$ diminishes. As a result, $\Gamma_{\perp}(z, B)/\Gamma_{\perp}(z, 0)$ grows with $B$ for $B \leq B_m$ and decreases otherwise. $B_m$ corresponds to the magnetic field strength that maximizes the integral in Eq. (\ref{SEPerp_NF}). The approximated value of $B_m$ can be calculated numerically by solving the equation ${k_{||}^{\textrm{max}}}_{\!\!\!\!\!\!\!r}\ \ = {k_{||}^{\textrm{max}}}_{\!\!\!\!\!\!\!\rho}\ $. In other words, $B_m$ satisfies
\begin{equation}
|\sigma_{xx}(\omega_0, B_m)| \simeq \varepsilon_0 \omega_0 z[\varepsilon_{s}(\omega_0)/\varepsilon_0+1]\, .
\label{MaximoCampoMagnetico}
\end{equation}
In Fig. (\ref{SEGraphene5}) the vertical dashed lines highlight the position of the peak of $\Gamma_{\perp}(z, B)/\Gamma_{\perp}(z, 0)$ obtained through Eq. (\ref{MaximoCampoMagnetico}). For $z = 100$ nm we have $B_m \simeq 5.05$ T ($B_m \simeq 8.54$ T) for $\mu_c = 115$ meV ($\mu_c = 150$ meV). For $z = 1\ \mu$m Eq. (\ref{MaximoCampoMagnetico}) gives  $B_m = 1.67$ T ($B_m = 2.44$ T) for $\mu_c = 115$ meV ($\mu_c = 150$ meV). From these plots it is clear that the approximated numerical results for $B_m$ given by the previous equation yield a reasonable estimate for the magnetic field that maximizes the SE rate. Finally, Fig. \ref{SEGraphene4} depicts $B_m$ as a function of the emitter-wall separation for $\mu_c = 115$ meV and $T = 4$ K. Note that $B_m \approx \textrm{constant}$ for some distance ranges and has strong variations with $z$ between consecutive plateaus. This is again a manifestation of the discrete Landau energy levels in graphene.
\vspace{10pt}
\begin{figure}[!ht]
\centering
\includegraphics[scale=0.55]{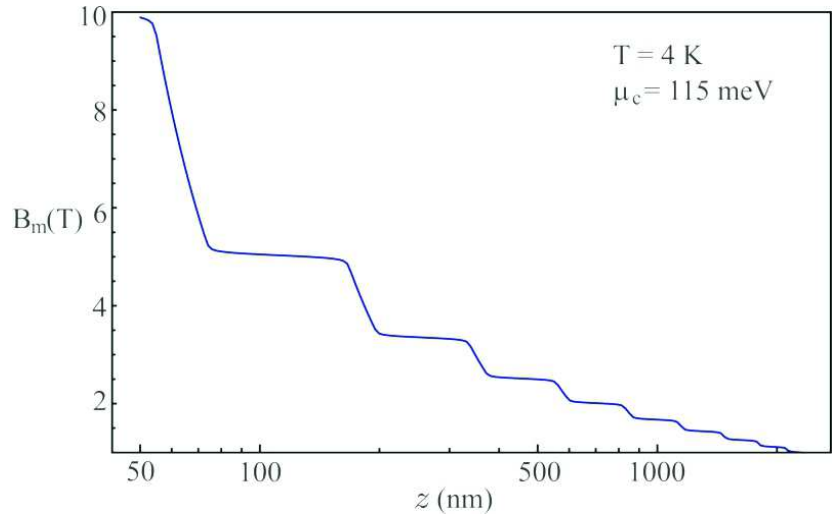}
\vspace{10pt}
\caption{Magnetic field strength $B_m$ that maximizes $\Gamma_{\perp}(z, B)/\Gamma_{\perp}(z, 0)$ as a function of the emitter-graphene distance $z$ for $T = 4$ K and $\mu_c = 115$ meV.}
\label{SEGraphene4}
\end{figure}

Despite the fact that our discussion has been mainly restricted to the effects of the magnetic field on $\Gamma_{\perp}(z, B)$, similar results also hold for $\Gamma_{||}(z, B)$. For completeness we show in Fig. \ref{SEGraphene7}(a)  a 3D plot of the relative spontaneous emission $\Delta\Gamma_{||}(\mu_c,B)/\Gamma_{||}(\mu_c, 0)$ as a function of both $\mu_c$ and $B$ for $z = 2\ \mu$m. The system is at temperature $T = 4$ K. This plot endorses our early conclusions regarding the strong control that can be achieved on the radiative properties of quantum emitters by using graphene-based materials. A comment is in order here. If the emitter's transition frequency is close or above $2\mu_c/\hbar$ the interband transitions in graphene should be taken into account in the expressions of the conductivity. In our case, the interband contributions may be relevant only for $\mu_c \lesssim \hbar\omega_0/2 \sim 1.5$ meV.  Furthermore, Fig. \ref{SEGraphene7}(a) shows that the SE rate can also be controlled by keeping the external magnetic field strength constant and changing the graphene's chemical potential. This could be an easier way to tailor the emitter's lifetime since $\mu_c$ can be altered by applying a gate voltage on the graphene sheet \cite{Graphene1, Graphene2}. Figure  \ref{SEGraphene7}(b) is a top view of panel (a) and shows a contour plot  of $\Delta\Gamma_{||}(\mu_c,B)/\Gamma_{||}(\mu_c, 0)$ as a function of $\mu_c$ and $B$. In this panel we also highlight the curves $B_n = \mu_c^2/(2ne\hbar v_f^2)$, $n =1\, ,  2\, , 3\, , ...$ where the discontinuities in the SE rate take place, namely, the magnetic field $B_n$ necessary for the crossing of the $n$-th Landau level through a given $\mu_c$ is $\mu_c^2/(2ne\hbar v_f^2)$.
\vspace{10pt}
\begin{figure}[!ht]
\centering
\includegraphics[scale=0.77]{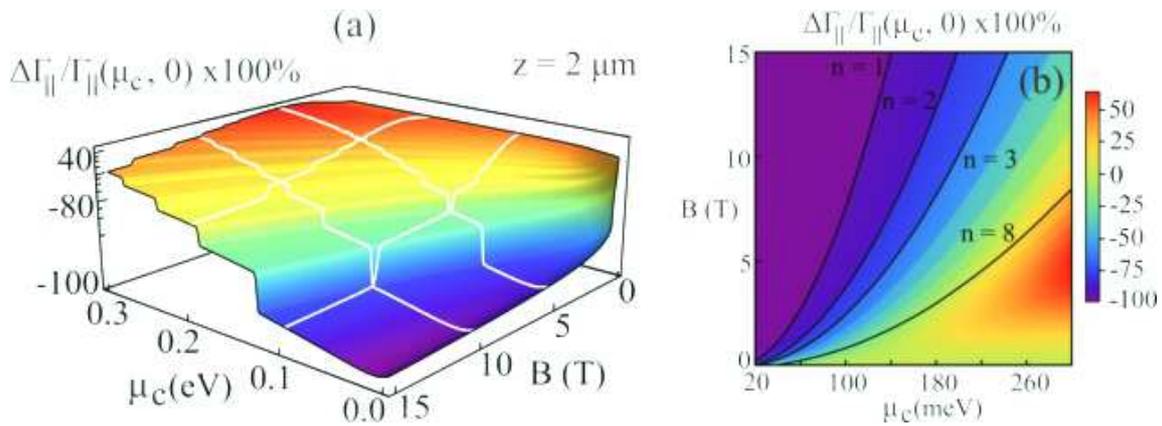}
\vspace{10pt}
\caption{{\bf (a)} 3D plot of the relative SE rate $\Delta\Gamma_{||}(\mu_c,B)/\Gamma_{||}(\mu_c, 0)$ as a function of both graphene chemical potential $\mu_c$ and external magnetic field $B$ for $z = 2\, \mu$m and $T = 4$ K. {\bf (b)} Contour plot of $\Delta\Gamma_{||}(\mu_c,B)/\Gamma_{||}(\mu_c, 0)$ as a function of $\mu_c$ and $B$ for the same parameters as in panel (a). The solid black lines are plots of $B_n = \mu_c^2/(2ne\hbar v_f^2)$ for $n = 1\, , 2\, , 3$, and $8$.}
\label{SEGraphene7}
\end{figure}
\pagebreak
\vspace{10pt}
\begin{figure}[!ht]
\centering
\includegraphics[scale=0.78]{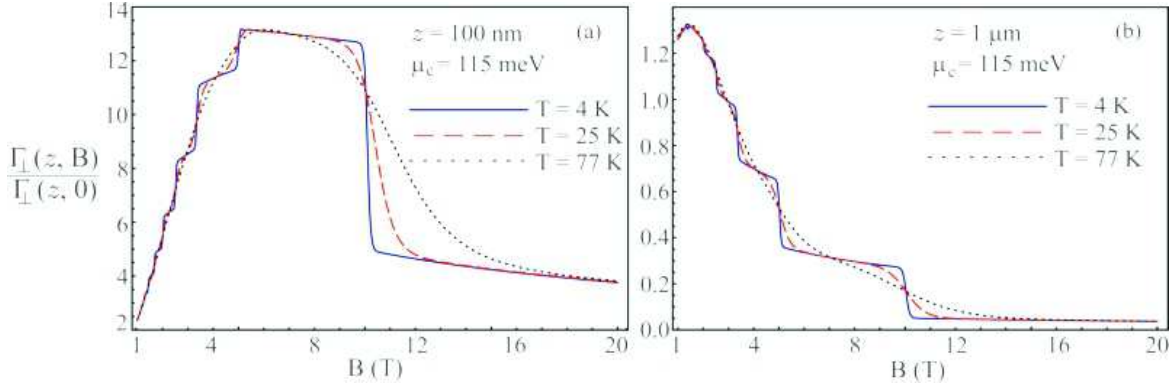}
\vspace{10pt}
\caption{Normalized spontaneous emission rate $\Gamma_{\perp}(z, B)/\Gamma_{\perp}(z, B = 0)$ as a function of $B$ for {\bf (a)} $z = 100$ nm, and {\bf (b)} $z = 1\ \mu$m. The temperatures used are $T = 4$ K (blue solid line), $T = 25$ K (red dashed line), and $T = 77$ K (black dotted line). The chemical potential is $\mu_c = 115$ meV. {\bf (b)} Spontaneous emission rate $\Gamma_{\perp}(z, B)$ normalized by $\Gamma_{\perp}(z, 0)$ as a function of $B$ for {\bf (a)} $z = 100$ nm, and {\bf (b)} $z = 1\ \mu$m. The blue solid line and the red dashed one correspond to $\mu_c = 115$ meV and $\mu_c = 150$ meV, respectively. The vertical dashed lines highlight the position of the peak of $\Gamma_{\perp}(z, B)/\Gamma_{\perp}(z, 0)$ obtained through Eq. (\ref{MaximoCampoMagnetico}).}
\label{SEGraphene8}
\end{figure}

We have also studied  thermal effects on the SE rate of a two-level quantum emitter. In Fig. \ref{SEGraphene8} we plot $\Gamma_{\perp}(z, B)/\Gamma_{\perp}(z, 0)$ as a function of $B$ for $\mu_c = 115$ meV,  $z = 100$ nm (left panel) and $z = 1\ \mu$m (right panel). Different curves correspond to distinct values of temperature of the graphene-coated half-space, namely,  $T = 4$ K (blue solid line), \linebreak $T = 25$ K (red dashed line), and $T = 77$ K (black dotted line). Notice that the main effect of the temperature is to smooth  $\Gamma_{\perp}(z, B)$ as a function of the magnetic field strength. Indeed, we notice that even for temperatures as low as $T = 25$ K the discontinuities in the atomic lifetime are not present anymore. This is a consequence of the fact that the Fermi-Dirac distribution becomes a smoother function of energy near the chemical potential as the temperature increases. Therefore, the step-like function approximation for $n_{F}(E)$ fails. It means that even if a given LL crosses $\mu_c$ there will be a nonvanishing probability that it remains filled. Therefore, the allowed intraband and interband transitions are not suddenly quenched as the magnetic field intensity is increased. The reader should also note that, in this case,  we must use Eqs. (\ref{Conductivity1}) and (\ref{Conductivity2}) instead of Eqs. (\ref{LongitudinalIntra}), (\ref{LongitudinalInter}), (\ref{TransversalIntra}) and  (\ref{TransversalInter}). Finally, it is worth mentioning that differently from stimulated emission the SE is generally not affected by changes in the temperature of the objects in the vicinities of the quantum emitter. However here we conclude that the temperature may be explored as another external agent to tailor the radiative properties of the quantum emitter. This is related to the fact that both $\sigma_{xx}$ and $\sigma_{xy}$ are sensitive to variations in the temperature. Consequently, the boundary conditions at $z = 0$ may be manipulated which allow us to control the coated substrate reflection coefficients and the EM field modes.

\vspace{10pt}

In this chapter we have investigated the SE rate of a two-level atom in the vicinities of a plasmonic cloak. We have analyzed the dependence of the SE rate on the distance between the atom and the plasmonic cloak.  We concluded that there is a substantial reduction of the SE rate for a large range of distances, even for small separations between the atom and the cloak. We have also investigated the dependence of the SE rate on the geometrical parameters of the cloak as well as on its material parameters, such as the electrical permittivities.  We found that the strong suppression of the SE rate is robust against the variation of both the geometrical and material parameters of the cloak, even taking into account realistic ohmic losses. In addition, we have also studied the SE rate of a quantum emitter in the vicinities of a graphene-coated wall under the influence of an external magnetic field $B$. We have demonstrated that, in the near-field regime, the atomic lifetime can be strongly affected by changes in $B$, in the chemical potential $\mu_c$ as well as in the temperature $T$ of the environment. Particularly, the SE rate presents several discontinuities as a function of $B$ for low temperatures, a direct consequence of the discrete Landau energy levels in graphene. Furthermore, we have also investigated the different decay channels in the near-field regime; our results show that the magnetic field allows us to manipulate possible mechanisms of quantum emission by using external agents. We hope that our findings demonstrate that cloaked objects as well as graphene-based materials may play a fundamental role in the development of new photonic devices.

\end{chapter}

\begin{chapter}{Dispersive interaction between an atom and arbitrary objects}
\label{cap7}

\begin{flushright}
{\it
Although the effect is small an experimental confirmation\\
seems not unfeasible and might be of a certain interest.
}

{\sc H. B. Casimir}
\end{flushright}

\hspace{5 mm}  It has been known for a long time that quantum fluctuations give rise to interactions between neutral but polarizable objects that do not possess any permanent electric or magnetic multipoles. In this Chapter we review the basic features about dispersive interactions between atoms and neutral bodies. Particularly,  we present the Eberlein and Zietal's method \cite{Eberlein-Zietal-2007} that allows for the calculation of dispersive interactions between an atom and conducting surfaces of arbitrary shape in the short distance regime. We use this method to calculate the interaction between an atom and an ellipsoid. We also present a very general expression for  the dispersive interaction between an atom and an anisotropic flat surface, valid for any regime of distances and temperatures. We use this result to investigate the dispersive Casimir-Polder interaction between a Rubidium atom and a suspended graphene sheet subjected to an external magnetic field ${\bf B}$. We demonstrate that this physical system allows for an unprecedented control of dispersive interactions at micro and nanoscales.
%Indeed, we show that the application of an external magnetic field can induce a $80\%$ reduction of the Casimir-Polder energy relative to its value without the field. We also show that sharp discontinuities emerge in the Casimir-Polder interaction energy for certain values of the applied magnetic field at low temperatures. Moreover, for sufficiently large distances these discontinuities show up as a plateau-like pattern with a quantized Casimir-Polder interaction energy, in a phenomenon that can be explained in terms of the quantum Hall effect. In addition, we point out the importance of thermal effects in the Casimir-Polder interaction, which we show that must be taken into account even for considerably short distances. In this case, the discontinuities in the atom-graphene dispersive interaction do not occur, which by no means prevents the tuning of the interaction in $\sim 50 \%$ by the application of the external magnetic field.
%

\section{Introduction}
\label{DispersiveInteractions}

\hspace{5mm} Intermolecular forces have been studied for approximately three centuries. Since molecules in a real gas condense into liquids and freeze into solids, it is natural to expect that there should exist attractive intermolecular forces, a conclusion that had already been achieved by Newton at the end of the 17th century \cite{Israelachvili-2011}. The phenomenon of capillarity - the property that allows liquids to adhere to the walls of a recipient in spite of external forces like gravity - was studied by the first time by Clairaut who suggested in 1743 that this phenomenon could be explained if the forces between the molecules of the liquid and those of a tube of glass were of a different nature from the intermolecular forces between the molecules of the liquid themselves \cite{Margenau-Kestner-1969}. This same phenomenon was considered later on by Laplace, in 1805, and by Gauss, in 1830. Many others  were also involved in the study of intermolecular forces as, for instance, Maxwell and Boltzmann.  A more complete list of references can be found in  Israelachvili's book \cite{Israelachvili-2011}.

Following a different approach, J. D. van der Waals suggested in his dissertation~\cite{VanDerWaals-1873} an equation of state for real gases, given for one mol of gas by $(P + a/V^2)(V - b) = RT$, where $P$, $V$ and $T$ are respectively the pressure, volume and absolute temperature of the gas, $R$ is the universal constant of gases, and $a$ and $b$ are two adjustable parameters. Parameter $b$ was introduced to take into account the finite volumes of the molecules  and the term $a/V^2$ would be  related to the existence of an attractive intermolecular force. These attractive forces are called  van der Waals forces. In fact, we must distinguish three types of van der Waals forces: the orientation force, the induction force and the dispersion force.

Orientation forces occur between two molecules with permanent electric dipoles. It is possible to show that for two randomly oriented dipoles ${\bf d}_1$ and ${\bf d}_2$ the van der Waals interaction energy between them is \cite{Keesom-1915,Keesom-1920}
\begin{equation}
 U_{\textrm{or}}(r) = -\, \mbox{$\dfrac{2d_1^2  d_2^2}{3 k_B T (4\pi\varepsilon_0)^2 r^6}$}\, , \ \ \ k_B T \gg \mbox{$\dfrac{d_1d_2}{4\pi\varepsilon_0 r^3}$}\, ,
\end{equation}
where $r$ is the distance between the two dipoles, $T$ is the temperature of the system, and $k_B$ is the Boltzmann constant. Although there are as many configurations that give rise to attractive forces as configurations that give rise to repulsive forces, Boltzmann weight ($e^{-{\cal E}/k_B T}$) favours the lower energies which correspond to ``attractive configurations". Clearly, as $T$ increases indefinitely all configurations become equally available, leading to a vanishing force.

Induction forces occur between a non-polar but polarizable molecule and another one that has a permanent electric dipole (or a higher multipole). Evidences that  non-polar molecules indeed exist led Debye \cite{Debye-1920,Debye-1921} and others to consider this kind of forces. The permanent dipole of one molecule  induces a dipole in the non-polar one, leading to a behavior similar to the previous dipole-dipole interaction. Hence, if $d_1$ is the magnitude of the dipole moment of the polar molecule then, apart from a numerical factor, the interaction energy between this molecule and a non-polar one will be\cite{Milonni-1994, WiltonAJP}
\begin{equation}
U_{\textrm{ind}}(r) \sim -\dfrac{\alpha_2 d_1^2}{r^6}\, ,
\end{equation}
with $\alpha_2$ being its static polarizability. The induction force does not disappear for high temperatures, since the orientation of dipoles 1 and 2 are not independent. Indeed, the induced dipole is parallel (for isotropic molecules) to the field generated by dipole 1 at the position of molecule 2, which explains the attractive character of the induction van der Waals interaction.

The above two types of van der Waals forces do not explain the attraction between atoms or non-polar molecules, like those that occur for instance in noble gases. The explanation for this kind of forces was provided only after the advent of Quantum Mechanics  \cite{Eisenschitz-London-1930, London-1930}. Due to quantum fluctuations, the charge and current distributions in an atom fluctuate and, consequently, instantaneous dipoles exist and give rise to an electromagnetic interaction. These fluctuations are ultimately related to the Heisenberg uncertainty principle. The first formal derivation of the interaction energy between two neutral atoms was carried out by Eisenschitz and London \cite{Eisenschitz-London-1930, London-1930}. Considering two Hydrogen atoms these authors were able to show that the interaction energy ought to be given by \cite{Eisenschitz-London-1930, London-1930,CohenEtAl-1973,Mahanty-1977}
\begin{equation}
U_{\textrm{disp}}(r) = -\mbox{$\dfrac{3}{4}$}
\mbox{$\dfrac{\hbar\omega_0\alpha(0)^2}{(4\pi\varepsilon_0)^2 r^6}$}\, ,
\end{equation}
where $\alpha(\omega)$ is the atomic polarizability of Hydrogen and $\omega_0$  is the dominant transition frequency for the interaction. Since the dynamical polarizability $\alpha(\omega)$ is related to the permittivity $\varepsilon(\omega)$, these forces were called by London dispersive van der Waals forces \cite{London-1930}.

Experiments with colloids, made at the Phillips laboratories in the first half of the 1940's by Verwey  and Overbeek, showed that if London's result was used there would be discrepancies between theoretical predictions and experimental data \cite{Verwey-Overbeek-1948}. They noticed that in order to retrieve agreement between experimental data and theory the dispersion interaction between two atoms should fall for large distances faster than $1/r^6$. Further, Overbeek conjectured that such a change in the force law was due to retardation effects of the electromagnetic interaction. These effects become important as the time elapsed by light to propagate from one atom to the other is of the order of characteristic times of the atoms. Assuming there is a dominant transition frequency $\omega_0$, retardation effects become relevant for $r/c \gtrsim 1/\omega_0$. Therefore, we distinguish two regimes for dispersion interactions: the non-retarded or short distance regime and the (asymptotically) retarded or large distance regime. The latter is valid for $r\gg \lambda_0 = 2\pi c/\omega_0$ while the former is valid for $a_0\ll r\ll \lambda_0$, with $a_0$ being the Bohr radius. The influence of retardation effects on the London-van der Waals forces was first reported by Casimir and Polder  in the late 40's \cite{Casimir-Polder-1946,Casimir-Polder-1948}. They showed that for $r\gg \lambda_0$ the dispersive Casimir-Polder energy between two atoms in the long-distance regime can be cast as
\begin{equation}
U_{\textrm{ret}}(r) = -\mbox{$\dfrac{23\hbar c}{4\pi}$}
\mbox{$\dfrac{\alpha_1\alpha_2}{(4\pi\epsilon_0)^2 r^7}$}\, ,
\end{equation}
where $\alpha_1$ and $\alpha_2$ are the static polarizabilities of atoms 1 and 2, respectively. We can understand, qualitatively, why retardation effects weaken the interaction between two atoms as follows. When the time taken for the electric field created by the fluctuating dipole of atom $1$ to reach atom $2$ and return to atom $1$ is of the order of the period of the fluctuating dipole itself, the instantaneous dipole of atom $1$ will have changed substantially from its original value so that the mutual configuration of both atoms is less correlated and less favourably disposed for an attractive interaction.

In contrast to the Coulomb interaction, which obeys the superposition principle, dispersive van der Waals interactions are not pairwise additive \cite{Axilrod-Teller-1943, Israelachvili-2011, Margenau-Kestner-1969,Langbein-1974,Milonni-1994,Farina-Santos-Tort-1999}, as first noticed by Axilrod and Teller \cite{Axilrod-Teller-1943}. This fact must be taken into account in the computation of the van der Waals force between an atom and a macroscopic body. A pairwise integration with London or Casimir and Polder forces would be justified only for diluted bodies. Non-additivity effects on the energy of a system may be positive or negative and they can be very important \cite{Israelachvili-2011}.

Although we shall not discuss any experiment on  dispersive forces between atoms and macroscopic surfaces, we shall mention a few of them. In 1993, a remarkable experiment was done by Sukenik {\it et al} \cite{SukenikEtAl-1993}, in which for the first time the change in the power law between retarded and non-retarded regimes was observed directly with atoms. In 1996, Landragin {\it et al} \cite{LandraginEtAl-1996} measured the van der Waals force in an atomic mirror based on evanescent waves. In 2001, quantum reflection was used to measure dispersive forces by Shimizu \cite{Shimizu-2001}. A short but valuable description of these experiments can be found in the nice paper by Dalibard \cite{Dalibard-2002}. In 2005, due to the Casimir-Polder potential, modifications in the center-of-mass oscillation frequency of a Bose Einstein condensate made of Rubidium atoms trapped in the vicinities of a silica surface were observed \cite{Harber-2005}. Further, in 2007, the same technique was used to measure thermal effects in the interaction \cite{Obrecht-2007}. More recently the interaction energy between atoms and surfaces for temperatures in the 500-1000 K range and distances $\sim 100$ nm have been reported \cite{Laliotis}.

Dispersive forces appear not only in different areas of Physics, as atomic and molecular physics, condensed matter physics and quantum field theory, but also in Engineering, Chemistry and Biology \cite{Parsegian-Book, Milton04, Lamoreaux05, CasimirLivro, Mostepanenko-Book-2009, vdW-Chemistry}. In QFT it is closely connected to the Casimir effect
\cite{Elizalde-Romeo-1991,Farina-BJP-2006, Milonni-1994,Proceedings-Leipzig-1998,Milton-Book-2004,Mostepanenko-Book-2009}. There are even more surprising situations where dispersive forces play an important role, like in the adhesion of geckos \cite{AutumnEtAl-Gecko-2002,LeeEtAl-Nature-2007} or as an important element in the generation of electric potentials in thunderstorms \cite{Lamoreaux-PhysicsToday-2007}. The interested reader can found more information about the history of intermolecular forces and calculations of the dispersive van der Waals interactions between two atoms at any separation in Refs.~\cite{Milton-AJP, Buhmann-Welsch-2007, Rowlinson-Book-2002, Holstein-2001,Craig-Thiru-1998,Salam-2010}.

This chapter is organized as follows: in Section {\bf \ref{InteractionTheoryEberlein}} we present the Eberlein-Zietal methot for calculation of dispersive forces between atoms and conducting surfaces of arbitrary shape in the short distance regime. As selected examples we calculate the van der Waals interaction between an atom and a grounded/isolated sphere or grounded ellipsoid. In Section {\bf \ref{DispersiveInteractionTheory}} we study the dispersive interaction between an atom and a flat surface with arbitrary optical properties in any regime of distances and temperature. Finally, in Section {\bf \ref{AtomGrapheneForce}} we use the formulation of the previous section to investigate the dispersive force between an atom and a suspended graphene sheet.

\section{van der Waals interaction between an atom and conducting objects of arbitrary shapes}
\label{InteractionTheoryEberlein}

\hspace{5mm} In this section we shall be concerned only with non-retarded dispersive forces, which do not demand quantization of the electromagnetic field. Most of the methods and results presented in this section and in the following subsections were published in Ref. \cite{WiltonAJP}. Actually, non-retarded dispersive forces can be computed quantum mechanically.  Particularly, we shall focus our attention on the van der Waals interaction energy between an atom and a perfectly conducting surface, which will be calculated using the method introduced by Eberlein and Zietal \cite{Eberlein-Zietal-2007} in 2007. As it will become clear, this method has the advantage of relating the quantum problem to a corresponding classical one in electrostatics so that all we need is to compute the appropriate Green function. Interestingly, the Eberlein and Zietal method is very convenient when the corresponding electrostatic problem admits a known image solution~\cite{ReinaldoEtAl-2011,ReinaldoEtAl-Proceeding-2012,WiltonAJP}.

In order to compute the van der Waals interaction  between two neutral but polarizable atoms, we first need  to determine the interaction Hamiltonian operator to be used in the perturbative quantum mechanical calculation of their interaction. In this case, since we are in the short distance regime this is achieved simply by computing the coulomb interaction between all charges of one atom with all charges of the other atom \cite{CohenEtAl-1973,WiltonAJP}. If the distance between the atoms is supposed to be small enough so that retardation effects can be neglected, but still large enough when compared to the Bohr radius, a Taylor expansion can be done leading to a dipole-dipole interaction as the dominant term. However, for the case of a neutral but polarizable atom interacting with a macroscopic body (a perfect conducting surface in our case), we have to take into account a huge number of pairwise coulomb interactions. Therefore, it is convenient to introduce the electrostatic potential $\Phi$ and volume charge density $\rho$ and write the Coulomb interaction hamiltonian between the atom and the conducting body in terms of the electrostatic potential, namely,
\begin{equation}
\label{UCoulomb}
U_{\textrm{Coul}} = \frac{1}{2}\int \rho({\bf r}) \Phi({\bf r})\, d{\bf r}\, .
\end{equation}

The electrostatic potential $\Phi({\bf r})$  satisfies the Poisson equation $\nabla^2\Phi(\mathbf{r}) = -\rho(\mathbf{r})/\varepsilon_0$, subjected to proper boundary condition on the surface $S$ of the object. For a grounded surface the BC imposed on $\Phi$ is given by $ \Phi(\mathbf{r})\Big|_{\mathbf{r}\in\mathcal{S}} = 0$. The electrostatic energy of the configuration is then given by (\ref{UCoulomb}). Solutions of the Poisson equation can be obtained from the Green function method \cite{Byron1992}, where the Green function $G({\bf r},{\bf r}^{\,\prime})$ satisfies
\begin{equation}
\label{green}
\nabla^2G(\mathbf{r},\mathbf{r}') = -\delta(\mathbf{r}-\mathbf{r}') \, .
\end{equation}
Thus, a general solution for the potential can be written as
\begin{equation}
\label{phig}
\Phi(\mathbf{r}) = \frac{1}{\varepsilon_0}\int G(\mathbf{r},\mathbf{r}') \rho (\mathbf{r}') d\mathbf{r}' \, .
\end{equation}
To guarantee that the electrostatic potential obeys the above mentioned BC it suffices to impose  $ G(\mathbf{r},\mathbf{r}')\Big|_{\mathbf{r}\in\mathcal{S}}=0$. In terms of the Green function $G({\bf r},{\bf r}^{\,\prime})$ the electrostatic energy given by (\ref{UCoulomb}) takes the form
\begin{equation}\label{energyg}
U_{\textrm{Coul}} =
\frac{1}{2\varepsilon_0}\int\!\! d\mathbf{r}\, d\mathbf{r}'\;\rho(\mathbf{r}) G(\mathbf{r},\mathbf{r}')\rho(\mathbf{r}')  \, .
\end{equation}

Since equation (\ref{green}) is nothing but the Poisson equation for a point charge at position $\mathbf{r}'$ (except for a  multiplicative constant factor) a particular solution of (\ref{green}) is readly obtained,
$G_p({\bf r},{\bf r}^{\,\prime}) = \mbox{\large$\frac{1}{4\pi|\mathbf{r}-\mathbf{r}'|}$}$.
However, this solution does not obey the correct boundary condition at the surface of the conducting objects. To adjust the BC, we add to this particular solution a solution of the homogeneous equation and write
\begin{equation}\label{G}
G(\mathbf{r},\mathbf{r}')=\frac{1}{4\pi|\mathbf{r}-\mathbf{r}'|} + G_H(\mathbf{r},\mathbf{r}') \, ,
\end{equation}
where $G_H({\bf r},{\bf r}^{\,\prime})$  satisfies Laplace equation, namely, $ \nabla^2G_H(\mathbf{r},\mathbf{r}')=0 $. It is straightforward to show that the BC satisfied by $G_H({\bf r},{\bf r}^{\,\prime})$ is given by
\begin{equation}
\label{cch}
\left[\frac{1}{4\pi|\mathbf{r}-\mathbf{r}'|} + G_H(\mathbf{r},\mathbf{r}')\right]_{\mathbf{r}\in S}=0 \, .
\end{equation}

Let us now consider the charge density to be used in our problem. Regarding the atom as an electric dipole ${\bf d}$  with the positive point charge at position ${\bf r_0}$ and the negative one at position ${\bf r_0} + {\bf h}$ (in a moment we shall take the limit ${\bf h}\rightarrow {\bf 0}$), we write $ \rho (\mathbf{r}) =  q \Bigl[\delta(\mathbf{r} - \mathbf{r}_0) - \delta\Bigl(\mathbf{r} - (\mathbf{r}_0 + \mathbf{h})\Bigr)\Bigr]$\footnote{\label{note13} The contribution to the electrostatic energy coming from the term $\frac{1}{2}\oint_{\cal{S}}\sigma({\bf r})\Phi({\bf r})dA = \frac{1}{2}Q\Phi_0$, where $Q$ is the total charge of the conducting object and $\Phi_0$ is the electrostatic potential at the surface, vanishes. Indeed, for a grounded object $\Phi_0 = 0$ whereas for an isolated but neutral body $Q = 0$.}. By using this expression for $\rho$ together with Eqs. (\ref{energyg}) and (\ref{G}), and taking the limit
$\mathbf{h}\rightarrow 0$, with $q\mathbf{h} \rightarrow \mathbf{d}$, we obtain
\begin{eqnarray}\label{taylor1}
U_{\textrm{Coul}} =
\dfrac{q^2}{2\varepsilon_0} \lim\limits_{\mathbf{h}\rightarrow 0\atop{ q\mathbf{h}=\mathbf{d}}}
\!\!\!\!\!\!&\Big\{&\!\!\!\!\!\!G_H(\mathbf{r}_0+\mathbf{h},\mathbf{r}_0+\mathbf{h})- G_H(\mathbf{r}_0+\mathbf{h},\mathbf{r}_0)\cr
&-&\!\!\!G_H(\mathbf{r}_0,\mathbf{r}_0+\mathbf{h})+G_H(\mathbf{r}_0,\mathbf{r}_0) \Big\} \, ,
\end{eqnarray}
where we have already subtracted the divergent self-interaction terms~\cite{Eberlein-Zietal-2007}. After making a Taylor expansion and generalizing the result to quantum mechanics it can be shown that the desired  interaction Hamiltonian operator between the atom and the conducting surface is given by \cite{Eberlein-Zietal-2007,WiltonAJP}
\begin{eqnarray}
 \hat{\cal{H}}_{\textrm{int}}= \frac{1}{2\varepsilon_0}(\mathbf{\hat{d}}\cdot\nabla')(\mathbf{\hat{d}}\cdot\nabla)G_H(\mathbf{r},\mathbf{r}')\bigg|_{\mathbf{r}=\mathbf{r}'=\mathbf{r}_0} \, , \label{taylor2}
\end{eqnarray}
where the atomic dipole moment that appears in (\ref{taylor2}) is a quantum operator.

In first order of perturbation theory the desired non-retarded interaction energy between the atom and the conducting surface is just the quantum expectation value of the Hamiltonian operator given by Eq. (\ref{taylor2}). We shall always work with an orthonormal basis, for which the expectation value of the aforementioned Hamiltonian\linebreak operator gives~\cite{WiltonAJP}
\begin{equation}\label{eberlein}
U^{\textrm{NR}}({\bf r}_0) =\langle \hat{\cal{H}}_{\textrm{int}} \rangle =  \frac{1}{2\varepsilon_0}\sum\limits_{m=1}^3\langle \hat{d}_m^2\rangle\nabla_m\nabla_m^{\,\prime} G_H(\mathbf{r},\mathbf{r}')\big|_{\mathbf{r}=\mathbf{r}'=\mathbf{r}_0} \, ,
\end{equation}
which is precisely the expression obtained by Eberlein and Zietal~\cite{Eberlein-Zietal-2007}.

This method has the advantage of relating the quantum problem to a corresponding classical one in electrostatics. Its remarkable simplicity relies on the fact that to obtain the non-retarded van der Waals interaction energy of an atom near any conducting surface we must  find only the homogeneous solution of Laplace equation, $G_H(\mathbf{r},\mathbf{r}')$, corresponding to that geometry and satisfying the BC given by Eq. (\ref{cch}). It is interesting to note that, except for constants, the equations satisfied by $G_H(\mathbf{r},\mathbf{r}')$ are those that yield the electrostatic potential of the image charges for the problem of a point charge at position $\mathbf{r}'$ in the presence of surface $\mathcal{S}$. Therefore, we see that the image method is a very useful tool to find the homogeneous solution $G_H(\mathbf{r},\mathbf{r}')$  which, in turn, is the only function needed in the Eberlein and Zietal's method to obtain the (quantum) non-retarded dispersive interaction between an atom and a conducting surface $\mathcal{S}$ of arbitrary shape. We should emphasize, however, that the use of the image method is not mandatory. In fact, in the pioneering work on this method  \cite{Eberlein-Zietal-2007}, the authors discussed the problems of an atom interacting with an infinite conducting semi-plane, and an atom interacting with an infinite conducting cylinder using the Green function method. Also, in subsequent papers, this method was employed without the use of the image method, as in the calculation of the non-retarded interaction between an atom and a dielectric slab \cite{Contreras-Eberlein-2009}, a disc \cite{Eberlein2012}, and a plane with a circular hole \cite{Eberlein-Zietal-2011, Milton-2011}. The latter as well as the atom-disc system were discussed with the aid of the more complicated Sommerfeld's image method \cite{Sommerfeld-1896} in a recent article \cite{ReinaldoEtAl-2011}. Finally, it is worth mentioning that the Eberlein and Zietal's method remains valid even in the case of a not-grounded isolated conducting object. Even in this case, the  Eberlein and Zietal's formula (\ref{eberlein}) is valid and $G_H({\bf r},{\bf r}^{\,\prime})$ still coincides with the electrostatic potential created by the image charges of the problem \cite{WiltonAJP}.

\subsection{Atom-sphere interaction}

\hspace{5mm} We now illustrate the Eberlein and Zietal's method in the simple case of an atom near an isolated but neutral conducting sphere. The atom-sphere system had been investigated before by many authors \cite{Marvin-Toigo-1982,Jhe-Kim-1995,Jhe-Kim-1995-2,BuhmannEtAl-2004} for spheres with different properties and recently has been a subject of great interest \cite{Sambale-Buhmann-Scheel-2010,EllingsenEtAl-2011,EllingsenEtAl-2012, Milton-2012}. Here, the corresponding electrostatic problem we need to solve is that of a point charge $q$ at position $\mathbf{r}'=(x',y',z')$ in the presence of the conducting sphere. The image method \cite{Griffiths1999} tells us that in this case we have to put an image charge $q_{i} = -Rq/r^{\,\prime}$ at position $\mathbf{r}^{\,\prime}_{i} = R^2\mathbf{r}'/r'^2$, and an extra image charge $-q_i$ at the origin (${\bf r}^\prime_i = {\bf 0}$), as  sketched in Fig. \ref{atesat}.
\vspace{10pt}
\begin{figure}[!ht]
\centering
\includegraphics[scale=0.4]{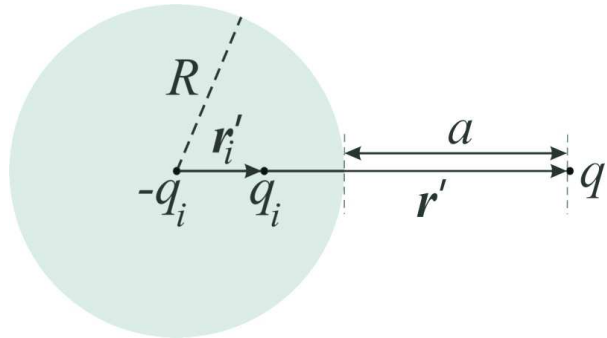}
\vspace{10pt}
\caption{Point charge $q$ near an isolated but neutral conducting sphere of radius $R$ and their images  $q_i$ and $-q_i$.}
\label{atesat}
\end{figure}

It is straightforward to show that the homogeneous Green function for this\linebreak problem reads \cite{Griffiths1999, WiltonAJP}
\begin{equation}
\label{GreenFunctionSphere}
	G_H(\mathbf{r},\mathbf{r}') =
-\dfrac{R}{4\pi|\mathbf{r}'||\mathbf{r}-\mathbf{r}^{\,\prime}_{i}|}+\dfrac{R}{4\pi|\mathbf{r}||\mathbf{r}'|} \, .
\end{equation}

Substituting (\ref{GreenFunctionSphere}) in (\ref{eberlein}) and assuming for the sake of simplicity an isotropic atom, $\langle \hat{d}_x^2\rangle = \langle \hat{d}_y^2\rangle = \langle \hat{d}_z^2\rangle = \langle {\bf \hat{d}}^2\rangle/3$,  it is possible to show that the dispersive interaction in the non-retarded regime is given by  \cite{WiltonAJP}
\begin{equation}\label{AtomoEsferaIsolada-aR}
U\!_{\textrm{ais}}^{\textrm{ NR}}(a,R) = -\dfrac{\langle {\bf \hat{d}}^2\rangle}{24\pi\varepsilon_0\, a^3}
 \left\{ \dfrac{4}{(2 + a/R)^3} +  \dfrac{a/R}{(2 + a/R)^2} - \dfrac{a^3/R^3}{(1 + a/R)^4}\right\}\, ,
\end{equation}
where $a$ is  the smallest distance between the atom and the sphere. It is important to remark that the second term on the right hand side of (\ref{AtomoEsferaIsolada-aR}) is, in absolute value, greater than the third one. Therefore, the interaction of an atom with an isolated but neutral conducting sphere is always attractive.

The only difference between the Green function of grounded and isolated spheres is the term $R/{4\pi|\mathbf{r}||\mathbf{r}'|} $ present in (\ref{GreenFunctionSphere}). As a consequence, the atom-grounded sphere interaction can be easily achieved just by omitting the last term in (\ref{AtomoEsferaIsolada-aR}) \cite{WiltonAJP}
\begin{equation}\label{AtomoEsfera-aR}
U\!_{\textrm{\textrm{ags}}}^{\textrm{ NR}}(a,R) = -\dfrac{\langle {\bf \hat{d}}^2\rangle}{24\pi\varepsilon_0\, a^3}
 \left\{ \dfrac{4}{(2 + a/R)^3} +  \dfrac{a/R}{(2 + a/R)^2} \right\}\, .
\end{equation}
Hence,  we conclude that the attraction is stronger in the case of a grounded sphere than in the case of an isolated sphere. Physically this can be understood as a consequence of the charge acquired by the grounded sphere. In addition, note that if in the previous equation we take the limit  $R\rightarrow\infty$, with finite  $a$, both equation (\ref{AtomoEsferaIsolada-aR}) and (\ref{AtomoEsfera-aR}) will reproduce the non-retarded result for the atom-plane system, namely,
\begin{equation}
\label{Lennard-JonesEq}
U\!_{\textrm{ap}}^{\textrm{ NR}}(a) = - \dfrac{\langle {\bf \hat{d}}^2\rangle}{48\pi\varepsilon_0\, a^3}\, .
\end{equation}
This is the well-known interaction between an atom and an infinite conducting plane in the non-retarded regime firstly obtained in 1932 by Lennard-Jones\footnote{\label{note10} It is worth mentioning that in Lennard-Jones result a factor of $1/2$ is missing. This same $1/2$ factor is also missing in Ref. \cite{CohenEtAl-1973}. This numerical error follows from a misuse of the image method \cite{TaddeiEtAl-2009}.
%Recall that, for the case at hand, this method states that the force acting on a dipole near a perfectly conducting (infinite) plane is that exerted by the image dipole and, consequently, the electrostatic energy between the dipole and the conducting plane is given by $1/2$ times the interacting energy between the dipole and its image \cite{TaddeiEtAl-2009}.
}
\cite{Lennard-Jones-1932, CohenEtAl-1973,Dalibard-2002}.

We finish this section by taking the limit $R\rightarrow 0$, but with $4\pi\varepsilon_0R^3 \rightarrow \alpha_s$, where $\alpha_s$ is the (finite) polarizability of a very small conducting sphere. Thus, Eq. (\ref{AtomoEsferaIsolada-aR}) reduces to
\begin{equation}\label{PontoCondutor}
\lim_{R\rightarrow 0\atop{\alpha_s\, =\, 4\pi\varepsilon_0 R^3 }}
U\!_{\textrm{ais}}^{\textrm{ NR}}(a,R) = -\dfrac{\langle {\bf \hat{d}}^2\rangle}{24\pi\varepsilon_0\, a^3}
 \dfrac{4R^3}{a^3} \; =\;
 -\, \dfrac{\hbar\omega_{10}\alpha_a\alpha_s}{(4\pi\varepsilon_0)^2\, a^6}\, ,
\end{equation}
where in the last step we used the explicit expression of $\langle {\bf \hat{d}}^2\rangle$ in terms of the atomic polarizability $\alpha_a$ \cite{Milonni-1994} and assumed that the transition from the fundamental state to the first excited one is dominant (transition frequency $\omega_{10}$). This result is a London-like dipole-dipole interaction\cite{London-1930, CohenEtAl-1973, Milonni-1994}, as expected.
\subsection{Atom-prolate ellipsoid interaction}

\hspace{5mm} Here, we apply the Eberlein and Zietal's method in a more complex situation, namely, the non-retarded interaction between an atom and a grounded prolate ellipsoid with arbitrary eccentricity, as depicted in Fig. \ref{Ellipsoid}.  In this case, it is not convenient to solve the corresponding electrostatic problem using the image method\footnote{\label{note15} As far as we know, the exact solution of this problem via the image method is lacking in the literature.}. As a consequence, in order to obtain $G_{H}({\bf r}, {\bf r}')$ we must solve the Laplace equation using other techniques, such as separation of variables. Due to the symmetry of the problem it is convenient to use prolate spheroidal coordinates $(\xi, \eta, \varphi)$, which are related to the cartesian ones by~\cite{Lebedev, Straton}
\begin{eqnarray}
x\!\!\! &=&\!\!\! f\sqrt{(1-\eta^2)(\xi^2-1)}\cos\varphi\, , \\
y\!\!\! &=&\!\!\! f\sqrt{(1-\eta^2)(\xi^2-1)}\sin\varphi\, , \\
z\!\!\! &=&\!\!\! f\xi\eta\, ,
\end{eqnarray}
where $1\leq \xi < \infty$, $-1\leq \eta \leq 1$, $0 \leq \varphi \leq 2\pi$, and $f$ is a scale factor. Actually, it is possible to show that $\xi =$ constant ($\eta=$ constant) defines the surface of an ellipsoid (double-sheeted hyperboloid) of revolution with foci at $z = \pm f$.
\vspace{10pt}
\begin{figure}[!ht]
\centering
\includegraphics[scale=0.35]{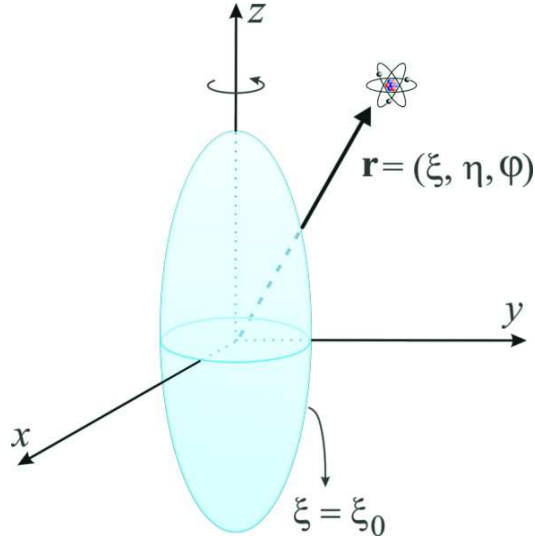}
\vspace{10pt}
\caption{Atom near a grounded conducting prolate ellipsoid. The position of the atom is given in prolate spheroidal coordinates by $(\xi, \eta, \varphi)$. The ellipsoid surface is characterized by $\xi = \xi_0 = $ constant.}
\label{Ellipsoid}
\end{figure}

Let us consider that the conducting surface of the ellipsoid is defined by $\xi = \xi_0$.  The general solution of the Laplace equation in prolate spheroidal coordinates in the region outside the ellipsoid can be cast into the form~\cite{Lebedev}
\begin{eqnarray}
G_H({\bf r},{\bf r}') = \sum_{n=0}^{\infty}\sum_{m=0}^{n} P_n^m(\eta) Q_n^m(\xi) \left[\alpha_{nm}^{H}({\bf r}')\cos(m\varphi) + \beta_{nm}^{H}({\bf r}')\sin(m\varphi) \right]\, , \ \ \xi \geq \xi_0\, ,
\end{eqnarray}
where $P_n^m(x)$ and  $Q_n^m(x) $ are the associate Legendre polynomials of first and second kind~\cite{Abramowitz},  $\alpha_{nm}^{H}({\bf r}')$, and $\beta_{nm}^{H}({\bf r}')$ are coefficients to be determined. By enforcing that the BC in Eq. (\ref{cch}) must be satisfied at $\xi = \xi_0$ and using that~\cite{Lebedev, Straton}
\begin{eqnarray}
\dfrac{(4\pi)^{-1}}{|{\bf r} - {\bf r}'|} =\!\! \sum_{n=0}^{\infty}\sum_{m=0}^{n} P_n^m(\eta) P_n^m(\xi)\Big[\alpha_{nm}^{<}({\bf r}')\cos(m\varphi)
+\beta_{nm}^{<}({\bf r}')\sin(m\varphi) \Big], \ \xi_0 \leq \xi < \xi' ,
\end{eqnarray}
where
\begin{eqnarray}
\alpha_{nm}^{<}({\bf r}')\!\!\! &=&\!\!\! \dfrac{1}{4\pi f}(-1)^m(2-\delta_{m,0}) (2n+1)\left[\dfrac{(n-m)!}{(n+m)!}\right]^2 P_n^m(\eta') Q_n^m(\xi') \cos(m\varphi')\, ,\\
\beta_{nm}^{<}({\bf r}')\!\!\! &=&\!\!\! \dfrac{1}{4\pi f}(-1)^m(2-\delta_{m,0}) (2n+1)\left[\dfrac{(n-m)!}{(n+m)!}\right]^2 P_n^m(\eta') Q_n^m(\xi') \sin(m\varphi')\, ,
\end{eqnarray}
it can be shown that
\begin{eqnarray}
\alpha_{nm}^{H}({\bf r}') = -\dfrac{P_n^m(\xi_0)}{Q_n^m(\xi_0)} \alpha_{nm}^{<}({\bf r}')\, , \ \ \textrm{and} \ \  \beta_{nm}^{H}({\bf r}') = -\dfrac{P_n^m(\xi_0)}{Q_n^m(\xi_0)} \beta_{nm}^{<}({\bf r}')\, .
\end{eqnarray}

The previous equations allow us to obtain the interaction energy between a conducting prolate ellipsoid and an atom at any position of space. However, for the sake of simplicity we shall be concerned here  with the case of an atom located on the $z-$axis and mainly polarizable in this direction ($\langle d_z^2\rangle \gg \langle d_x^2\rangle, \langle d_y^2\rangle$). In this situation it can be shown that the non-retarded interaction in the considered problem can be put into the form
\begin{eqnarray}
\label{atom-ellipsoid-eberlein-energy}
U_{\textrm{ape}}^{\textrm{NR}}(a, e) = -\dfrac{\langle d_z^2\rangle }{8\pi\varepsilon_0a^3}  \dfrac{(a^3/b_0^3)}{e^3} \sum_{n=0}^{\infty} \dfrac{(2n+1)P_n(1/e)}{Q_n(1/e)} \left. \left[\dfrac{dQ_n(\xi)}{d\xi} \right]^2\right|_{\xi = \left[(a/b_0)+1\right]/e}\, .
\end{eqnarray}
where $e $ and $b_0$ are the eccentricity and the major semi-axes of the ellipsoid, and $a = z - b_0$ is the smallest distance between the atom and the ellipsoid.
\vspace{10pt}
\begin{figure}[!ht]
\centering
\includegraphics[scale=0.77]{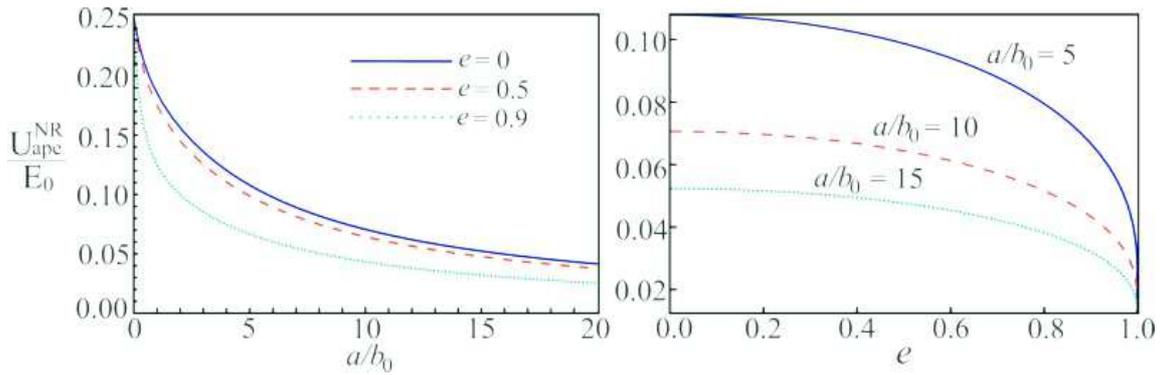}
\vspace{10pt}
\caption{van der Waals interaction energy between an atom and a prolate ellipsoid as a function of {\bf (a)} $a/b_0$ and  {\bf (b)} $e$. In panel (a) different curves correspond to $e = 0$ (blue solid line), $e = 0.5$ (red dashed line), and $e = 0.9$ (green dotted line). In panel (b) the different plots are for  $a/b_0 = 5$ (blue solid line),  $a/b_0 = 10$ (red dashed line),  $a/b_0 = 15$ (green dotted line). In all cases the atom is located along the $z-$axes and it is mainly polarizable in this same direction.}
\label{AtomEllipsoidInteraction}
\end{figure}

In Fig. \ref{AtomEllipsoidInteraction}(a) we plot $U_{\textrm{ape}}^{\textrm{NR}}(a, e)/E_0$, where $E_0 = -\langle d_z^2\rangle/8\pi\varepsilon_0a^3$, as a function of $a/b_0$ for different eccentricities: $e = 0$ (blue solid line), $e = 0.5$ (red dashed line), and $e = 0.9$ (green dotted line). Note that the interaction energy presents a monotonic behavior as function of $a/b_0$ and for $a/b_0 \rightarrow 0$ all plots approach $0.25$. This is due to the fact that for very small atom-ellipsoid separations the effects of the curvature of the conducting body are negligible and the interaction can be well described by the atom-wall result\footnote{\label{note12} Note that for $a/b_0 \rightarrow 0$ we have $U_{\textrm{ape}}^{\textrm{NR}}(a, e)\rightarrow -\langle d_z^2\rangle/32\pi\varepsilon_0a^3$. This result is different from that in Eq. (\ref{Lennard-JonesEq}). Nevertheless, the reader should remember that in Eq. (\ref{Lennard-JonesEq}) we considered an isotropic atom whilst here the atom is supposed to be polarizable only along the $z-$ direction.}. Furthermore, for a fixed value of $a/b_0$ we note that the interaction energy decreases as the ellipsoid becomes more needle-like. This is confirmed in Fig. \ref{AtomEllipsoidInteraction}(b) where $U_{\textrm{ape}}^{\textrm{NR}}(a, e)/E_0$ is shown as a function of $e$ for $a/b_0 = 5$ (blue solid curve), $a/b_0 = 10$ (red dashed curve), and $a/b_0 = 15$ (green dotted curve). As it is clear the dispersive energy has a maximum for spherical objects ($e = 0 $) and decreases strongly with $e$ for eccentricities above $\sim 0.9$.

We leave to the reader to show that in the limit $e \rightarrow 0$ the atom-sphere interaction is retrieved and that for very thin ellipsoids ($e\rightarrow 1$) the interaction energy can be rewritten conveniently as
\begin{eqnarray}
U_{\textrm{an}}^{\textrm{NR}} = -\left[\dfrac{\langle d_z^2\rangle}{8\pi\varepsilon_0b_0^3}\right] \times \left[\dfrac{3\alpha_{\textrm{needle}}}{4\pi \varepsilon_0 b_0^3}\right]
\sum_{n=0}^{\infty} (2n+1) \left. \left[\dfrac{dQ_n(\xi)}{d\xi} \right]^2\right|_{\xi = (a/b_0)+1}\, ,
\end{eqnarray}
where
\begin{equation}
\dfrac{\alpha_{\textrm{needle}}}{4\pi\varepsilon_0} = \dfrac{2a_0^2b_0}{3}\dfrac{e^2}{(1-e^2)\log\left(\dfrac{1+e}{1-e}\right)}
\end{equation}
is the polarizability of a needle with semi-axis $b_0$ and  $a_0 \ll b_0$ \cite{Landau}.

\section{Casimir-Polder interaction between an atom and a dispersive wall}
\label{DispersiveInteractionTheory}
\hspace{5mm}  In spite of being a simple method to obtain the qualitative behavior of the dispersive interaction between an atom and a surface with arbitrary shape in the short distance regime, the Eberlein and Zietal's method applies only to very idealized case of bodies that can be approximated by perfect conductors. However, in order to make realistic quantitative predictions about the dispersive force between an atom and a body we must take into account   the dispersive character of the involved materials, thermal effects and possible retardation effects,  among others \cite{Milonni-1994,Milton-Book-2004,Mostepanenko-Book-2009,Buhmann-Welsch-2007}. Such effects have been first considered in the seminal paper of Lifshitz in 1956 \cite{Lifshitz-1956}, and generalized by Dzyaloshinskii, Lifshitz and Pitaevskii \cite{DLP-1961} a couple of years later, in the calculation of the dispersive interaction between two isotropic flat surfaces in any regime of distances and temperatures. In the case where the considered objects have more complicated geometries and/or optical properties the dispersive force can be calculated via a multiple\linebreak scattering approach \cite{Reynaud-1991, Lambrecht-2006, Milton-2008, Lambrecht-2014}.

Computations of dispersive interactions between atoms and arbitrary shape surfaces with general EM properties can be a challenging task. Therefore, in this section we shall concentrate in the dispersive interaction energy $U_{\textrm{T}}(z)$ between an atom and a dispersive flat half-space at any regimes of temperatures and distances. In order to obtain an appropriate expression for $U_{\textrm{T}}(z)$ we will start from the well known result for the interaction energy between two semi-infinite media with arbitrary optical properties \cite{Felipe-PRA-LANL}, namely,
\begin{equation}
\label{LifshitzAnisotropy}
U_{\textrm{L}} = k_B T {\sum_{l=0}^\infty}' \int \dfrac{d{\bf k}_{||}}{(2\pi)^2} \textrm{Tr} \log [\mathbb{I} - \mathfrak{R}_1\mathfrak{R}_2 e^{-2\kappa_l d}]\, ,
\end{equation}
where the prime denotes that the term $l =0$ must be weighted by a factor $1/2$, $k_B$ is the Boltzmann constant, $T$ is the absolute temperature of the system, $\textrm{Tr} $ represents the trace, $\kappa_l = \sqrt{k_{||}^2 + \xi_l^2/c^2}$, $\xi_l = 2\pi l k_B T/\hbar$ $(n = 0,\ 1,\ 2,\ 3...)$ are the so called Matsubara's frequencies and $d$ is the separation distance between the walls. Besides, $\mathfrak{R}_1$ and $\mathfrak{R}_2$ are the reflection matrices of the two semi-spaces and can be cast as
\begin{equation}
\mathfrak{R}_1 = \left(
\begin{tabular}{cc}
$r_{1}^{\textrm{TE,TE}}$ & $r_1^{\textrm{TE,TM}}$ \\
$r_{1}^{\textrm{TM,TE}}$ & $r_1^{\textrm{TM,TM}}$
\end{tabular}
\right)\ , \ \ \ \textrm{and} \ \ \
\mathfrak{R}_2 = \left(
\begin{tabular}{cc}
$r_{2}^{\textrm{TE,TE}}$ & $r_2^{\textrm{TE,TM}}$ \\
$r_{2}^{\textrm{TM,TE}}$ & $r_2^{\textrm{TM,TM}}$
\end{tabular}
\right)\ ,
\end{equation}
with $r_k^{\textrm{i, j}}$ $( k =1, 2 \textrm { and } \textrm{i, j} = \textrm{TE, TM})$ being the surfaces reflection coefficients. Finally, all optical properties must be evaluated at the imaginary frequencies $i\xi_l$.

Equation (\ref{LifshitzAnisotropy}) cannot be obtained by pairwise integration over the intermolecular potentials between atoms of media 1 and 2 since dispersive interactions are not additive.  However, in the case where one of the media, let us say medium 2, is made of a rarefied isotropic material the nonadditivity effects become negligible and Eq. (\ref{LifshitzAnisotropy}) can be written as
\begin{equation}
\label{CP1}
U_{\textrm{L}}  = \int_{d}^{\infty} U_{\textrm{T}}(z) dz\, ,
\end{equation}
with $U_{\textrm{T}}(z)$ being the interaction energy between an isotropic atom of medium 2 at a distance $z$ above the interface of medium 1 as illustrated in Fig. \ref{CPFigure}.
\vspace{10pt}
\begin{figure}[!ht]
\centering
\includegraphics[scale=0.4]{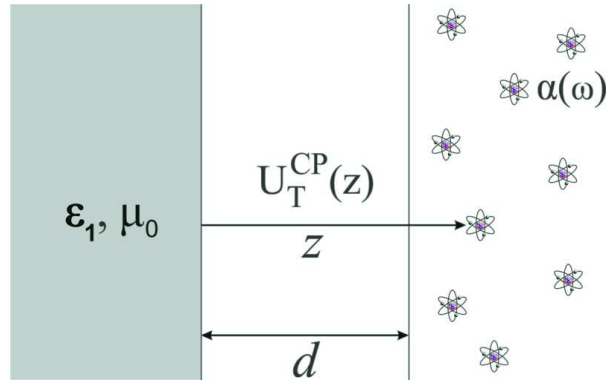}
\vspace{10pt}
\caption{In the limit of very low densities the dispersive interaction energy between two half-spaces corresponds to the sum over the interaction between all atoms of medium 2 and medium 1.}
\label{CPFigure}
\end{figure}

Let us suppose that medium 2 is made of isotropic atoms, thus, $r_2^{\textrm{TE,TM}} = r_{2}^{\textrm{TM,TE}} = 0$ and the diagonal reflection coefficients are given by Eq. (\ref{FresnelCoefficients}). In addition, in the limit of very low density the permittivity $\varepsilon_2(i\xi)$ of medium 2 can be conveniently written in terms of the atomic polarizability $\alpha(i\xi)$ as \cite{Griffiths1999}
\begin{equation}
\varepsilon_2(i\xi) \simeq \varepsilon_0 + N \alpha(i\xi)\, ,
\end{equation}
where $N$ is the number of atoms per unit of volume. In this case,  up to first order in $N\alpha/\varepsilon_0$, $\mathfrak{R}_2$ is given by\cite{Milonni-1994}
\begin{equation}
\label{ReflectionMatrixApproximated}
\mathfrak{R}_2(i\xi) \simeq \dfrac{N\alpha(i\xi)}{2\varepsilon_0}\left(
\begin{tabular}{cc}
$-\dfrac{\xi^2}{2c^2\kappa^2}$ & 0\\
0 & $1 - \dfrac{\xi^2}{2c^2\kappa^2}$
\end{tabular}\right)
\end{equation}

Substituting last equation into Eq. (\ref{LifshitzAnisotropy}) and expanding the logarithm in powers of $\alpha$ it is straightforward to show that %
\begin{equation}
\label{LifshitzAnisotropy2}
U_{\textrm{L}} \simeq N \int_d^{\infty} dz \left\{\dfrac{k_B T}{\varepsilon_0 c^2} {\sum_{l=0}^\infty}' \xi_l^2 \alpha(i\xi_l)\int \dfrac{d{\bf k}_{||}}{(2\pi)^2} \dfrac{e^{-2\kappa_lz}}{2\kappa_l}\left[r_1^{\textrm{TE,TE}} - r_1^{\textrm{TM,TM}} \left(1+\dfrac{2c^2k_{||}^2}{\xi_l^2} \right) \right]\right\}\, .
\end{equation}

Since in the low density limit the non additivity effects become negligible, a comparison  between (\ref{CP1}) and (\ref{LifshitzAnisotropy2}) allows us to identify the term in braces in last equation as the Casimir-Polder interaction between an isotropic atom and a flat surface in any regime of temperature and distances \cite{Contreras-2010, Mostepanenko-Book-2009, MostepanenkoRMP},
\begin{equation}
\label{CP2}
U_{\textrm{T}}(z) = \dfrac{k_B T}{\varepsilon_0 c^2} {\sum_{l=0}^\infty}' \xi_l^2 \alpha(i\xi_l)\int \dfrac{d{\bf k}_{||}}{(2\pi)^2} \dfrac{e^{-2\kappa_lz}}{2\kappa_l}\left[r^{\textrm{TE,TE}} - r^{\textrm{TM,TM}} \left(1+\dfrac{2c^2k_{||}^2}{\xi_l^2} \right) \right]\, .
\end{equation}
In the limit $T \to 0$ the sum in the Matsubara's frequencies in  Eq. (\ref{CP2}) can be replaced by an integral over frequencies so that in the low temperature regime the CP energy can be cast as
\begin{equation}
\label{CP3}
U_{\textrm{0}}(z) = \dfrac{\hbar}{\varepsilon_0 c^2} \int_0^{\infty}\dfrac{d\xi}{2\pi} \xi^2 \alpha(i\xi)\int \dfrac{d{\bf k}_{||}}{(2\pi)^2} \dfrac{e^{-2\kappa z}}{2\kappa}\left[r^{\textrm{TE,TE}} - r^{\textrm{TM,TM}} \left(1+\dfrac{2c^2k_{||}^2}{\xi^2} \right) \right]\, ,
\end{equation}
where $\kappa = \sqrt{k_{||}^2 + \xi^2/c^2}$.

If the material under consideration is a perfect mirror the above equation can be simplified by using that $r^{\textrm{TE,TE}} = -1$ and $r^{\textrm{TM,TM}} = 1$. Hence, at zero temperature the interaction between an atom and a conducting surface $U_{\textrm{0}}^{\textrm{perf}}(z)$ is given by \cite{MostepanenkoRMP}
\begin{equation}
\label{CPConductor}
U_{\textrm{0}}^{\textrm{perf}}(z) = -\dfrac{\hbar}{(4\pi)^2\varepsilon_0z^3} \int_0^{\infty} d\xi \alpha(i\xi) e^{-2\xi z/c} \left[1 + \dfrac{2\xi z}{c} + 2 \left(\dfrac{\xi z}{c}\right)^2 \right]\, .
\end{equation}
This expression is valid for any atom-wall distance and allows us to show that in the case of a perfect mirror the dispersive force $F_{\textrm{0}}^{\textrm{perf}}(z) = -dU_{\textrm{0}}^{\textrm{perf}}(z)/dz$ is always attractive. In the regime of very short distances, $z \ll \lambda_0$ where $\lambda_0$ is a typical transition wavelength of the atom, retardation effects are negligible and the interaction between the atom and the mirror is almost instantaneous. In this regime of distances, also called van der Waals regime, the exponential factor $e^{-2\xi z/c}$ in the above integrand can be replaced by $1$ and the polynomial by its static value since the scale of decay of $\alpha(i\xi)$ is typically $\xi \sim c/\lambda_0 \ll c/z$. Therefore, the expression for the interaction energy can be simplified to \cite{Milonni-1994}
\begin{equation}
\label{vdWallsPlane}
U_{\textrm{0}}^{\textrm{perf, vdW}}(z \ll \lambda_0) \simeq  -\dfrac{\hbar}{(4\pi)^2\varepsilon_0z^3} \int_0^{\infty} \alpha(i\xi)d \xi\, .
\end{equation}
Note that this equation could also be obtained by taking the limit $c \to \infty$ in Eq. (\ref{CPConductor}). Moreover, in this case the energy falls with $z^{-3}$ as expected [see Eq. (\ref{Lennard-JonesEq})]. On the other hand, in the so called Casimir-Polder regime $z \gg \lambda_0$ the retardation effects dominate the atom-wall interaction. In this situation the exponential factor decays faster than the polarizability so that we can replace $\alpha(i\xi)$ by $\alpha (0)$. Hence, in the retarded regime the interaction energy reads \cite{Milonni-1994}
\begin{equation}
\label{CPPlane}
U_{\textrm{0}}^{\textrm{perf, CP}}(z \gg \lambda 0) \simeq  -\dfrac{3\hbar c\alpha(0)}{32\pi^2\varepsilon_0z^4}\, ,
\end{equation}
where we ought to note that in the CP regime and for $T\to 0$ the dispersive energy \linebreak falls with $z^{-4}$.

It is worth mentioning that thermal effects usually become relevant for distances $z \sim \lambda_{\textrm{th}} := \hbar c/k_B T$ though they may be negligible in the van der Waals regime. Indeed, if instead of using (\ref{CP3}) we use (\ref{CP2}) to obtain the atom-wall interaction at a temperature $T \neq 0$, it is possible to show that the results for short atom-plane separations remain unchanged whereas in the Casimir-Polder regime the dispersive interaction energy \linebreak is given by \cite{Mostepanenko-Book-2009}
\begin{equation}
U_{\textrm{T}}^{\textrm{perf, CP}}(z) \simeq -\dfrac{k_BT \alpha(0)}{16\pi\varepsilon_0z^3}\, .
\end{equation}
Note that, the main effect of considering a finite temperature on the dispersive interaction is to give a dependence of the force with the same power law in $z$ at both short and long distances. Besides, note that in the CP regime the difference between the results at $T = 0$ and $T \neq 0$ for a fixed distance $z$ increases with the temperature of the system. These results in both regimes of temperatures and distances are summarized in Table (\ref{TableCP}).
\vspace{10pt}
\begin{table}[!ht]
\centering
 \caption{Dependence of the dispersive interaction energy $U_{\textrm{T}}^{\textrm{perf}}(z)$ between an isotropic atom and a perfect mirror with $z$ in the van der Waals and Casimir-Polder regimes at zero and finite temperatures. For the sake of simplicity we defined $I_{\alpha} = \int_{0}^{\infty} \alpha(i\xi)d\xi$.}
  \vspace{10pt}
  {\renewcommand{\arraystretch}{2}
\begin{tabular}{|c|c|c|}
\hline
                      & $T = 0$ & $T \neq 0$ \\ \hline
van der Waals regime  & $ -\dfrac{\hbar I_{\alpha}}{(4\pi)^2\varepsilon_0z^3}  $& $-\dfrac{\hbar I_{\alpha}}{(4\pi)^2\varepsilon_0z^3}$\\ \hline
Casimir-Polder regime &  $-\dfrac{3\hbar c\alpha(0)}{32\pi^2\varepsilon_0z^4} $ & $ -\dfrac{k_BT \alpha(0)}{16\pi\varepsilon_0z^3}$ \\ \hline
 \end{tabular}
 }
 \label{TableCP}
\end{table}

Finally, it is worth mentioning that even if we take into account that the half-space is actually a dispersive material described by permittivity $\varepsilon(i\xi)$ and with reflection coefficients given by the Fresnel Eq. (\ref{FresnelCoefficients}), the aforementioned features for the CP interaction will remain qualitatively unaltered, as shown in Fig. \ref{CPFigure2}. In this figure we show the dispersive interaction calculated with Eq.  (\ref{CP2}) between a Rubidium atom and a flat substrate made of Silicon Carbide as a function of distance $z$. The reflection coefficients of the substrate are given by Eq. (\ref{FresnelCoefficients}) and the electric permittivity of SiC is given by Eq. (\ref{sic}).  The Rubidium atom is a convenient choice, since there are experimental data on its complex electric polarizability $\alpha(\omega)$ for a wide range of frequencies \cite{Rubidium}. As it is clear from Eq. (\ref{CP2}), we actually need the polarizability evaluated at imaginary frequencies, which can be obtained from Kramers-Kronig relations provided one has the data for $\textrm{Im} \, \alpha(\omega)$ \cite{Jackson}. In Fig. \ref{CPFigure2} all results are normalized by $U_{\textrm{0}}^{\textrm{perf, CP}}(z) \propto z^{-4}$. The results for zero temperature are given by the blue solid line. Notice that for distances below $10$ nm the curve is approximately a straight line, $U_{\textrm{0}}^{\textrm{SiC}}(z)/ U_{\textrm{0}}^{\textrm{perf, CP}}(z) \propto z$, indicating that the atom-SiC substrate distance is within the van der Waals regime, $U_{\textrm{0}}^{\textrm{SiC, vdW}}(z) \propto z^{-3}$. On the other hand, for distances $z \gtrsim 1\ \mu$m we have $U_{\textrm{0}}^{\textrm{SiC}}(z)/ U_{\textrm{0}}^{\textrm{perf, CP}}(z) \simeq $ constant $\simeq 0.65$, showing that $U_{\textrm{0}}^{\textrm{SiC, CP}}(z) \propto z^{-4}$ in the Casimir-Polder regime. Note that $U_{\textrm{0}}^{\textrm{SiC}}(z)/ U_{\textrm{0}}^{\textrm{perf, CP}}(z)$ does not approach $1$ as the atom-substrate distance increases due to the fact that SiC behaves as a dielectric and the normalization is with respect to the perfect conductor case. The analytical results for $U_{\textrm{0}}^{\textrm{SiC, vdW}}(z)$ and $U_{\textrm{0}}^{\textrm{SiC, CP}}(z)$ [normalized by $U_{\textrm{0}}^{\textrm{perf, CP}}(z)$] are given in Fig. \ref{CPFigure2} by the dashed green and black lines, respectively. In Fig. \ref{CPFigure2} we also show the results for $T = 300$ K. In this case the thermal wavelength is $\lambda_{\textrm{th}} \simeq 7.6\ \mu$m. As we mentioned before in the van der Waals regime the results at zero and nonvanishing  temperatures coincide asymptotically. However, for atom-wall distances comparable to $\lambda_{\textrm{th}}$, thermal effects become relevant and the results at zero and finite temperature are quite different. Particularly, in the CP regime we note that $U_{\textrm{300}}^{\textrm{SiC, CP}}(z)/ U_{\textrm{0}}^{\textrm{perf, CP}}(z) \propto z$ so that $U_{\textrm{300}}^{\textrm{SiC, CP}}(z) \propto z^{-3}$. In Fig. \ref{CPFigure2} the dashed gray curve represents a plot of the analytical results for the atom-dielectric interaction at large separations and $T = 300$ K.
\vspace{5pt}
\begin{figure}[!ht]
\centering
\includegraphics[scale=0.6]{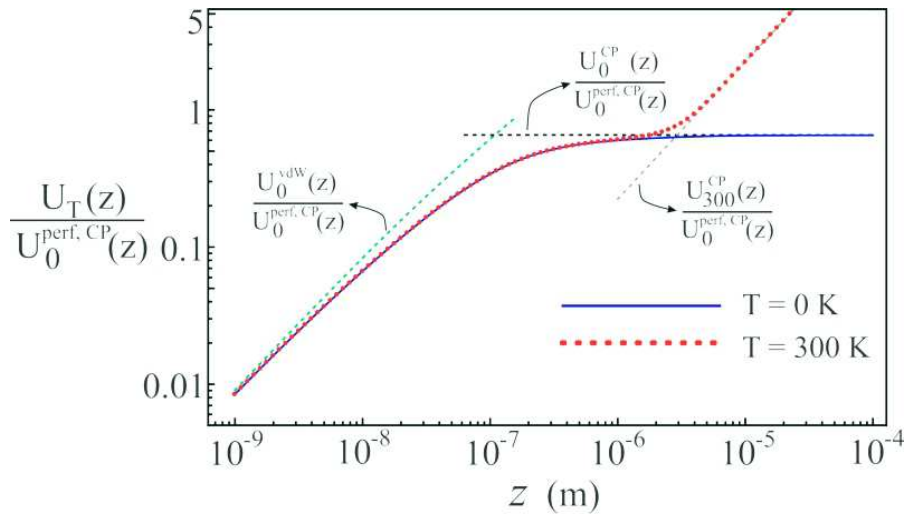}
\vspace{5pt}
\caption{Dispersive interaction $U_{\textrm{T}}^{\textrm{SiC}}$ [normalized by $U_{\textrm{0}}^{\textrm{perf, CP}}$] between a Rubidium atom and a substrate made of Silicon Carbide as a function of the mutual distance. The blue solid curve shows the result for $T = 0$ K whilst the red dotted one is for $T = 300$ K. The dashed green, black and gray lines present the asymptotic behaviors of the interaction energy in the van der Waals and Casimir-Polder regimes.}
\label{CPFigure2}
\end{figure}
%

%In the next section we will apply the previous formalism to investigate the dispersive interaction between an atom and a suspended graphene sheet.
%
\section{Tailoring the dispersive interaction between an atom and a graphene sheet}
\label{AtomGrapheneForce}

\hspace{5mm}  Recently, great  attention has been devoted to dispersive interactions in carbon nanostructures, such as graphene sheets. These systems are specially appealing since graphene possesses unique mechanical, electrical, and optical properties \cite{Graphene1, Graphene2, Graphene3}. Recently, dispersive interactions between graphene sheets and/or material walls have been investigated~\cite{Bordag2009,Bordag2012,Drosdoff,Fialkovsky,Galina2013,Gomez,Macdonald_PRL,Sernelius,Svetovoy, Klimchitskaya-Mostepanenko-14, Lilia2010, Lilia2012-1, Lilia2012-2,Lilia2013}, as well as the Casimir-Polder interaction between atoms and graphene \cite{Chaichian,Eberlein,Churkin,Galina2014,Judd,Sofia-Scheel-13, Sofia-Scheel-13E}. In particular, the impact of graphene coating on the atom-plate interaction has been calculated for different atomic species and substrates; in some cases this results in modifications of the order of $20\%$ in the strength of the interaction~\cite{Galina2014}. Furthermore, results on the possibility of shielding the dispersive interaction with the aid of graphene sheets have been reported~\cite{Sofia-Scheel-13}.

However, the possibility of controlling the Casimir-Polder interaction between an atom and a graphene sheet by an external agent has never been envisaged so far. The possibility of varying the atom-graphene interaction without changing the physical system would be extremely appealing for both experiments and applications. With this motivation, and exploring the extraordinary magneto-optical response of graphene, we investigate the Casimir-Polder interaction between a Rubdium atom and a suspended graphene sheet under the influence of an external magnetic field ${\bf B}$ \cite{Cysne-2014}. We show that just by changing the applied magnetic field, the atom-graphene interaction can be greatly reduced, up to $80\%$  of its value without the field. Furthermore, we demonstrate that at low temperatures ($T\simeq 4$ K) the Casimir-Polder energy exhibits sharp discontinuities at certain values of $B$, that is a manifestation of the discrete Landau energy levels in graphene. As the distance $z$ between the atom and the graphene sheet grows to $z \gtrsim 1 \ \mu$m, these discontinuities form a plateau-like pattern with quantized values for the Casimir-Polder energy.  We also show that at room temperature ($T\simeq 300$ K) thermal effects must be taken into account even for considerably short distances. Moreover, in this case the discontinuities in the atom-graphene interaction do not occur, although the Casimir-Polder energy can still be tuned in $\sim 50 \%$ by applying an external magnetic field. Most of the results presented in the following were published in Ref. \cite{Cysne-2014}.
\vspace{10pt}
\begin{figure}[!ht]
  \centering
  \includegraphics[scale = 0.55]{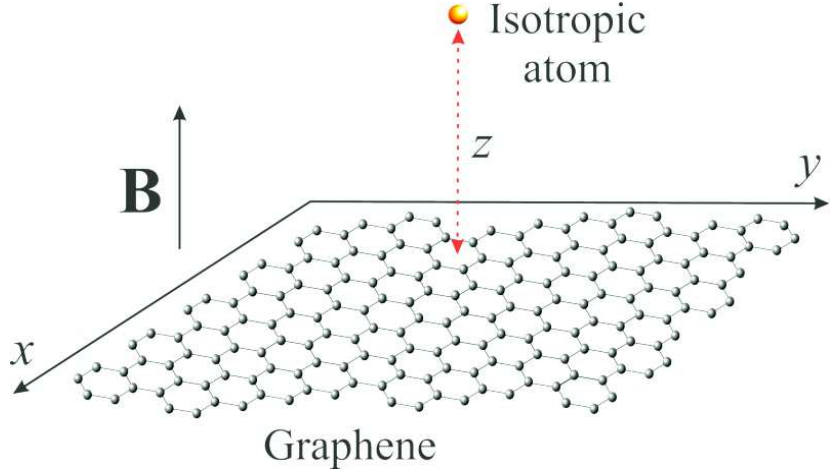}
  \vspace{10pt}
  \caption{Suspended graphene and an isotropic atom at a distance $z$, in presence of a uniform and static magnetic field perpendicular to the graphene sheet.}
  \label{Casimir1}
\end{figure}
\pagebreak

Let us consider that an isotropic atom is placed at a distance $z$ above a suspended graphene sheet biased by a static magnetic field  ${\bf B} = B \hat{{\bf z}}$, as depicted in Fig.~\ref{Casimir1}. \linebreak The whole system is assumed to be in thermal equilibrium at temperature $T$. When $B \neq 0$ the optical properties of graphene can be well described in terms of an homogeneous but anisotropic two-dimensional conductivity tensor. In this case, the Casimir-Polder (CP) energy interaction can be calculated by means of Eq. (\ref{CP2}). In the case of suspended graphene, the reflection coefficients $r^{\textrm{TE, TE}}_{\textrm{s}}(\textbf{k},i\xi, B)$ and $r^{\textrm{TM, TM}}_{\textrm{s}}(\textbf{k},i\xi, B)$ can be obtained from Eqs. (\ref{ReflectionCoefficients_SS_Exact}) and (\ref{ReflectionCoefficients_PP_Exact}) by setting $k_{z1} = k_{z0}$~\cite{Cysne-2014},
\begin{eqnarray}
\label{ReflectionCoefficients_SS_Casimir}
\hspace{0pt}r^{\textrm{TE, TE}}_{\textrm{s}}({\bf k}_{||},i\xi,B)\!\!\! &=&\!\!\!  - \dfrac{2\sigma_{xx}(i\xi, B) \mathfrak{S}^{\textrm{H}} +Z_0^2[\sigma_{xx}(i\xi, B)^2+\sigma_{xy}(i\xi, B)^2]} {\Delta({\bf k}_{||},i\xi, B)}\, ,  \\
\label{ReflectionCoefficients_PP_Casimir}
\hspace{0pt} r^{\textrm{TM, TM}}_{\textrm{s}}({\bf k}_{||},i\xi,B)\!\!\! &=&\!\!\!  \dfrac{2\sigma_{xx}(i\xi, B) \mathfrak{S}^{\textrm{E}} + Z_0^2[\sigma_{xx}(i\xi, B)^2+\sigma_{xy}(i\xi, B)^2]} {\Delta({\bf k}_{||},i\xi, B)}\, ,  \\
\Delta({\bf k}_{||},i\xi, B) = [2\!\!\! &+&\!\!\! \mathfrak{S}^{\textrm{H}} \sigma_{xx}(i\xi, B)][2+ \mathfrak{S}^{\textrm{E}} \sigma_{xx}(i\xi,B)] +[Z_0\sigma_{xy}(i\xi, B)]^2  \, .
\end{eqnarray}
where now $\mathfrak{S}^H = \xi \mu_0 / \kappa$, and $\mathfrak{S}^E = \kappa/(\xi \epsilon_0)$.

We consider a Rubidium atom since there are experimental data on its complex electric polarizability $\alpha(\omega)$ for a wide range of frequencies \cite{Rubidium}. At this point it is worth mentioning that the Rubibium ground state polarizability is strongly dominated by the D-line transitions \cite{Rubidium2}, which occur around 380 THz. Even though Zeeman shifts $\sim$0.05 THz/T (strong magnetic field regime) do exist at these transitions \cite{Rubidium3}, we have checked that they are negligible for the calculations of the dispersive interaction for the values of $B$ used throughout the thesis.

Our first results are summarized in Fig. \ref{Casimir2}, where we depict the CP energy of our setup as a function of the atom-graphene distance $z$ and magnetic field $B$ for\linebreak $T = 4$ K (normalized by its corresponding value for $B=0$ T). The hallmark of this plot is, surely, the great amount of discontinuities shown by the energy as a function of $B$ for all distances considered. These drops show up even more clearly in Figs. \ref{Casimir22}(a) and \ref{Casimir22}(b), where we take two cuts of Fig.~\ref{Casimir2} at two different fixed values of $z$, namely, $z=100$ nm and $z=1$ $\mu$m, and present them as 2-D plots.   Such discontinuities are directly linked to the discrete Landau levels brought about by the application of an external magnetic field on the system \cite{Cysne-2014}. %Indeed,  the crossing of the $n$-th level through the chemical potential  sharply quenches the $M_{n} \leftrightarrow M_{n+1}$ transition; at the same time that it gives birth to the $M_{n-1} \leftrightarrow M_{n}$ one \cite{Graphene3, Gusynin1, Gusynin2}, in a process that changes the conductivity, and thus the interaction energy, discontinuously.  The fact that the CP energy always drops down at a discontinuity as we increase $B$ may be understood by recalling the behavior of the relativistic Landau levels $\propto \sqrt{n}$. This square-root growth implies that the \linebreak $n-1$ $\leftrightarrow$ $n$ gap is wider than the $n$ $\leftrightarrow$ $n+1$ one, making the transition weaker, hence reducing the overall conductivity.
\vspace{10pt}
\begin{figure}[!ht]
  \centering
  \includegraphics[scale = 0.6]{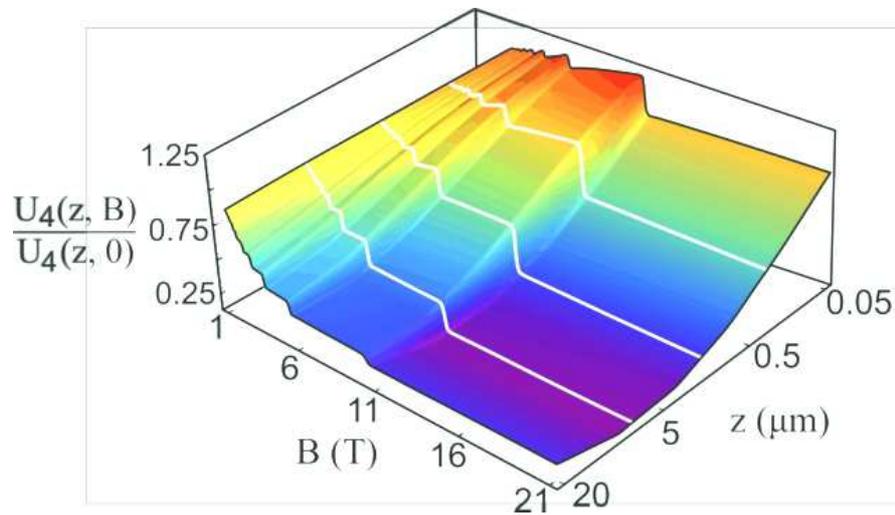}
  \vspace{10pt}
  \caption{Casimir-Polder energy of a Rubidium atom in front of a graphene sheet subjected to a magnetic field $B$, as a function of $B$ and the distance $z$. In all plots $\mu_c = 115$ meV, $T = 4$ K and we have normalized $U_4(z,B)$ by the energy in the absence of magnetic field $U_4(z,0)$.}
\label{Casimir2}
\end{figure}

\pagebreak
Figure \ref{Casimir22} also reveals that a flattening of the steps in the CP energy between drops occurs as $B$ increases. However, if on the one hand for $z=100$ nm only the very last step is really flat, on the other hand many plateau-like steps exist for $z=1$ $\mu$m. This result is connected to the electrostatic limit of the conductivity: for large distances the exponential factor in (\ref{CP2}) strongly suppress all contributions coming from $l \neq 0$, whereas for $l=0$ and large magnetic fields $\sigma_{xx} \rightarrow 0$ and $\sigma_{xy}$
%, where $N$ is an integer
turns to be a discrete function of $B$ \cite{Gusynin1, Gusynin2}. 
Therefore, in the limit of very large distances (of the order of micrometers) only the Hall conductivity contributes to $U_{4}(z, B)$, and the CP energy becomes almost quantized~\cite{Macdonald_PRL}. Furthermore, one should note the striking reduction in the force as we sweep through different values of $B$. While for $z = 100$ nm this reduction can be as hight as $45\%$, one can get up to an impressive $80\%$ decrease in the CP interaction for $z=1$ $\mu$m and  $B \gtrsim$ 10 T, with huge drops in between. Finally, it should be remarked that for $B \gtrsim$ 10 T the CP interaction is practically insensitive to changes in the magnetic field, regardless of the atom-graphene distance. 
%In this regime the discontinuities in the CP energy do not occur any longer. This effect has its origins in the fact that there is a critical value of the magnetic field $B_c$ (in the present case, $B_c \sim 10$ T) for which the transition $M_0 \rightarrow M_1$ is dominant since all Landau levels, except $M_0$, are above the chemical potential. The value of the critical magnetic field is given by  $B_c = \mu_c^2/ 2\hbar e v_F^2 $. 
Altogether, our findings suggest that the atom-graphene is a particularly suited system for investigation of the effects of external magnetic fields on CP forces, and may pave the way for an active modulation of dispersion forces in general.

We should mention that in order to calculate the dispersive interaction between the Rubidium atom and the graphene sheet the whole EM spectrum must be considered. Indeed, whereas in the SE process only EM field modes with $\omega \simeq \omega_0$ are significant, in the CP interaction several  frequencies may play a fundamental role.  Particularly, even frequencies above the interband threshold ($\omega\gtrsim 2\mu_c/\hbar$) will give in general a nonvanishing  contribution to the integrals in Eq. (\ref{CP2}). As a consequence, the interband transitions in graphene  cannot be neglected in the expressions of $\sigma_{xx}$ and $\sigma_{xy}$. In Figs. \ref{Casimir22}(a) and \ref{Casimir22}(b) we plot the contributions $U_{\textrm{4}}^{\textrm{intra}}$ and $U_{\textrm{4}}^{\textrm{inter}}$ of intraband (blue dashed line) and interband (red dot-dashed line) transitions in graphene to the CP energy. As the system is held at a very low temperature ($T = 4$ K) these plots can be obtained by separately considering that the longitudinal and transversal conductivities are given by Eqs. (\ref{LongitudinalIntra}) and   (\ref{TransversalIntra}) or Eqs. (\ref{LongitudinalInter}) and (\ref{TransversalInter}), respectively. Note that though the intraband transitions are dominant for the two distances considered in the plots, interband transitions cannot be neglected since they contribute up to 30\% of the total interaction energy. Moreover, for $B > B_c \sim 10$ T we note that $U_{\textrm{4}}^{\textrm{inter}} \simeq 0 \Rightarrow U_{\textrm{4}} \simeq  U_{\textrm{4}}^{\textrm{intra}}$. This is due to the fact that for these values of magnetic field the $M_0 \rightarrow M_1$ transition dominates the conductivity in graphene and we have classified transitions involving the LL $n = 0$ as intraband-like.
\vspace{10pt}
\begin{figure}[!ht]
  \centering
  \includegraphics[scale = 0.77]{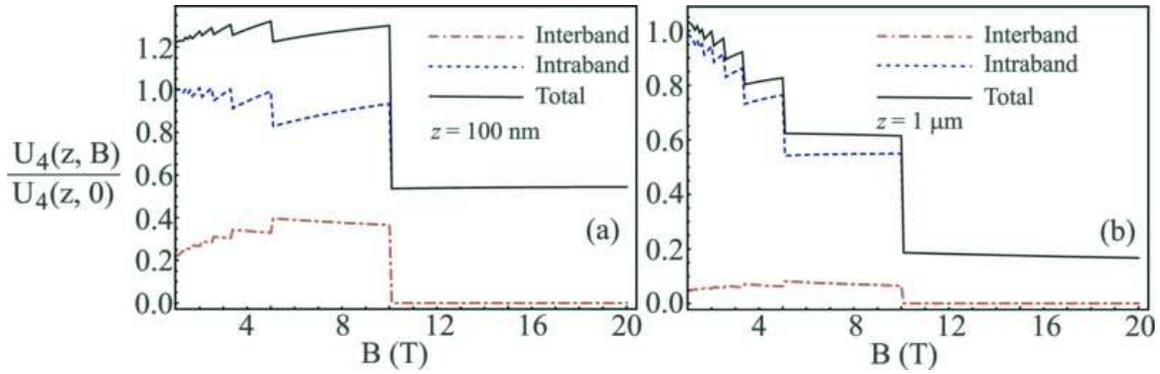}
  \vspace{10pt}
  \caption{Casimir-Polder energy [normalized by $U_4(z,0)$] as a function of $B$ for two fixed distances: {\bf (a)} $z = 100$ nm, and {\bf (b)} $z = 1$ $\mu$m. Both panels show the contributions of the intraband $U_{\textrm{4}}^{\textrm{intra}}$ (blue dashed line) and interband $U_{\textrm{4}}^{\textrm{inter}}$ (red dot-dashed line) transition in graphene to the total (black solid line) CP interaction. The material parameters are the same as in Fig.~\ref{Casimir2}.}
\label{Casimir22}
\end{figure}

Note in Fig. \ref{Casimir22} that $U_{\textrm{4}}^{\textrm{inter}}$ is more significative at $z = 100$ nm than at $z = 1\ \mu$m for $B < B_c$. The effects of distance on the relative contributions of intraband and interband transitions to the overall CP interaction energy are more evident in Fig. \ref{Casimir23}. In this figure we plot both $U_{\textrm{4}}^{\textrm{intra}}$ (red dashed curve) and $U_{\textrm{4}}^{\textrm{inter}}$ (black dot-dashed curve) as a function of $z$ for $B = 5$ T and all other parameters exactly the same as in Fig. \ref{Casimir22}. Interestingly, the relative contribution of $U_{\textrm{4}}^{\textrm{inter}}$ to $U_{\textrm{4}}$ (blue solid line) increases as the atom gets closer to the graphene sheet. For $z \lesssim 1\ \mu$m we note that $U_{\textrm{4}}^{\textrm{inter}}/U_{\textrm{4}} \gtrsim 0.1$. On the other hand, for large atom-graphene separations the interband transitions in graphene are almost negligible for the CP energy for the material parameters chosen. These results are in agreement with our previous discussion that for large distances the static limit (low frequencies) dominates the integrals in Eq. (\ref{CP2}). Consequently, interband transitions are not expected to be relevant to the dispersive force at such distances once they are mainly affected by high frequency EM field modes ($\omega \gtrsim 2\mu_c/\hbar \sim 3.5 \times 10^{14}$ rad/s).
\vspace{10pt}
\begin{figure}[!ht]
  \centering
  \includegraphics[scale = 0.6]{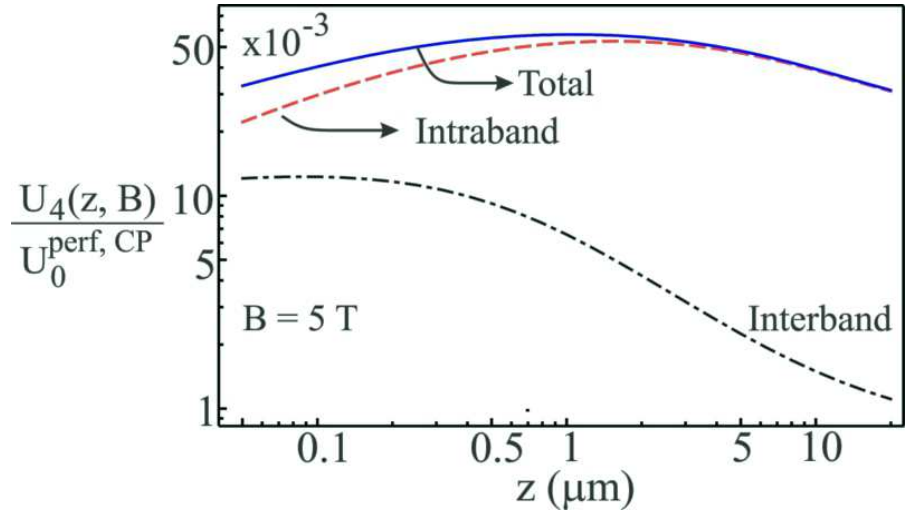}
  \vspace{10pt}
  \caption{%{\bf (a)} The Casimir-Polder energy between a Rubidium atom and a suspended graphene sheet subjected to a magnetic field $B$, as a function of distance $z$ for $B = 0$ T (blue solid line), $B = 5$ T (red dashed line), $B = 10$ T (green dotted line), and $B = 15$ T (black dot-dashed line). {\bf(b)}
  Contributions of the intraband (red dashed curve) and interband (black dot-dashed curve) graphene transitions to the total Casimir-Polder energy (blue solid curve) for the system depicted in Fig. \ref{Casimir1} as a function of $z$ for $B = 5$ T. We set $\mu_c = 115$ meV, $T = 4$ K and normalized the plotted functions by $U_0^{\textrm{perf, CP}}$.}
\label{Casimir23}
\end{figure}

In order to investigate thermal effects, in Fig.~\ref{Casimir4}(a) we present the CP energy as a function of both $z$ and $B$ at room temperature. The most distinctive aspect of Fig.~\ref{Casimir4}(a) is the complete absence of discontinuities that characterize the behavior of the CP energy at low temperatures. At $T = 300$ K the Fermi-Dirac distribution is a quite smooth function of the energy levels, allowing for a partial filling of many Landau levels. Hence the effects of the crossing between these levels and the graphene's chemical potential is hardly noticed, resulting in a smooth CP energy profile. Another important aspect of Fig.~\ref{Casimir4}(a) is that the CP energy becomes essentially independent of $B$ for $z \gtrsim 2$ $\mu$m.  For this set of parameters, the system is already in the thermal regime, where the CP energy is essentially dominated by the electrostatic conductivity. %
\vspace{10pt}
\begin{figure}[!ht]
  \centering
  \includegraphics[scale = 0.75]{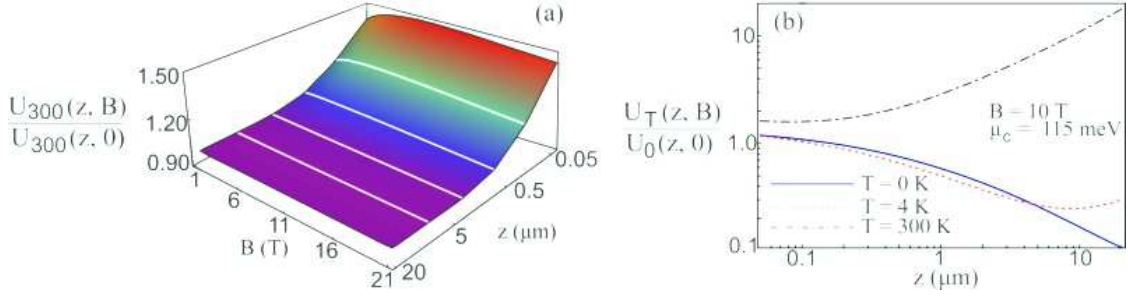}
  \vspace{10pt}
  \caption{{\bf (a)} Casimir-Polder interaction energy $U_{300}(z,B)$ [normalized by $U_{300}(z,0)$] between a Rubidium atom and a graphene sheet as a function of both distance and external magnetic field strength for $T = 300$ K. {\bf (b)} Casimir-Polder energy [normalized by $U_{0}(z,0)$] as a function of the mutual distance between atom and graphene sheet for $B = 10$ T and $T = 0$ K (solid line), $T = 4$ K (dashed line), and $T = 300 K$ (dot-dashed line). In both panels (a) and (b) the graphene chemical potential is $\mu_c = 115$ meV.}
    \label{Casimir4}
\end{figure}
In this regime, the CP energy is very weakly affected by variations in $B$ due to the already discussed exponential suppression of the $l \geq 1$  terms in Eq. (\ref{CP2}). In this case, the zero frequency contribution dominates the integrals in Eq. (\ref{CP2}). However, taking the limit $\xi \rightarrow 0$ in Eqs. (\ref{ReflectionCoefficients_SS_Casimir}) and (\ref{ReflectionCoefficients_SS_Casimir}) leads to $r^{\textrm{TE, TE}}_{\textrm{s}}({\bf k}_{||},i\xi,B) \rightarrow 0$ and $r^{\textrm{TM, TM}}_{\textrm{s}}({\bf k}_{||},i\xi,B) \rightarrow 1$ regardless of the value of $B$. This would explain the low influence of the magnetic field on $U_{\textrm{T}}$ for very large atom-graphene separations.  We emphasize, however, that the absence of discontinuities does not prevent us from tuning the CP interaction between a Rb atom and a graphene sheet at least at short distances. This tunability can be achieved even for relatively modest magnetic fields, as the value of the CP energy can increase up to 50 $\%$ (compared to the case where $B = 0$ T) by applying a magnetic field of $B=5$ T for $z = 50$ nm. For $B=5$ T and $z = 100$ nm, the variation in the interaction can still be as high as 30$\%$. In Fig.~\ref{Casimir4}(b) the CP energy is calculated for $B=10$ T and for different temperature values, $T = 0$, $4$, and $300$ K, all normalized by the zero-temperature, zero-field energy value $U_{\textrm{0}}(z,0)$. Figure~\ref{Casimir4}(b) reveals that thermal corrections are relevant even for low temperatures, and for a broad range of distances: we have a 10-20\% variation in the relative difference of $U_{\textrm{4}}(z,10)$ and $U_{\textrm{0}}(z,10)$ in the 1-10 $\mu$m interval, which is in the ballpark of recent/current experiments' precision. Besides, Fig.~\ref{Casimir4}(b) demonstrates that at room temperature not only the thermal effects are absolutely dominant in the micrometer range, but also they play an important role even for small distances. Indeed, at $z = 100$ nm the relative difference between $U_{\textrm{300}}(z,10)$ and $U_{\textrm{0}}(z,10)$ is $\sim 45\%$ and at $z = 1$ $\mu$m it is $\sim 400 \%$; so in the latter approximately $80 \%$ of the CP energy come from the thermal contribution. We conclude that, at room temperature, these effects should be taken into account for a wide range of distances between the atom and the graphene sheet.

In Fig. \ref{Casimir5} we plot the CP energy as a function of the magnetic field for $z = 1\ \mu$m and distinct temperatures of the system, namely, $T = 10$ K (red dashed line), $T = 20$ K (green dotted line), and $T = 40$ K (black dot-dashed line). As the temperature of the system increases we note that a smoothing of the discontinuities of $U_{\textrm{T}}$ takes place. As it is clear, even for temperatures of a few dozen Kelvin the discontinuities in the interaction are not present. Besides, note that the effects of increasing $T$ is not just to smoothen $U_{\textrm{T}}$ as function of $B$, but also enhance the interaction energy. This result is not surprising since in Eq. (\ref{CP2}) several functions depend on $T$ through the Matsubara's frequencies and not only the Fermi-Dirac distribution.
\vspace{10pt}
\begin{figure}[!ht]
  \centering
  \includegraphics[scale = 0.6]{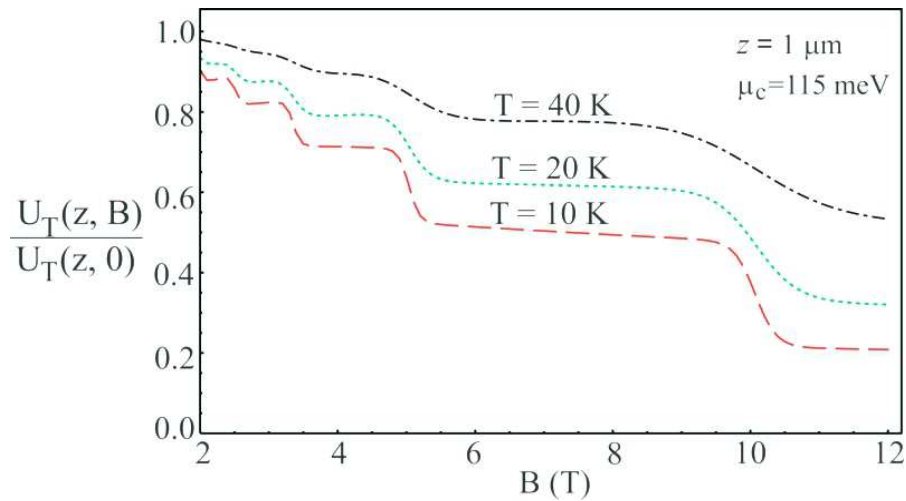}
  \vspace{10pt}
  \caption{{\bf (a)} Casimir-Polder interaction energy $U_{T}(z,B)$ [normalized by $U_{T}(z,0)$] between a Rubidium atom and a graphene sheet as a function of $B$ for $z = 1\ \mu$m and several temperatures of the system, namely, $T = 10$ K (red dashed line), $T = 20$ K (green dotted line), and $T = 40$ K (black dot-dashed line). All the other material parameters are the same as in Fig. \ref{Casimir4}.}
    \label{Casimir5}
\end{figure}

\vspace{30pt}

In this chapter we have discussed the main features of dispersive interactions between atoms and dispersive surfaces. We have presented a method for computing the van der Waals interaction between an atom and a perfect conducting surface of arbitrary shape. We have applied this method to compute the dispersive interaction between an atom and a sphere or ellipsoid. In addition, we derived a expression for computing the interaction energy between an atom and a semi-infinite flat wall valid in any regime of distances and temperatures and for materials with arbitrary optical properties. In particular, we used this expression so as to investigate the dispersive Casimir-Polder interaction between a Rubidium atom and a suspended graphene sheet subjected to an external magnetic field ${\bf B}$.  We showed that just by changing the applied magnetic field, this interaction can be reduced up to $80\%$  of its value in the absence of the field. Further, due to the discrete Landau energy levels in graphene, we showed that for low temperatures the Casimir-Polder interaction energy acquires sharp discontinuities at given values of $B$ and that these discontinuities approach a plateau-like pattern with a quantized Casimir-Polder interaction energy as the distance between the atom and the graphene sheet increases. Finally, we showed that at room temperature thermal effects must be taken into account even for considerably short distances. In this case, the discontinuities in the atom-graphene dispersive interaction do not occur, although the interaction can still be tuned in $\sim 50 \%$ by applying an external magnetic field.

\end{chapter}

\begin{chapter}{Near-field heat transfer in composite media}
\label{cap8}

\begin{flushright}
{\it
The theory of heat will hereafter form one\\
of the most important branches of general physics.
}

{\sc J. B. Fourier}
\end{flushright}

\hspace{5 mm} Energy transfer via electromagnetic radiation is a well-known phenomenon that has led to the development of several technologies. Nowadays, this area has been attracting renewed interest due to the recent advances in manipulating the structure of the EM field modes in the near-field. In this chapter, we study near-field heat transfer (NFHT) between a semi-infinite medium and a nanoparticle. Using concepts of fluctuation electrodynamics we derive a convenient expression for calculating the amount of energy exchanged between these bodies. We apply the developed formalism to investigate NFHT in the case of a nanoparticle made of composite materials.   For metallic inclusions embedded in a dielectric host-medium we show that, in the effective medium approximation, heat transfer at the near-field can be greatly enhanced by considering composite media, being maximal at the insulator-metal (percolation) phase transition.

\section{Introduction}
\label{Intro8}

\hspace{5mm} For billions of years thermal radiation has been of fundamental relevance for the development of life on Earth. Solar EM radiation is not only one of the main existing sources of energy but it is also crucial for various biologic processes such as photosynthesis. The modern history of thermal radiation begins with the discovery of infrared light by W. Herschel~\cite{Herschel} in 1800. In the following decades, studies carried out by G. Kirchhoff~\cite{Kirchhoff1} on absorption and emission lines of hot gases led to the well known Kirchhoff's law: the absorptivity of a body in thermal equilibrium with the environment is equal to its emissivity~\cite{Kirchhoff2}. Further, J. Stefan~\cite{Stefan} and L. Boltzmann~\cite{Boltzmann} have proposed that the total radiance emitted by a blackbody (an idealized concept introduced by Kirchhoff for an object that absorbs all EM radiation at all frequencies and at all incidence angles) should be proportional to the fourth power of its temperature $T$. Despite describing properly the dependence of the blackbody radiance with $T$, Stefan and Boltzmann were not able to provide the spectral distribution of emission. The correct description of the spectrum emitted by a blackbody is given by Planck's spectral distribution, derived by Max Planck in 1900~\cite{Planck}. The dominant emission wavelength $\lambda_{\textrm{d}}$ at temperature $T$ can be calculated using the Wien's law~\cite{Wien1, Wien2}.

Planck's distribution has accurately described far-field emission for more than a century. Nevertheless, recent works on radiative energy transfer between bodies separated by sub-wavelength distances have revealed that Planck's law  may fail~\cite{Greffet2005, Basu2009, Dorofeyev2011,Raschke2013,BiehsIR,Volokitin2007}. This is due to the fact that Planck's result is based on the assumption that only propagating waves contribute to the energy exchange. Besides, the relative phases between distinct waves are not taken into account so that the energy transfer process is incoherent. Of course that for distances $d \gg \lambda_{\textrm{d}}$ (far-field) this approximation holds, since the coherence length of blackbody radiation is of the same order of magnitude as $\lambda_{\textrm{d}}$ \cite{Donges}. However, since the seminal work by Polder and van Hove~\cite{PvH1971} it is known that for $d \ll \lambda_{\textrm{d}}$ (near-field) the bodies' evanescent waves contribute to the heat transfer, sometimes even dwarfing the propagating contribution. Indeed, it can be shown that this so called near-field heat transfer can vastly surpass the blackbody limit of radiative transfer by orders of magnitude, radically changing the landscape of possibilities in the arena of out-of-equilibrium phenomena~\cite{Greffet2005, Basu2009, Dorofeyev2011,Raschke2013,BiehsIR,Volokitin2007}. Furthermore, in the near-field the EM spectrum presents a nearly monochromatic profile, with a peak that is orders of magnitude higher than the far-field typical values \cite{Mulet2002, GreffetNature, Carminati1999}.

Recently,  numerous works have been carried out to investigate, both theoretically and experimentally, the physics involved in the NFHT process. On the theoretical front, the formulation of the NFHT in terms of scattering matrices~\cite{Narayanaswamy2008,Bimonte2009,Messina2011,Kruger2011, Kruger2012} opened new venues for investigating the effects of non-trivial geometries, as did the more numerical oriented approaches of fluctuating surface currents \cite{Rodriguez2013} and finite-difference time-domain (FDTD) computations~\cite{Rodriguez2012}. As selected (but by no means exhausting) examples, we can highlight studies of the heat transfer for the sphere-plate configuration~\cite{Otey2011}, between gratings~\cite{Guerout2012,Lussange2012}, between particles and surfaces~\cite{ChapuisEtAl2008,HuthEtAl2010}, tips (small dipole) and surfaces~\cite{PBA2013}, and various shapes~\cite{Rodriguez2013}. There was also great activity regarding the material properties in the NFHT, like hyperbolic materials~\cite{Biehs2012}, porous media~\cite{Felipe-2011}, photonic crystals~\cite{PBA2010}, and graphene-based materials \cite{Volokitin2011, Ognjen, Lim, MessinaGraphene}. On the experimental side, several groups carried out measurements of NFHT for different geometries,  such as tip-surface~\cite{Kittel}, sphere-plate~\cite{NarayaEtAl2008, NarayaEtAl2009, RousseauEtAl2009}, and plate-plate~\cite{HuEtAl2008,Ottens2011}, all in fairly good agreement \linebreak with  theoretical predictions.

All this development in the field of NFHT has naturally led to investigations of possible applications. Among many ideas, there has been studies in thermal imaging~\cite{BiehsEtAl2008}, thermal rectification and control \cite{Fan2010,Zwol2011,BiehsRosaAPL,BiehsPBA2013}, and optimization of thermophotovoltaic cells \cite{Narayanaswamy2003,Laroche2006}, all of which take advantage of the large increase of the heat flux brought forth by the near field. As a result, enhancing the process of NFHT is crucial for the development of new and/or optimized applications. Indeed, there are some recent proposals in this direction (see {\em e.g.} Refs.~\cite{Guerout2012,Worbes2013}), where enhancements up to a factor of a few tens have been reported. The aim of this chapter is to introduce a novel approach to enhance the heat transfer in the near field by exploiting the versatile material properties of composite media \cite{WiltonPRB2014}.  To this end we investigate the NFHT between a semi-infinite dielectric medium and metallic nanoparticles, with various concentrations and geometries, embedded in dielectric hosts. Applying homogenization techniques, we demonstrate that the NFHT is strongly enhanced in composite media if compared to the case where homogeneous media are considered. In particular we show that NFHT is maximal precisely at the insulator-metal (percolation) transition. We also demonstrate that at the percolation transition more modes effectively contribute to the heat flux, widening the transfer frequency band.

In the next section we investigate the NFHT process between a nanoparticle and a semi-infinite space in the scope of the fluctuating electrodynamics. Among other things we discuss the role played by frustrated TIR modes and  surface waves in the problem. In Section {\bf \ref{NFHTCompositeMaterials}} we discuss the effect of considering composite materials for NFHT and we present our main results.

\section{NFHT between a nanoparticle and a half-space}

\hspace{5mm} Let us consider the system depicted in Fig. \ref{HTPerc1}. The half-space $z<0$ is composed of an isotropic and homogeneous (bulk) material of local dielectric constant $\epsilon_B (\omega)$, at temperature $T_B$. The upper medium $z>0$ is vacuum and a sphere of radius $a$ and temperature $T_S$ is located at ${\bf r}_s =  d\hat{{\bf z}}$. For the sake of simplicity we consider in this thesis that $T_S = 0$ K. In order to describe the energy transfer process we follow Rytov's stochastic description of electrodynamics\cite{Rytov1, Rytov2}. In this description, the heat transfer process is governed by fluctuating currents, with zero mean value, in the bulk.  The currents in the bulk, in local thermal equilibrium, induce electromagnetic fields that eventually illuminate the sphere. Once the Poynting vector depends on the correlation function of these random currents, there exists a nonvanishing net energy exchanged between the considered bodies.
\vspace{10pt}
\begin{figure}[!ht]
  \centering
  \includegraphics[scale = 0.5]{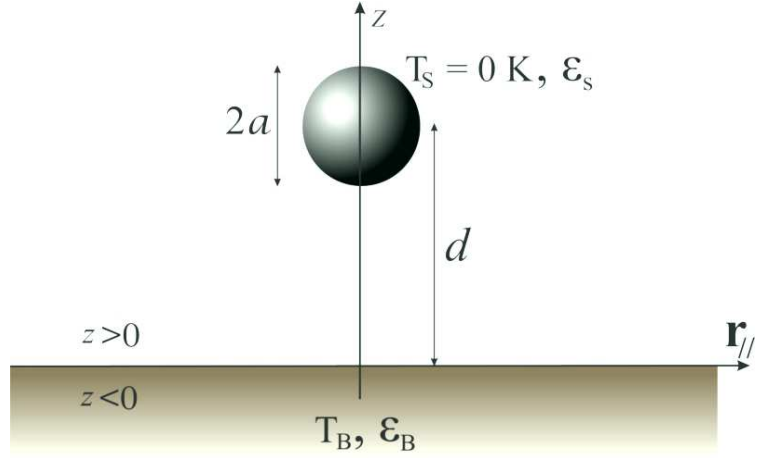}
  \vspace{10pt}
  \caption{An isotropic and homogeneous sphere of radius $a$ and dielectric constant $\epsilon_s(\omega)$ at position ${\bf r}_s = d \hat{{\bf z}}$ is held at temperature $T_S = 0$ K in the vicinities of a half-space with a local dielectric constant $\epsilon_B(\omega)$ and temperature $T_B$.}
  \label{HTPerc1}
\end{figure}

In the case where the relevant wavelengths to NFHT are much larger than $a$, and $d$ is at least of order of a few radii, the electromagnetic response of the sphere can be described in terms of its electric and magnetic dipoles~\cite{ChapuisEtAl2008,HuthEtAl2010,magneticdipole, Kruger2011, Kruger2012}. Consequently, the mean power absorbed by the sphere can be cast into the form~\cite{magneticdipole}
\begin{eqnarray}
\label{Power1}
{\cal {P}}_{\textrm{abs}}^{\textrm{total}}= \left\langle \int {\bf J}({\bf r}, t) \cdot {\bf E}({\bf r}, t) dV \right\rangle
                                 = \left\langle \dfrac{\partial {\bf p}({\bf r}_s, t)}{\partial t} \cdot {\bf E}({\bf r}_s, t) - {\bf m}({\bf r}_s, t) \cdot \dfrac{\partial {\bf B}({\bf r}_s, t)}{\partial t} \right\rangle\, ,
\end{eqnarray}
where ${\bf E}$ and ${\bf B}$ are the electric and magnetic fields impinging on the particle, $\langle ... \rangle$ denotes statistical average over bulk current fluctuations, and we have used that the current in the sphere is given by \cite{Jackson}
\begin{eqnarray}
{\bf J}({\bf r}, t) = \dfrac{\partial {\bf P}({\bf r},t)}{\partial t} + \nabla \times {\bf M}({\bf r}, t)\, ,
\end{eqnarray}
with ${\bf P}({\bf r},t) = {\bf p}({\bf r},t) \delta({\bf r} - {\bf r}_s)$ and ${\bf M}({\bf r},t) = {\bf m}({\bf r},t) \delta({\bf r} - {\bf r}_s)$. Besides, ${\bf p}({\bf r}_s, t)$ and ${\bf m}({\bf r}_s, t)$ are the sphere's electric and magnetic dipoles, respectively. Here we do not take into account the diamagnetic response of the material so that the magnetic dipole moment is due to eddy currents in the composite particle~\cite{ChapuisEtAl2008,HuthEtAl2010,magneticdipole, Kruger2011, Kruger2012}. In order to proceed we express the EM field as well as the electric and magnetic dipoles in terms of their Fourier components
\begin{eqnarray}
\label{FourierFields1}
{\bf E}({\bf r}, t) \!\!\! &=&\!\!\! \int_{-\infty}^{\infty} {\bf \mathcal{E}} ({\bf r}, \omega) e^{-i\omega t} d\omega\, , \ \ \ \
{\bf B}({\bf r}, t) = \int_{-\infty}^{\infty} {\bf \mathfrak{B}} ({\bf r}, \omega) e^{-i\omega t} d\omega\, , \\
\label{FourierDipoles1}
{\bf p}({\bf r}_s, t) \!\!\! &=&\!\!\! \int_{-\infty}^{\infty} {\bf \mathfrak{p}} ({\bf r}_s, \omega) e^{-i\omega t} d\omega\, , \ \ \ \
{\bf m}({\bf r}_s, t) = \int_{-\infty}^{\infty} {\bf \mathfrak{m}} ({\bf r}_s, \omega) e^{-i\omega t} d\omega\, .
\end{eqnarray}

Since the fields in the time-domain are real functions we must enforce that \linebreak ${\bf \mathcal{E}}^{*}({\bf r}, \omega) = {\bf \mathcal{E}} ({\bf r}, -\omega) $, ${\bf \mathfrak{B}}^{*}({\bf r}, \omega) = {\bf \mathfrak{B}} ({\bf r}, -\omega)$, ${\bf \mathfrak{p}}^{*}({\bf r}_s, \omega) = {\bf \mathfrak{p}} ({\bf r}_s, -\omega)$, and \linebreak ${\bf \mathfrak{m}}^{*}({\bf r}_s, \omega) = {\bf \mathfrak{m}} ({\bf r}_s, -\omega)$. Furthermore, the relations between the  impinging fields on the particle and the electric and magnetic dipoles can be properly expressed in the Fourier space as \cite{ChapuisEtAl2008, HuthEtAl2010}
\begin{equation}
\label{Polarizabilities1}
{\bf \mathfrak{p}} ({\bf r}_s, \omega) = \varepsilon_0 \alpha_E(\omega) {\bf \mathcal{E}} ({\bf r}_s, \omega)\, , \ \ \textrm{and} \ \
{\bf \mathfrak{m}} ({\bf r}_s, \omega) =  \dfrac{\alpha_H(\omega)}{\mu_0} {\bf \mathfrak{B}}({\bf r}_s, \omega) = \alpha_H(\omega) {\bf \mathcal{H}}({\bf r}_s, \omega)\, ,
\end{equation}
where $\alpha_E(\omega)$ and $\alpha_H(\omega)$ are the sphere's electric and magnetic polarizabilities, respectively. Substituting (\ref{FourierFields1}), (\ref{FourierDipoles1}), and (\ref{Polarizabilities1}) in (\ref{Power1}) and expressing all quantities as functions of positive frequencies we can show that the mean absorbed power can \linebreak be written as
\begin{eqnarray}
\label{Power2}
{\cal {P}}_{\textrm{abs}}^{\textrm{total}}\!\!\! &=&\!\!\! \varepsilon_0 \textrm{Im}\left\{\int_0^{\infty}\!\!\! d\omega  \int_0^{\infty}\!\!\! d\omega' \ \omega\alpha_E(\omega)\right. \cr
\!\!\! &\times& \!\!\! \left.
 \left[\left\langle {\bf \mathcal{E}} ({\bf r}_s, \omega)\cdot {\bf \mathcal{E}} ({\bf r}_s, \omega')  \right\rangle e^{-i(\omega+\omega')t} +  \left\langle {\bf \mathcal{E}} ({\bf r}_s, \omega)\cdot{\bf \mathcal{E}}^{*} ({\bf r}_s, \omega')  \right\rangle e^{-i(\omega-\omega')t} \right] \right\} \cr
 \!\!\!&+&\!\!\!  (1/\mu_0)\textrm{Im}\left\{\int_0^{\infty}\!\!\! d\omega  \int_0^{\infty}\!\!\! d\omega' \ \omega'\alpha_H(\omega)\right. \cr
 \!\!\! &\times&\!\!\! \left.
 \left[\left\langle {\bf \mathfrak{B}} ({\bf r}_s, \omega)\cdot {\bf \mathfrak{B}} ({\bf r}_s, \omega')  \right\rangle e^{-i(\omega+\omega')t} +  \left\langle {\bf \mathfrak{B}} ({\bf r}_s, \omega)\cdot{\bf \mathfrak{B}}^{*} ({\bf r}_s, \omega')  \right\rangle e^{-i(\omega-\omega')t} \right] \right\}
\end{eqnarray}

The correlation functions in the previous equation cannot be evaluated without any further information about $(i)$ the statistics of the fluctuating currents ${\bf j}^f({\bf r}, \omega)$, and $(ii)$ the relationship between these currents and their generated EM fields.  In Rytov's stochastic electrodynamics formalism the first requirement is overcome via the application of the fluctuation dissipation theorem \cite{Greffet2005, Basu2009, Dorofeyev2011,Raschke2013,BiehsIR,Volokitin2007, Carminati1999, Mulet2001}
\begin{eqnarray}
\label{FluctuationDissipation}
\langle j_m^f({\bf r},  \omega)  {j_n^{f}}^{*}\!({\bf r'},  \omega') \rangle\!\! =\! 2\pi\varepsilon_0 \omega \Theta(\omega, T) \textrm{Im}[\epsilon_B(\omega)] \delta_{m,n} \delta({\bf r} - {\bf r}') \delta(\omega-\omega')\!\, ,
\end{eqnarray}
 $\langle  {j_m^{f}}^{*}\!({\bf r},  \omega)j_n^f({\bf r'},  \omega') \rangle = \langle j_m^f({\bf r},  \omega)  {j_n^{f}}^{*}\!({\bf r'},  \omega') \rangle$, and $\langle  {j_m^{f}}\!({\bf r},  \omega)j_n^f({\bf r'},  \omega') \rangle = \langle {j_m^f}^{*}({\bf r},  \omega)  {j_n^{f}}^{*}\!({\bf r'},  \omega') \rangle = 0$. Besides, $\Theta(\omega, T) =  \hbar \omega/[\exp{(\hbar\omega/k_BT)}-1]$ is the mean energy of a quantum oscillator in thermal equilibrium at temperature $T$.
%Due to the presence of $\hbar$ in Eq. (\ref{FluctuationDissipation}) it is clear that the dissipation theorem is a quantum mechanical relation. Once the stochastic EM fields are classical functions, Rytov's stochastic electrodynamics should be faced as a semi-classical approach.
Furthermore, the Fourier components of the EM field may be expressed in terms of ${\bf j}^f({\bf r}, \omega)$ through the use of Green's dyadic functions of the problem $\mathbb{G}({\bf r}, {\bf r}'; \omega)$ \cite{BiehsIR, Carminati1999, Mulet2001}
\begin{eqnarray}
\label{FluctuatingField}
{\bf \mathcal{E}} ({\bf r}, \omega) = i\omega \mu_0\int_{\cal{R}} \mathbb{G}({\bf r}, {\bf r}'; \omega) \cdot {\bf j}^f({\bf r}', \omega) d{\bf r}'\, ,
\end{eqnarray}
where the integral extends over the whole region occupied by the semi-infinite medium, and a similar expression holds for the magnetic field. Together, Eqs. (\ref{FluctuationDissipation}) and (\ref{FluctuatingField}) allow us to demonstrate that $\left\langle {\bf \mathcal{E}} ({\bf r}, \omega)\cdot {\bf \mathcal{E}} ({\bf r}, \omega')  \right\rangle = \left\langle {\bf \mathfrak{B}} ({\bf r}, \omega)\cdot {\bf \mathfrak{B}} ({\bf r}, \omega')  \right\rangle = 0$ and \linebreak $\left\langle {\bf \mathcal{E}} ({\bf r}, \omega)\cdot {\bf \mathcal{E}}^{*}({\bf r}, \omega')  \right\rangle = \left\langle {\bf \mathfrak{B}} ({\bf r}, \omega)\cdot {\bf \mathfrak{B}}^{*}({\bf r}, \omega')  \right\rangle \propto \delta(\omega-\omega')$. As a consequence, Eq. (\ref{Power2}) can be greatly simplified and the mean power radiated by the bulk and absorbed by the composite particle can be cast as~\cite{ChapuisEtAl2008,HuthEtAl2010}
\begin{eqnarray}
{\cal {P}}_{\textrm{abs}}^{\textrm{total}} = \int_{0}^{\infty} d\omega \left[{\cal {P}}_{\textrm{abs}}^{E}(\omega) + {\cal {P}}_{\textrm{abs}}^{H}(\omega)\right]\, ,
\end{eqnarray}
where
\begin{eqnarray}
{\cal {P}}_{\textrm{abs}}^{E}(\omega) = \omega\textrm{Im}[\alpha_E(\omega)] \varepsilon_0\langle |{\bf \mathcal{E}}|^2 \rangle\, , \ \ \textrm{and} \ \
{\cal {P}}_{\textrm{abs}}^{H}(\omega) = \omega \textrm{Im}[\alpha_H(\omega)] \mu_0\langle |{\bf \mathcal{H}}|^2 \rangle\, ,
\label{pabs}
\end{eqnarray}
are the electric and magnetic contributions to the absorbed spectral power. The correlation function $\langle |{\bf \mathcal{E}}|^2 \rangle$ can be calculated using Eqs. (\ref{FluctuationDissipation}) and (\ref{FluctuatingField}) together with the explicit expression for the electric Green's dyadic of two semi-infinite media sharing a plane interface at $z = 0$, given in Eqs. (\ref{GreenFunctionSplit}) - (\ref{ReflectionMatrix}). After a lengthy calculation we can demonstrate that \cite{HuthEtAl2010}
\begin{eqnarray}
\label{AbsorbedSpectralElectric}
{\cal {P}}_{\textrm{abs}}^{E}(\omega)= \dfrac{\omega^2}{4\pi^2c^2}\textrm{Im}[\alpha_E(\omega)] \Theta(\omega, T_B)
\left\{\int_0^{\omega/c}\dfrac{k_{||}}{\xi}
\Big[2-|r^{\textrm{TE, TE}}|^2-|r^{\textrm{TM, TM}}|^2 \Big] dk_{||} \right. \cr
+ \left.  2\int_{\omega/c}^{\infty} \dfrac{k_{||}e^{-2\zeta d}}{\zeta}
\left[\textrm{Im}(r^{\textrm{TE, TE}}) + \left(\dfrac{2k_{||}^2c^2}{\omega^2} - 1 \right) \textrm{Im}(r^{\textrm{TM, TM}})
\right] dk_{||} \right\}\, .
\end{eqnarray}
\pagebreak
The magnetic spectral absorbed power can obtained from the previous equation by interchanging $\alpha_E \leftrightarrow \alpha_H$ and $r^{\textrm{TE, TE}} \leftrightarrow r^{\textrm{TM, TM}}$,
\begin{eqnarray}
\label{AbsorbedSpectralMagnetic}
{\cal {P}}_{\textrm{abs}}^{H}(\omega)= \dfrac{\omega^2}{4\pi^2c^2}\textrm{Im}[\alpha_H(\omega)] \Theta(\omega, T_B)
\left\{\int_0^{\omega/c}\dfrac{k_{||}}{\xi}
\Big[2-|r^{\textrm{TE, TE}}|^2-|r^{\textrm{TM, TM}}|^2 \Big] dk_{||} \right. \cr
+ \left.  2\int_{\omega/c}^{\infty} \dfrac{k_{||}e^{-2\zeta d}}{\zeta}
\left[\textrm{Im}(r^{\textrm{TM, TM}}) + \left(\dfrac{2k_{||}^2c^2}{\omega^2} - 1 \right) \textrm{Im}(r^{\textrm{TE, TE}})
\right] dk_{||} \right\}\, .
\end{eqnarray}

Note that for propagating modes ($k_{||} \leq \omega/c $) the heat transfer do not depend on the distance $d$ between the particle and the dielectric half-space. Moreover, the propagating electric and magnetic spectral powers differ only by a multiplicative factor given by the imaginary part of the polarizabilities of the sphere. As a consequence, in the far-field the relative amounts of energy absorbed through either electric or magnetic dipoles depends exclusively on the strengths of $\textrm{Im}[\alpha_E(\omega)]$ and $\textrm{Im}[\alpha_H(\omega)]$.

On the other hand, the evanescent EM modes ($k_{||} > \omega/c $) contribution to heat transfer is strongly affected by the sphere-wall separation. Besides, the relative contributions of the electric and magnetic channels is not determined only by $\textrm{Im}[\alpha_E(\omega)]$ and $\textrm{Im}[\alpha_H(\omega)]$. As a result, the magnetic [electric] contribution to the energy exchange may dominate the transfer process even if $\alpha_E(\omega) > \alpha_H(\omega)$ [$\alpha_H(\omega) > \alpha_E(\omega)$].

In order to calculate the absorbed power we still have to specify the electric and magnetic polarizabilities. For the case of a spherical object
they can be calculated via Mie scattering theory provided $x := \omega a/c \ll 1$. Indeed, it is possible to show that $\alpha_E(\omega)$ and $\alpha_H(\omega)$ are given by~\cite{ChapuisEtAl2008,HuthEtAl2010,magneticdipole, Kruger2011, Kruger2012}
\begin{eqnarray}
\label{PolarizabilidadeEexata}
\alpha_E(\omega)\!\!\! &=&\!\!\! 2\pi a^3 \dfrac{(2y^2+x^2)(\sin{y}-y\cos{y})-x^2y^2\sin{y}}
{(y^2-x^2)[\sin{y}-y\cos{y}]-x^2y^2\sin{y}}\, , \\ \cr
\label{PolarizabilidadeHexata}
\alpha_H(\omega)\!\!\! &=&\!\!\! \dfrac{\pi a^3}{3} \left[\dfrac{(x^2-6)}{y^2}(y^2+3y\cot{y}-3)-\dfrac{2x^2}{5} \right],
\end{eqnarray}
where $y = \sqrt{\epsilon_{s}(\omega)}x$. In case the skin depth $\delta = c/\textrm{Im}[\sqrt{\epsilon_S(\omega)}\omega]$ is also much larger than the radius of the sphere, $|y| \ll 1$ , the previous equations can be simplified to
\begin{eqnarray}
\label{PolarizabilidadeCM}
\alpha_E^{CM}(\omega) = 4\pi a^3 \dfrac{\epsilon_s(\omega)-1}{\epsilon_s(\omega)+2}\, , \ \ \textrm{and} \ \
\alpha_H^{CM}(\omega) = \dfrac{2\pi}{15} a^3 \left(\dfrac{\omega a}{c}\right)^2\ [\epsilon_s(\omega)-1]\, .
\end{eqnarray}
Clearly, the expression for $\alpha_E^{CM}(\omega)$ corresponds to the well known Classius-Mossoti formula. Usually the condition $|y|\ll 1$ cannot be satisfied for all relevant frequencies for the NFHT, so that Eq. (\ref{PolarizabilidadeCM}) is of limited utility.

Let us now briefly discuss the different ways in which  thermal energy can be exchanged between the semi-infinite medium and the sphere. For a distance $d \gg \lambda_{\textrm{th}}$ only the first integrals in (\ref{AbsorbedSpectralElectric}) and (\ref{AbsorbedSpectralMagnetic}) contribute to the heat transfer. In this case, for a fixed frequency the relevant EM field modes are limited to a circle of radius $\omega/c$ in ${\bf k}_{||}-$space, as sketched in Fig.~\ref{HTPercSiCEvanescenteModes}(a). Nevertheless, if the distance between the sphere and the substrate decreases up to values much smaller than the thermal wavelength, TIR waves may become frustrated and hence may tunnel through the wall-sphere gap, effectively contributing to the heat flux\cite{Greffet2005, Basu2009, Dorofeyev2011,Raschke2013,Volokitin2007, BiehsIR}. As we have discussed in chapter {\bf \ref{cap6}}, TIR modes are expected to show up for $\omega/c < k_{||} < \omega\sqrt{\epsilon_B(\omega)}/c$. Hence, though still restricted to the interval $k_{||} < \omega\sqrt{\epsilon_B(\omega)}/c$ the number of modes that contribute to the heat transfer is larger than in the far-field regime, enhancing the absorbed power by the  sphere at frequency $\omega$. Another kind of evanescent waves responsible for an astounding enhancement of the NFHT are the so called surface phonon polaritons (SPhP) \cite{Maier_Plasmonics}. SPhP are surface waves originated owing to the coupling between the EM field and lattice vibrations (phonons) of dielectrics. They are characterized for being EM waves confined to propagate along the surface of the material with their amplitudes decaying exponentially in the perpendicular direction. For an isotropic, homogeneous, and non-magnetic half-space only TM SPhP do exist \cite{Maier_Plasmonics}. In this case,  the dispersion relation of the SPhP is given by the pole of the TM Fresnel reflection coefficient in Eq. (\ref{FresnelCoefficients}), to wit \cite{Greffet2005, Maier_Plasmonics}
\begin{equation}
\label{DispersionPlasmons}
 k_{||}^{\textrm{SPhP}} = \dfrac{\omega}{c} \sqrt{\dfrac{\epsilon_{B}(\omega)}{\epsilon_{B}(\omega)+1}}\, .
 \end{equation}
\vspace{10pt}
\begin{figure}[!ht]
  \centering
  \includegraphics[scale = 0.73]{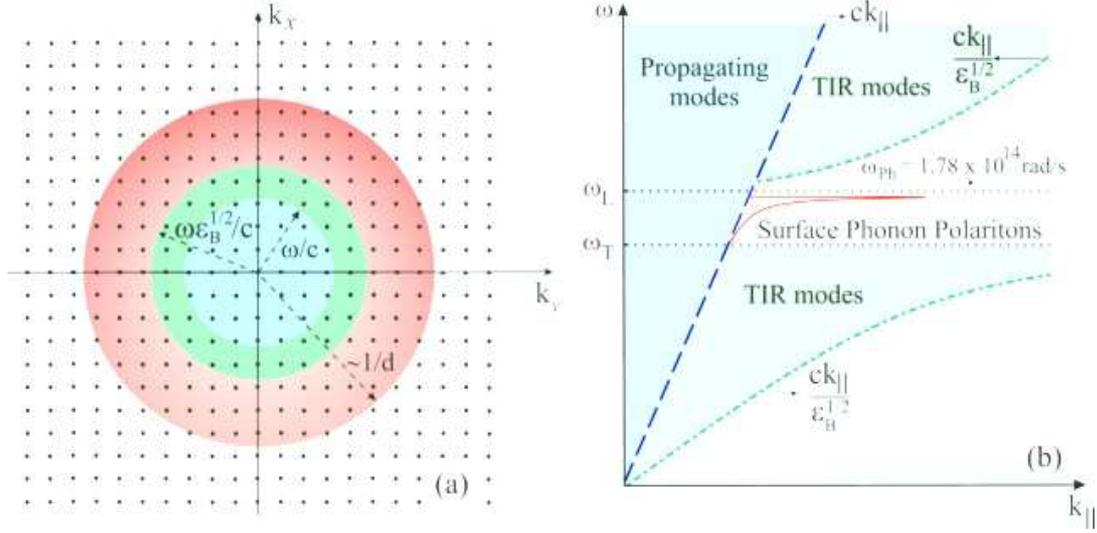}
  \vspace{10pt}
  \caption{{\bf (a)} Schematic representation in ${\bf k}_{||}-$space of the number of EM field modes that contribute to the heat flux for a fixed value of frequency $\omega$. The blue, green, and red areas show the propagating, frustrated TIR, and surface phonon polariton modes, respectively. {\bf (b)} Sketch of the $k_{||}-\omega$ plane of the regions where propagating, frustrated TIR, and surface phonon polariton waves due to a semi-infinite substrate made of SiC are expected to contribute to the NFHT in the situation depicted in Fig. \ref{HTPerc1}.}
  \label{HTPercSiCEvanescenteModes}
\end{figure}
It is important to mention that the above relation is valid only in the situation of \linebreak $\epsilon_B(\omega)<0$. Note that the dispersion relation in Eq. (\ref{DispersionPlasmons}) has a resonance for $\textrm{Re}[\epsilon_B({\omega})] = -1$ provided $\textrm{Im}[\epsilon_B(\omega)] \ll \textrm{Re}[\epsilon_B(\omega)]$. Surface waves become dominant in the NFHT process if $d \ll \delta^{\textrm{SPhP}}$, where $\delta^{\textrm{SPhP}} = 1/\textrm{Im}[\sqrt{\omega^2/c^2 - ({k_{||}^{\textrm{SPhP}}})^2}]$ is the penetration depth of surface waves in vacuum. In this regime of distances the approximation \linebreak $e^{-2\zeta d} \sim e^{-2 k_{||} d}$ in Eqs. (\ref{AbsorbedSpectralElectric}) and (\ref{AbsorbedSpectralMagnetic}) is valid. Hence, only surface modes with $ k_{||} \lesssim 1/d$ effectively contribute to  the absorbed power, as presented in Fig. \ref{HTPercSiCEvanescenteModes}(a). As a result, the number of EM field modes contributing to the NFHT increases as the sphere-wall separation diminishes. It is worth mentioning that SPhP at the surface of the sphere may also be important to the NFHT process. In this case the plasmon resonance in the sphere takes place for $\textrm{Re}[\epsilon_s(\omega)] = -2$ \cite{Mulet2001}.  However, we have checked that for the materials of the sphere chosen in the following section, this resonance do not contribute significantly to NFHT. Therefore, the surface waves of the semi-infinite medium govern the transfer process in the extreme near-field regime.  In Fig. \ref{HTPercSiCEvanescenteModes}(b) we show in the $k_{||}-\omega$ plane the regions where propagating, frustrated TIR, and SPhP modes are expected to contribute to the absorbed power for a semi-infinite medium made of silicon carbide. We model the optical response of SiC by means of its permittivity, given by Eq. (\ref{sic}).  The blue shaded area above $\omega = k_{||}c$ corresponds to propagating modes whereas the green areas show the regions where frustrated TIR modes are expected to exist. In the range $\omega_T \leq \omega \leq \omega_L$ the real part of the dielectric constant of SiC is negative, so that SPhP may contribute to the NFHT. The red curve in Fig. \ref{HTPercSiCEvanescenteModes}(b) presents $\textrm{Re}[k_{||}^{\textrm{SPhP}}]$ in this frequency range. Note that a peak in the dispersion relation emerges around $\omega_{\textrm{Ph}} \simeq 1.78 \times 10^{14}$ rad/s. This corresponds to the aforementioned SPhP resonance. Finally, we mention that in Fig.~\ref{HTPercSiCEvanescenteModes}(b) the weight of $\textrm{Im}[\alpha_E(\omega)]$ and $\textrm{Im}[\alpha_H(\omega)]$ for the total absorbed power is not considered. The relevance of propagating waves, frustrated TIR modes or SPhP to ${\cal {P}}_{\textrm{abs}}^{E}(\omega)$ and ${\cal {P}}_{\textrm{abs}}^{H}(\omega)$ depend also on $\textrm{Im}[\alpha_E(\omega)]$ and $\textrm{Im}[\alpha_H(\omega)]$ at frequency $\omega$. It means that even if a propagating or evanescent mode is allowed to exist for a given frequency $\omega$ and transverse wavevector ${\bf k}_{||}$ it may give a negligible contribution to the NFHT depending on $\Theta(\omega, T)$, $\textrm{Im}[\alpha_E(\omega)]$, and $\textrm{Im}[\alpha_H(\omega)]$.
\section{NFHT with composite materials}
\label{NFHTCompositeMaterials}

\hspace{5mm} The main goal of this section is to introduce a novel approach to enhance the heat transfer in the near-field by exploiting the versatile material properties of composite media \cite{WiltonPRB2014}.  Most of the results presented here were published in Ref. \cite{WiltonPRB2014}. To this end we investigate the NFHT between a semi-infinite dielectric medium and a composed (small) sphere with metallic nanoparticles embedded in a dielectric host medium. For the sake of simplicity we will use Eqs. (\ref{PolarizabilidadeEexata}) and (\ref{PolarizabilidadeHexata}) to obtain the electric and magnetic polarizabilities of the sphere, with the dielectric constant calculated through effective \linebreak medium theories \cite{Choy1999}.

Effective medium theories allow us to determine the effective dielectric constant $\epsilon_e$ of a composite medium as a function of the constituent dielectric constants and shapes as well as of the fractional volumes characterizing the mixture~\cite{Lagarkov1996, Brouers1986,Goncharenko2004,Choy1999}. One of the most important and successful effective medium approaches is the Bruggeman Effective Medium Theory (BEMT), which is the simplest analytical model that predicts an insulator-metal transition at a critical concentration of metallic particles in the dielectric host \cite{Choy1999, Sahimi1993}. BEMT treats the dielectric host medium and the metallic inclusions symmetrically, and it is based on the following assumptions: $(i)$ the grains are assumed to be randomly oriented spheroidal particles, and $(ii)$ they are embedded in an homogeneous effective medium of dielectric constant $\epsilon_e$  that will be determined self-consistently. If a quasi-static electromagnetic field ${\bf E}_0$ impinges on such an inhomogeneous medium, the electric field ${\bf E}_m^{\textrm{in}}$ inside the metallic grains ($\epsilon_m$), and the field ${\bf E}_{hm}^{\textrm{in}}$ inside the dielectric grains ($\epsilon_{hm}$) read\cite{Landau, Lagarkov1996}
\begin{eqnarray}
{\bf E}_m^{\textrm{in}}\!\!\! &=&\!\!\! \left[\dfrac{1}{3} \dfrac{\epsilon_e}{\epsilon_e + L(\epsilon_m - \epsilon_e)}
+ \dfrac{2}{3}  \dfrac{2\epsilon_e}{2\epsilon_e + (1-L)(\epsilon_m - \epsilon_e)}\right]{\bf E}_0\, ,\\
{\bf E}_{hm}^{\textrm{in}}\!\!\! &=&\!\!\! \left[\dfrac{1}{3} \dfrac{\epsilon_e}{\epsilon_e + L(\epsilon_{hm} - \epsilon_e)}
+ \dfrac{2}{3}  \dfrac{2\epsilon_e}{2\epsilon_e + (1-L)(\epsilon_{hm} - \epsilon_e)}\right]{\bf E}_0\, ,
\end{eqnarray}
where $0\leq L \leq 1$ is the depolarization factor of the spheroidal inclusions. For an arbitrary ellipsoidal particle characterized by the semi-axes $a_0\, , \ b_0\, ,\ c_0$,  the depolarization factor $L_i$ ($i = a_0, \, b_0, \, c_0$) is given by \cite{Landau}
\begin{eqnarray}
L_i = \dfrac{a_0b_0c_0}{2} \int_0^{\infty} \dfrac{ds}{(s+i^2)\sqrt{(s+a_0^2)(s+b_0^2)(s+c_0^2)}}\, ,
\end{eqnarray}
where the relation $L_{a_0}+L_{b_0}+L_{c_0}=1$ holds. In this work we consider spheroidal inclusions (ellipsoids with $b_0 = c_0$) such that $L_{a_0} = L$ and $L_{b_0} = L_{c_0} = (1-L)/2$. In this case, the depolarization factor can be written in terms of the eccentricity of the \linebreak ellipsoids as \cite{Landau,HuthEtAl2010}
\begin{eqnarray}
\label{DepoalrizationFactor}
L = \left\{ \!\!\!
\begin{tabular}{c}
$\dfrac{1-e^2}{2e^3}\left[\ln\left(\dfrac{1+e}{1-e}\right) -2e \right]\, , \ \ b_0<a_0\, ,$ \\ \\
$\dfrac{1+e^2}{e^3}\left[e-\arctan(e) \right]\, , \ \ \ \ b_0>a_0\, ,$
\end{tabular}
\right.
\end{eqnarray}
where $e = \sqrt{1-b_0^2/a_0^2}$ ($e = \sqrt{b_0^2/a_0^2-1}$) for prolate (oblate) ellipsoids.

In the BEMT the displacement vector $\langle {\bf D} \rangle$  is calculated by averaging the individual contributions of metallic and dielectric grains. The effective dielectric constant in the BEMT is then defined through

\begin{eqnarray}
\langle{\bf D} \rangle = \varepsilon_0\epsilon_e \langle {\bf E}\rangle \Longrightarrow
 f\varepsilon_0\epsilon_m {\bf E}_m^{\textrm{in}} + (1-f) \varepsilon_0\epsilon_{hm}{\bf E}_{hm}^{\textrm{in}} = \varepsilon_0\epsilon_e f {\bf E}_m^{\textrm{in}} + \varepsilon_0\epsilon_{e} (1-f) {\bf E}_{hm}^{\textrm{in}} \, ,
 \label{epsilonefetivo}
\end{eqnarray}
where $f$ ($0\leq f \leq 1$) is the volume filling factor for the metallic inclusions. Therefore, provided the size of the inclusions is much smaller than the relevant wavelengths (both in vacuum and inside the materials), the effective dielectric constant $\epsilon_{e}(\omega, f, L)$ is given, in the Bruggeman effective medium approach, by~\cite{Lagarkov1996, Brouers1986,Goncharenko2004,Choy1999, Sahimi1993},
\begin{eqnarray}
(1-f)\!\!\!\!\!\!\!\!\!\!\!\!&&\!\!\!\left\{\dfrac{\epsilon_{hm} - \epsilon_{e}}{\epsilon_{e} + L(\epsilon_{hm}-\epsilon_{e})}
+ \dfrac{4(\epsilon_{hm} - \epsilon_{e})}{2\epsilon_{e} + (1-L)(\epsilon_{hm}-\epsilon_{e})}\right\}\cr
\!\!\!\!\!\!&+&f\!\left\{\dfrac{\epsilon_{m} - \epsilon_{e}}{\epsilon_{e} + L(\epsilon_{m}-\epsilon_{e})}
+ \dfrac{4(\epsilon_{m} - \epsilon_{e})}{2\epsilon_{e} + (1-L)(\epsilon_{m}-\epsilon_{e})}\right\}=0 \, \label{BEMT}\, .
\end{eqnarray}
The previous equation can also be obtained by enforcing that the average over all directions of the scattered Poynting vector vanishes. Besides, note that Eq. (\ref{BEMT}) has several roots but only the one with $\textrm{Im}(\epsilon_e) \geq 0$ is physical since we are assuming passive materials.

With the expression of $\epsilon_e$ in hand we are able to evaluate the electric and magnetic polarizabilities of the composite sphere by setting $\epsilon_s = \epsilon_e$. In Fig.~\ref{HTPerc5} $\textrm{Im}[\alpha_E]$ is shown, within the BEMT, as a function of both frequency $\omega$ and filling factor $f$ for two different values of $L$: ${\bf (a)}\ L=0.1$  (needle-like grains with $e\simeq 0.95$) and ${\bf (b)} \ L=1/3$ (spherical grains), for an inhomogeneous  sphere of radius $a = 50$ nm with Copper inclusions randomly distributed inside a Polystyrene host sphere\footnote{\label{note14} We assume that the metallic grains can be well described in terms of a Drude model. The parameters characterizing all materials used in this chapter can be found in Appendix {\bf \ref{apendicec}}.}. Fig.~\ref{HTPerc5} reveals that, for every frequency, the maximal value of $\textrm{Im}[\alpha_E]$ occurs at (a) $f_{m} \simeq 0.25$ and\linebreak (b) $f_m \simeq 1/3$. Interestingly, these values of $f_m$ correspond precisely to the percolation threshold $f_c$ predicted by the BEMT~\cite{Goncharenko2004,Brouers1986,Lagarkov1996}, to wit
\begin{equation}
f_{c}(L) = \frac{L(5-3L)}{(1+ 9L)}.
\label{bruggeman}
\end{equation}
The percolation threshold $f_c$ corresponds to a critical value in the filling factor for which the composite medium undergoes an insulator-conductor transition and the system exhibits a dramatic change in its electrical and optical properties~\cite{Goncharenko2004,Brouers1986,Lagarkov1996,Sarychev2000}. This critical filling factor is calculated by taking $\omega \rightarrow 0$ in Eq. (\ref{BEMT}). In this limit, $\epsilon_m \gg \epsilon_{hm}$, $\textrm{Im}[\epsilon_m] \gg \textrm{Re}[\epsilon_m]$, and $\textrm{Im}[\epsilon_{hm}] \ll \textrm{Re}[\epsilon_{hm}]$ provided the host medium does not have a resonance near $\omega = 0$. Consequently, $\epsilon_m$ ($\epsilon_{hm}$) may be approximated by a pure imaginary (real) function. Besides, if $f<f_c$ $(f\geq f_c)$ the effective medium behaves as a dielectric-like (metal-like) material so that $\textrm{Re}[\epsilon_e(\omega)] > 0$ ($\textrm{Re}[\epsilon_e(\omega)] <0$)  in the low frequency regime. Hence, the critical threshold filling factor can be obtained by the condition $\textrm{Re}[\epsilon_e(\omega)] = 0$. Finally, we have verified that the results in Fig. ~\ref{HTPerc5} apply for every value of $L$ (only two distinctive examples are shown in Fig.~\ref{HTPerc5}). Figure~\ref{HTPerc5} also shows that, for a given $f$, $\textrm{Im} (\alpha_E)$ weakly depends on frequency, as it remains approximately constant as we vary $\omega$.
\vspace{10pt}
\begin{figure}[!ht]
  \centering
  \includegraphics[scale = 0.73]{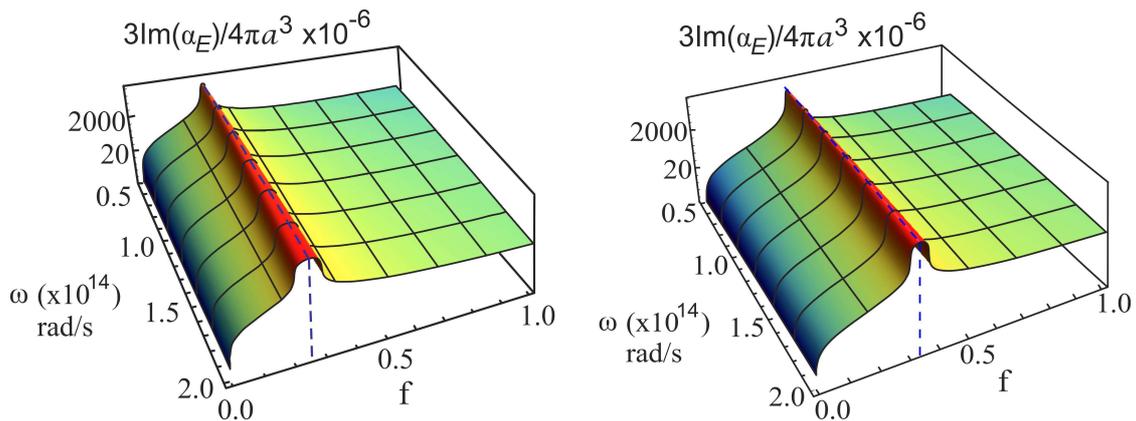}
  \vspace{10pt}
  \caption{Imaginary part of the electric polarizability of an inhomogeneous sphere made of copper inclusions randomly distributed in a polystyrene spherical host as a function of frequency $\omega$ and volume fraction $f$ for two different values of the depolarization factor $L$: {\bf (a)} $L=0.1 \ \  (e\simeq 0.95)$ and {\bf (b)} $L=1/3 \ \ (e = 0)$. The particle radius is $50$ nm.}
  \label{HTPerc5}
\end{figure}

In Fig.~\ref{HTPerc6} $\textrm{Im} (\alpha_H)$ is shown as a function of both frequency $\omega$ and filling factor $f$ for the same parameters of Fig.~\ref{HTPerc5}. As it occurs for $\textrm{Im} (\alpha_E)$, $\textrm{Im} (\alpha_H)$ is weakly dependent on frequency for fixed $f$. Note that a local maximum in $\textrm{Im} (\alpha_H)$ exists at $f = f_{c}$. As we will see in a moment the joint contribution of the maxima of $\textrm{Im} (\alpha_E)$ and $\textrm{Im} (\alpha_H)$ at $f_{c}$ will ultimately lead to a large enhancement of the heat transfer between the composite sphere and the substrate, mainly in the near-field regime~\cite{WiltonPRB2014}.
\vspace{10pt}
\begin{figure}[!ht]
  \centering
  \includegraphics[scale = 0.73]{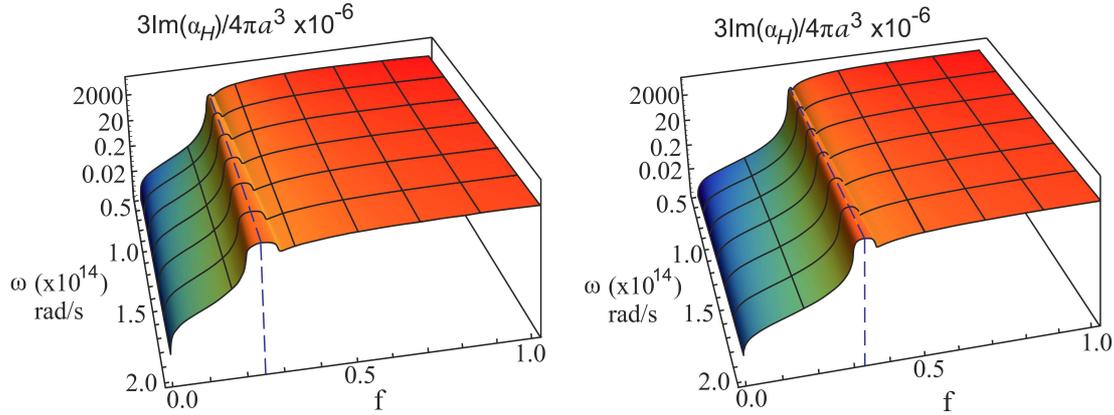}
  \vspace{10pt}
   \caption{Imaginary Part of the magnetic polarizability of an inhomogeneous sphere made of copper inclusions randomly distributed in a polystyrene spherical host as a function of frequency $\omega$ and volume fraction $f$ for two different values of the depolarization factor $L$: {\bf (a)} $L=0.1 \ \  (e\simeq 0.95)$ and {\bf (b)} $L=1/3 \ \ (e = 0)$. The particle radius is $50$ nm.}
  \label{HTPerc6}
\end{figure}
\subsection{The role of the percolation transition in NFHT}

\hspace{5mm} Let us now investigate NFHT between a composite sphere and a semi-infinite medium. In the following calculations we take the substrate to be made of Silicon Carbide and a composite medium of randomly dispersed spheroidal Copper nanoparticles in a host sphere of Polystyrene $(C_8H_8)_n$. The dispersive models for these materials are well known and were taken from the references\footref{note14}~\cite{Palik, Ordal, Hough1980}. The sphere's radius is $a = 50$ nm and the distance between the particle and the half-space is fixed at $d = 200$ nm. We set the temperature of the bulk as $T_B = 300$ K. We have verified that for the materials and geometric parameters chosen the applicability of the dipole approximation is guaranteed, and contributions from higher multipoles and multiple scattering, which are not taken into account in Eq.~(\ref{pabs}), are negligible to NFHT. Besides, the relevant wavelengths (in vacuum and in the materials) for the NFHT process are much larger that the radius $a$ of the sphere,  so the BEMT can be used.
\vspace{10pt}
\begin{figure}[!ht]
  \centering
  \includegraphics[scale = 0.7]{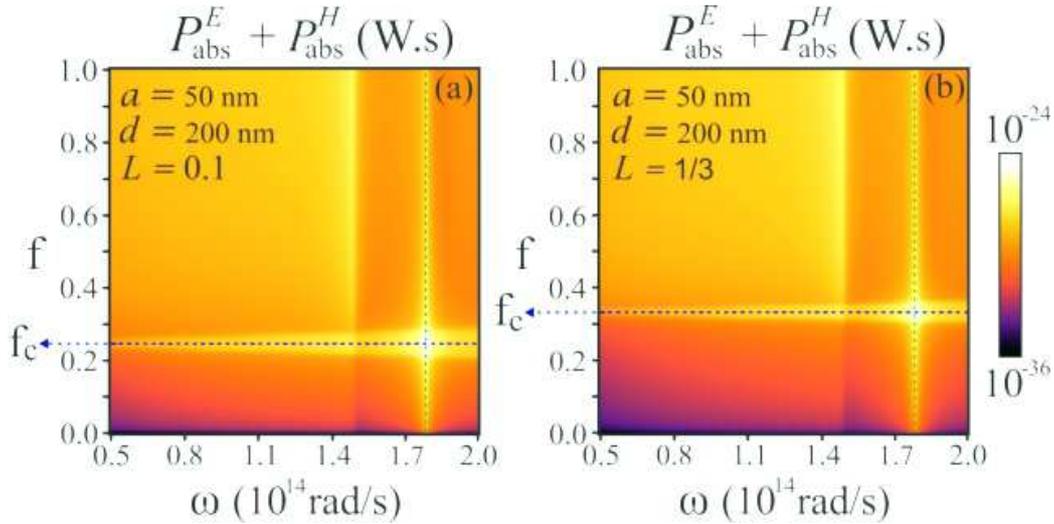}
  \vspace{10pt}
  \caption{Spectral mean power absorbed by a Polysterene particle with embedded Copper inclusions as a function of frequency and filling factor for a fixed distance between the nanoparticle and the SiC medium and two different values of the depolarization factor, {\bf (a)} $L = 0.1$ \ ($e \simeq 0.95$) and {\bf (b)} $L=1/3 \ (e = 0)$. In both cases, the horizontal dashed lines correspond to percolation threshold $f_{c}$ predicted by the Bruggeman effective medium theory whereas the vertical dashed lines correspond to the position of the phonon polariton resonance for SiC.}
  \label{HTPerc2}
\end{figure}

In Fig.~\ref{HTPerc2} the mean power absorbed by the particle ${\cal {P}}_{\textrm{abs}}^{E} + {\cal {P}}_{\textrm{abs}}^{H}$ between $\omega$ and  $\omega + d \omega $ is shown as a function of frequency and the volume fraction $f$ for two different values of the depolarization factor $L$, which encodes all the information related to the microgeometry of the inclusions: $L=0.1$ (needle-like particles with eccentricity $e \simeq 0.95$) and $L=1/3$ (spherical inclusions, $e = 0$). In both cases, there is a strong enhancement in ${\cal {P}}_{\textrm{abs}}^{E} + {\cal {P}}_{\textrm{abs}}^{H}$ due to the excitation of surface phonon polaritons in the bulk that occur at $\textrm{Re} \left[{\varepsilon}_{B}(\omega_{\textrm{Ph}})\right] = -1$ (regardless of the value of $f$), related to a peak in the density of states at $\omega_{\textrm{Ph}}$~\cite{Greffet2005}. For SiC, $\omega_{\textrm{Ph}} \approx 1.787 \times10^{14}$ rad/s \cite{Greffet2005}, as shown by the vertical dashed lines. Also, it is clear from Fig.~\ref{HTPerc2} that there exists a value of the volume fraction $f$ for which the absorbed power by the particle is maximal. For $L=0.1$ this peak occurs at $f_{m} \simeq 0.25$ whereas for $L=1/3$ it shows up for $f_{m} = 1/3$. It is also important to emphasize that: {\it (i)} there is a broadening of the spectral heat flux at  $f_{m}$, {\it i.e.} more modes effectively contribute to the NFHT process; {\it (ii)} for any frequency the maximal enhancement in ${\cal {P}}_{\textrm{abs}}^{E} + {\cal {P}}_{\textrm{abs}}^{H}$ occurs at $f_{m}$, as it can be seen from Fig.~\ref{HTPerc2} for both spheres and needle-like particles. Remarkably, these values of $f_{m}$ correspond exactly to the percolation threshold $f_{c}$ predicted by the BEMT~\cite{Brouers1986,Lagarkov1996,Goncharenko2004}, as discussed previously. There is also a peak in ${\cal {P}}_{\textrm{abs}}^{E} + {\cal {P}}_{\textrm{abs}}^{H}$ that shows up at the SiC resonance frequency, $\omega_0 \simeq 1.49\times 10^{14}$ rad/s, and it is significantly more pronounced for $f\geq f_c$. This peak is related to the magnetic dipole contribution to the absorbed power, which is proportional to $\textrm{Im} \left[\varepsilon_B(\omega) \right]$ in the near-field. In the present system its contribution is more relevant for metallic spheres than for dielectric ones.
%, {\it i.e.} above the percolation threshold, $ f \geq f_c$.

To further investigate the effects of percolation on the NFHT process, in Fig.~\ref{HTPerc3}(a) we depict the total power absorbed by the composite particle $P^{\textrm{total}}_{\textrm{abs}} (f, L, d)$ [normalized by its value for a corresponding homogeneous metallic sphere $P^{\textrm{total}}_{\textrm{abs}} (1, L, d)$], calculated as function of the volume fraction $f$ for $L=0.1$ ($e = 0.95$) and $L=1/3$ ($e = 0$). From Fig.~\ref{HTPerc3}(a) it is clear that $P^{\textrm{total}}_{\textrm{abs}}$ is maximal at the percolation threshold $f_{c}$ for the two inclusion geometries, confirming that NFHT is greatly enhanced at the percolation critical point. It is also very important to stress that the simple fact of considering a composite particle, even for inclusion concentration far from $f_{c}$, often enhances the NFHT process if compared to the case where the materials involved are homogeneous. Indeed, the enhancement factor in $P^{\textrm{total}}_{\textrm{abs}}$ due the inclusion of copper nanoparticles can be as high as $15$ if compared to the case of an homogeneous particle made of Copper and $10^{5}$ if compared to the one of an homogeneous Polystyrene particle. This result unambiguously demonstrates that composite media can largely outperform homogeneous media in NFHT, which therefore may find novel applications and optimize heat transfer at the nanoscale. We also point out that since both $\textrm{Im}[\alpha_E]$ and $\textrm{Im}[\alpha_H]$ present a maximum at $f_c$ the enhancement in the heat transfer occurs even in the far-field regime. Nevertheless, the effects of the percolation transition are more relevant in the near-field regime. Indeed, in Fig. \ref{HTPerc3}(b) we plot ${\cal{P}_{\textrm{abs}}^{\textrm{total}}}(f_c, L, d)/{\cal{P}_{\textrm{abs}}^{\textrm{total}}}(1, L, d) $  as a function of $d$ for the case of spherical (red solid line) and needle-like (blue dashed line) Copper inclusions embedded in  a Polystyrene host medium. Note that for distances $d \gtrsim 1\ \mu$m the largest enhancement factor (compared to the case of a copper homogeneous sphere) is of the order of $3.5$ for $L=0.1$ and $2.5$ for $L = 1/3$. If the sphere-wall distance is decreased up to values smaller than $1\ \mu$m both frustrated TIR and SPhP may contribute to the NFHT. In this case we note from the plots that the enhancement factor increases as the sphere gets closer \linebreak to the substrate.
%This result definitively shows that the near-field develops a fundamental role in the problem, otherwise the enhancement factor in the absorbed power would not depend on $d$.
%
\vspace{10pt}
\begin{figure}[!ht]
  \centering
  \includegraphics[scale = 0.77]{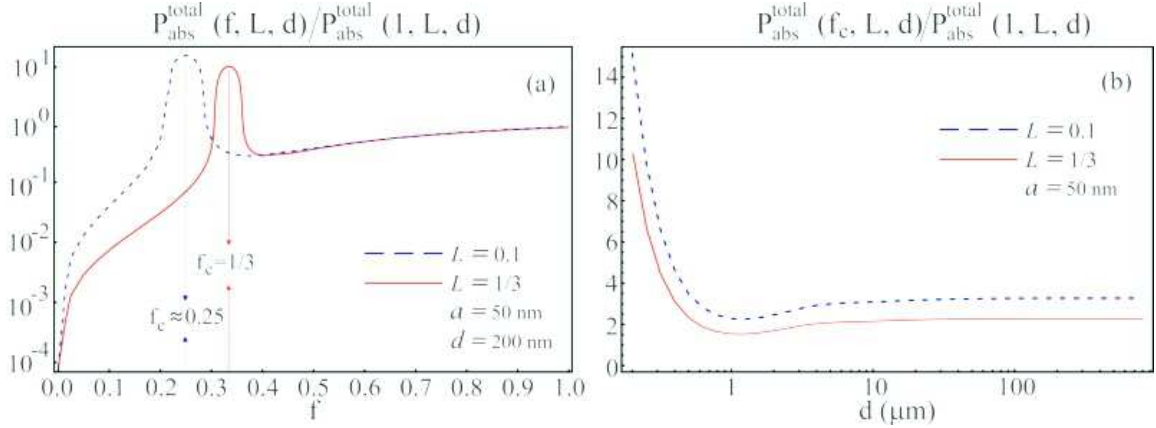}
  \vspace{10pt}
 \caption{{\bf (a)} Total power absorbed by the composite particle [normalized by its value for a corresponding homogeneous metallic sphere $P^{\textrm{total}}_{\textrm{abs}} (1, L, d)$] as a function of $f$ for $L = 0.1$ (blue dashed line) and $L=1/3$ (solid red line). The vertical arrows highlight that the values of maximum heat transfer occur precisely at the percolation threshold $f_{c}$ given by Eq.~(\ref{bruggeman}). {\bf (b)}${\cal{P}_{\textrm{abs}}^{\textrm{total}}}(f_c, L, d)/{\cal{P}_{\textrm{abs}}^{\textrm{total}}}(1, L, d) $  as a function of the distance $d$. The red solid (blue dashed) curve corresponds to spherical (needle-like) inclusions in a polystyrene host medium. All other parameters are the same as in Fig.~\ref{HTPerc2}.}
  \label{HTPerc3}
\end{figure}

The dependence of $P^{\textrm{total}}_{\textrm{abs}}$ on the shape of the Copper inclusions is investigated in Fig.~\ref{HTPerc4}(a), where $P^{\textrm{total}}_{\textrm{abs}} (f,L,d)$ is calculated as a function of both $f$ and the depolarization factor $L$, which only depends on the geometry of the inclusions~\cite{Goncharenko2004,Brouers1986,Lagarkov1996}. Figure~\ref{HTPerc4}(a) reveals that the maximal enhancement in $P^{\textrm{total}}_{\textrm{abs}} (f,L,d)$ occurs at $f_{c}$ not only for the two particular inclusion geometries considered above (spheres and needle-like particles) but for all possible spheroids. Indeed, the value of the filling factor that leads to maximal $P^{\textrm{total}}_{\textrm{abs}} (f,L,d)$ corresponds precisely to the prediction of the percolation threshold $f_{c}$ of the BEMT [Eq.~(\ref{bruggeman})] for all $L$, demonstrating the robustness of our findings against the variation of the shape of identical inclusions.  For $L \gtrsim 0.7$ (oblate spheroids) the global maximum in $P^{\textrm{total}}_{\textrm{abs}} (f,L,d)$ becomes more broadly distributed around the percolation threshold; nevertheless, on average, it still occurs at $f_{c}$. For Copper inclusions, the ratio between the total power absorbed by the particle at the percolation threshold, $P^{\textrm{total}}_{\textrm{abs}} (f=f_{c},L,d)$, and its value for an homogeneous sphere made of the same material of the inclusions, $P^{\textrm{total}}_{\textrm{abs}} (f=1,L,d)$ for a distance $d = 200$ nm is 15.4 for $L=0.1$ (needle-like inclusions), and 10.4 for $L=1/3$ (spherical inclusions). It is worth mentioning that the enhancement factors at $f_c$, relative to the corresponding homogeneous sphere, are at least 5 times larger than the previous results obtained for composite systems~\cite{Felipe-2011}. The values of the ratio $P^{\textrm{total}}_{\textrm{abs}} (f=f_{c},L,d)/P^{\textrm{total}}_{\textrm{abs}} (f=1,L,d)$ for several metals is shown in Table \ref{Tab1}. For all investigated metallic materials the enhancement in NFHT is maximal at $f_{c}$, for every $L$, a fact that suggests that our findings are independent of the metals of choice.
\vspace{10pt}
\begin{figure}[!ht]
  \centering
  \includegraphics[scale = 0.7]{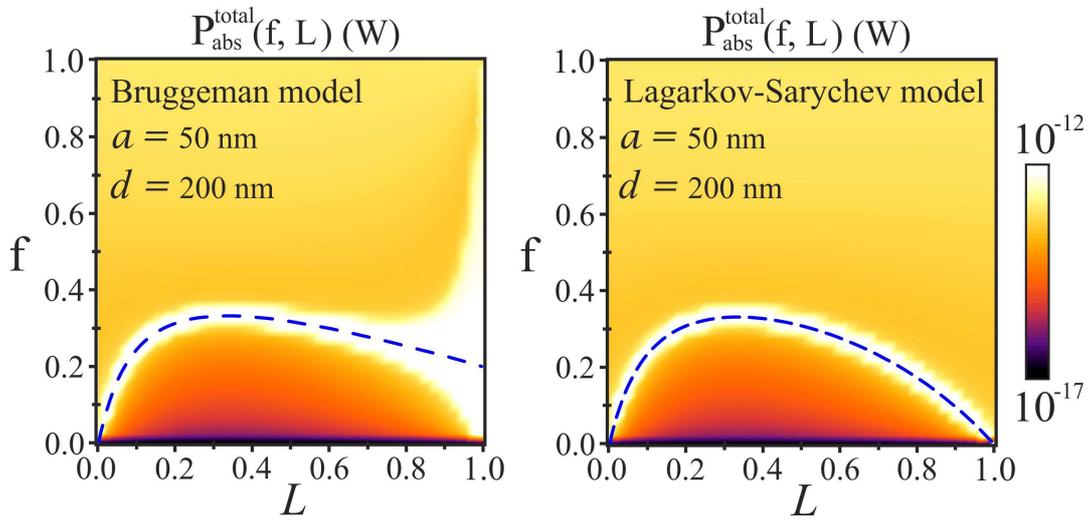}
  \vspace{10pt}
  \caption{Contour plot of $P^{\textrm{total}}_{\textrm{abs}}$ as a function of both the depolarization factor $L$ and filling factor $f$. The dashed blue curve corresponds to the critical filling factor $f_{c}$ that determines the percolation threshold for {\bf (a)} Bruggeman effective medium theory and {\bf (b)} Lagarkov-Sarychev model. The other numerical parameters are the same as in Fig.~\ref{HTPerc2}.}
  \label{HTPerc4}
\end{figure}
\vspace{10pt}
\begin{table}[!ht]
\centering
\caption{Ratio $P^{\textrm{total}}_{\textrm{abs}} (f_{c},L,d)/P^{\textrm{total}}_{\textrm{abs}} (1,L,d)$, for several metals and for $L=0.1 \ (e\simeq0.95)$ (needle-like inclusions) and for $L=1/3 \  (e = 0) $ (spherical inclusions). The first and third columns refers to the Bruggemann (B) model, whereas the other two refers to the Lagarkov-Sarychev (LS) model.}
\vspace{10pt}
\begin{tabular}{|c|c|c|c|c|}
\cline{2-5}
\multicolumn{1}{c|}{} & \multicolumn{2}{c|}{$L=0.1$} & \multicolumn{2}{c|}{$L=1/3$} \\ \cline{2-5}
\multicolumn{1}{c|}{} & B & LS & B & LS\\ \hline
Titanium & 28.7 & 30.4 & 20.7 & 20.7\\ \hline
Copper & 15.4 & 16.2 & 10.4 & 10.4\\ \hline
Vanadium  & 7.0 & 7.6 & 5.0& 5.0\\ \hline
Silver & 5.1 & 5.8 & 3.7  & 3.7\\ \hline
Gold & 3.6 & 4.1 & 2.6 & 2.6\\ \hline
\end{tabular}
\label{Tab1}
\end{table}

In order to test the robustness of our results against modifications of the effective medium theory, in Fig.~\ref{HTPerc4}(b) we depict, as a function of $L$ and $f$, the total power $P^{\textrm{total}}_{\textrm{abs}}$ absorbed by an inhomogeneous particle with its effective electric permittivity being obtained by means of an alternative homogenization technique, namely the one proposed by Lagarkov and Sarychev in Ref.~\cite{Lagarkov1996}. In the Lagarkov-Sarychev approach the effective dielectric constant $\epsilon_{e}$ of the composite material is obtained following a self-consistent prescription that is analogous to the one of BEMT, but under the assumption that the dielectric and metallic grains are not treated symmetrically. Rather, the host medium is supposed to have spherical symmetry, whereas the metallic inclusions are considered to be spheroidal particles; this approach leads to the following expression to $\epsilon_e$~\cite{Lagarkov1996}
\begin{eqnarray}
9(1-f)\left\{\dfrac{\epsilon_{hm} - \epsilon_{e}}{2\epsilon_{e} + \epsilon_{hm}}\right\} +
f\left\{\dfrac{\epsilon_{i} - \epsilon_{e}}{\epsilon_{e} + L(\epsilon_{i}-\epsilon_{e})}
+ \dfrac{4(\epsilon_{i} - \epsilon_{e})}{2\epsilon_{e} + (1-L)(\epsilon_{i}-\epsilon_{e})}\right\}=0\, .
\end{eqnarray}
This effective dielectric constant is known to give more accurate results for $f_{c}$ than the BEMT in the regime of small $L$ ($L \ll 1 $). Similarly to the BEMT, the percolation threshold $f_{c}$ is calculated by taking the limit $\omega \rightarrow 0$ in the previous equation. Therefore, in the Lagarkov-Sarychev effective medium theory $f_c(L)$ is given by \cite{Lagarkov1996}
\begin{eqnarray}
f_c(L) = \dfrac{9L(1-L)}{2+15L-9L^2}\, .
\end{eqnarray}
Figure~\ref{HTPerc4}(b) reveals that, using this alternative effective medium prescription, the maximal value for $P^{\textrm{total}}_{\textrm{abs}}$ again occurs at the percolation threshold for all $L$, as it happens within the BEMT. This fact suggests that the maximal enhancement of the NFHT in inhomogeneous media at the percolation threshold is, at least to a certain degree, independent of the effective medium theory used. Finally, it should be mentioned that we have also investigated NFHT between an homogeneous SiC sphere and a semi-infinite composite medium. In this case we have verified that our main conclusion also holds, {\it i.e.} the NFHT is maximal at the percolation threshold.

In order to understand the physical mechanism leading to the strong enhancement of the NFHT at the percolation threshold $f_{c}$, we recall that the metal-insulator transition associated with percolation is a geometric phase transition where current and electric field fluctuations are expected to be large~\cite{Sarychev2000}. In composite mixtures of metallic grains embedded in dielectric hosts, these strong fluctuations at $f_{c}$ induce a local electric field concentration (``hot spots")~\cite{Losquin2013,Caze2013} at the edge of metal clusters. In addition, for an ideally loss-free ($\textrm{Im} [\epsilon_{hm}] = \textrm{Im} [\epsilon_{i}]=0$) inhomogeneous mixture of metallic grains embedded in a dielectric,  the effective dielectric constant $\epsilon_{e}$ at the percolation critical point is mainly imaginary ($\textrm{Im} [\epsilon_{e}] \gg \textrm{Re} [\epsilon_{e}$]) so that the composite medium is highly absorptive~\cite{Gadenne1998}.  Hence the electric fields localized at the ``hot spots", and consequently the electromagnetic energy stored in the medium, are expected to increase unlimitedly at $f_{c}$. These local fields of course remain finite due to unavoidable losses but are still very large at $f_{c}$, resulting in maximal absorption by the composite medium and explaining why there is a peak in $P^{\textrm{total}}_{\textrm{abs}}$  precisely at the percolation threshold. Furthermore, in our particular system where we have considered realistic material losses, we have verified that the relation $\textrm{Im} [\epsilon_{e}] \gg \textrm{Re} [\epsilon_{e}]$ still holds at $f_{c}$. Finally, it is important to emphasize that the previous arguments for the enhancement of NFHT at $f_{c}$ rely on important critical properties of the percolation phase transition, which are known to be independent of the details of the effective medium model of choice~\cite{Sarychev2000}. This reasoning, together with the fact that our results are found to be independent of the investigated homogenization techniques, provides evidence that our findings should hold even beyond the effective medium approximation.

\vspace{20pt}
In conclusion, we have investigated the near-field heat transfer between a half-space and a nanoparticle made of composite materials. Using the Bruggeman effective medium theory, we show that heat transfer between the nanoparticle and the half-space is largely enhanced by the fact that the particle contains randomly distributed inclusions; the enhancement factor can be as large as thirty if one compares it to the case of an homogeneous metallic sphere. We also demonstrate that heat transfer is maximal at the percolation threshold in the nanoparticle for all possible spheroids, a result we show to be robust against material losses. We argue that this effect is related to the critical properties of the percolation phase transition, such as enhancement of fluctuations of currents and electromagnetic fields inside the particle, which are known to be universal and independent of the details of the effective medium model. Our findings suggest that composite media can be used as a new, versatile material platform, of easy fabrication, to tailor and optimize near-field heat transfer at the nanoscale.

\end{chapter}

\clearpage\phantomsection
\addcontentsline{toc}{chapter}{Conclusions and final considerations}

\begin{chapter}*{Conclusions and final considerations}
\label{cap9}

%\begin{flushright}
%{\it
%Science never solves a problem without creating ten more.
%}

%{\sc G. B. Shaw}
%\end{flushright}

\hspace{5mm}   In spite of the notable progress in harnessing light-matter interactions in recent years, the development of practical and tunable photonic devices for light manipulation still remains a challenge. In this thesis, we presented new approaches to circunvent some existing difficulties for an active control of electromagnetic radiation at nanoscale.

The first part of the thesis was focused in the analysis of the flow of light in situations that do not demand quantization of the EM field. Specifically, we studied the physical mechanisms of invisibility cloaks, with emphasis in devices designed via the scattering cancellation technique. In this method the EM scattering is almost suppressed by using plasmonic coatings capable to cancel the dominant multipoles induced on the cloaked object. In chapter {\bf \ref{cap4}} we have shown that magneto-optical materials under the influence of static and uniform magnetic fields may represent a new material platform for dynamical control of invisibility cloaks \cite{kortkamp2013-2, kortkampPRL, kortkampJOSAA}.  Particularly, by investigating EM scattering by a dielectric cylinder coated with a magneto-optical shell we demonstrated that the magnetic field can play the role of an external agent capable to actively tune the operation of the invisibility device. Our results suggest that the application of ${\bf B}$ may drastically reduce the scattering cross section for all observation angles.  Conversely, a magnetic field can suppress invisibility in a system originally designed to act as an invisibility cloak.   Moreover, we have shown that this tuning mechanism allows not only for a switchable cloak but also leads to a broadening of the operation frequency band of cloaking devices. Besides, we demonstrated that cloaking typically occurs in frequency ranges where the absorption cross section is minimal. This serendipitous result suggests that magneto-optical cloaks could circumvent one of the major problems of many cloaking devices, whose performance is typically decreased due to unavoidable material losses. Furthermore, we have demonstrated that the external magnetic field allows to control the angular distribution of scattered radiation. This result may be relevant in the fabrication of photonic devices where a high control of the direction of light flow is required.

The second part of the thesis was devoted to the study of light-matter interactions mediated by quantum fluctuations of the vacuum electromagnetic field. In particular, we have investigated spontaneous emission and dispersive interactions. After obtaining suitable expressions for calculating the lifetime of two-level quantum emitters in terms of both EM field modes and dyadic Green's function, we have studied the influence of materials with unusual optical properties on the lifetime of these emitters.   The effects of the environment on the atomic lifetime is known as Purcell effect, and is directly related to the boundary conditions that must be satisfied by the vacuum EM field. It is well established that whenever bodies are brought to the neighborhood of a quantum emitter,  its decay rate is in general altered. However, in chapter {\bf \ref{cap6}} we showed that if the objects in the vicinity of the atomic system are properly coated with invisibility cloaks, the Purcell effect can be suppressed provided the emitter's transition frequency is within the operation frequency range of the device \cite{KortKamp2013}. Specifically, we have investigated the SE rate of a two-level atom placed in the vicinities of a spherical plasmonic cloak. We have analyzed the dependence of the SE rate on the distance between the atom and the plasmonic cloak.  We concluded that there is a substantial reduction of the SE rate for a large range of distances, even in the near-field regime. We investigated the dependence of the SE rate on the geometrical parameters of the cloak as well as on its material parameters.  We found that the strong suppression of the SE rate is robust against the variation of both the geometrical and material parameters of the cloak.
%, even taking into account realistic ohmic losses.

In addition to the SE of an atom near an invisibility cloak, we have investigated in chapter {\bf \ref{cap6}} the influence of the magneto-optical properties of a graphene-coated  wall under the influence of a uniform magnetic field on the lifetime of quantum emitters.  We have demonstrated that, in the near-field regime, the SE rate can be strongly affected by the application of $B$ and by varying the chemical potential and temperature of graphene.  For moderate magnetic fields and distances between the emitter and the graphene sheet (in the range $1 - 10\ \mu$m) a suppression of the Purcell effect as high as 98\%  in relation to the case $B = 0$ T can be achieved. We have also demonstrated that, as a consequence of the discrete Landau energy levels in graphene, the SE rate exhibits several discontinuities as a function of $B$ for low temperatures. Besides, we have shown that, depending on materials and geometric parameters, the decay rate could be enhanced due to the presence of the external magnetic field. In the quasi-static regime we calculated the value of $B$ that maximizes the SE rate.  Furthermore, we have investigated the different decay channels in the near-field regime. We have proved that, for the material parameters chosen, there are three mechanisms of quantum emission, namely: the emitter can emit a photon into $(i)$ a propagating mode, $(ii)$ a total internal reflection, or $(iii)$ the energy can be lost to the substrate via a lossy surface wave. We showed that the magnetic field applied on the system could not only control the total decay rate but also allow us to manipulate the decay channels in the system.

In chapter {\bf \ref{cap7}} we have discussed the main features of dispersive interactions between atoms and dispersive materials. In the first part we developed a convenient method  for computing the van der Waals interaction between an atom and a perfect conducting surface of arbitrary shape. This method has the advantage of mapping a quantum problem into an electrostatic one. As a consequence, we showed that the image method could be a very useful tool to determine the non-retarded dispersive interaction between an atom and a conducting surface \cite{WiltonAJP}. As applications of this technique, we have analyzed the van der Waals interaction between an atom and a grounded/isolated sphere \cite{WiltonAJP} or a prolate ellipsoid. In the second part of chapter {\bf \ref{cap7}} we have considered dispersive interactions in a simpler geometry, namely, the atom-wall. In this case we took into account other relevant effects, such as the dispersive characteristics of the materials, optical anisotropy as well as thermal effects. In particular, we investigated the dispersive Casimir-Polder interaction between a Rubidium atom and a suspended graphene sheet subjected to an external magnetic field ${\bf B}$ \cite{Cysne-2014}.  We showed that by changing the applied magnetic field, this interaction can be reduced up to $80\%$  of its value in the absence of the field. Further, due to the magneto-optical properties of graphene, we showed that for low temperatures the Casimir-Polder interaction energy acquires sharp discontinuities at given values of $B$.  These discontinuities creates a plateau-like pattern with a quantized Casimir-Polder interaction energy as the distance between the atom and the graphene sheet increases. Besides, we have investigated the role of intraband and interband transitions in graphene in the dispersive interaction. Moreover, we showed that at room temperature thermal effects must be taken into account even for considerably short distances. In this case, the discontinuities in the atom-graphene dispersive interaction do not exist, although the interaction can still be tuned in $\sim 50 \%$ by applying an external magnetic field.

The third part of the thesis was dedicated to the study of radiative heat transfer in the near-field regime. In this case the electromagnetic fields are treated as stochastic quantities generated by random currents in local thermal equilibrium. We concluded that if the distances between the bodies are much smaller than the typical thermal wavelengths, frustrated total internal reflection modes and surface phonon polariton waves can play an important role in the heat transfer process. As a consequence the total energy exchanged between the bodies can exceed the blackbody limit by several orders of magnitude. In chapter {\bf \ref{cap8}} we have been particularly interested in the near-field heat transfer between a semi-infinite medium and a composite nanoparticle.  Using the Bruggeman effective medium theory, we showed that heat transfer between the nanoparticle and the half-space is largely enhanced by the fact that the particle contains randomly distributed metallic inclusions in a dielectric host medium \cite{WiltonPRB2014}.  We also demonstrated that heat transfer is maximal at the percolation transition for all possible mettalic spheroids. We explain this effect in terms of the critical properties of the percolation phase transition, such as enhanced fluctuations of currents and electromagnetic fields inside the particle, which are known to be universal of the effective medium model. Our findings suggest that composite media can be used as a new, versatile material platform to tailor and optimize near-field heat transfer at the nanoscale.

Altogether, we believe that our findings may pave the way for novel, alternative applications involving magneto-optical media, graphene-based materials, and nanostructured materials in disruptive photonic technologies.

\end{chapter}

%%%%%%%%%%%%%%%%%%%%%%%%%%%%%%%%%%%%%%%%%%%%%%%%%%%%%%%%%%%%%%%%%%%%%%%%%

%%%%%%%%%%%%%%%%%%%%%%%%%%% Bibliografia %%%%%%%%%%%%%%%%%%%%%%%%%%%%%%%%
\newpage
\phantomsection
\addcontentsline{toc}{chapter}{Bibliography}
\def\bibindent{2em}

%%%%%%%%%%%%%%%%%%%%%%%%%%%%% Apêndices %%%%%%%%%%%%%%%%%%%%%%%%%%%%%%%%%

\appendix
\begin{chapter}{List of publications during Ph.D. thesis work}
\label{apendiced}

\begin{itemize}
\item Kort-Kamp, W. J. M.; Rosa, F. S. S.; Pinheiro, F. A.; Farina, C.\\
 {\it Achieving invisibility with a tunable cloaking device}.\\
 Proceedings of the 7th International Congress on Advanced Electromagnetic Materials in Microwaves and Optics, p. 328 - 330,  2013.

\item Kort-Kamp, W. J. M.; Rosa, F. S. S.; Pinheiro, F. A.; Farina, C.\\
 {\it Tuning plasmonic cloaks with an external magnetic field}.\\
  Physical Review Letters,  vol. 111, p. 215504-1 - 215504-5, 2013.

\item Kort-Kamp, W. J. M.; Rosa, F. S. S.; Pinheiro, F. A.; Farina, C.\\
 {\it Spontaneous emission in the presence of a spherical plasmonic metamaterial}.\\
  Physical Review A, vol. 87, p. 023837-1 - 023837-7,  2013.
  
 \item De Melo E Souza, R.; Kort-Kamp, W. J. M.; Sigaud, C.; Farina, C.\\
 {\it Image method in the calculation of the van der Waals force between an atom\\ and a conducting surface}.\\
  American Journal of Physics, vol. 81, p. 366 - 376, 2013.
\pagebreak
  
 \item Kort-Kamp, W. J. M.; Rosa, F. S. S.; Pinheiro, F. A.; Farina, C.\\
 {\it Molding the flow of light with a magnetic field: plasmonic cloaking\\ and directional scattering}. \\
 Journal of the Optical Society of America A, vol. 31, p. 1969 - 1976, 2014.

\item Cysne, T.; Kort-Kamp, W. J. M.; Oliver, D. ; Pinheiro, F. A;\\ Rosa, F. S. S.; Farina, C.\\
 {\it Tuning the Casimir-Polder interaction via magneto-optical effects in graphene}.\\
  Physical Review A, vol. 90, p. 052511-1 -  052511-5, 2014.

\item Kort-Kamp, W. J. M.; Caneda, P. I.; Rosa, F. S. S.; Pinheiro, F. A.\\
 {\it Enhancing near-field heat transfer in composite media:\\ effects of the percolation transition}.\\
  Physical Review B, vol. 90, p. 140202(R)-1 - 140202(R)-5, 2014.
\end{itemize}

\end{chapter}

\begin{chapter}{Drude-Lorentz model for dispersion in magneto-optical materials}
\label{apendicea}

\hspace{5 mm} Let us consider an arbitrarily polarized EM wave propagating in a generic direction inside a magneto-optical medium. The system is under the influence of an external static and uniform magnetic field ${\bf B}$. By assumption the interaction between charges and fields in this model is given by the Lorentz force ${\bf F} = -e[{\bf E} + {\bf v} \times ({\bf B} + \hat{{\bf k}}\times {\bf E}/c)]$, where $-e\ (e>0)$ is the electric charge of the particles and $\hat{{\bf k}}$ is the unit wave vector. Denoting by ${\bf r}$ the displacement of a particle in relation to its equilibrium position, we can write the following equation of motion
\begin{eqnarray}
\label{MotionEquation}
m\ddot{{\bf r}} = -\dfrac{m}{\tau} \dot{{\bf r}} - m \omega_0^2 {\bf r} - e({\bf E} + \dot{{\bf r}} \times {\bf B})\, ,
\end{eqnarray}
where $m$ is the particle mass, $1/\tau$ and $\omega_0$ are the damping constant and resonance frequency, respectively. Besides, in order to obtain the above equation we neglected terms ${\cal{O}}(v/c)$ by supposing that the velocities of the charges in the material are nonrelativistic. Furthermore, we assumed that $|{\bf r}| \ll \lambda$, where $\lambda$ is the incident wavelength. Hence, we can neglect the spatial variation of the EM field over the region where the motion of the particle takes place. This means that in the previous equation the electric field is evaluated at the equilibrium position. Actually, we will work with quantities such as the polarization of the medium, which are volume densities calculated over a volume $\delta V$. Therefore, the condition $|{\bf r}|^3 \ll \delta V \ll \lambda^3$ has to be satisfied so that there are several atoms within $\delta V$ and the spatial variation of the field can be neglected.

If we suppose that the static uniform magnetic field is applied along the $z$-direction, we can rewrite Eq. (\ref{MotionEquation}) as a set of three coupled equations of motion
\begin{eqnarray}
\ddot{x}\!\!\! &+&\!\!\! \dfrac{1}{\tau} \dot{x} + \omega_0^2 x + \omega_c(B) \dot{y}= -\dfrac{e}{m} E_x\, , \\
\ddot{y}\!\!\! &+&\!\!\! \dfrac{1}{\tau} \dot{y} + \omega_0^2 y - \omega_c(B) \dot{x}= -\dfrac{e}{m} E_y\, , \\
\ddot{z}\!\!\! &+&\!\!\!\dfrac{1}{\tau} \dot{z} + \omega_0^2 z = -\dfrac{e}{m} E_z\, ,
\end{eqnarray}
where $\omega_c(B) = eB/m$. Considering now a temporal harmonic dependence $e^{-i\omega t}$ for the EM field and writing $x(t) = x_0 e^{-i\omega t}$, $y(t) = y_0 e^{-i\omega t} $, and $z(t) = z_0 e^{-i\omega t}$ we obtain the following equations for the oscillation amplitudes
\begin{eqnarray}
\label{x0Eq}
(\omega_0^2 \!\!\! &-&\!\!\! \omega^2 -i\omega/\tau)x_0 - i\omega_c(B)\omega y = -\dfrac{e}{m} E_{0x}\, , \\
\label{y0Eq}
(\omega_0^2 \!\!\! &-&\!\!\! \omega^2 -i\omega/\tau)y_0 + i\omega_c(B)\omega x = -\dfrac{e}{m} E_{0y}\, , \\
\label{z0Eq}
(\omega_0^2 \!\!\! &-&\!\!\! \omega^2 -i\omega/\tau)z_0 = -\dfrac{e}{m} E_{0z}\, ,
\end{eqnarray}
where $E_{0x}$, $E_{0y}$, and $E_{0z}$ are the amplitudes of the oscillating electric field along directions $x$, $y$, and $z$, respectively.

While the solution for $z_0$ follows directly from Eq. (\ref{z0Eq}), we ought to decouple Eqs. (\ref{x0Eq}) and (\ref{y0Eq}) so as to obtain $x_0$ and $y_0$. We leave for the interested reader to show that this can be accomplished through the use of the complex variable $\xi_{\pm} = x_0 \pm iy_0$. After straightforward algebric manipulations it is possible to show that the solution of the previous equations can be cast as
\begin{eqnarray}
x_0\!\!\! &=&\!\!\! -\dfrac{e}{m}\left[\dfrac{(\omega_0^2 - \omega^2 - i\omega/\tau)E_{0x} + i\omega_c(B)\omega E_{0y}}
{(\omega_0^2 - \omega^2 - i\omega/\tau)^2 - \omega_c(B)^2 \omega^2}\right]\, , \\
y_0\!\!\! &=&\!\!\! -\dfrac{e}{m}\left[\dfrac{-i\omega_c(B)\omega E_{0x} +  (\omega_0^2 - \omega^2 - i\omega/\tau)E_{0y}}
{(\omega_0^2 - \omega^2 - i\omega/\tau)^2 - \omega_c(B)^2 \omega^2}\right]\, , \\
z_0\!\!\! &=&\!\!\! -\dfrac{e}{m}\dfrac{E_{0z}}{\omega_0^2 - \omega^2 - i\omega/\tau}\, .
\end{eqnarray}

The polarization induced by the EM wave within a volume $\delta V$ of the magneto-optical material can be written as  ${\bf P}(t) = {\bf P}_0 e^{-i\omega t}$, where ${\bf P}_0 = -eN(x_0\hat{{\bf x}}+ y_0\hat{{\bf y}}+ z_0\hat{{\bf z}})$ with $N$ being the number of charged particles per volume. Consequently, we have
\begin{eqnarray}
\label{a1}
{\bf P}_0 = \varepsilon_0 \Omega^2\!\!\! &\Bigg\{&\!\!\! \left[\dfrac{(\omega_0^2 - \omega^2 - i\omega/\tau)E_{0x} + i\omega_c(B)\omega E_{0y}} {(\omega_0^2 - \omega^2 - i\omega/\tau)^2 - \omega_c(B)^2 \omega^2}\right] \hat{{\bf x}} \cr
&+&\!\!\! \left[\dfrac{-i\omega_c(B)\omega E_{0x} +  (\omega_0^2 - \omega^2 - i\omega/\tau)E_{0y}} {(\omega_0^2 - \omega^2 - i\omega/\tau)^2 - \omega_c(B)^2 \omega^2}\right] \hat{{\bf y}} \cr
&+&\!\!\! \left[\dfrac{E_{0z}}{\omega_0^2 - \omega^2 - i\omega/\tau}\right] \hat{{\bf z}}\Bigg\}\, ,
\end{eqnarray}
where we defined the oscillating strength frequency $\Omega^2 = Ne^2/m\varepsilon_0$.

On the other side, ${\bf P}_0 = ($\mbox{{\mathversion{bold}${\varepsilon}$}}$ - \mathbb{I} \varepsilon_0){\bf E}_0 = P_{0x}\hat{{\bf x}} + P_{0y}\hat{{\bf y}} + P_{0z}\hat{{\bf z}}$, so that the cartesian components of the polarization vector can be cast into the form
\begin{eqnarray}
\label{a2}
P_{0x} &=& (\varepsilon_{xx} - \varepsilon_0)E_{0x} + \varepsilon_{xy}E_{0y} + \varepsilon_{xz} E_{0z}\, , \\
\label{a3}
P_{0y} &=& \varepsilon_{yx} E_{0x} + (\varepsilon_{yy} - \varepsilon_0)E_{0y} + \varepsilon_{yz} E_{0z}\, , \\
\label{a4}
P_{0z} &=& \varepsilon_{zx} E_{0x} + \varepsilon_{zy} E_{0y} + (\varepsilon_{zz} - \varepsilon_0) E_{0z}\, .
\end{eqnarray}

Comparing Eqs. (\ref{a1}), (\ref{a2}), (\ref{a3}) and (\ref{a4}) it is straightforward to show that the dispersive expressions for the electric permittivity tensor elements of magneto-optical materials described by the Drude-Lorentz model are \cite{King2009}
\begin{eqnarray}
\varepsilon_{xx}(\omega,B) = \varepsilon_{yy}(\omega,B) = \varepsilon_0\left[1 - \dfrac{\Omega^2(\omega^2+i\omega/\tau - \omega_0^2)}{(\omega^2+i\omega/\tau - \omega_0^2)^2 - \omega^2\omega_c(B)^2} \right]\, ,
\end{eqnarray}
\begin{eqnarray}
\varepsilon_{xy}(\omega,B) = -\varepsilon_{yx}(\omega,B) = \varepsilon_0 \dfrac{i\Omega^2\omega\omega_c(B)}{(\omega^2+i\omega/\tau - \omega_0^2)^2 - \omega^2\omega_c(B)^2}\, ,
\end{eqnarray}
\begin{eqnarray}
\varepsilon_{zz}(\omega)= \varepsilon_0\left[1 -
\dfrac{\Omega^2}{(\omega^2+i\omega/\tau - \omega_0^2)} \right]\, ,
\end{eqnarray}
and $\varepsilon_{xz}(\omega,B) = \varepsilon_{zx}(\omega,B) = \varepsilon_{yz}(\omega,B) = \varepsilon_{zy}(\omega,B) = 0$.
\end{chapter}

\begin{chapter}{Graphene's reflection coefficients}
\label{apendiceb}

\hspace{5 mm}  Let us consider the situation depicted in Fig. \ref{ReflectionCoefficientsFig}. An incoming arbitrarily polarized wave propagating in vacuum impinges on the surface of an homogeneous medium occupying the half-space $z\leq0$ coated by a graphene layer at $z = 0$. For the sake of simplicity we consider here that the substrate is a non-magnetic material, {\it i. e. }$\mu_1(\omega) = \mu_0$.  The impinging electric and magnetic fields can be conveniently expressed as
\begin{eqnarray}
\label{EletricoIncidente}
{\bf E}_0\!\!\! &=&\!\!\! \left[E_0^{\textrm{TE}} \mbox{{\mathversion{bold}${\epsilon}$}}_{\textrm{TE}}^{+}  + E_0^{\textrm{TM}} \mbox{{\mathversion{bold}${\epsilon}$}}_{\textrm{TM}_0}^{+}  \right] e^{i({\bf k}_0^+ \cdot {\bf r} - \omega t)}\, , \\
\label{MagneticoIncidente}
{\bf H}_0\!\!\! &=&\!\!\! \sqrt{\dfrac{\varepsilon_0}{\mu_0}} \left[ E_0^{\textrm{TM}}  \mbox{{\mathversion{bold}${\epsilon}$}}_{\textrm{TE}}^{+}  - E_0^{\textrm{TE}} \mbox{{\mathversion{bold}${\epsilon}$}}_{\textrm{TM}_0}^{+}  \right] e^{i({\bf k}_0^+ \cdot {\bf r} - \omega t)} \, ,
\end{eqnarray}
where $E_0^{\textrm{TE}}, \, E_0^{\textrm{TM}}$ are the transverse electric and transverse magnetic incoming amplitudes, respectively.
\vspace{10pt}
\begin{figure}[!ht]
\centering
\includegraphics[scale=0.5]{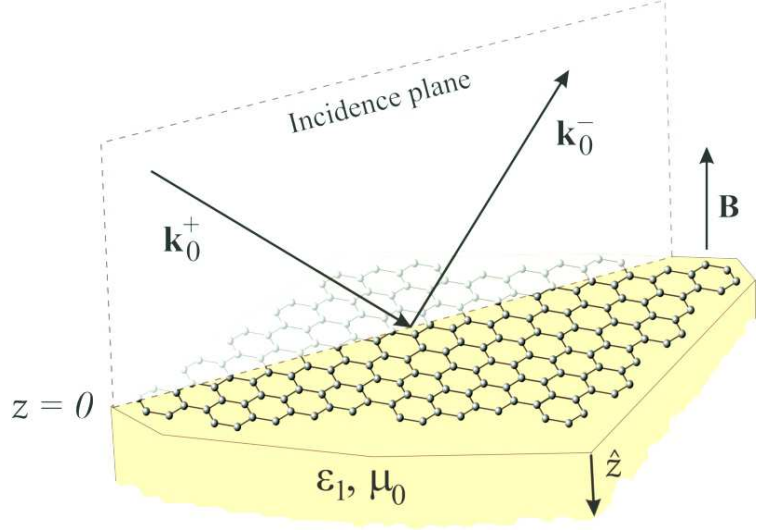}
\vspace{10pt}
\caption{Monochromatic plane wave impinging on a graphene-coated non magnetic wall of electric permittivity $\varepsilon_1(\omega)$.}
\label{ReflectionCoefficientsFig}
\end{figure}
Similarly, the reflected and transmitted fields are written as
\begin{eqnarray}
\label{EletricoRefletido}
{\bf E}_R\!\!\! &=&\!\!\! \left[E_R^{\textrm{TE}} \mbox{{\mathversion{bold}${\epsilon}$}}_{\textrm{TE}}^{-}  + E_R^{\textrm{TM}} \mbox{{\mathversion{bold}${\epsilon}$}}_{\textrm{TM}_0}^{-}  \right] e^{i({\bf k}_0^- \cdot {\bf r} - \omega t)}\, , \\
\label{MagneticoRefletido}
{\bf H}_R\!\!\! &=&\!\!\! \sqrt{\dfrac{\varepsilon_0}{\mu_0}} \left[ E_R^{\textrm{TM}}  \mbox{{\mathversion{bold}${\epsilon}$}}_{\textrm{TE}}^{-}  - E_R^{\textrm{TE}} \mbox{{\mathversion{bold}${\epsilon}$}}_{\textrm{TM}_0}^{-}  \right] e^{i({\bf k}_0^- \cdot {\bf r} - \omega t)} \, ,
\end{eqnarray}
and
\begin{eqnarray}
\label{EletricoTransmitido}
{\bf E}_T\!\!\! &=&\!\!\! \left[E_T^{\textrm{TE}} \mbox{{\mathversion{bold}${\epsilon}$}}_{\textrm{TE}}^{+}  + E_T^{\textrm{TM}} \mbox{{\mathversion{bold}${\epsilon}$}}_{\textrm{TM}_1}^{+}  \right] e^{i({\bf k}_1^+ \cdot {\bf r} - \omega t)}\, , \\
\label{MagneticoTransmitido}
{\bf H}_T\!\!\! &=&\!\!\! \sqrt{\dfrac{\varepsilon_1}{\mu_0}} \left[ E_T^{\textrm{TM}}  \mbox{{\mathversion{bold}${\epsilon}$}}_{\textrm{TE}}^{+}  - E_T^{\textrm{TE}} \mbox{{\mathversion{bold}${\epsilon}$}}_{\textrm{TM}_1}^{+}  \right] e^{i({\bf k}_1^+ \cdot {\bf r} - \omega t)} \, ,
\end{eqnarray}
where the polarization vectors are given by
\begin{eqnarray}
\mbox{{\mathversion{bold}${\epsilon}$}}_{\textrm{TE}}^{+} = \mbox{{\mathversion{bold}${\epsilon}$}}_{\textrm{TE}}^{-} = {\bf \hat{z}}\times{\bf\hat{k}}_{||}\, , \ \
\textrm{and} \ \  \mbox{{\mathversion{bold}${\epsilon}$}}_{{\textrm{TM}}_i}^{\pm} = \mbox{{\mathversion{bold}${\epsilon}$}}_{\textrm{TE}}^{\pm} \times {\bf\hat{k}}_i^{\pm} \, ,
\end{eqnarray}
with
\begin{equation}
{\bf\hat{k}}_i^{\pm} = \dfrac{{\bf k}_i}{|{\bf k}_i|} = \dfrac{{\bf k}_{||} \pm k_{zi}{\bf\hat{z}}}{\sqrt{k_{||}^2+k_{zi}^2}}\, ,
\end{equation}
where $i = 0,\ 1$.

Our problem consists in determining the reflected $E_R^{\textrm{TE, TM}}$ and transmitted $E_T^{\textrm{TE, TM}}$ amplitudes so that we can calculate the reflection and transmission coefficients
\begin{equation}
r^{\textrm{i, j}} = \dfrac{E_R^{\textrm{i}}}{E_0^{\textrm{j}}}\ \ \ \textrm{and} \ \ \  t^{\textrm{i, j}} = \dfrac{E_T^{\textrm{i}}}{E_0^{\textrm{j}}}\, .
\end{equation}
where $\textrm{(i, j)} = \textrm{(TE, TM)}$.

The reflected and transmitted amplitudes are obtained by solving Maxwell's equations and forcing the appropriate boundary conditions on the interface at $z = 0$.  Particularly, we model the graphene sheet as a surface current ${\bf K} = \mbox{{\mathversion{bold}${\sigma}$}}\cdot{{\bf E}_T}|_{z = 0}$, where $\mbox{{\mathversion{bold}${\sigma}$}} = \sigma_{xx}({\bf\hat{x}}{\bf\hat{x}} + {\bf\hat{y}}{\bf\hat{y}}) + \sigma_{xy}({\bf\hat{x}}{\bf\hat{y}} - {\bf\hat{y}}{\bf\hat{x}})$ is the graphene conductivity. Hence,  the boundary conditions that must be satisfied by the EM field at $z = 0$ are
\begin{eqnarray}
\label{CC1}
{\bf \hat{z}} \times \left[{\bf E}_T - {\bf E}_R - {\bf E}_0 \right]\!\!\! &=&\!\!\! {\bf 0}\, , \\
\label{CC2}
{\bf \hat{z}} \times \left[{\bf H}_T - {\bf H}_R - {\bf H}_0 \right]\!\!\! &=&\!\!\! {\bf K} = \mbox{{\mathversion{bold}${\sigma}$}}\cdot{{\bf E}_T}\, .
\end{eqnarray}

Using, Eqs. (\ref{EletricoIncidente})-(\ref{EletricoTransmitido}), and  (\ref{MagneticoTransmitido}) in (\ref{CC1}) and (\ref{CC2}) it is straightforward to show that the reflected and transmitted amplitudes satisfy the following equations
\begin{eqnarray}
&(i)&\ \ E_T^{\textrm{TE}} = E_0^{\textrm{TE}} + E_R^{\textrm{TE}}\, , \\
&(ii)&\ \ \dfrac{k_{z1}}{k_1} E_T^{\textrm{TM}} = \dfrac{k_{z0}}{k_0} \left[E_0^{\textrm{TM}} - E_R^{\textrm{TM}} \right]\, , \\
&(iii)&\ \ \left[\dfrac{k_{1}}{k_{z1}} \sigma_{xx} + \sqrt{\dfrac{\varepsilon_1}{\mu_0}}\right] E_T^{\textrm{TE}} - \sigma_{xy} E_T^{\textrm{TM}} = \sqrt{\dfrac{\varepsilon_0}{\mu_0}}\dfrac{k_1k_{z0}}{k_0k_{z1}} \left[E_0^{\textrm{TE}} - E_R^{\textrm{TE}} \right]\, ,     \\
&(iv)&\ \ \left[\dfrac{k_{z1}}{k_1} \sigma_{xx} + \sqrt{\dfrac{\varepsilon_1}{\mu_0}}\right] E_T^{\textrm{TM}} + \sigma_{xy} E_T^{\textrm{TE}} = \sqrt{\dfrac{\varepsilon_0}{\mu_0}}\left[E_0^{\textrm{TM}} + E_R^{\textrm{TM}} \right]\, .
\end{eqnarray}

If now we consider separately the cases of TE and TM incidence it is possible to show that the reflection coefficients are given by

\begin{eqnarray}
\label{ReflectionCoefficients_SS}
\!\!\!\!\!\!\!\!\!\!r^{\textrm{TE, TE}} \!\!\! &=& \!\!\!-\dfrac{\Delta_{+}^{E}\Delta_{-}^{H}+Z_0^2k_{z_0}k_{z_1} \sigma_{xy}^2}{\Delta_{+}^{E}\Delta_{+}^{H}+Z_0^2k_{z_0}k_{z_1} \sigma_{xy}^2}\, ,\\ \cr
\label{ReflectionCoefficients_PP}
\!\!\!\!\!\!\!\!\!\!r^{\textrm{TM, TM}} \!\!\!&=& \!\!\! \dfrac{\Delta_{-}^{E}\Delta_{+}^{H}+Z_0^2k_{z_0}k_{z_1} \sigma_{xy}^2}{\Delta_{+}^{E}\Delta_{+}^{H}+Z_0^2k_{z_0}k_{z_1} \sigma_{xy}^2}\, , \\ \cr
\label{ReflectionCoefficients_SP}
\!\!\!\!\!\!\!\!\!\!r^{\textrm{TE, TM}} \!\!\! &=& \!\!\! r^{\textrm{TM, TE}} = \dfrac{2k_{z_0}k_{z_1}Z_0\sigma_{xy}}{\Delta_{+}^{E}\Delta_{+}^{H}+Z_0^2k_{z_0}k_{z_1} \sigma_{xy}^2}\, , %\\ \cr
%\label{TransmissionCoefficients_SS}
%\!\!\!\!\!\!\!\!\!\!t^{\textrm{TE, TE}} \!\!\! &=& \!\!\!\dfrac{2k_{z0}\Delta_{+}^{E}}{\Delta_{+}^{E}\Delta_{+}^{H}+Z_0^2k_{z_0}k_{z_1} \sigma_{xy}^2}\, ,\\ \cr
%\label{TransmissionCoefficients_PP}
%\!\!\!\!\!\!\!\!\!\!t^{\textrm{TM, TM}} \!\!\!&=& \!\!\!\dfrac{2\sqrt{\varepsilon_1/\varepsilon_0}k_{z0}\Delta_{+}^{H}}{\Delta_{+}^{E}\Delta_{+}^{H}+Z_0^2k_{z_0}k_{z_1} \sigma_{xy}^2}\, , \\ \cr
%\label{TransmissionCoefficients_SP}
%\!\!\!\!\!\!\!\!\!\!t^{\textrm{TE, TM}} \!\!\! &=& \!\!\! r^{\textrm{TE, TM}}\, ,  \ \ \ \textrm{and} \ \ \ t^{\textrm{TM, TE}} = -\sqrt{\dfrac{\varepsilon_1}{\varepsilon_0}}\dfrac{k_{z0}}{k_{z1}} r^{\textrm{TM, TE}}\, ,
\end{eqnarray}
where
\begin{eqnarray}
\Delta_{\pm}^{i} = (k_{z_1}\delta_{iE}+k_{z_0}\delta_{iH})(\mathfrak{S}^i  \sigma_{xx}\pm 1) + [\varepsilon_1(\omega)/\varepsilon_0]k_{z_0} \delta_{iE} + k_{z_1} \delta_{iH} \, ,
\end{eqnarray}
with $i = (E, \, H)$,  $\mathfrak{S}^H = \omega \mu_0 / k_{z0}$, $\mathfrak{S}^E = k_{z0}/(\omega \epsilon_0)$, $Z_0=\sqrt{\mu_0/\epsilon_0}$ is the vacuum impedance, $k_{z0} = \sqrt{(\omega/c)^2 - k_{||}^2}$, and ${k_z}_1 = \sqrt{\mu_0\varepsilon_{1}\omega^2 - k_{||}^2}$.

\end{chapter}

\begin{chapter}{Optical data of the materials used in chapter {\bf \ref{cap8}}}
\label{apendicec}

%\hspace{5 mm} The electrical permittivity $\varepsilon_{B}$ of Silicon Carbide (SiC) is well described by following dispersive model \cite{Palik}
%
%\begin{equation}
%\dfrac{\varepsilon_{\textrm{SiC}}(\omega)}{\varepsilon_0} = \epsilon_{\infty}\left(1 + \dfrac{\omega_L^2 - \omega_T^2}{\omega_T^2 - \omega^2 - i \omega/{\tau_{\,}}_{\textrm{SiC}}}\right),
%\end{equation}
%
%where $\varepsilon_{\infty} = 6.7$,  $\omega_L = 182.7\times10^{12}$ rad/s, $\omega_T = 149.5\times10^{12}$ rad/s, and $1/{\tau_{\,}}_{\textrm{SiC}} = 0.9\times10^{12}$ rad/s.

%For the polystyrene the electric permittivity $\varepsilon_{\textrm{Py}}$ reads \cite{Hough1980}
%
%\begin{equation}
%\dfrac{\varepsilon_{\textrm{Py}}(\omega)}{\varepsilon_0} = 1 + \dfrac{\omega_{p1}^2}{\omega_{r1}^2 - \omega^2 - i\omega/\tau_{1}} +\dfrac{\omega_{p2}^2}{\omega_{r2}^2 - \omega^2 - i\omega/\tau_{2}}\, ,
%\end{equation}
%
%with material parameters, including losses, given by  $\omega_{p1} = 1.11\times10^{14}\  \textrm{rad/s}\, ,$ \linebreak $\omega_{r1} = 5.54\times10^{14}\  \textrm{rad/s}\, , \ \ \  \omega_{p2} = 1.96\times10^{16}\  \textrm{rad/s}\, , \ \ \omega_{r2} = 1.35\times10^{16}\  \textrm{rad/s}\, , \ \ \ \textrm{and}$ \linebreak $1/\tau_1 = 1/\tau_2 = 0.1 \times 10^{12}\  \textrm{rad/s} $

\hspace{5mm} Throughout chapter {\bf \ref{cap8}} the dielectric constant of the metals is assumed to be described by the Drude model \cite{Ordal}
\begin{equation}
\dfrac{{\varepsilon_{\! }}_{\textrm{m}}(\omega)}{\varepsilon_0} = 1 - \dfrac{{\omega_{\! }}_{\textrm{m}}^{\ 2}}{\omega^2 + i\omega/{\tau_{\! }}_{\textrm{m}}}\, ,
\end{equation}
where the material parameters for metallic inclusions of Titanium (Ti), Copper (Cu), Vanadium (V), Silver (Ag), and Gold (Au) are given in Table \ref{TabMaterials}.
\vspace{10pt}
\begin{table}[!ht]
\centering
\caption{Material parameters used in the dispersive Drude model for metallic materials.}
\vspace{10pt}
\begin{tabular}{|c|c|c|}
\hline
 & $\omega_{\textrm{m}}$ (rad/s) & $1/\tau_{\textrm{m}}$ (rad/s) \\ \hline
Titanium &  $3.83 \times 10^{15}$  & $7.19 \times 10^{13}$  \\ \hline
Copper & $1.12 \times 10^{16}$ & $1.38 \times 10^{13}$ \\ \hline
Vanadium   & $7.84 \times 10^{15}$ & $9.26 \times 10^{13}$\\ \hline
Silver & $1.37 \times 10^{16}$ & $2.73 \times 10^{13}$ \\ \hline
Gold & $1.37 \times 10^{16}$  & $4.05 \times 10^{13}$ \\ \hline
\end{tabular}
\label{TabMaterials}
\end{table}

The dielectric constant of  Polystyrene and Silicon Carbide are the same as in Eqs. (\ref{Poly}) and (\ref{sic}), respectively.

\end{chapter}

%\input{apendiced.tex}

%%%%%%%%%%%%%%%%%%%%%%%%%%%%%%%%%%%%%%%%%%%%%%%%%%%%%%%%%%%%%%%%%%%%%%%%%

\end{document}